\documentclass[a4paper,11pt]{article}
\usepackage{pos}

\usepackage{lineno}
\usepackage{aas_macros}

\usepackage{placeins}
\usepackage{bm}
\usepackage{graphicx, color, hepunits}
\usepackage{multirow}
\usepackage[compat=1.0.0]{tikz-feynman}
\usepackage{slashed}
\usepackage{textcomp}
 \usepackage{hyperref} %Automatically links \label and \ref commands; Always load last
\hypersetup{
    colorlinks=true,       % false: boxed links; true: colored links
    linkcolor=blue,        % color of internal links
    citecolor=blue,        % color of links to bibliography
    filecolor=magenta,     % color of file links
    urlcolor=blue          % color of external links
}
\usepackage[utf8]{inputenc}
\usepackage[english]{babel}

\newcommand{\Ps}[1]{\slashed{p_{#1}}}
\newcommand{\qs}{\slashed{q}}
\newcommand{\ks}{\slashed{k}}

\newcommand{\es}[2] {\begin{equation} \label{#1} \begin{split} #2 \end{split} \end{equation}}

\title{TASI Lectures on the Particle Physics and Astrophysics of Dark Matter}

\author[a,b]{Benjamin R. Safdi}

\affiliation[a]{Berkeley Center for Theoretical Physics, \\ University of California, Berkeley, CA 94720, U.S.A.}

\affiliation[b]{Physics Division, \\ Lawrence Berkeley National Laboratory, Berkeley, CA 94720, U.S.A.}

\emailAdd{brsafdi@berkeley.edu}

\FullConference{%
  TASI 2022 -- ``Ten Years After the Higgs Discovery: Particle Physics Now and Future," \\
  June 6 - July 1 2022\\
  Boulder, CO USA
}

\abstract{
These lecture notes on the particle physics and astrophysics of dark matter (DM) were delivered at TASI 2022 ``Ten Years After the Higgs Discovery: Particle Physics Now and Future."  The focus of these lecture notes, aimed at the level of advanced graduate students and beginning postdocs, is on indirect ({\it i.e.}, astrophysical and cosmological) probes of particle DM models.  While DM models and indirect detection are broadly discussed, the examples of weakly interacting massive particles (WIMPs) and axions are worked out in detail.   
The topics covered include: the role of DM in the cosmology and astrophysics of structure formation, including DM density profiles in galaxies, general constraints on particle DM models, the theory of minimal DM, with the higgsino as a relevant and illustrative example, indirect detection with gamma-rays, including with the upcoming Cherenkov Telescope Array, axions as a solution to the strong-{\it CP} problem and a DM candidate, including discussions of possible ultraviolet completions and of axion string cosmology, and astrophysical probes of axions such as with isocurvature perturbations, $N_{\rm eff}$, black hole superradiance, radio telescopes, spectral modulations, stellar polarization, and stellar cooling, amongst other topics.  Example \texttt{Jupyter} notebooks are provided that walk the reader through relevant analyses, including an example statistical analysis of a DM annihilation search towards the Segue I dwarf galaxy with gamma-ray data from the {\it Fermi} Large Area Telescope that is relevant for DM explanations of the {\it Fermi} Galactic Center Excess.  We also provide an introduction to frequentist statistics for particle and astro-particle physics.  These lecture notes are meant to be pedagogical, with the focus on explaining the underlying physical principles and analysis techniques that are set to play crucial roles in the search for particle DM in the coming decade.
}

\tableofcontents

\begin{document}

\maketitle

\section{Introduction}
Cosmological observations indicate that around 27\% of the current energy density in our Universe is in the form of cold dark matter (DM)~\cite{Planck:2018vyg}.  However, the microscopic nature of DM remains unknown.  What is abundantly clear is that the existence of DM necessitates beyond the Standard Model (SM) physics.  There is no candidate for cold DM within the SM -- for example, the SM neutrinos would be too hot -- and moreover the abundance of ordinary matter is constrained by big bang nucleosynthesis (BBN).  Primordial black holes could potentially explain the DM, but even in that case new physics is needed for their early universe formation.

On the other hand, many well motivated extensions to the SM naturally contain cold DM candidates.  For example, extensions to the SM motivated by naturalness considerations, such as supersymmetry, often contain a stable, neutral relic with weak-scale mass and interaction cross-sections that can acquire the correct relic DM abundance in the early universe through thermal freezeout.   Such DM candidates are known as weakly interacting massive particles (WIMPs), and they are the subject of extensive collider, direct, and indirect searches today.  A particularly motivated WIMP candidate that will be probed by indirect searches in the coming years is the thermal higgsino, which emerges in the context of supersymmetry.  The quantum chromodynamics (QCD) axion is another example of a motivated DM candidate. It was originally introduced to solve the strong-{\it CP} problem related to the absence of the neutron electric dipole moment (EDM), and only later was it realized that the axion could also acquire the correct relic abundance to explain the observed DM.  Axions are now understood to arise generically in the context of string theory compactifications, and they may play important roles in the context of quantum gravity. They are being searched for extensively in the laboratory, but astrophysical probes are complementary and indeed necessary in order to make sure that the full parameter space is covered in the coming years.   

Axions and WIMPs are but two examples of a plethora of possibilities for particle DM that have been proposed in the nearly 100 years since the discovery of DM in our Universe.  What these and other models have in common, however, is that they may leave unique signatures in precision astrophysical and cosmological observables, which we may leverage to confirm their existence or rule them out as models of nature.  In these lecture notes, which are based upon lectures titled ``Astrophysical Probes of Dark Matter" given at the TASI 2022 summer school on ``Ten Years After the Higgs Discovery: Particle Physics Now and Future", we review some of the most promising DM models and their astrophysical probes, concentrating in particular on probes that will be relevant in the coming decade. 

\subsection{Historical context circa 2022}

The search for DM, and beyond the SM (BSM) physics more generally, is going through a period of introspection.  As I describe below, in light of a sequence of null results and theoretical insights over the past few decades, efforts to find BSM physics may not appear especially promising. However, as I argue here, I believe that the contrary is true: we are set to embark on an incredibly exciting next few decades of searches for new physics, in the dark sector in particular, which stand a real chance of revolutionizing our understanding of the Universe.  

We begin, however, by acknowledging and describing the reasons why the 
 search for new physics may appear bleak at present.  WIMP DM with the naive weak-scale cross-section (assuming $Z$-mediated, spin-independent scattering) has long been ruled out.  Large, underground direct detection experiments have now searched far below the original target cross-sections, with no signs of new physics having yet appeared.  Similarly, indirect searches for {\it e.g.} gamma-rays from WIMP DM annihilation have not turned up any definitive signs of new physics, though there are a few anomalies such as the {\it Fermi} Galactic Center (GC) Excess (GCE) that remain outstanding and that we discuss more in these lecture  notes.  WIMP DM could have also shown up in colliders such as the Large Hadron Collider (LHC), where it could be pair produced and leave missing energy signatures, but it did not.  Beyond the WIMP paradigm, sterile neutrino DM (with $\sim$keV scale sterile neutrino masses) is increasingly constrained by indirect searches for DM decay in $X$-rays and small-scale structure observations.  In fact, until recently, mismatches between the expected DM halo profiles of small galaxies and the expected numbers of small galaxies with observations seemed to point towards a model of DM that could modify the shapes and abundances of halos on astrophysical scales, such as DM with sizable self interactions (simply referred to as self-interacting DM), warm DM, or fuzzy DM.  Yet with more data and better modeling of  cosmological structure formation with baryons the necessity for any modification to the standard collisionless cold DM paradigm has been diminished.  Axions have also been subject to a few, narrow null results over the past few years from experiments such as the Axion Dark Matter eXperiment (ADMX), though as we discuss further relative to WIMPs axions are still remarkably unconstrained at present. 

As dire as the situation seems for DM, it only appears worse when viewed through the lens of the broader search for BSM physics.  Foremost, the observation of the non-zero cosmological constant is deeply troubling, as a small but non-zero value for the cosmological constant defies explanation within any of the standard, deterministic theories, which naively predict a much larger value.  One of the most compelling explanations for the small but non-zero value of the cosmological constant was made prior to its discovery by Weinberg~\cite{Weinberg:1987dv}, who noted that a cosmological constant near the value that would later be observed appears necessary to have the correct conditions for structure formation as we know it and thus life as we know it.  Weinberg's work helped solidify the anthropic principle, which is the possibility that we live in a multiverse with different patches, or universes, that have different values of the fundamental constants (such as the cosmological constant). From the point of view of the full theory in the multiverse these parameters may be chosen dynamically, but us low-energy 4D observers may never be able to probe that high-scale dynamics, and we can also not access the other universes.  We may simply be living in the patch we are living in, with a small but non-zero cosmological constant, because we would not be able to exist anywhere else in the multiverse. 

The cosmological constant casts a long shadow on searches for BSM physics. Consider, for example, the electroweak hierarchy problem related to the light Higgs mass, which in quantum field theory receives quadratically-divergent radiative corrections that should naturally push it towards the ultraviolet (UV) cut-off of the theory.  To solve the hierarchy problem, one needs new physics to appear near the electroweak scale, and in the LHC era it is clear that such new physics needs to be higher in mass than naively expected given null results so far. Completely natural solutions to the hierarchy problem appear unlikely unless we have fundamentally misunderstood key aspects of how nature works~\cite{Craig:2022uua}.  In the shadow of the cosmological constant problem, we cannot help but wonder if perhaps the Higgs mass is light because it needs to be light in order to have the complicated particle dynamics necessarily to eventually lead to life like us that contemplate such problems.  This is scary not just because it means we may never know the solution to the hierarchy problem, but also because without new physics at the electroweak scale, there is less motivation for WIMP DM, since the WIMP miracle connects the DM candidate with the assumption of new physics at the electroweak scale.  If we give up on the WIMP, then perhaps we should turn to the axion. But remember that the axion was also introduced to solve a fine tuning problem -- the strong-{\it CP} problem -- and perhaps this problem is solved anthropically also! 

 While the above paragraphs give you a taste of the types of existential questions that particle physics are grappling with at present, I strongly suspect that our ability to improve our fundamental understanding of nature is not nearly as dire and depressing as described above.  In these lecture notes, on the contrary, I hope to instill a sense of optimism that the next decade, roughly, will be one where we stand to learn deep aspects about how nature works and, perhaps, discover new BSM particles that open the door to paradigm-shifting UV physics.  My optimism comes foremost from the fact that DM exists and is definitive proof of BSM physics. One cannot use anthropic arguments to ``explain away" the existence of matter in our Universe.  It is true that DM may interact only through Planck-scale suppressed operators, but even in that case we may be able to detect the particle nature of DM by, for example, observing the rare DM decays induced by the Planck-scale operators.  Secondly, as I argue in these lectures notes, WIMP DM remains a well-motivated paradigm. In addition to broadly discussing the WIMP paradigm, I work out in detail the specific example of the nearly-pure higgsino. This model is in many ways the most WIMP-like of all WIMP models (with a capital $W$), since it places the DM in the fundamental representation of the electroweak force. The higgsino mass is predicted to be around 1 TeV. With a small mass splitting between the two neutral Majorana states -- expected in the context of UV completions -- the theory is unobservable in direct detection experiments since the tree-level $Z$-exchange process is cut off. However, as I discuss, the next generation of indirect detection searches will discover or rule out this particle.  The same is true for other WIMP benchmark models, such as the wino. 
 Axions are similarly facing an extremely exciting decade. From a theoretical point of view they only appear better motivated in recent years, with connections between string theory and axions emerging more concretely. (Also, the anthropic solution to the strong-{\it CP} problem does not appear to work, as we explain, since the CP-violating parameter has little effect on any physical processes.) While it is true that ADMX has ruled out small slivers of parameter space for the axion, the vast majority of the axion's parameter space remains open.  Most exciting, through a combination of astrophysical and laboratory probes the situation for the axion will change drastically in the coming decade or two.  Assuming the theorists and experimentalists are able to do the necessary work, we will rule out or confirm the existence of the QCD axion, in addition to potentially discovering axion-like particles over large regions of mass and coupling parameter space.

 While these lecture notes focus on WIMPs and axions, there are many other compelling paradigms for DM at present.  Broadly speaking, one can divide searches for DM into two categories: model-driven approaches and signature-driven efforts.  WIMP and axion searches are examples of model-driven approaches.  We have specific UV models in mind for DM, motivated by other considerations ({\it e.g.}, the electroweak hierarchy problem and the strong-{\it CP} problem); we are then forced to come up with laboratory and astrophysical probes of these models to confirm them or rule them out as particles of nature. Other well-motivated models in this category include, for example, sterile neutrino DM and asymmetric DM (see~\cite{Green:2022hhj} and references therein).  Signature-driven approaches, on the other hand, accept our ignorance of the dark sector and work to parameterize the possible ways that DM, whatever it may be, could show up in the laboratory or in astrophysics. Self-interacting DM is a prime example, where the strength of self interactions discussed is necessarily at the level of the current sensitivity of halo probes. Fuzzy DM, which modifies the structure of halos on astrophysical scale as well, is another example of a signature-driven DM model.  Then, there are DM efforts that do not so cleanly divide between these two categories but rather come from attempts to classify the ways in which particle DM could come to acquire the necessary relic abundance.  Freeze-in DM, strongly-interacting massive particle (SIMP) DM, and dark photon DM are all examples of efforts in this direction, and each of these efforts has helped catalyze laboratory and astrophysical efforts to search for DM. Keep in mind that while these lecture notes focus on axions and WIMPs, they are but two possibilities in a broad class of ongoing efforts to understand the particle nature of DM.  On the other hand, they do illustrate many of the general techniques that are being applied to astrophysical and cosmological searches for DM, and so by studying these models in detail, one is well prepared to approach other DM models as well.     

 At present, it appears that there is a good chance we stand at a very unique junction in history where we know of the existence of DM, but we do not know {\it what} it is. It is my sincere hope that this epoch is short-lived and that soon the wave-function of possibilities for the nature of DM collapses and the identity of DM is known forevermore.  Depending on what DM ends up being, it is very likely that it will provide a portal through which we can understand physics at higher energy scales and better contextualize the SM and our place in the Universe.

\subsection{A user's guide to these lecture notes} 

These lecture notes focus primarily on the indirect signatures of DM, with a specific focus on axions and WIMPs.  The theoretical foundations for these models are reviewed as well, as are the more general principles behind DM from a cosmological and astrophysical perspective.  These lecture notes are meant to be pedagogical, which means that the focus is on explaining key concepts and not on (i) getting all of the ${\mathcal O}(1)$ factors correct, and   (ii) completely summarizing the literature. For a more complete literature review of any of the topics discussed here, I recommend following the citation trails from the recent Snowmass 2021 process, starting with this Report~\cite{Green:2022hhj}.  The ${\mathcal O}(1)$  factors, on the other hand, may be worked out by the careful reader or found in the original works.  Note, also, that these lecture notes almost completely avoid mention of DM direct detection. The reason is not that laboratory efforts for DM identification are not promising, but rather that the scope of these lecture notes is limited to only discuss DM models and their astrophysical and cosmological probes.  There are many other great sets of lecture notes out there that discuss laboratory probes of DM (see, {\it e.g.}, the TASI notes~\cite{Lisanti:2016jxe,Hook:2018dlk,Lin:2019uvt}).
These lecture notes are aimed at the level of advanced graduate students and postdocs, though certain aspects may be of broader interest to more advanced researchers.

For junior researchers just getting started some of the biggest (and most annoying) hurdles can be practical. How do I actually integrate that differential equation numerically? What are the ways that I can perform a statistical analysis given a model and data set?  How do I then make the plots look nice? To help shorten the learning curve with some of these everyday tasks that nearly all junior researchers need to face in this field, a supplementary data directory is supplied through Google Drive and Google Colaboratory (Colab) at the following address~\cite{suppData}.\footnote{
See these \href{https://drive.google.com/drive/folders/1Cry-Q90fAeXFqEoV_pqeF4sXwEGSZDYu?usp=sharing}{Colab  Jupyter Notebooks.}
}  The Drive contains a number of example \texttt{Jupyter} notebooks (\texttt{python}), which may be run remotely through Colab, that recreate plots and analyses described in these lecture notes, in addition to the supplementary data set for the example statistical analysis in Sec.~\ref{sec:segue}.  The Colab notebooks may be run ``out-of-the-box" by anyone using a web browser through Google cloud computing with no local installations. 

The remainder of these lecture notes are organized as follows. Sec.~\ref{sec:DM} discusses the cosmological and astrophysical roles of DM.  General considerations for particle DM models are reviewed in Sec.~\ref{sec:gen}.  The WIMP paradigm is then introduced in Sec.~\ref{sec:4}, while Sec.~\ref{sec:indirect} discusses indirect detection schemes for annihilating and decaying DM.  Axions as solutions to the strong-{\it CP} problem, along with their ultraviolet completions and DM production mechanisms, are described in Sec.~\ref{sec:axions}. 
 Indirect detection searches for axions are described in Sec.~\ref{sec:axions_indirect}.  There are then three Appendices: App.~\ref{sec:cosmo} gives an overview of the standard cosmological picture; App.~\ref{sec:stats} provides an introduction to frequentist statistics; App.~\ref{app:xsec} gives additional details for the calculation of the higgsino annihilation cross-section.   

\section{The structure of dark matter in galaxies}
\label{sec:DM}

One can make the argument that DM is a crucial building block to life in our Universe.  The reason is that the small-amplitude perturbations in the density of matter imprinted by quantum fluctuations during inflation later collapsed, during the matter-domination epoch, to form gravitationally bound structures such as galaxies and galaxy clusters. Since DM is more abundant than ordinary matter by around a factor of five, it is mostly responsible for forming the deep potential wells that pulled in the ordinary matter (referred to as ``baryons" in cosmology), which then further collapsed, through dissipation, to form the visible components of galaxies in all their complexity.  Without DM, the baryons alone would not have been able to form deep enough potential wells to pull in sufficient matter to form galaxies like our Milky Way.  This line of reasoning may cause one to wonder if perhaps the reason the ratio of energy density in ordinary matter to DM is so close to unity  
is due to some kind of anthropic selection. That is, perhaps in other parts of the ``multiverse" there is more or less DM, relative to baryons, but those regions do not have the correct conditions to support ``observers" like us to contemplate their existence. Such line of reasoning may seem philosophical, but it is worthwhile to contemplate such possibilities because this may just be how the Universe works.\footnote{The coincidence between the DM and baryon energy densities could also emerge naturally in the underlying theory of DM, as is the case in {\it e.g.} asymmetric DM~\cite{Kaplan:2009ag,Petraki:2013wwa}, with the dynamics that set the baryon abundance tied to the dynamics that set the DM abundance. } We will return to this idea when discussing axion DM.

One crucial feature that differentiates DM and baryons is that DM, at least in the standard cosmological paradigm called $\Lambda$CDM (cold DM plus a cosmological constant), does not have strong (or perhaps any) non-gravitational self interactions and, especially important, does not have dissipative pathways. Ordinary matter, on the other hand, does have self interactions and does have dissipative pathways by which it may cool.  For example, hot gas clouds emit radiation, thanks to the existence of our massless photon, which allows them to lose energy and collapse under their gravity. As the gas clouds collapse by shedding energy to radiation, they begin to rotate faster in order to conserve angular momentum. The fast rotation can collapse the spinning gas clouds down to compact disks, which is what is observed in a  variety of astrophysical systems ranging from Saturn's rings to the solar system as a whole to spiral galaxies like our own Milky Way.  However, while the gas in Milky Way like galaxies collapses due to dissipation to form compact disks, the DM halo remains much larger in extent and much more spherically symmetric. 

Numerical cosmological simulations, such as $N$-body simulations that discretize the DM density distribution into $N$ massive particles that interact gravitationally, confirm the picture of hierarchical structure formation leading to large, spherically symmetric DM halos surrounding galaxies (see, {\it e.g.},~\cite{Angulo:2021kes} for a review of large-scale cosmological simulations).  Inflation generates a nearly conformal spectrum of density perturbations across spatial scales. As these scales fall within the horizon during the matter dominated epoch they begin to grow.  Most cosmological simulations start by using perturbation theory to evolve the perturbations until a redshift $z \sim 100$, beyond which point the perturbations may begin to become nonlinear and thus numerical methods are necessary.  DM-only simulations discretize the cold DM fluid into particles and solve the gravitational dynamics of these particles in an expanding spacetime. Zoom-in simulations are a common technique for achieving higher resolution around the resulting halos of interest at low redshift. For zoom-in simulations, one starts with a coarse simulation, identifies structures of interest at late time ({\it e.g.}, $z = 0$), traces back to find which region of the initial states collapsed down to form the halo of interest, and then simulate that region alone with with higher particle number and thus higher resolution.

DM-only $N$-body simulations do not have baryons, and so after forming DM halos at late times these halos must be artificially ``populated" by galaxies.  Realistically, the ordinary matter is pulled into the DM halos during their formation, leading to galaxy formation. Hydrodynamic simulations attempt to capture these dynamics (see, for example,~\cite{Hopkins:2017ycn} and references therein).  Hydrodynamic simulations include both discretized DM particles and gas that is evolved using a hydrodynamic solver.  The gas is subject to cooling, self-interactions, and gravity. If various thresholds are met, the gas can also undergo star formation, which means that some fraction of the mass within a volume is converted to stars, which are co-evolved in the dynamics.  Stars have many important feedback mechanisms that can affect the gravitational dynamics. For example, stars can undergo supernova, which eject energy and heat up the surrounding gas. These feedback processes are also included in the hydrodynamic code and can have important implications for the DM profiles of halos. Unfortunately, a first-principles approach to modeling {\it e.g.} star-formation and supernova is not realistic.  Instead, these complicated events are handled phenomenologically and are controlled by parameters that must be tuned to data. The many free parameters in modern hydrodynamic simulations can make it difficult to separate potential signs of new physics (in, {\it e.g.} DM self interactions) from baryonic physics, but on the other hand modern hydrodynamic simulations are incredibly successful in modeling the diversity of behavior seen in real galaxies (see~\cite{Hopkins:2017ycn}).

Since baryons are a small fraction of the total matter in the Universe, it might seem strange that they can strongly affect the shapes of DM halos. However, it is the case that hydrodynamic simulations find important differences in DM halo shapes relative to pure DM-only simulations, as we discuss further below. Heuristically, one can understand the importance of baryons as follows. First, for many DM searches we are interested in the centers of galaxies, since the DM densities rise towards the GCs. For example, DM annihilation signals scale as the DM density squared. Since the baryons collapse through dissipation, they are more compact than the DM halos, and in the centers of galaxies like the Milky Way they can dominate the gravitational potentials. This means that at the centers of galaxies the DM acts as tracer particles in the potential set by the baryons. Baryons can undergo adiabatic contraction, whereby the potential well slowly collapses towards the origin; this process pulls in the DM and can form cuspier profiles.  On the other hand, baryons can also have non-adiabatic processes such as supernova, which heat up the gas and eject material, causing the gravitational potential to flatten in the inner part of the galaxy. If this happens non adiabatically, then the DM will have too much kinetic energy and also ``evaporate" from the central region, coring the otherwise cuspy DM profile. For Milky Way size galaxies, both adiabatic contraction and baryon-induced cores can take place simultaneously, leading to highly non-trivial DM density profiles within roughly the radius of the galactic bulge where baryons dominate the potential.  To complicate the situation even more, one galaxy can be completely different at $z= 0 $ from another because of, for example, their merger histories. Understanding the interplay between baryonic physics and DM halos in a cosmological context is a difficult but extremely important problem, since DM halos are invoked in a broad range of astrophysical probes of particle DM across multiple DM models.

DM halos, like that which hosts our own Milky Way, form through hierarchical structure formation.  Roughly, the picture is that a density perturbation on some length scale $k^{-1}$ will become dynamical and begin to rapidly collapse, assuming we are in the epoch of matter domination, when the mode ``enters the horizon" ({\it i.e.}, $k > H$, with $H$ the Hubble parameter at that epoch).  Since $H$ becomes smaller as the Universe expands, this implies that the highest-$k$ modes (smallest scale perturbations) collapse first. This, in-turn, means that small DM halos are formed before large ones. After and during their formation the small DM halos merge together to form larger halos, such that the largest DM halos are composed of the mergers of many smaller halos. Some of the subhalos may survive until today as so-called DM substructure, but many of them are completely disrupted, with the DM becoming virialized within the host halo.  A full study of structure formation is well beyond the scope of these lectures, but note that much progress can be made analytically by formalizing the picture described above using Press-Schechter approach~\cite{1974ApJ...187..425P}.

For our purposes, the observables of primary interest are the DM density profiles of galaxies. These DM density profiles may be extracted from numerical simulations or, at the other extreme, inferred non-parametrically from data, such as stellar kinematic data, that traces the gravitational potential.  A very successful way forward  is to combine these two approaches and to use numerical simulations to motivate parametric DM density profiles, whose model parameters may be constrained in individual systems using real kinematic data.  We describe a few of the more common spherically-symmetric parametric DM density profiles in Sec.~\ref{sec:density}.  In Sec.~\ref{sec:DM_MW} we then discuss the DM density profile for the Milky Way, which is one of the most important DM halos since it is that which hosts our galaxy.   In Sec.~\ref{sec:DM_halo_other} we discuss DM density profiles in a range of other galaxies of interest, such as nearby dwarf galaxies and clusters.

\subsection{Parametric dark matter density profiles}
\label{sec:density}

Let us suppose that you are starting an indirect detection project for your favorite model.  Perhaps you are thinking about a WIMP, a supermassive decaying DM candidate, a sterile neutrino, an axion, a black hole, or something completely different; regardless, at some point in the project you will likely need to write down a DM density profile for either the Milky Way galaxy or another DM halo, perhaps a Milky Way dwarf galaxy or a much larger galaxy cluster.  The go-to model for modeling DM density profiles is that of Navarro-Frenk-White (NFW)~\cite{Navarro:1995iw,Navarro:1996gj}:
\es{eq:NFW}{
\rho_{\rm NFW} = {\rho_s \over r / r_s \left( 1+ r / r_s\right)^2 } \,,
}
where $r_s$ is called the scale radius, $\rho_s$ sets the level of the DM density, and $r$ is the distance from the galaxy center.  This profile is spherically symmetric about the GC. The NFW DM profile has been found to be a good description of DM halos in cosmological DM-only $N$-body simulations across a range of scales, ranging from dwarf galaxy sizes to the larger clusters.  The NFW profile is purely phenomenological; there is no first-principle derivation of this profile in the context of hierarchical mergers.  With that said, in addition to describing the results of $N$-body simulations well, the NFW profile has also proven to be very successful in describing actual halo data (see, {\it e.g.},~\cite{Klypin:2001xu}), with a few notable potential discrepancies that we will discuss in more detail.  Note that the NFW profile is ``cuspy" in that it diverges as $r \to 0$. 

The mass enclosed from the NFW profile within some radius $r$, which we define as $M_{\rm NFW}(r)$, logarithmically diverges at large $r \gg r_s$.  However, at very large $r$ eventually the expression in~\eqref{eq:NFW} will fall well below the critical energy density of the universe $\rho_c$, at which point it is clear that the halo profile in~\eqref{eq:NFW} is no longer valid. It is customary to cut-off the DM halo profile at a radius $r_{\rm max}$ defined by 
\es{eq:M_NFW}{
{M_{\rm NFW}(r_{\rm max})   \over {4 \over 3} \pi r_{\rm max}^3 }= \Delta_c \rho_c \,,
}
where $\Delta_c$ is a threshold quantity relative to the critical density.

Before continuing, let us clarify two points about this discussion. First, the concepts we are describing also apply for DM profiles beyond the NFW profile, though we will illustrate them here for the NFW model. Second, we will assume here that we are looking at halos at redshift $z = 0$; going to higher $z$ the expressions written here need to be modified in relatively straightforward ways (see, {\it e.g.},~\cite{Lisanti:2017qoz} for an overview), but high-$z$ halos are often less useful in DM indirect detection than low-$z$ halos.   

There are various choices of $\Delta_c$ that are common in the literature, as the exact point at which the DM halo merges with the ambient energy density is slightly ambiguous. One principled choice is that appropriate for virial quantities: $\Delta_v = 18 \pi^2 - 82 x - 39 x^2$, where $x = \Omega_m - 1$, with $\Omega_m$ the energy density fraction in matter.  The virial radius is $r_v$, defined using~\eqref{eq:M_NFW} with $\Delta_c \to \Delta_v$ and $r_{\rm max} \to r_v$, has the property that within the virial radius the virial theorem holds for the bound DM, assuming the halo formed from the collapse of a spherically expanding top-hat perturbation~\cite{1974ApJ...187..425P,1980lssu.book.....P,Bryan:1997dn}.  Numerically, $\Delta_v \sim 200$, which motivates using the ``200" convention, where we simply define $\Delta_c = \Delta_{200} \equiv 200$ with $r_{200}$ defined analogously through~\eqref{eq:M_NFW}.  Use caution: some references use $r_{200}$, while others use $r_{v}$.  The {\it concentration parameter} $c_v$ is defined as $c_v = r_v / r_s$ or alternatively $c_{200} = r_{200} / r_s$.  The total mass enclosed within $r_v$ is given by
\es{}{
M_{\rm NFW}(r_v) \equiv M_v = 4 \pi \rho_0 r_s^3 \left( \log(1+c_v) - {c_v \over 1 + c_v} \right) \,,
}
with an analogous relation holding for the total mass within $r_{200}$, referred to as $M_{200}$.  

Halos are specified by {\it e.g.} their virial masses and concentration parameters.  The concentration parameters can be measured directly in cosmological $N$-body simulations and also derived using semi-analytic techniques based off of the initial spectrum of density perturbations and the picture of hierarchical mergers (see, {\it e.g.},~\cite{2012MNRAS.423.3018P,Diemer:2014gba,Correa:2015dva,Ludlow:2016ifl}).  For example, Ref.~\cite{Correa:2015dva} provides a fitting function to their semi-analytic halo-mass relation, which is claimed to extend down to arbitrarily small masses since it is based off of physical arguments and not purely phenomenological fitting functions, of the form
\es{eq:c_M_relation}{
\log_{10} c_{200}(M_{200}) \approx \alpha + \beta \log_{10} {M_{200} \over M_\odot} \left[ 1 + \gamma \left(\log_{10} {M_{200} \over M_\odot}\right)^2 \right] \,,
}
where $\alpha \approx 1.3991$, $\beta \approx -0.02572$, and $\gamma \approx 0.00481$.  Of course, this does not mean that every halo of a given mass will have the concentration parameter given above; the above formula is meant to describe the mean expected concentration over an ensemble of halos. $N$-body simulations observe a large amount of scatter; for example, in the \texttt{DarkSky} simulations the typical scatter is $\sim$0.15 in $\log_{10} c_{200}(M_{200})$ for Milky Way mass halos and above~\cite{Skillman:2014qca,Lisanti:2017qoz}.

There are a number of parametric halo profiles found in the literature beyond the NFW profile.  For example, some studies ({\it e.g.},~\cite{He:2019svf}) suggest that the generalized NFW (gNFW) profile provides a better description of cosmological hydrodynamic simulations that include baryons. The gNFW profile has an extra parameter $\gamma_{\rm gNFW}$ and is given by 
\es{}{
\rho_{\rm gNFW} = {\rho_s \over (r / r_s)^{\gamma_{\rm gNFW}} \left( 1+ r / r_s\right)^{3 - \gamma_{\rm gNFW}} } \,.
}   
Note that when an annihilating DM model is fit to the {\it Fermi} GCE the spatial morphology is consistent with a gNFW profile with $\gamma_{\rm gNFW} \approx 1.25$~\cite{Fermi-LAT:2017opo}.  Another common spherically-symmetric DM density profile discussed in the literature is the Einasto profile~\cite{Graham:2005xx,2012A&A...540A..70R,2016MNRAS.457.4340K}, which has been claimed to provide a better fit to simulation data for halos in some DM-only $N$-body simulations relative to the NFW profile.  The profile is given by 
\es{}{
\rho_{\rm ein}(r) = \rho_s \exp \left[ - {2 \over \alpha_s} \left( \left( {r \over r_s} \right)^{\alpha_s} - 1 \right) \right] \,, 
}
where the three model parameters are $\rho_s$, $r_s$, and $\alpha_s$.  Typical values for $\alpha_s$ found in simulations are $\sim$0.2. The parameter $r_s$ is typically similar to the NFW scale radius.  Lastly, another parametric DM profile discussed frequently is the Burkert profile~\cite{Burkert:1995yz,Burkert:2000di,Salucci:2000ps}:
\es{}{
\rho_{\rm burk}(r) = \rho_0 {r_0^3 \over (r + r_0) (r^2 + r_0^2) } \,,
}
where $\rho_0$ and $r_0$ are the model parameters.  Here, $r_0$ corresponds to a core size for the central DM density.  The Burkert profile was introduced to explain apparent cores in the DM density profiles of spiral galaxies.  Some analyses of the rotation curve of the Milky Way even suggest $r_0 \sim 10$ kpc~\cite{Nesti:2013uwa}.

\subsection{Dark matter in the Milky Way}  
\label{sec:DM_MW}

The DM halo of the Milky Way plays an especially important role in DM indirect detection.  Unfortunately, it is exceptionally difficult to determine the DM density profile in the inner regions of the Galaxy, which is typically the region most important for indirect searches, since the gravitational potential becomes increasingly baryon dominated as one moves towards the GC. If the potential is dominated by baryons, it becomes very difficult to infer sub-dominant affects on {\it e.g.} stellar dynamics from the DM contribution. Thus, typically efforts to model the Milky Way's DM profile measure bulk properties of the DM halo, well away from the GC, and then extrapolate to the inner halo using one of the phenomenological profiles discussed above, which are validated on simulations. Thus, in this framework simulations play an especially important role, and constructing more accurate hydrodynamic simulations that model the DM profile in the baryon-dominated inner regions of galaxies is crucial.  Unfortunately, though, current simulations (see, {\it e.g.},~\cite{Hopkins:2017ycn}) point to a large diversity of halo profiles in the inner regions of galaxies for halos of otherwise similar masses, complicating efforts to precisely predict the halo shape in the inner kpc of the Galaxy by extrapolating from simulations.

With the caveats above aside, let us discuss the measured values for the Milky Way. Ref.~\cite{Cautun:2019eaf} performed an analysis combining the {\it Gaia} rotation curve data of stars~\cite{2019ApJ...871..120E} with estimates of the total mass of the Milky Way from studies of the motion of its satellite galaxies and globular clusters~\cite{2019ApJ...873..118W,Callingham:2018vcf} to infer the virial mass $M_{\rm 200} = 0.82_{-0.18}^{+0.09} \times 10^{12}$ $M_\odot$, with a concentration parameter $c_{200} = 13.31_{-2.68}^{+3.60}$ for a fit of the NFW profile.  While the precise values of these quantities are debated in detail and the subject of ongoing research, the numbers quoted above provide good benchmark quantities for studies of the Milky Way halo using the NFW profile. Note that these central values correspond to $r_s \approx 15$ kpc, $\rho_0 \approx 1.1 \times 10^7$ $M_\odot/$kpc$^3$, and $r_{200} \approx 200$ kpc.  Note that the mean prediction for the concentration for a halo of this mass from~\eqref{eq:c_M_relation} is  $c_{200} \sim 8$, and thus this measurement is consistent with the expected concentration mass relation within the predicted scatter.  The distance between the Sun and the GC is measured to be $r_\odot \approx 8.23 \pm 0.12$ kpc~\cite{2023MNRAS.519..948L}.  This implies a local DM density at the solar location of $\rho_\odot \approx 8.6 \times 10^{-3}$ $M_\odot/$pc$^3 \approx 0.33$ GeV/cm$^3$.  The local DM density is the discussion of intense discussion and debate because it plays an important role in anchoring the DM density profile for indirect detection but also because it normalizes the signal strength in DM direct detection.  The local DM density can be inferred from global data ({\it e.g.}, rotation-curve data) or locally by looking at the velocity dispersion and density profiles of stars in the local neighborhood. The different measurements tend to lie in the range $\sim$$0.2 - 0.6$ GeV/cm$^3$, but out-of-equilibrium behavior of the galaxy in addition to systematic uncertainties associated with modeling baryonic contributions to the potential are hindering efforts to achieve more precise and accurate measurements (see~\cite{Read:2014qva,ParticleDataGroup:2022pth} for discussions).   

Ref.~\cite{Pieri:2009je} found that Milky Way-like halos in the DM-only {\it Via Lactea II} and {\it Aquarius} simulations could be well-modeled by Einasto profiles with $r_s \approx 20$ kpc and $\alpha \approx 0.17$. Normalizing to a common local DM density of $\rho_\odot = 0.33$ GeV/cm$^3$, in order to compare with the fiducial NFW profile described above, implies a normalization parameter of $\rho_s \approx 2 \times 10^3$ $M_\odot/$kpc$^3$.  Ref.~\cite{Nesti:2013uwa} fit the Burkert profile to a variety of Milky Way stellar kinematic data and found a large best-fit core radius $r_0 \approx 9.26$ kpc.  Such a large core would radically affect DM indirect searches in the Galaxy, since the DM density profile would essentially be constant within the solar radius. On the other hand, recent hydrodynamic simulations do not appear to find such large cores in simulations of Milky Way size galaxies (see, {\it e.g.},~\cite{2020MNRAS.497.2393L} for a discussion in the context of the FIRE-2 simulations).  In fact, the Milky Way mass halos are found to be more concentrated because of baryonic feedback, though DM cores may still form at smaller scales $\sim$0.5 - 2 kpc.  Even closer to the GC one reaches the radius of influence of the central supermassive black hole Sagittarius A* ($\sim$3 pc), where the gravitational dynamics are dominated by the potential of the black hole.  Analytic arguments and simulations suggest that the DM may develop a kinematic cusp within the radius of influence, where the DM density may rise sharply (perhaps as $\rho_{\rm DM} \sim r^{-1.5}$) towards the event horizon (see, {\it e.g.},~\cite{Merritt:2006mt}).  Needless to say, DM studies that involve searches in the inner regions of the Galaxy should be regarded with a healthy degree of skepticism, given the large and increasing uncertainties as one approaches the GC. 

\begin{figure}[htb]  
\begin{center}
\includegraphics[width=0.75\textwidth]{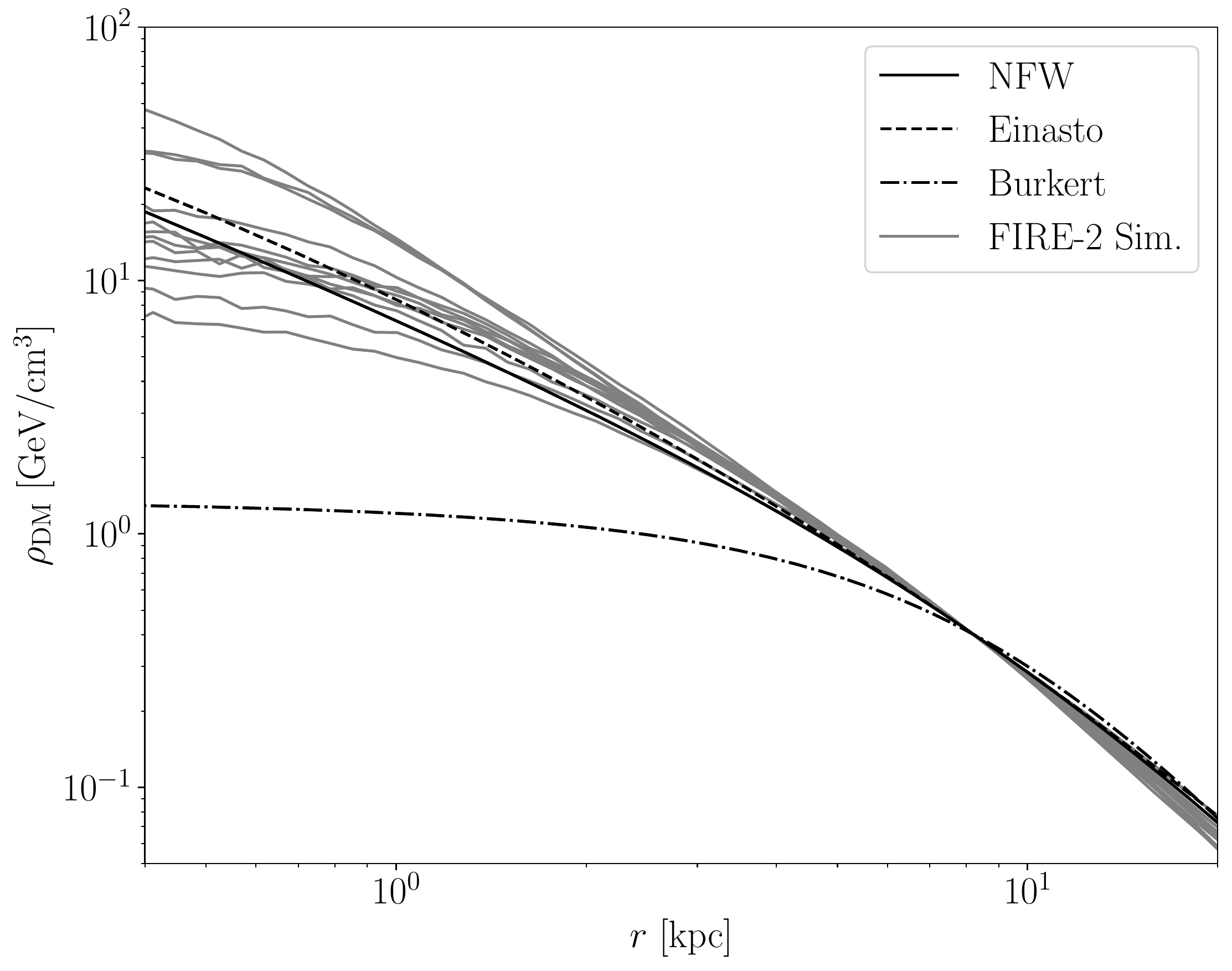}
\caption{Spherically-symmetric DM density profiles for the Milky Way. All of these density profiles assume the local DM density of $\rho_\odot = 0.4$ GeV/cm$^3$. For these examples, the NFW profile has $r_s = 15$ kpc, while the Einasto profile has $r_s = 20$ kpc and $\alpha = 0.17$. The Burkert profile is illustrated with a core radius of $r_0 = 9.26$ kpc.  Twelve non-parametric profiles are also illustrated from Milky Way analogue galaxies in the FIRE-2 hydrodynamic simulations. These profiles show a combination of adiabatic contraction, which enhances the DM density relative to {\it e.g.} pure NFW, and also central cores, which flattens the innermost DM density relative to NFW.  For some analogue galaxies one effect is more important than the other, giving a reasonable spread between the profiles of otherwise similar halos, though none of the galaxies is found to be as cored as the example Burkert profile.  This figure may be reproduced using \href{https://colab.research.google.com/drive/1MZ6ZeyGYmXlx1pXpeh3NrSani54_zBhr?usp=sharing}{this Colab Jupyter Notebook.}
}
\label{fig:DM}
\end{center}
\end{figure}

In Fig.~\ref{fig:DM} we compare the DM profiles in the Milky Way between the fiducial NFW, Einasto, and Burkert profiles discussed above, though for the sake of comparison we normalize all of the profiles to the common local DM density $\rho_\odot = 0.4$ GeV$/$cm$^3$ at the solar location, which is fixed at $r_\odot = 8.23$ kpc.  While the NFW and Einasto profiles are relatively similar, the large Burkert core predicts significantly less DM at the inner radii.
In addition, we show the DM density profiles, normalized the same way as discussed above, from the 12 Milky Way analogue galaxies in the recent FIRE-2 hydrodynamic simulations~\cite{Hopkins:2017ycn,2020MNRAS.497.2393L,2022MNRAS.513...55M}.  These 12 galaxies have similar DM and baryonic masses to those of the Milky Way,  The effects of adiabatic contraction are visible in most of the galaxies, as seen in Fig.~\ref{fig:DM}, with the DM density generally enhanced relative to that of the NFW and Einasto profiles at a few kpc from the GC. Closer to the GC, however, the FIRE-2 profiles may develop modest cores at scales $\sim$0.5 - 2 kpc~\cite{2020MNRAS.497.2393L}.  These simulations, while far from definitive in terms of the Milky Way's DM profile, highlights both the diversity of possibilities and the possible complexity of baryonic feedback on halo profiles, simultaneously inducing inner cores through baryonic feedback and enhancing the densities at moderate distances through adiabatic contraction.  On the other hand, the Burkert profile appears to be an outlier with respect to the FIRE-2 halo profiles, with its inner DM densities much lower than found in those simulations.

\subsection{Dark Matter in nearby galaxies and halos}
\label{sec:DM_halo_other}

Some of the strongest and most robust constraints to-date on annihilating DM models come from Milky Way satellites including classical and ultra-faint dwarf galaxies. These are low-mass DM subhalos that live as substructure within the bulk DM halo of the galaxy. In the modern picture of hierarchical structure formation such substructure is expected, since the galaxy is made up from the merger of smaller DM halos.  Many of these halos are disrupted and go on to form the virialized DM halo of the Milky Way, but many others are predicted to still survive today as gravitationally bound subhalos. 

DM subhalos are predicted to exist to (almost) arbitrarily small mass scales. The less massive subhalos formed earlier, which is why they are more compact (see~\eqref{eq:c_M_relation}).  The minimum subhalo mass is usually set by the DM mass contained within a Hubble volume at the cosmological epoch where DM starts to behave like a cold, collisionless fluid. For WIMP DM models this can correspond to $M_{\rm vir} \sim 10^{-6}$ $M_\odot$ or even smaller halo masses. (Exceptions to this rule include DM models that are warm or have self interactions or other non-trivial particle physics properties, such as fuzzy DM, that wash out structure on small scales.) However, these ultra-low-mass halos do not form deep enough potential wells to pull in a sufficient  mass of baryons to form stars. After all, a $10^{-6}$ DM halo will certainly not be able  to accrete enough baryons to form a $\sim$1$M_\odot$ mass star! Ultra-faint dwarfs are the smallest known DM halos that form stars. They have stellar masses today of around $M_{\bm \star} \sim 10^2 - 10^5$ $M_\odot$~\cite{Bullock:2017xww}, with $M_{\bm \star}$ denoting the stellar mass.  Despite their small number of stars, ultra-faint dwarfs are surrounded by massive DM halos.  Low-mass halos ($M_{v} \lesssim 10^{11}$ $M_\odot$) have a roughly power-law relation between the stellar mass $M_{\bm \star}$ and the halo mass $M_{v}$: 
\es{}{
M_{\bm \star} \sim10^8 \, \, M_\odot \left( {M_v \over 10^{11} \, \, M_\odot} \right)^2 \,,
}
 though there for a fixed $M_v$ cosmological simulations and observations find a significant spread around this approximate, mean prediction~\cite{2013ApJ...770...57B,Garrison-Kimmel:2016szj,2019MNRAS.488.3143B}. (The spread is more pronounced at low $M_v$.) Note that at low halo masses the ratio $M_{\bm \star} / M_v$ becomes increasingly small, which implies that the halos become increasingly DM dominated. However, the ratio $M_{\bm \star} / M_v$ peakes around $M_v \sim 10^{12}$ $M_\odot$; above roughly this value the ratio $M_{\bm \star} / M_v$ decreases again going to cluster scales. 
 
 Slightly more massive than the ultra-faint dwarfs, the so-called {\it classical dwarfs} have $M_{\bm \star} \sim 10^5 - 10^7$ $M_\odot$~\cite{Bullock:2017xww}.  Continuing the classification, there are {\it bright dwarfs} with $M_{\bm \star} \sim 10^7 - 10^9$ $M_\odot$ and then Milky Way type galaxies with $M_{\bm \star} \sim 10^{10} - 10^{11}$  $M_\odot$. (The Milky Way has $M_{\bm \star} \approx 6 \times 10^{10}$ $M_\odot$~\cite{2011MNRAS.414.2446M}).  In the context of hydrodynamic simulations it is found that baryonic feedback does not strongly affect the DM density profiles of ultra-faint dwarf galaxies and the low-mass end of classical dwarfs (see, {\it e.g.},~\cite{2020MNRAS.497.2393L}).  Bright dwarfs, on the other hand, are strongly affected by baryons, which can produce sizable cores in the inner parts of the galaxies. Milky Way mass halos are more complicated, with -- as discussed above -- the possibility of adiabatic contraction increasing the DM density at moderate distances from the GC and then possibly smaller cores developing in the very inner parts of the galaxy.  Baryonic feedback appears to produce the most significant cores for bright dwarf scales galaxies, with $M_{\bm \star} / M_v \sim {\rm few} \times 10^{-3}$~\cite{2020MNRAS.497.2393L}.  For the ultra-light dwarf galaxies, however, using a DM profile motivated from DM-only $N$-body simulations, such as Einasto or NFW, is likely a good approximation to reality, assuming the subhalo cores are not tidally disrupted by interactions with {\it e.g.} their hosts.
 
 Let us consider now some of the Milky Way's ultra-faint dwarf spheroidal galaxies (dSphs). Here spheroidal means that the stellar content is approximately spherically distributed, as opposed to in {\it e.g.} an elliptical galaxy.  In Tab.~\ref{table:dSphs} we give the best-fit properties of a sample of ultra-faint dSphs from~\cite{2019MNRAS.482.3480P}, with best-fit parameters quoted in~\cite{DiMauro:2022hue}.  The general method for determining the DM profiles from these systems, which is what~\cite{2019MNRAS.482.3480P} follows (but see~\cite{Nguyen:2022ldb} for a new machine learning approach), is to use Jeans modeling. One treats the stars as tracers in the gravitational potential and then models their phase-space distribution using the collisionless Boltzmann equation in conjunction with the Poisson equation for gravity, which in this case is sourced completely by the DM halo. The output of the Jeans equation is compared to stellar kinematic data, namely the radial velocity dispersion and the radial stellar density profile, in order to constrain the model parameters of the DM density profile. Since these systems essentially behave as DM-only halos, an NFW profile is appropriate, which is what~\cite{2019MNRAS.482.3480P}  assumed (with best-fit parameters displayed in Tab.~\ref{table:dSphs}). Note that there are many more ultra-faint dSphs than given in Tab.~\ref{table:dSphs}, but we have chosen to only show here six of the brightest as ranked by their predicted DM annihilation signatures, which we discuss shortly.

\begin{table}[h!]
\centering
\begin{tabular}{||c | c c  c c c cc||} 
 \hline
dSph & $M_{\rm halo}$ [$M_\odot$] & $\rho_s$ [${M_\odot \over {\rm kpc}^3}$] & $r_s$ [kpc] & $r_t$ [kpc] & $d$ [kpc] & ${\bar {\mathcal J}}$ [${\rm{GeV}^2 \over {\rm cm}^5}$] & ${\bar {\mathcal D}}$ [${{\rm GeV} \over \rm{cm}^2}$]  \\[0.5ex] 
 \hline\hline
 Willman I & $2.7 \times 10^8$ & $10^{8.25}$ & $0.45$ & $3.74$ & 38 & $1.6\times 10^{23}$ & $1.5 \times 10^{22}$ \\ 
Ursa Major II & $8.5 \times 10^8$ & $10^{7.61}$ & $1.25$ & $5.29$ & 35 & $1.2 \times 10^{23}$ & $2.0 \times 10^{22}$ \\ 
Ursa Major I & $6.6 \times 10^8$ & $10^{7.43}$ & $1.12$ & $9.91$ & 97 & $7.2 \times 10^{21}$ & $5.5 \times 10^{21}$ \\ 
Tucana II & $9.5 \times 10^8$ & $10^{7.31}$ & $1.65$ & $6.74$ & 57 & $3.1 \times 10^{22}$ & $1.2 \times 10^{22}$ \\ 
Segue I & $1.9 \times 10^7$ & $10^{8.30}$ & $0.19$ & $1.14$ & 23 & $3.6 \times 10^{22}$ & $5.1 \times 10^{21}$ \\ 
Reticulum II &1.$7 \times 10^8$ & $10^{7.55}$ & $0.83$ & $2.62$ & 32 & $3.6 \times 10^{22}$ & $9.4 \times 10^{21}$ \\ [1ex] 
 \hline
\end{tabular}
\caption{A table of some of the top five Milky Way dSphs for indirect searches for DM (particularly annihilation).  The tables of halo masses and NFW parameters, along with the tidal radius $r_t$ that serves as a density cut-off radius, are taken from~\cite{2019MNRAS.482.3480P}.  The distances $d$ are also given, as are the calculated ${\bar {\mathcal J}}$ and ${\bar {\mathcal D}}$ factors averaged over a $0.5^\circ$ ROI around the halo center. (These quantities are discussed in Sec.~\ref{sec:indirect}.)  Note that uncertainties on the dSph parameters are not shown but are important to account for in a real analysis.  The ${\bar {\mathcal J}}$ and ${\bar {\mathcal D}}$ factors in this table may be reproduced using \href{https://colab.research.google.com/drive/1MZ6ZeyGYmXlx1pXpeh3NrSani54_zBhr?usp=sharing}{this Colab Jupyter Notebook.}
}
\label{table:dSphs}
\end{table}

In Tab.~\ref{table:dSphs} we also give the so-called tidal radius $r_t$.  Following~\cite{Read:2018pft,DiMauro:2022hue}, the NFW profile for the subhalos is cut off at the tidal radius (see~\cite{DiMauro:2022hue} for an explicit formula for the tidal radius). The tidal radius is, roughly, the radius beyond which the gravitational force that pulls a tracer particle towards the center of the subhalo is overcome by the tidal gravitational force of the host galaxy that wants to strip the particle out of the subhalo. Thus, the density of DM outside of the subhalo should be heavily reduced relative to what one would find in field halos, which are DM halos that exist in isolation and are not subhalos living in the potential well of a larger halo. The closer a subhalo comes to the center of the galaxy, the more DM that is tidally stripped and the smaller $r_t$.  

Lastly, we note that DM halos do likely exist within galaxies like the MW that are simply not massive enough to form stars; as mentioned previously, subhalos could have masses extending all the way down to moon or even asteroid scales. Moreover, the number of subhalos is expected to rise sharply with decreasing halo mass. Based off of simulations and analytic arguments, the number of subhalos $N_{\rm sh}$ is expected to depend on halo mass $M_{\rm halo}$ as, roughly, $dN_{\rm sh} / M_{\rm halo} \sim 1/M_{\rm halo}^2$ (see, {\it e.g.},~\cite{Ando:2019xlm} and references therein). These subhalos can play important roles in many different types of particle DM searches. For example, they can provide a mechanism for ``boosting" the annihilation signal of extragalactic DM halos, such as those surrounding large galaxy clusters.

Having discussed the smallest DM halos, let us briefly comment on some of the largest.  Going away from the Milky Way, the first massive DM halo that is encountered is that surrounding the Andromeda galaxy (also called M31). M31 is relatively similar to the Milky Way; for example, it has a virial mass $M_v \sim 10^{12}$ $M_\odot$.  It is at a distance of around 750 kpc. M31 plays an important role in many searches for particle DM, including annihilating and decaying scenarios. Going further, one of the next most important systems for DM searches is the Virgo cluster.  Virgo has a mass $M_v \sim 10^{15}$ $M_\odot$.  Like all clusters, Virgo has many constituent galaxies.  It is at a distance of around 16 Mpc from the Milky Way. Note that the DM halos of both M31 and Virgo are relatively extended on the sky. The virial radius of M31 (Virgo) extends an angle of $2.6^\circ$ ($1.2^\circ$) on the sky from the center of the galaxy~\cite{Lisanti:2017qlb}.  See Ref.~\cite{Lisanti:2017qlb} for a complete list of the local galaxy groups, their DM properties, and descriptions of their relative importance for DM decay and annihilation signals.  

\section{General considerations for particle dark matter models}
\label{sec:gen}

Before discussing specific particle DM models and their indirect detection strategies, let us consider a few general constraints that any particle DM models should satisfy to be consistent with the $\Lambda$CDM paradigm. 

\subsection{Bounds on the dark matter mass}

First, we motivate a bound on fermionic DM candidates known as the Tremaine-Gunn bound~\cite{1979PhRvL..42..407T}.  The key to the Tremaine-Gunn bound is to note that we can only put one fermion per quantum state, and as we decrease the fermion mass we necessarily need to pack the fermions closer together. To illustrate this point, imagine that we have a constant-density sphere of mass $M$ and radius $R$ made up of fermionic DM with mass $m_\chi$. The Fermi energy is given by 
\es{}{
E_f = {1 \over 2 m_\chi} \left( n_f 3 \pi^2 \right)^{2/3} = {1 \over 2 m_\chi} \left( {9 \pi \over 4} {M \over m_\chi R^3} \right)^{2/3}  \,,
}
where $n_f = M / m_\chi / (4/3 \pi R^3)$ is the fermion number density.  The Fermi velocity $v_f$ is given by the relation $E_f = {1 \over 2} m_\chi v_f^2$, which implies
\es{}{
v_f = {1 \over m_\chi} \left( {9 \pi \over 4} {M \over m_\chi R^3} \right)^{1/3} \,.
}
Most of the states have velocities near the Fermi velocity. If, however, the Fermi velocity surpasses the escape velocity $v_{\rm esc}$ at the edge of the sphere 
\es{eq:vesc}{
v_{\rm esc} = \sqrt{ {2 G M \over R} }\,,
}
then the sphere will be unstable and evaporate due to the Fermi degeneracy pressure. Let us now approximate a typical dwarf galaxy as a sphere of mass $M \sim 5 \times 10^7$ $M_\odot$ and size $R \sim 2.5$ kpc. Requiring the Fermi velocity to be less than the escape velocity restricts $m_\chi \gtrsim 100$ eV, which is close to what one finds from a more careful analysis~\cite{1979PhRvL..42..407T,Boyarsky:2008ju}.

If the DM is less in mass than roughly a keV, then it needs to be a boson because the quantum occupation numbers will necessarily be greater than unity in dense systems like dwarf galaxies. In fact, when we discuss axion DM in Sec.~\ref{sec:axions} this observation will be critical to our treatment of DM as a classical field in that case.  For now, however, let us try to determine the absolute lower bound on the DM mass, assuming that the DM is bosonic. A key insight here comes from the uncertainty principle $\Delta x \Delta p \gtrsim 1/2$. Let us use this relation in the context of our dwarf galaxy illustration above, modeled as a sphere of radius $R \sim 1$ kpc and mass $M \sim 5 \times 10^7$ $M_\odot$. Given the size $\Delta x \sim R \sim 1$ kpc, the velocity of the particles, from the uncertainty principle, cannot be determined to greater precision than
\es{}{
v \gtrsim {1 \over R\, m_\chi} \sim 20 \, \,{ {\rm km} \over {\rm s}} {10^{-22} \, \, {\rm eV} \over m_\chi} \,,
}
with again $m_\chi$ the DM mass. Referring back to~\eqref{eq:vesc}, however, we estimate the escape velocity of this system to be $v_{\rm esc} \sim 20$ km/s.  Thus, if $m_\chi \lesssim 10^{-22}$ eV, the quantum pressure, induced from the uncertainty principle, will not allow for the formation of DM structures as compact as observed dwarf galaxies. Indeed, pseudo-scalar DM models with $m_\chi \sim 10^{-22}$ eV are referred to as fuzzy DM, and historically fuzzy DM has been invoked to try to explain small-scale structure anomalies. More recently, however, the upper bound on the DM mass has been pushed up by some orders of magnitude, such that today pseudo-scalar DM with masses roughly below ${\rm few} \times 10^{-19}$ eV appear excluded~\cite{Dalal:2022rmp}.  We will discuss such ultra-light DM candidates more later in these lecture notes.

Having bounded the DM mass from below, let us now try to bound it from above. Consider a DM point particle of mass $m_\chi$. We may formally calculate the Schwarzschild radius of such as particle as $r_s = 2 G m_\chi$.  On the other hand, the uncertainty principle implies that a particle cannot be localized to better than, roughly, a length scale smaller than the Compton wavelength $r_c = {1 \over m_\chi}$.  If $m_\chi$ is small, then the Compton wavelength is much larger than the Schwarzchild radius, in which case we cannot localize the particle within its own Schwarzchild radius. On the other hand, if $m_\chi \gtrsim G^{-1/2} \sim 10^{19}$ GeV, which is the Planck scale, then we can localize a particle within its own Schwarzchild radius. This implies that the particle should be thought of as a finite-size black hole instead of as a point particle.  This reasoning motivates only thinking about fundamental point particles at masses below the Planck scale. Larger-mass DM candidates may exist, but they must be spatially extended at sizes beyond or equal to the Schwarzchild radius. Examples of such massive DM candidates include primordial black holes and bound-state DM candidates. 

Let us now consider another observational property of DM: it is cold. Here, cold means that DM particles have small primordial velocities relative to those acquired during gravitational collapse.  If the DM were to have a primordial velocity dispersion, then it would wash out structure on small spatial scales because of free-streaming. This is simply the statement that if I create an initial overdensity of matter with a finite velocity dispersion, the DM will free-stream away and wash out the overdensity to an extent determined by the initial overdensity size and the magnitude of the velocity dispersion.  Here, we estimate the magnitude of this effect by making the very rough and not completely correct assumption that a thermal DM candidate, of the type discussed more in the next section, freezes out semi-relativistically ($v \sim 1$) at the temperature $T = m_\chi$ where it begins to become non-relativistic.  If this occurs during the radiation dominated epoch then the co-moving horizon size at decoupling is $(R H)^{-1} \sim \left( {T_0 \over T} {T^2 \over m_{\rm pl}} \right)^{-1} \sim {m_{\rm pl} \over T_0 m_\chi} \sim 50 \, \, {\rm kpc} \, {1 \, \, {\rm keV} \over m_\chi}$, with $T_0$ the temperature of the cosmic microwave background (CMB) today and $m_{\rm pl}$ the reduced Planck mass.  We expect structure to be washed out on smaller spatial scales.  Note that the mass contained within the horizon at decoupling is given by the critical energy density today times a volume of radius the co-moving horizon size, which is $M_{\rm min} \sim 10^{8} \, \, M_\odot \left( {1 \, \, {\rm keV} \over m_\chi} \right)^3$.  We expect to not form DM subhalos at or below, approximately, masses $M_{\rm min}$. Given the existence of dwarf galaxies less massive than this, however, we can roughly motivate a limit constraining thermal DM candidates to have masses at or above the keV scale, regardless of whether or not the DM candidate is a boson or fermion.  This argument is very rough -- in particular it neglects the fact that after chemical decoupling the DM is kept in thermal equilibrium through elastic scattering processes for much longer (see, {\it e.g.},~\cite{Bringmann:2006mu}), but it roughly reproduces the result of more careful analyses (see, for example,~\cite{DES:2020fxi}) that constrain thermal DM candidates to have mass more than around 7 keV. 

DM candidates with a finite initial velocity dispersion at the level that affects observable structures, such as thermal candidates with masses near the keV scale and keV-scale sterile neutrinos, are called warm DM candidates. Historically, just like fuzzy DM, warm DM has been invoked to solve a number of small-scale structure anomalies, though today the necessity of such modifications to the $\Lambda$CDM paradigm are less compelling and the limits on warm DM are stronger~\cite{DES:2020fxi}.  On the other hand, note that axions have significantly smaller masses, much less than the eV scale, and are still viable DM candidates. The reason is that the axion relic density is produced non-thermally, as we discuss further in Sec.~\ref{sec:axions}. Thus, DM candidates do exist with masses below the keV scale, but they must be bosons and also have non-thermal production mechanisms in the early universe.  

\subsection{Bounds on the dark matter lifetime}

Next, let us consider constraints arising from the fact that DM, while produced in the early universe and probed at the epoch of CMB decoupling through the CMB power spectrum, still needs to be around today. The age of the Universe is around $t_{\rm univ} \sim 4 \times 10^{17}$ s, implying that any viable DM model should have a decay rate (to either SM or dark radiation) of $\Gamma_\chi \lesssim 3 \times 10^{-18}$ s$^{-1}$.  This is a non-trivial constraint in light of the fact that quantum gravity is expected to violate global symmetries (we discuss this point in more detail in Sec.~\ref{sec:axions}).  
An illustrative example of this point is found in the context of glueball DM (see, {\it e.g.},~\cite{Carlson:1992fn,Faraggi:2000pv,Boddy:2014yra,Soni:2016gzf,Foster:2022ajl} for further discussions of this model and in particular discussions of how the glueball may obtain the correct relic abundance).  

The glueball DM model starts, in its simplest form, with a non-abelian hidden sector with gauge field strength $G^D_{\mu \nu}$ and dark fine structure constant $\alpha_D$.  Let us assume for simplicity that the dark sector is pure glue ({\it i.e.}, no light matter) and that the theory confines at the scale $\Lambda_D$.  For example, the dark gauge group could be $SU(N_D)$ for some $N_D \geq 2$.  Let us imagine that the dark gauge group is thermalized at some temperature $T_D$ in the early universe, which does not necessarily need to equal the SM temperature and could be significantly lower so long as the thermalization time scale between the two sectors is less than Hubble.  For $T_D \gg \Lambda_D$ the dark sector consists of free dark gluons, but for $T_D \ll \Lambda_D$ the dark sector is confined and the dark gluons condense into neutral (from the point of view of $G_D$) bosonic bound states of gluons called glueballs. The lightest glueball is typically the $0^{++}$ state, with $0$ referring to its spin and $++$ referring to its CP quantum numbers. The relation between the mass of this state, $m_{0^{++}}$, and the dark confinement scale is typically $m_{0^{++}} \sim {\rm few} \times \Lambda_D$, with the exact prefactor depending on the dark gauge group (see~\cite{Foster:2022ajl} for a discussion).  The lightest glueball can make up the observed DM, assuming the cosmology is arranged such that it acquires the correct relic abundance.

Generically, we expect higher dimensional operators in this theory of form
\es{eq:glueball_5}{
{\mathcal L} \supset {c \alpha_D \over 4 \pi} G^D_{\mu \nu} G^{D, \mu \nu} {H^\dagger H \over \Lambda^2} \,,
}
where $c$ is a constant of order unity, $H$ is the SM Higgs doublet, and $\Lambda$ is the mass scale where this operator emerges in the UV completion.  In particular, there is (naively) no gauge symmetry preventing this operator from appearing at the Planck scale ($\Lambda = m_{\rm pl}$), though the operator could also arise at a lower scale. Equation~\eqref{eq:glueball_5} allows for glueball decay, as may be estimated by taking $G^D_{\mu \nu} G^{D, \mu \nu} \to \Lambda_D^3 \chi$ in~\eqref{eq:glueball_5}, with $\chi$ the $0^{++}$ glueball. Working out the decay rate in the limit $m_{0^{++}} \gg 10^2$ GeV so that we can treat the Higgs as massless yields the result~\cite{Juknevich:2009gg}
\es{}{
\Gamma_\chi \approx  5 \times 10^{-24} \, \, {\rm s}^{-1}\, c^2\left( {m_{0^{++}} \over 10^6 \, \, {\rm GeV}} \right)^5 \left( {2 \times 10^{18} \, \, {\rm GeV} \over \Lambda} \right)^4 \,.
}
While this decay rate, for the default parameters above, implies that the DM lifetime is much longer than the age of the universe, this benchmark model is in fact ruled out by orders of magnitude. The reason is that the decay products from this decay leave observable signatures in high-energy neutrino and gamma-ray experiments~\cite{Cohen:2016uyg}.  In fact, constraints from the {\it Fermi} gamma-ray telescope for $m_{0^{++}} = 10^6$ GeV require $\Gamma_\chi \lesssim 10^{-28}$ s$^{-1}$.  The dark glueball model illustrates two general points: (i) unless there is a local symmetry that protects the DM candidate, slow decays to the SM can be expected, and (ii) often direct searches for the DM decay product give much stronger constraints on $\Gamma_\chi$ than found from requiring the DM lifetime be longer than the age of the Universe (decaying DM signatures are discussed more in Sec.~\ref{sec:indirect}). 

One way to avoid the decaying DM constraints in the context of the glueball model is to take $m_{0^{++}}$ to be much smaller than a PeV. On the other hand, as we discuss now, for significantly smaller confinement scales the glueball DM model is subject to self-interaction bounds. By dimensional analysis we may infer that below the confinement scale the DM effective Lagrangian should have quartic terms of the form ${\mathcal L} \supset \chi^4$, with coefficients order unity since after confinement there are no parametrically small coupling constants.  This implies that we can expect $\chi \chi \to \chi \chi$ scattering processes today with cross-section $\sigma_{\chi \chi \to \chi \chi} \sim 1 / m_{0^{++}}^2$.  Let us now consider how such scattering processes are constrained today.

\subsection{Bounds on the dark matter self interactions}

Perhaps the most canonical example of DM self interaction constraints arises from the bullet cluster~\cite{Markevitch:2003at,Robertson:2016xjh}.  The bullet cluster is a system around 1 Gpc away ($z \sim 0.3$) that is the merger of two galaxy clusters. The clusters have already passed through each other. One observes the following: the hot gas in the two clusters interacted strongly during the merger and is left at the interaction point (this is observed in $X$-rays); the stars did not interact and passed through unperturbed (this is observed in {\it e.g.} optical); the two DM halos also passed through unaffected, as is observed by noting through gravitational lensing that the matter associated with the two clusters are still clustered around the stars.  This means that the DM did not undergo appreciable self interactions during the crossing, as otherwise the DM -- like the gas -- would have also stopped to some degree around the interaction point.  To estimate the limit on $\sigma_{\chi \chi \to \chi \chi}$ from this observation let us model both of the clusters as having DM halos with sizes around a Mpc and masses around $5 \times 10^{14}$ $M_\odot$.  Consider a DM particle in one halo which transits through the other. The probability $p_{\chi \chi \to \chi \chi}$ that it undergoes scattering during the transit is estimated as $p_{\chi \chi \to \chi \chi} \sim n_\chi \sigma_{\chi \chi \to \chi \chi} L$, where $n_\chi \sim M_{\rm halo} / (4/3 \pi L^3) / m_\chi$ is an estimate for the number density in the halo and where $L$ is an estimate for the halo size. At the least, we want $p_{\chi \chi \to \chi \chi} < 1$, as otherwise almost every DM will have undergone scattering.  This requirement translates to $\sigma_{\chi \chi \to \chi \chi} / m_\chi \lesssim 10$ cm$^2$/g.  Modeling the bullet cluster more carefully leads to a comparable but slightly stronger result $\sigma_{\chi \chi \to \chi \chi} / m_\chi \lesssim 1$ cm$^2$/g (in more particle-physics-oriented units this is $\sigma_{\chi \chi \to \chi \chi} / m_\chi \lesssim \left({1 \over 0.06 \, \, {\rm GeV}} \right)^3$)~\cite{Markevitch:2003at,Robertson:2016xjh}.  In the context of our glueball model this implies that the glueball mass should be above around $50$ MeV in order to not be excluded by self-interaction bounds.

One interesting remark about the glueball model is that at first glance it appears like a doomsday scenario that would be impossible to confirm with data, since the dark sector may only talk to the SM through Planck-suppressed operators. On the other hand, as we have seen above, even nearly-secluded dark sectors can give discernible astrophysical signatures, either through their rare decays or through their self interactions.

To conclude the discussion of DM self interactions, let us note that it is possible to leave larger signatures of DM self interactions in small-scale DM halos such as dwarf galaxies and still be consistent with {\it e.g.} the bullet cluster bound by invoking velocity dependent self scattering processes (see~\cite{2021MNRAS.503..920C} and references therein). Velocity dependent scattering would be expected in a model with an effective long-range force ({\it i.e.}, a light mediator). In this case, the self-scattering cross-section would be larger in systems with lower velocity dispersions, such as dwarf galaxies, than in larger systems like clusters with higher velocity dispersions.

\section{WIMP dark matter}
\label{sec:4}

WIMPs are perhaps the most well-studied of all possible DM candidates at present. Moreover, despite some claims to the contrary they are most certainly not dead. In fact, some of the most well-motivated WIMP candidates, such as the thermal higgsino, have yet to be probed experimentally or observationally. As we discuss below, however, multiple advancements in indirect detection will cover qualitatively new parameter space, such as that of the higgsino and other WIMP benchmark scenarios, in the coming years. At the same time, current and future generations of direct detection experiments, such as the current LZ (see, {\it e.g.},~\cite{Akerib:2022ort} for a summary of current and future WIMP direct detection efforts),  will cover complementary parameter space to the indirect probes. In these lecture notes, however, we focus on indirect searches (see~\cite{Schumann:2019eaa} and ~\cite{Boveia:2018yeb} for reviews that discuss laboratory and collider probes of WIMPs).  We begin with a brief review of thermal freeze-out, before discussing the minimal DM candidates such as the higgsino, and then turning to indirect probes.

\subsection{Thermal freeze-out of WIMP DM and implications for present-day annihilation rates}
\label{sec:freeze_out}

Before starting this section, it is a good idea to review App.~\ref{sec:cosmo}. That section provides a brief review of cosmology and the early universe.  We will be making use of a few key points from that section during the radiation-dominated era, in particular:
\begin{itemize}
\item The Hubble parameter $H = \dot R / R$, with $R(t)$ the scale factor, scales as $H \sim T^2 / m_{\rm pl}$ in radiation domination (see~\eqref{eq:Hubble_rad}).
\item The number density $n_{\rm eq}$ of a relativistic particle in equilibrium ($T\gg m$, with $m$ the particle mass) scales with temperature as $n_{\rm eq} \propto T^3$ (see~\eqref{eq:n_rel}).
\item The number density of a non-relativistic particle in equilibrium ($T\ll m$) scales with temperature as $n_{\rm eq} \propto (m \, T)^{3/2} e^{-m/T}$ (see~\eqref{eq:n_NR}).
\item The co-moving entropy density $s$, defined in~\eqref{eq:entropy}, is conserved: ${d \over dt} \big(R^3 s \big)= 0$.
\end{itemize}

\begin{figure}
\centering
\begin{tikzpicture}
  \begin{feynman}
    \vertex (a){DM};
    \vertex [right=2cm of a] (b);
    \vertex [below=2cm of b] (c);
    \vertex [below=2cm of a] (d){DM};
    \vertex [below=0.7cm of b, blob]  (e) {};
    \vertex [right=2cm of b] (f){SM};
    \vertex [below=0.66cm of f] (j){$\vdots$};
    \vertex [below=1.33cm of f] (h){SM};
    \vertex [right=2cm of c] (i){SM};
    \diagram* {
      (a) --  (e),
      (e) -- (d),
      (e) -- (f),
      (e) -- (h),
      (e) -- (i)
    };
  \end{feynman}
\end{tikzpicture}%
\hspace{3cm}
\begin{tikzpicture}
  \begin{feynman}
    \vertex (a){SM};
    \vertex [right=2cm of a] (b);
    \vertex [below=2cm of b] (c);
    \vertex [below=2cm of a] (d){SM};
    \vertex [below=0.7cm of b, blob]  (e) {};
    \vertex [right=2cm of b] (f){DM};
    \vertex [below=0.66cm of f] (j){$\vdots$};
    \vertex [below=1.33cm of f] (h){DM};
    \vertex [right=2cm of c] (i){DM};
    \diagram* {
      (a) --  (e),
      (e) -- (d),
      (e) -- (f),
      (e) -- (h),
      (e) -- (i)
    };
  \end{feynman}
\end{tikzpicture}
\caption{In the early universe thermal DM  candidates like WIMPs are kept in equilibrium with the SM by scattering processes like those shown above, where {\it e.g.} two DM particles may annihilate into any number of SM particles, along with the reverse process, or two SM particles may come together to form any number of DM particles. These diagrams can also be read vertically. For simplicity, in these lecture notes we restrict ourselves to models where the most relevant scattering processes are two-to-two. }
\label{fig:feyn}
\end{figure}
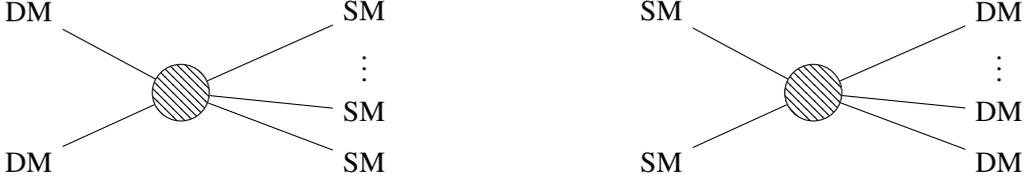

Let us now imagine that at some high temperature $T \gg m_\chi$ our massive DM candidate $\chi$, with mass $m_\chi$ and anti-particle $\bar \chi$, is in thermal equilibrium with the SM plasma. Let us assume that there is a two-body annihilation channel, where $\chi$ and $\bar \chi$ may annihilate to any number of SM final states but also SM final states may come together to form $\chi$-$\bar \chi$ pairs (see Fig.~\ref{fig:feyn}). (Note that in the examples we consider the annihilation will be 2-to-2.)
Let the rate of DM annihilation be given by $\Gamma_{\chi \bar \chi \to {\rm SM}}$. Then -- neglecting the inverse process for now -- the Boltzmann equation describing the evolution of the DM number density $n_\chi$ is approximately
\es{eq:Boltz_0}{
{d (R^3 n_\chi) \over dt} = - \Gamma_{\chi \bar \chi \to {\rm SM}} n_\chi R^3 \, \qquad \Longrightarrow \qquad
 {d n_\chi \over dt} + 3 H n_\chi = - \Gamma_{\chi \bar \chi \to {\rm SM}} n_\chi \,.
}
Note that the right hand side describes the depletion of $\chi$ particles through annihilation processes, while the factor of $R^3$ simply accounts for dilution from the expansion of the universe.  Of course, $\chi$ particles may also be created through SM annihilation processes, which implies that a term should appear on the right hand side of~\eqref{eq:Boltz_0} with the opposite sign.  

The exact form for the so-called ``collision terms" appearing on the right hand side of the Boltzmann equation may be derived in detail by starting with the Boltzmann equation for the six-dimensional phase-space distribution, writing down the collision term in terms of the phase-space factors of all participating scattering particles with the associated cross-sections, and then integrating over momentum (see, {\it e.g.},~\cite{Kolb:1990vq,Hooper:2009zm,Lisanti:2016jxe,Slatyer:2017sev,Hooper:2018kfv,Lin:2019uvt} for more detailed reviews).  However, we may argue for the form of this term through the following logic.  First, let us consider the expected form of $ \Gamma_{\chi \bar \chi \to {\rm SM}}$. This rate is given by $\Gamma_{\chi \bar \chi \to {\rm SM}} = n_\chi \langle \sigma_{\rm \chi \bar \chi \to {\rm SM}} v\rangle$, since a given DM particle, traveling with velocity $v$, needs to find another DM particle, with number density $n_\chi$, in order to annihilate with cross-section $\sigma_{\rm \chi \bar \chi \to {\rm SM}}$.\footnote{For identical annihilating particles, it is more appropriate to write the collision term as ${n_\chi^2 \over 2} \langle \sigma_{\rm \chi \bar \chi \to {\rm SM}} v\rangle \times 2$, with the factor of $1/2$ a symmetry factor for identical particles but the extra factor of two in the numerator arising since DM annihilation removes two particles at once.}  
Now note that when the DM is in thermal equilibrium with the SM, the DM production term should exactly cancel the DM depletion term (the right-hand-side of~\eqref{eq:Boltz_0}) such that the total number density of $\chi$ particles follows the equilibrium expectation $n_\chi \propto 1/R^3$. This allows us to hypothesize (this hypothesis may again be verified in detail by thinking more carefully about the collision terms)
\es{eq:Boltz_1}{
 {d n_\chi \over dt} + 3 H n_\chi = - \langle \sigma v \rangle (n_\chi^2 - n_{\rm eq}^2) \,,
}
where for simplicity we have dropped the label on the annihilation cross-section $\sigma$ and where $n_{\rm eq}$ denotes number density that a particle would have in equilibrium at time $t$ with mass $m_\chi$.  The right-hand-side above cancels trivially if $n_\chi = n_{\rm eq}$.  

The best way to compute how $n_\chi$ freezes out is to numerically evolve the non-linear differential equation in~\eqref{eq:Boltz_1}.  We will turn to this approach shortly. First, though, let us try to understand the parametrics of non-relativistic freeze-out. We expect $n_\chi$ to follow the equilibrium number density $n_{\rm eq}$ until the processes that keep $\chi$ in equilibrium are too slow relative to Hubble expansion, which is when $\langle \sigma v \rangle n_{\rm eq} \lesssim H$.  We will later show that this occurs for $x_{\rm fo} \sim 10$, with $x \equiv m_\chi / T$, roughly independent of $m_\chi$ and $\langle \sigma v \rangle$.  This value of the freeze-out temperature is  expected since $n_{\rm eq}$ drops exponentially for $x > 1$ (see~\eqref{eq:n_NR}), and so it is not surprising that $x_{\rm fo}$ is slightly larger than unity.  Below, we will be slightly approximate and solve for $\langle \sigma v \rangle$ in terms of the freeze-out temperature $T_{\rm fo}$ and the number density at freeze-out, $n_{\rm eq}^{\rm fo}$:
\es{}{
H_{\rm fo} \sim {T_{\rm fo}^2 \over m_{\rm pl}} \,, \qquad \Longrightarrow \qquad \langle \sigma v \rangle \sim {T_{\rm fo}^2 \over m_{\rm pl} n_{\rm eq}^{\rm fo}} \,.
}
Note that we expect $n_{\rm eq}^{\rm fo} \ll T_{\rm fo}^3$, since the number density is exponentially suppressed in the non-relativistic regime. To make progress, we use the fact that after freeze-out the number density will redshift like a gas of non-interacting particles: in particular, $n_\chi \propto 1/R^3$. This allows us to work backwards in the following sense. At matter radiation equality, with temperature $T_{\rm MRE}$, the energy density in DM is given by $\rho_\chi^{\rm MRE} = n_\chi^{\rm MRE} m_\chi \sim T_{\rm MRE}^4$, with the latter equality equating the matter energy density to that in radiation.  Thus, for a given $m_\chi$ we know the value of $n_\chi^{\rm MRE}$ to match the observed DM density.  Moreover,
\es{}{
 \left( R (T_{\rm fo}) \over R (T_{\rm MRE}) \right)^3 = \left( {g_{*s}(T_{\rm MRE}) \over g_{*s}(T_{\rm fo}) } \right) \left({T_{\rm MRE} \over T_{\rm fo}}\right)^3 \,,
 } 
 since the co-moving entropy density is conserved, with the entropy degrees of freedom $g_{*s}$ defined in App.~\ref{sec:cosmo}.  For a TeV thermal relic we may expect $g_{*s}$ to be around 10 - 30 times larger at $T_{\rm fo}$ than $T_{\rm MRE}$, but for simplicity we approximate the two $g_{*s}$ factors to be the same here for a back-of-the-envelop estimate. 
 This implies that the value of $n_{\rm eq}^{\rm fo}$ that we need to have to end up with the correct number density at matter radiation equality is approximately $n_{\rm eq}^{\rm fo} \approx n_\chi^{\rm MRE} (T_{\rm fo} / T_{\rm MRE} )^3$.  Putting everything together we then can compute that 
\es{eq:sigma_v_thermal}{
\langle \sigma v \rangle \sim {T_{\rm fo}^2 \over m_{\rm pl} n_{\rm eq}^{\rm fo}}  \sim {x_{\rm fo} \over m_{\rm pl} T_{\rm MRE}} \,. 
}
Remarkably, this gives a number for the velocity-averaged annihilation cross-section $\langle \sigma v \rangle$ in order to obtain the correct DM abundance that is independent of $m_\chi$, at least to the extent that $x_{\rm fo}$ is independent of $m_\chi$. Given that $T_{\rm MRE} \sim 1$ eV, we find that the correct DM abundance is obtained -- roughly -- for $\langle \sigma v \rangle \sim 10^{-9} \, \, {\rm GeV}^{-2} \sim 10^{-26} \, \, {\rm cm}^3 / {\rm s}$.  Doing the calculation more precisely leads to the result $\langle \sigma v \rangle \sim 2 \times 10^{-26} \,\, {\rm cm}^3 / {\rm s}$, with minor mass dependence for electroweak scale masses~\cite{Steigman:2012nb}.  

Let us now check our assumption that $x_{\rm fo} \sim 10$. Recall that we want to solve the following equation for $T_{\rm fo}$, using the fact that the equilibrium number density is exponentially suppressed for a non-relativistic particle~\eqref{eq:n_NR}:
\es{}{
{T_{\rm fo}^2 \over m_{\rm pl}} \sim \langle \sigma v \rangle n_{\rm eq}^{\rm fo} \sim \langle \sigma v \rangle  \left( {m_\chi T_{\rm fo}  \over 2 \pi} \right)^{3/2} e^{- m_\chi / T_{\rm fo}} \,,
}
where we have left off the possible spin degrees of freedom factor for the DM number density. We may find an approximate solution to this equation by noting that the only rapidly varying part of the equation above is the exponential. So, let us take $T_{\rm fo} = m_\chi$ everywhere except in the exponential, which allows us to approximate 
\es{}{
x_{\rm fo} \equiv {m_\chi \over T_{\rm fo}} &\sim  \log\left( m_\chi m_{\rm pl} \langle \sigma v \rangle \right) - {3 \over 2} \log 2 \pi \\
&\approx 25 \, \qquad (m_\chi = 1 \, \, {\rm TeV}) \,.
}
Note that the dependence of $x_{\rm fo}$ on $m_\chi$ is only logarithmic.

\begin{figure}[htb]  
\begin{center}
\includegraphics[width=0.75\textwidth]{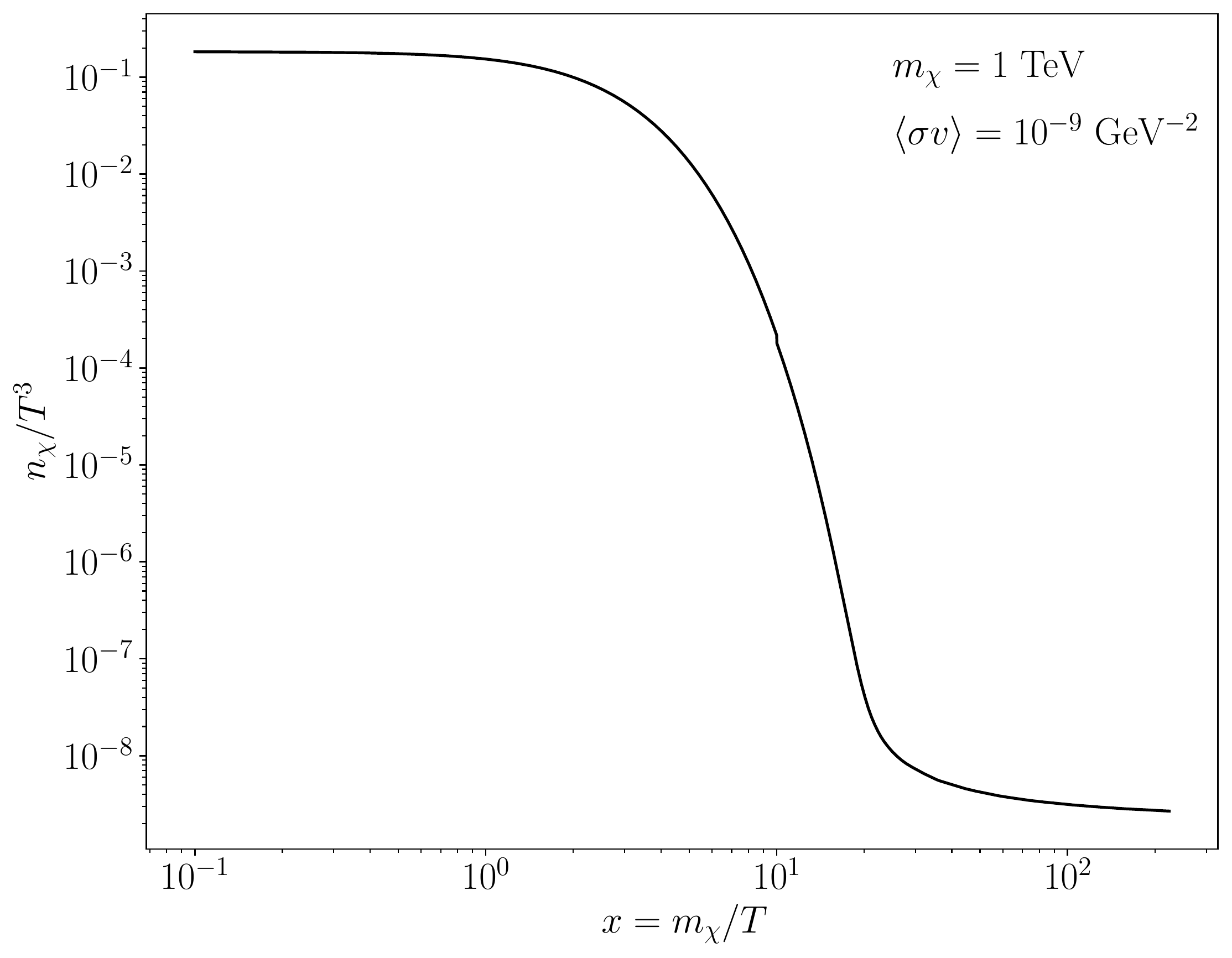}
\caption{The number density $n_\chi$ of a WIMP DM candidate during freeze-out for the indicated DM mass and thermal-averaged cross-section, shown as a function of $x = m_\chi /T$. This figure may be reproduced by solving the Boltzmann equation in~\eqref{eq:Boltz_1}, as illustrated in the supplementary Google Colab notebook. Note that when $\chi$ is in thermal equilibrium $n_\chi \propto T^3$, and this scaling is restored after the DM freezes out and simply dilutes with the expansion of the universe. For $x \sim 1$, the DM is exponentially depleted until the DM falls out of equilibrium. Larger values of the cross-section will keep the DM in equilibrium longer and thus deplete the DM abundance further.}
\label{fig:fo}
\end{center}
\end{figure}

While the above arguments are insightful and lead to a surprisingly accurate estimate of the correct freeze-out cross-section, to make more quantitative progress one should numerically solve the Boltzmann equation in~\eqref{eq:Boltz_1}. This is a non-linear ordinary differential equation.  In the Supplemental \texttt{jupyter} notebook provided through Colab we setup and solve this differential equation numerically.\footnote{
See \href{https://colab.research.google.com/drive/180GGT-iy83lQ4zizHgqXrpyQZe3oldD5?usp=sharing}{this Colab Jupyter Notebook.}
}  In Fig.~\ref{fig:fo} we show the output of the freeze-out calculation for a $m_\chi = 1$ TeV DM particle with $\langle \sigma v \rangle = 10^{-9}$ GeV$^{-2}$. As expected, the abundance is exponentially depleted for $x \gtrsim 1$ until the DM freezes out at $x_{\rm fo} \approx 25$.

One fascinating and powerful aspect of the WIMP DM paradigm is that, with a few caveats that we explain more below, if DM can annihilate in the early universe to set its abundance then it can still annihilate today, with a predictable rate. The annihilation rate in an infinitesimal volume of space at position ${\bm x}$, in units of number of annihilations per unit volume per unit time, will be $\Gamma_{\rm ann}({\bm x}) =( \rho_{\rm DM}({\bm x})^2 / m_\chi^2) \langle \sigma v \rangle$, with an additional symmetry factor of $1/2$ for identical annihilating particles.   Here, $\rho_{\rm DM}({\bm x})$ is the DM density today at position ${\bm x}$.  Naively, we may simply take $\langle \sigma v \rangle \approx 2 \times 10^{-26}$ cm$^3$/s and have a concrete prediction for $\Gamma_{\rm ann}$ given $m_\chi$.  There are a few reasons why this identification may be more complicated, however, which we partially enumerate below.  After describing these caveats we then discuss how to predict the DM annihilation signal in the Milky Way and nearby galaxies.

\subsubsection{Caveats to the standard WIMP annihilation prediction}

One key point to remember is that the relative velocity $v$ of the DM today is much smaller than the relative velocity of the DM at the epoch of freeze-out. For $v \ll 1$ we may write $\sigma v \sim v^{2 L}$, where $L = 0, 1, 2, \dots$ is the orbital angular momentum quantum number of the two initial annihilating states. While the s-wave ($L = 0$) annihilation typically dominates for this reason, there may be models where the s-wave cross-section vanishes and the first non-trivial annihilation cross-section occurs as p-wave ($L = 1$) or even higher $L$. (For an extended discussion of this point and models that realize p-wave annihilation, see~\cite{Kumar:2013iva}.)  Note that at freeze-out $v \sim 1/\sqrt{x_{\rm fo}} \sim 0.3$, while in the Milky Way today $v \sim 10^{-3}$ and in smaller dwarf galaxies $v \sim 10^{-4}$.  Thus, for p-wave annihilation models the annihilation rates are tens of thousands of times weaker today, at best, relative to the annihilation rates for models where s-wave annihilation is allowed.  Conversely, in some DM models the annihilation rate is enhanced at low velocities, a process known as Sommerfeld enhancement (see~\cite{Hisano:2004ds} for an extended discussion).  In the presence of a long-range force, the two annihilating DM particles may attract each other before annihilation, which enhances the cross-section by a factor $\sim$$\alpha_D / v$, with $\alpha_D$ the dark fine structure constant of the long-range force. This enhancement does not extend to arbitrarily small $v$; instead, the enhancement flattens when the relative momentum of the DM particles falls below the mediator mass, at which point the force may no longer be treated as long range.  Additional Sommerfeld resonances may also be possible, with further enhancements, depending on the mediator and DM masses. These resonances play a very important role in probing the wino DM model, which we will discuss in more detail shortly~\cite{Fan:2013faa,Cohen:2013ama}.

Coannihilations~\cite{Griest:1990kh} may also complicate the relation between the annihilation cross-section that sets the DM abundance and the DM annihilation cross-section today.  Coannihilations happen when there are multiple particles that freeze-out together with nearly degenerate masses, though the DM particle is the lightest among them.  After freeze-out the slightly heavier states decay to the DM state and perhaps small amounts of SM radiation. The DM abundance in this case is then set by the mutual evolution of the ensemble of particles. Coannihilations play an important role in determining the abundances of higgsinos and winos, which we discuss in more detail.

\subsection{Minimal dark matter models and the higgsino}

One of the most attractive aspects of the WIMP DM paradigm is the so-called ``WIMP miracle," which is the observation that the annihilation cross-section $\langle \sigma v \rangle \approx 2 \times 10^{-26}$ cm$^3$/s is naturally obtained for tree-level annihilations with electroweak scale interactions and electroweak scale masses. This begs the question of whether the DM can be in a representation of $SU(2)_L \times U(1)_Y$. The answer is yes, though of course the DM itself must be electrically neutral, and indeed this is the case in many supersymmetric realizations of WIMP DM.  Before discussing supersymmetry, however, we can simply talk about so-called ``minimal DM models"~\cite{Cirelli:2005uq}, which are the minimal models of WIMP DM where the DM is embedded into a representation of $SU(2)_L$. 

Let us first discuss the fermionic DM minimal model where the DM is in the fundamental representation of $SU(2)_L$. In some sense, this is the most minimal of the minimal models, since it is the lowest non-trivial representation, but also it turns out this model corresponds to the higgsino, which we argue is perhaps the most interesting WIMP DM model still standing.

Let us modify the Lagrangian of the SM to include a spin-$1/2$, Dirac fermion $\chi$ in the fundamental representation of $SU(2)_L$:
\es{}{
{\mathcal L} = {\mathcal L}_{\rm SM} + \bar \chi (i \slashed{D} + m_\chi) \chi \,,
}
where $D_\mu$ is the $SU(2)_L \times U(1)_Y$ covariant derivative, and $ {\mathcal L}_{\rm SM} $ is the SM Lagrangian.  Let us write the doublet as $\chi = ( \chi_+, \chi_0)^T$. Recall that the generator of electromagnetism is given by $Q = T_3 + Y$, where $T_3 = {1 \over 2} \sigma_3$, with $\sigma_3$ the third Pauli matrix, and $Y$ the identity matrix times the hypercharge $Y$ of $\chi$. We want an electrically neutral component of $\chi$ to be our DM candidate, so we chose $Y = 1/2$ such that $\chi_0$ is electrically neutral while $\chi_+$ has charge $+1$.

At tree level $\chi_+$ and $\chi_0$ have the same mass $m_\chi$.  This poses a problem, since if the two states are exactly degenerate then freeze-out would set a non-trivial abundance of $\chi_+$, which would be stable and have a relic density today. Such a DM candidate is strictly forbidden. Luckily, the electroweak symmetry is spontaneously broken, which means that $\chi_+$ and $\chi_0$ do not necessarily have the same mass at the quantum level. Indeed, at one-loop electroweak gauge bosons induce the mass splitting~\cite{Cirelli:2005uq} 
\es{}{
\Delta M \equiv m_+ - m_0 \approx \left( 1 + {2 Y \over \cos \theta_w} \right) \alpha_2 M_W \sin^2 {\theta_w \over 2} \approx 360 \, \, {\rm MeV} \,,
}
where $\theta_w$ is the Weinberg angle ($\sin^2\theta_w \approx 0.23$), $M_W\approx 80$ GeV is the $W$-boson mass, and $\alpha_2 = \alpha_{\rm EM} / \sin^2(\theta_w) \approx 0.034$ is the $SU(2)_L$ fine structure constant.  Thus, the heavier, charged state, with mass $m_+$, decays to charged SM fermions and the neutral, lighter state $\chi_0$, which is the DM candidate. However, one may verify that the lifetime of $\chi_+$ is much longer than the time-scale of freeze-out. Thus, for the purpose of the freeze-out calculations one may assume that the two Dirac states are degenerate, calculate the abundance of both $\chi_+$ and $\chi_0$, and then assume that the number density of $\chi_+$ is transferred to $\chi_0$ number density by subsequent decays of the charged particles.  Indeed, for many purposes we may simply ignore the precise value of the splitting $\Delta M$. 

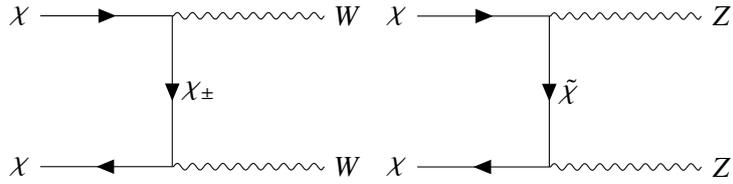
\begin{figure}[htb]  
\begin{center}
\begin{tikzpicture}
  \begin{feynman}
    \vertex (a){$\chi$};
    \vertex [right=2cm of a] (b);
    \vertex [below=2cm of b] (c);
    \vertex [below=2cm of a] (d){$\chi$};
    \vertex [right=2cm of b] (e){$W$};
    \vertex [right=2cm of c] (f){$W$};
    \diagram* {
      (a) -- [fermion] (b),
      (b) -- [fermion, edge label =\(\chi_\pm\)] (c),
      (c) -- [fermion] (d),
      (b) -- [photon] (e),
      (c) -- [photon] (f)
    };
  \end{feynman}
\end{tikzpicture}
\begin{tikzpicture}
  \begin{feynman}
    \vertex (a){$\chi$};
    \vertex [right=2cm of a] (b);
    \vertex [below=2cm of b] (c);
    \vertex [below=2cm of a] (d){$\chi$};
    \vertex [right=2cm of b] (e){$Z$};
    \vertex [right=2cm of c] (f){$Z$};
    \diagram* {
      (a) -- [fermion] (b),
      (b) -- [fermion, edge label =\(\tilde \chi\)] (c),
      (c) -- [fermion] (d),
      (b) -- [photon] (e),
      (c) -- [photon] (f)
    };
  \end{feynman}
\end{tikzpicture}
\end{center}
\caption{Annihilation channels for the higgsino DM candidate $\chi$, which is a neutral Majorana fermion, that are active both in the early universe and today for indirect detection.  Note that $\chi$ has a small mass splitting relative to the other neutral, Majorana higgsino state $\tilde \chi$. These neutral states in-turn are slightly lighter than the charged states $\chi_{\pm}$ that help make up the full $SU(2)_L$ multiplet for the higgsino.}
\label{fig:feyn_1}
\end{figure}

\begin{figure}
\begin{center}
\begin{tikzpicture}
  \begin{feynman}
    \vertex (a){$\chi$};
    \vertex [right=2cm of a] (b);
    \vertex [below=2cm of b] (c);
    \vertex [below=2cm of a] (d){$\chi_\pm$};
    \vertex [right=2cm of b] (e){$W$};
    \vertex [right=2cm of c] (f){$Z, \gamma$};
    \diagram* {
      (a) -- [fermion] (b),
      (b) -- [fermion, edge label =\(\chi_\pm\)] (c),
      (c) -- [fermion] (d),
      (b) -- [photon] (e),
      (c) -- [photon] (f)
    };
  \end{feynman}
\end{tikzpicture}
\begin{tikzpicture}
  \begin{feynman}
    \vertex (a){$\chi$};
    \vertex [right=2cm of a] (b);
    \vertex [below=2cm of b] (c);
    \vertex [below=2cm of a] (d){$\chi_\pm$};
    \vertex [right=2cm of b] (e){$Z$};
    \vertex [right=2cm of c] (f){$W$};
    \diagram* {
      (a) -- [fermion] (b),
      (b) -- [fermion, edge label =\(\tilde \chi\)] (c),
      (c) -- [fermion] (d),
      (b) -- [photon] (e),
      (c) -- [photon] (f)
    };
  \end{feynman}
\end{tikzpicture}%
\begin{tikzpicture}
  \begin{feynman}
    \vertex (a){$\chi$};
    \vertex [right=1.5cm of a] (b);
    \vertex [below=1cm of b] (c);
    \vertex [below=1cm of c] (d);
    \vertex [right=1cm of c] (e);
    \vertex [below=2cm of a] (f){$\chi_\pm$};
    \vertex [right=2.5cm of b] (g){$f$};
    \vertex [right=2.5cm of d] (h){$f$};
    \diagram* {
      (c) -- [fermion] (a),
      (c) -- [photon, edge label =\(W\)] (e),
      (f) -- [fermion] (c),
      (e) -- [fermion] (g),
      (h) -- [fermion] (e)
    };
  \end{feynman}
\end{tikzpicture}%
\begin{tikzpicture}
  \begin{feynman}
    \vertex (a){$\chi$};
    \vertex [right=1.5cm of a] (b);
    \vertex [below=1cm of b] (c);
    \vertex [below=1cm of c] (d);
    \vertex [right=1cm of c] (e);
    \vertex [below=2cm of a] (f){$\chi_\pm$};
    \vertex [right=2.5cm of b] (g){$W$};
    \vertex [right=2.5cm of d] (h){$H, Z, \gamma$};
    \diagram* {
      (c) -- [fermion] (a),
      (c) -- [photon, edge label =\(W\)] (e),
      (f) -- [fermion] (c),
      (e) -- [photon] (g),
      (h) -- [photon] (e)
    };
  \end{feynman}
\end{tikzpicture}
\end{center}
\caption{Co-annihilation channels involving $\chi$ and heavier, charged states $\chi_{\pm}$ only relevant in the early universe.}
\label{fig:feyn_3}
\end{figure}
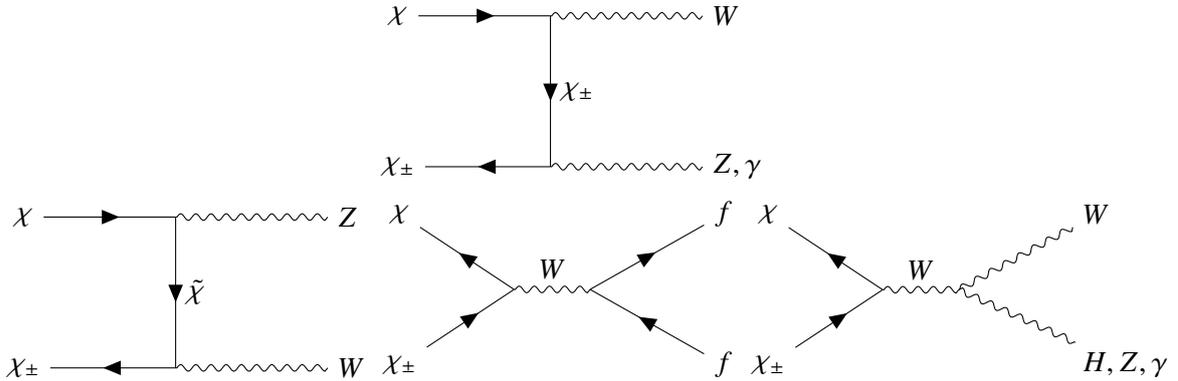

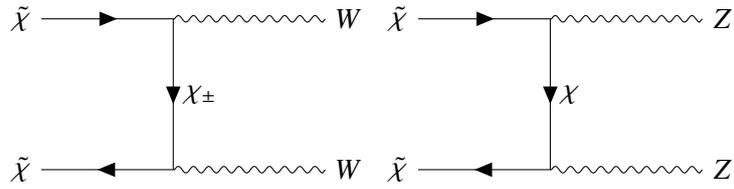
\begin{figure}
\begin{center}
\begin{tikzpicture}
  \begin{feynman}
    \vertex (a){$\tilde \chi$};
    \vertex [right=2cm of a] (b);
    \vertex [below=2cm of b] (c);
    \vertex [below=2cm of a] (d){$\tilde \chi$};
    \vertex [right=2cm of b] (e){$W$};
    \vertex [right=2cm of c] (f){$W$};
    \diagram* {
      (a) -- [fermion] (b),
      (b) -- [fermion, edge label =\(\chi_\pm\)] (c),
      (c) -- [fermion] (d),
      (b) -- [photon] (e),
      (c) -- [photon] (f)
    };
  \end{feynman}
\end{tikzpicture}
\begin{tikzpicture}
  \begin{feynman}
    \vertex (a){$\tilde \chi$};
    \vertex [right=2cm of a] (b);
    \vertex [below=2cm of b] (c);
    \vertex [below=2cm of a] (d){$\tilde \chi$};
    \vertex [right=2cm of b] (e){$Z$};
    \vertex [right=2cm of c] (f){$Z$};
    \diagram* {
      (a) -- [fermion] (b),
      (b) -- [fermion, edge label =\(\chi\)] (c),
      (c) -- [fermion] (d),
      (b) -- [photon] (e),
      (c) -- [photon] (f)
    };
  \end{feynman}
\end{tikzpicture}
\end{center}
\caption{Annihilation channels involving $\tilde \chi$ with itself.  These channels are relevant in the early universe but not today.}
\label{fig:feyn_4}
\end{figure}

\begin{figure}
\begin{center}
\begin{tikzpicture}
  \begin{feynman}
    \vertex (a){$\tilde \chi$};
    \vertex [right=2cm of a] (b);
    \vertex [below=2cm of b] (c);
    \vertex [below=2cm of a] (d){$\chi_\pm$};
    \vertex [right=2cm of b] (e){$W$};
    \vertex [right=2cm of c] (f){$Z, \gamma$};
    \diagram* {
      (a) -- [fermion] (b),
      (b) -- [fermion, edge label =\(\chi_\pm\)] (c),
      (c) -- [fermion] (d),
      (b) -- [photon] (e),
      (c) -- [photon] (f)
    };
  \end{feynman}
\end{tikzpicture}
\begin{tikzpicture}
  \begin{feynman}
    \vertex (a){$\tilde \chi$};
    \vertex [right=2cm of a] (b);
    \vertex [below=2cm of b] (c);
    \vertex [below=2cm of a] (d){$\chi_\pm$};
    \vertex [right=2cm of b] (e){$Z$};
    \vertex [right=2cm of c] (f){$W$};
    \diagram* {
      (a) -- [fermion] (b),
      (b) -- [fermion, edge label =\(\tilde \chi\)] (c),
      (c) -- [fermion] (d),
      (b) -- [photon] (e),
      (c) -- [photon] (f)
    };
  \end{feynman}
\end{tikzpicture}%
\begin{tikzpicture}
  \begin{feynman}
    \vertex (a){$\tilde \chi$};
    \vertex [right=1.5cm of a] (b);
    \vertex [below=1cm of b] (c);
    \vertex [below=1cm of c] (d);
    \vertex [right=1cm of c] (e);
    \vertex [below=2cm of a] (f){$\chi_\pm$};
    \vertex [right=2.5cm of b] (g){$f$};
    \vertex [right=2.5cm of d] (h){$f$};
    \diagram* {
      (c) -- [fermion] (a),
      (c) -- [photon, edge label =\(W\)] (e),
      (f) -- [fermion] (c),
      (e) -- [fermion] (g),
      (h) -- [fermion] (e)
    };
  \end{feynman}
\end{tikzpicture}%
\begin{tikzpicture}
  \begin{feynman}
    \vertex (a){$\tilde \chi$};
    \vertex [right=1.5cm of a] (b);
    \vertex [below=1cm of b] (c);
    \vertex [below=1cm of c] (d);
    \vertex [right=1cm of c] (e);
    \vertex [below=2cm of a] (f){$\chi_\pm$};
    \vertex [right=2.5cm of b] (g){$W$};
    \vertex [right=2.5cm of d] (h){$H, Z, \gamma$};
    \diagram* {
      (c) -- [fermion] (a),
      (c) -- [photon, edge label =\(W\)] (e),
      (f) -- [fermion] (c),
      (e) -- [photon] (g),
      (h) -- [photon] (e)
    };
  \end{feynman}
\end{tikzpicture}
\end{center}
\caption{Co-annihilation channels involving $\tilde \chi$ with $\chi_{\pm}$.  These channels are relevant in the early universe but not today.}
\label{fig:feyn_5}
\end{figure}
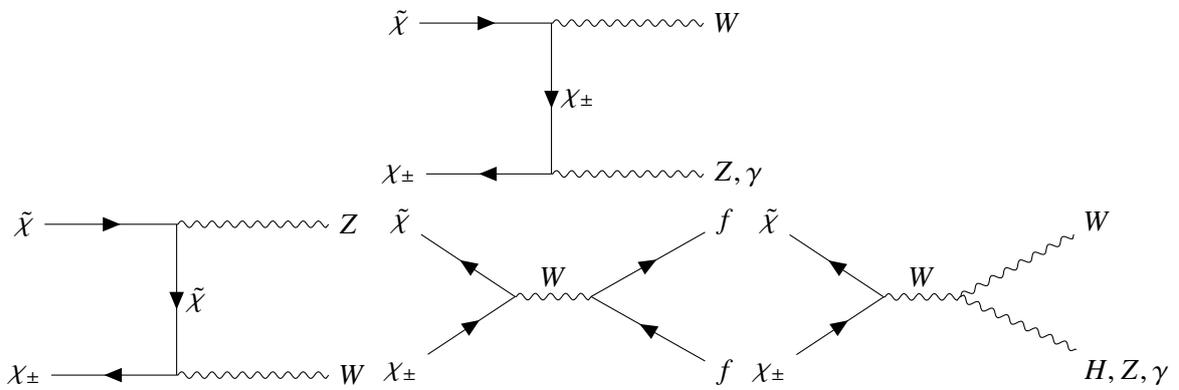

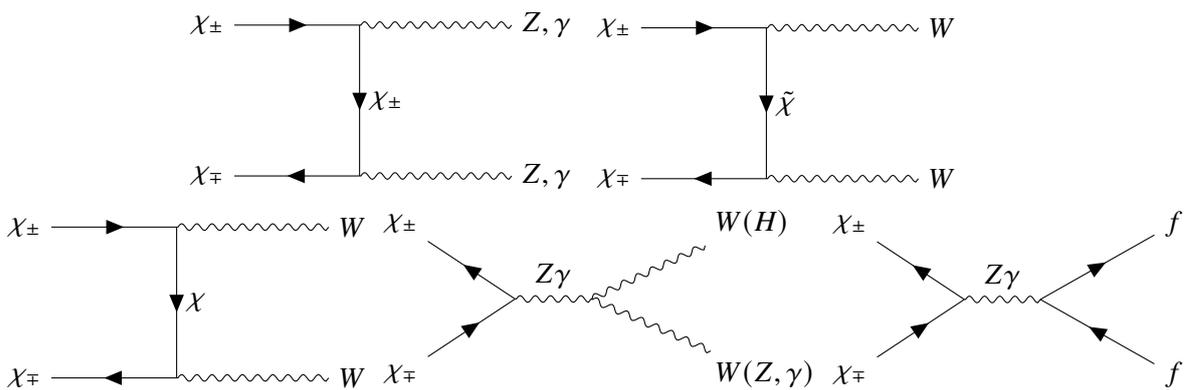
\begin{figure}
\begin{center}
\begin{tikzpicture}
  \begin{feynman}
    \vertex (a){$\chi_\pm$};
    \vertex [right=2cm of a] (b);
    \vertex [below=2cm of b] (c);
    \vertex [below=2cm of a] (d){$\chi_\mp$};
    \vertex [right=2cm of b] (e){$Z, \gamma$};
    \vertex [right=2cm of c] (f){$Z, \gamma$};
    \diagram* {
      (a) -- [fermion] (b),
      (b) -- [fermion, edge label =\(\chi_\pm\)] (c),
      (c) -- [fermion] (d),
      (b) -- [photon] (e),
      (c) -- [photon] (f)
    };
  \end{feynman}
\end{tikzpicture}
\begin{tikzpicture}
  \begin{feynman}
    \vertex (a){$\chi_\pm$};
    \vertex [right=2cm of a] (b);
    \vertex [below=2cm of b] (c);
    \vertex [below=2cm of a] (d){$\chi_\mp$};
    \vertex [right=2cm of b] (e){$W$};
    \vertex [right=2cm of c] (f){$W$};
    \diagram* {
      (a) -- [fermion] (b),
      (b) -- [fermion, edge label =\(\tilde \chi\)] (c),
      (c) -- [fermion] (d),
      (b) -- [photon] (e),
      (c) -- [photon] (f)
    };
  \end{feynman}
\end{tikzpicture}
\begin{tikzpicture}
  \begin{feynman}
    \vertex (a){$\chi_\pm$};
    \vertex [right=2cm of a] (b);
    \vertex [below=2cm of b] (c);
    \vertex [below=2cm of a] (d){$\chi_\mp$};
    \vertex [right=2cm of b] (e){$W$};
    \vertex [right=2cm of c] (f){$W$};
    \diagram* {
      (a) -- [fermion] (b),
      (b) -- [fermion, edge label =\(\chi\)] (c),
      (c) -- [fermion] (d),
      (b) -- [photon] (e),
      (c) -- [photon] (f)
    };
  \end{feynman}
\end{tikzpicture}%
\begin{tikzpicture}
  \begin{feynman}
    \vertex (a){$\chi_\pm$};
    \vertex [right=1.5cm of a] (b);
    \vertex [below=1cm of b] (c);
    \vertex [below=1cm of c] (d);
    \vertex [right=1cm of c] (e);
    \vertex [below=2cm of a] (f){$\chi_\mp$};
    \vertex [right=2.5cm of b] (g){$W (H)$};
    \vertex [right=2.5cm of d] (h){$W (Z,\gamma)$};
    \diagram* {
      (c) -- [fermion] (a),
      (c) -- [photon, edge label =\(Z \gamma\)] (e),
      (f) -- [fermion] (c),
      (e) -- [photon] (g),
      (e) -- [photon] (h)
    };
  \end{feynman}
\end{tikzpicture}%
\begin{tikzpicture}
  \begin{feynman}
    \vertex (a){$\chi_\pm$};
    \vertex [right=1.5cm of a] (b);
    \vertex [below=1cm of b] (c);
    \vertex [below=1cm of c] (d);
    \vertex [right=1cm of c] (e);
    \vertex [below=2cm of a] (f){$\chi_\mp$};
    \vertex [right=2.5cm of b] (g){$f$};
    \vertex [right=2.5cm of d] (h){$f$};
    \diagram* {
      (c) -- [fermion] (a),
      (c) -- [photon, edge label =\(Z \gamma\)] (e),
      (f) -- [fermion] (c),
      (e) -- [fermion] (g),
      (h) -- [fermion] (e)
    };
  \end{feynman}
\end{tikzpicture}
\end{center}
\caption{Annihilation channels involving $\chi_{+}$-$\chi_{-}$.  These channels are relevant in the early universe but not today.}
\label{fig:feyn_6}
\end{figure}

\begin{figure}
\begin{center}
\begin{tikzpicture}
  \begin{feynman}
    \vertex (a){$\chi_\pm$};
    \vertex [right=2cm of a] (b);
    \vertex [below=2cm of b] (c);
    \vertex [below=2cm of a] (d){$\chi_\pm$};
    \vertex [right=2cm of b] (e){$W$};
    \vertex [right=2cm of c] (f){$W$};
    \diagram* {
      (a) -- [fermion] (b),
      (b) -- [majorana, edge label =\(\tilde \chi\)] (c),
      (d) -- [fermion] (c),
      (b) -- [photon] (e),
      (c) -- [photon] (f)
    };
  \end{feynman}
\end{tikzpicture}
\begin{tikzpicture}
  \begin{feynman}
    \vertex (a){$\chi_\pm$};
    \vertex [right=2cm of a] (b);
    \vertex [below=2cm of b] (c);
    \vertex [below=2cm of a] (d){$\chi_\pm$};
    \vertex [right=2cm of b] (e){$W$};
    \vertex [right=2cm of c] (f){$W$};
    \diagram* {
      (a) -- [fermion] (b),
      (b) -- [majorana, edge label =\(\chi\)] (c),
      (d) -- [fermion] (c),
      (b) -- [photon] (e),
      (c) -- [photon] (f)
    };
  \end{feynman}
\end{tikzpicture}
\end{center}
\caption{Annihilation channels involving $\chi_{+}$-$\chi_{+}$ and $\chi_{-}$-$\chi_{-}$.  These channels are relevant in the early universe but not today.}
\label{fig:higgsino_ann}
\end{figure}
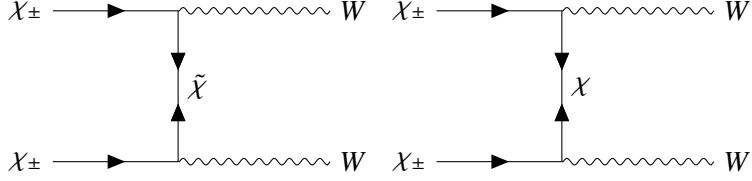

We will calculate below that to achieve the correct DM abundance the fermion $\chi$ should have a mass $\sim$1 TeV.\footnote{Since $m_\chi \gg v_{\rm ew}$, with $v_{\rm ew}$ the Higgs vacuum expectation value (VEV), it can be useful to make the approximation that $SU(2)_L$ is unbroken ({\it i.e.}, take $v_{\rm ew} \to 0$).}  
With that in mind, the tree-level Feynman diagrams that contribute to freeze-out, including co-annihilations ({\it i.e.}, annihilations between $\chi_+$ and $\chi_0$) are given in Figs.~\ref{fig:feyn_1},~\ref{fig:feyn_2},~\ref{fig:feyn_3},~\ref{fig:feyn_4},~\ref{fig:feyn_5},~\ref{fig:feyn_6}, and~\ref{fig:higgsino_ann}.  Note that we do not show diagrams which are related to those shown by $t$/$u$ channel exchanges.

\begin{figure}
\begin{center}
\begin{tikzpicture}
  \begin{feynman}
    \vertex (a){$\chi$};
    \vertex [right=2cm of a] (b);
    \vertex [below=2cm of b] (c);
    \vertex [below=2cm of a] (d){$\tilde\chi$};
    \vertex [right=2cm of b] (e){$W$};
    \vertex [right=2cm of c] (f){$W$};
    \diagram* {
      (a) -- [fermion] (b),
      (b) -- [fermion, edge label =\(\chi_\pm\)] (c),
      (c) -- [fermion] (d),
      (b) -- [photon] (e),
      (c) -- [photon] (f)
    };
  \end{feynman}
\end{tikzpicture}%
\begin{tikzpicture}
  \begin{feynman}
    \vertex (a){$\chi$};
    \vertex [right=1.5cm of a] (b);
    \vertex [below=1cm of b] (c);
    \vertex [below=1cm of c] (d);
    \vertex [right=1cm of c] (e);
    \vertex [below=2cm of a] (f){$\tilde\chi$};
    \vertex [right=2.5cm of b] (g){$W (H)$};
    \vertex [right=2.5cm of d] (h){$W (Z,\gamma)$};
    \diagram* {
      (c) -- [fermion] (a),
      (c) -- [photon, edge label =\(Z\)] (e),
      (f) -- [fermion] (c),
      (e) -- [photon] (g),
      (e) -- [photon] (h)
    };
  \end{feynman}
\end{tikzpicture}%
\begin{tikzpicture}
  \begin{feynman}
    \vertex (a){$\chi$};
    \vertex [right=1.5cm of a] (b);
    \vertex [below=1cm of b] (c);
    \vertex [below=1cm of c] (d);
    \vertex [right=1cm of c] (e);
    \vertex [below=2cm of a] (f){$\tilde\chi$};
    \vertex [right=2.5cm of b] (g){$f$};
    \vertex [right=2.5cm of d] (h){$f$};
    \diagram* {
      (c) -- [fermion] (a),
      (c) -- [photon, edge label =\(Z\)] (e),
      (f) -- [fermion] (c),
      (e) -- [fermion] (g),
      (h) -- [fermion] (e)
    };
  \end{feynman}
\end{tikzpicture}
\end{center}
\caption{Co-annihilation channels involving the DM candidate $\chi$ and the slightly heavier, neutral Majorana state $\tilde \chi$.  These channels are only relevant in the early universe. }
\label{fig:feyn_2}
\end{figure}
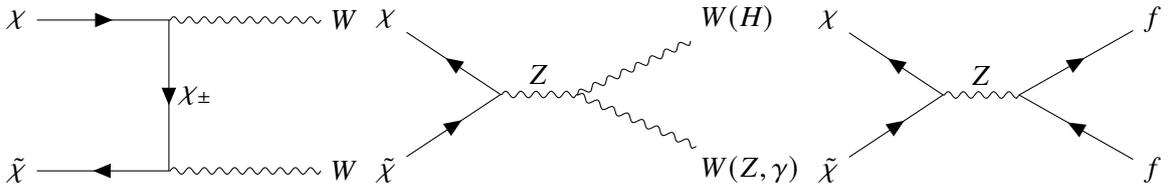

As we discuss further below, the neutral Dirac state $\chi_0$ must be split into two neutral Majorana states $\chi$ (the lighter, DM candidate) and $\tilde \chi$, which is slightly heavier. (Note that the Majorana $\chi$ should not be confused with our notation for the full Dirac $SU(2)_L$ multiplet $\chi$.)
The $f$ in the Feynman diagrams stand for generic SM fermions that are charged under the electroweak force, and all appropriate final state fermions should be summed over.  To include co-annihilations, we may assign the annihilating $\chi-\bar \chi$ pair $SU(2)_L$ indices $i$ and $j$, respectively, such that the annihilation cross-section is $\langle \sigma_{ij} v \rangle$~\cite{Griest:1990kh}. (Note that we are interested in the s-wave cross-section, obtained in the limit $v \to 0$.) Then, the relevant total annihilation cross-section, averaged over all channels, is $\langle \sigma v \rangle = {1 \over 16} \sum_{ij} \langle \sigma_{ij} v \rangle$. Additionally, we should average over the initial fermion spins and sum over final spins, polarizations, and gauge quantum numbers.  We also need to correctly account for the degrees of freedom in our Dirac $SU(2)_L$ doublet higgsino.  Full details of this calculation may be found in App.~\ref{app:xsec}.  The final annihilation cross-section evaluates to be~\cite{Cirelli:2005uq,Arkani-Hamed:2006wnf}
\es{eq:xsec_higgsino}{
    \langle  \sigma v \rangle  =  \sum_{i,j \in  \chi , \tilde \chi, \chi_+, \chi_-}  \frac{\langle \sigma v \rangle_{ij} }{16} = \frac{g^4}{512 \pi m_\chi^2} \left( 21 + 3 \tan^2 \theta_w + 11 \tan^4 \theta_w \right) \,. 
}
Let us now substitute this expression into~\eqref{eq:sigma_v_thermal} and solve for $m_\chi$. We find that the correct DM abundance is achieved for $m_\chi \sim 1$ TeV.  Refining this calculation yields $m_\chi = 1.08 \pm 0.02$ TeV~\cite{Bottaro:2022one}.

There is a fatal issue with the higgsino model described above (without a mass splitting between $\chi$ and $\tilde \chi$) that rules it out, in its minimal form, as a DM candidate: direct detection constraints exclude it by orders of magnitude.  Since the higgsino is charged under $U(1)_Y$, it has a large, spin-independent cross-section for scattering off of a nucleon of the order ${\mathcal \sigma}_{\rm SI} \sim G_F^2 M_{n}^2 \sim 10^{-40}$ cm$^2$, with $G_F$ the Fermi constant and $M_N$ the nucleon mass.  
 At $m_\chi \sim 1$ TeV the current constraints from the LZ experiment, with a Xe target, exclude spin-independent cross-sections larger than $\sim$$3 \times 10^{-46}$ cm$^2$.  Our model is excluded by more than 5 orders of magnitude!  The $Z$-exchange tree-level scattering process can be kinematically shut off, however, by inducing a small mass splitting between the two Majorana states that make up the Dirac fermion $\chi_0$.  To illustrate this point, consider a Dirac fermion $\chi_0$ that we write in the Weyl basis as $\chi = (\eta \, \, \bar \xi)^T$, where both $\eta$ and $\bar \xi$ are two-component Weyl spinors~\cite{Tucker-Smith:2001myb}. The normal Dirac mass term reads ${\mathcal L}_M = m_\chi \bar \chi_0 \chi_0 = m_\chi ( \xi \eta + {\rm h.c.})$. Now, suppose that we add in a Majorana mass term for $\eta$ only: ${\mathcal L}_\delta ={ \delta \over 2} \eta \eta + \, \, {\rm h.c.}$, where $\delta \ll m_\chi$. Then, diagonalizing the full mass matrix leads to two mass eigenstates: $\tilde \chi \approx {i \over \sqrt{2}} (\eta + \xi)$ and $\chi \approx {i \over \sqrt{2}} (\eta - \xi)$ (again, this Majorana $\chi$ should not be confused with our notation that also refers to the full $SU(2)$ Dirac multiplet as $\chi$). The $\chi$ has mass $m_- = m - \delta$, while the $\tilde \chi$ has mass $m_+ = m+\delta$. In the context of DM, the $\chi$ Majorana state becomes the DM, with the $\tilde \chi$ decaying to $\chi$ plus a small amount of SM radiation. Now consider the vector-like interaction that is used in $Z$-exchange: $\bar \chi_0 \gamma_\mu \chi_0 \approx i {\bar {\tilde \chi}} \bar \sigma_\mu \chi + \, \, {\rm h.c.}$. That is, the $Z$-exchange DM direct detection process is {\it inelastic} because the lighter $\chi$ state must up-scatter to the $\tilde \chi$ state.  The energy gap is $2 \delta$, while the energy available in the scattering process is $\sim$${1 \over 2} m_\chi v^2$, with $v \sim 300$ km/s.  Thus, we need $\delta \gtrsim 200$ keV in order to effectively shut-off the inelastic $Z$-exchange process.

In the context of supersymmetry the mass splitting $\delta$ may naturally arise from higher-dimension operators that effectively induce a small mixing between the higgsino states and heavier gaugino states. In order to understand this, we should first understand a bit more about how higgsinos arise in supersymmetry.  There are many excellent reviews of supersymmetry and the minimal supersymmetric SM (MSSM) (see,~{\it e.g.},~\cite{Martin:1997ns}); we will not review supersymmetry in any non-trivial detail here, but we simply remark on a few important points that are relevant to this discussion. First, note that the MSSM requires two Higgs doublets $H_u$ and $H_d$.  Both $H_u$ and $H_d$ are in the fundamental representation of $SU(2)_L$ but $H_u$ has $Y = {1 \over 2}$ while $H_d$ has $Y = -{1 \over 2}$. There are deep reasons why supersymmetry needs two Higgs fields. However, one mundane reason why we can immediately see that two Higgs fields are necessary is related to anomaly cancellation.  If we were to add a single, new Weyl fermion as the superpartner of the SM Higgs, then the SM would no longer be anomaly free. However, in the MSSM we have two new Weyl fermions $\tilde H_u$ and $\tilde H_d$ as superpartners of $H_u$ and $H_d$, and since these have opposite hypercharge they do not induce any additional anomaly contributions.  The higgsinos have a Dirac mass term, ${\mathcal L} \supset - \mu \tilde H_u \tilde H_d + \, \, {\rm h.c.}$, with all gauge and spacetime indices suppressed, which allows us to package $\tilde H_u$ and $\tilde H_d$ into a Dirac fermion, which we have been referring to as $\chi$.  At dimension five, there are two operators that can induce Majorana mass terms for $H_u$ and $H_d$ separately, which go as $\sim$$(H^\dagger)^2 \tilde H_u \tilde H_u$ and $\sim$$H^2 \tilde H_d \tilde H_d$, where indices are suppressed and $H$ is the SM Higgs field.  These operators are induced when integrating out the heavier bino and wino particles, which are the superpartners of the $U(1)_Y$ and $SU(2)_L$ gauge bosons, respectively. For example, suppose that the bino $\tilde B$ has the mass term ${\mathcal L} \supset - {M_b \over 2} \tilde B \tilde B + \, \, {\rm h.c.}$, with $M_b \gg m_\chi$. Since the Higgs is charged under $U(1)_Y$, the bino also has Yukawa interactions with the Higgs sector of the form ${\mathcal L} \sim H^\dagger \tilde H_u \tilde B + H \tilde H_d \tilde B$, where we have left off the dimensionless coefficients that  depend on {\it e.g.} the $U(1)_Y$ gauge coupling. Integrating out $\tilde B$ (the same is also true for integrating out the wino) generates the dimension-5 operators discussed above, suppressed by the scale $M_b$. Thus, we expect a mass splitting between the two Majorana higgsino states $\delta \sim {v_{\rm EW}^2 \over M_b}$, with here $M_b$ really indicating the smallest of the bino or wino masses.  Thus, as long as $M_b$ is less than roughly $10^8$ GeV, which is to be expected in models such as split SUSY and mini-Split~\cite{Wells:2003tf,Giudice:2004tc,Arkani-Hamed:2004ymt,Arvanitaki:2012ps} that raise the SUSY breaking scale and scalar superpartner masses above the electroweak scale while keeping fermion superpartners closer to the electroweak scale, the mass-splitting between the two mostly-higgsino Majorana eigenstates is sufficiently large to suppress the $Z$-mediated diagrams and therefore evade current direct detection constraints.~\cite{Nagata:2014wma}.
With the $Z$-exchange channel cut-off, direct detection must proceed through either higher-order loop processes or through the higher-dimensional operators that induce the Majorana mass splittings. Unless the scale of the higher-dimensional operators is low, both the spin dependent and spin independent direct detection rates are expected to be below the neutrino floor~\cite{Bottaro:2022one}.

As discovering a WIMP with a direct detection cross-section below the neutrino floor is remarkably challenging, indirect detection emerges as a promising discovery channel.
Other ways the higgsino could be detected, however, include from direct production at colliders or indirectly through electric dipole moments (see, {\it e.g.},~\cite{Co:2021ion,Bottaro:2022one}).  The indirect signatures of higgsino DM annihilation are straightforward to compute. The tree-level annihilation diagrams shown in Fig.~\ref{fig:feyn_1} are active today for DM annihilation in DM-rich environments like the GC of the Milky Way.  Furthermore, and this is a generic expectation for annihilating and decaying DM models~\cite{Foster:2022nva}, there are one-loop annihilation diagrams that give $\gamma \gamma$ and $\gamma Z$ final states. These final states, while loop suppressed, are especially useful because they lead to monochromatic gamma-ray signatures. Such signatures are more straightforward to search for than the continuum gamma-ray signatures that arise from DM annihilation to non-gamma-ray final states, like $W^+ W^-$, which then have to shower down to lower-energy, stable particles, creating a broad spectrum of gamma-rays in the process. In the following section we discuss the indirect signatures of DM annihilation in more detail.  

As a side note, let us suppose that instead of being in the fundamental representation of $SU(2)_L$ the DM is in the adjoint representation. In this representation the $SU(2)_L$ generator $T_3$  can be written as $T_3 = {\rm diag}(1,0,-1)$, such that if we write $\chi = (\chi_+, \chi_0, \chi_{-})$ and take $Y = 0$, then $\chi_{\pm}$ have charges $\pm 1$, respectively, while $\chi_0$ is a charge-neutral DM candidate. (A DM candidate is also found if $Y = 1$, though we will not discuss this case further.)  This particle has the exact same quantum numbers as the $W$-boson, and indeed it can be identified with the wino in the context of the MSSM.  As in the doublet case, the charged $\chi_{\pm}$ are slightly more massive than $\chi_0$ due to radiative corrections.  In this case, the annihilation cross-section is slightly larger than it was for the doublet, which naively increases the thermal DM mass to $\sim$2.4 TeV. However, at higher DM masses, approaching 3 TeV, Sommerfeld enhancement has an important effect on the DM relic abundance. Accounting for non-perturbative Sommerfeld enhancement increases the DM mass that produces the correct relic abundance for the adjoint model to $m_\chi \sim 3$ TeV (see, {\it e.g.},~\cite{Beneke:2016ync}).  Unlike the thermal higgsino, the thermal wino is (essentially) ruled out as a DM candidate by indirect detection~\cite{Cohen:2013ama,Fan:2013faa}. The reason that the wino is in strong tension with data while the higgsino is not is that the wino annihilation cross-section receives a large Sommerfeld enhancement due to its larger mass, while the higgsino annihilation cross-section today does not receive such an enhancement.  If the Milky Way  has a reasonably large DM core, however, the wino may still be viable.  

\section{Indirect detection of annihilating and decaying DM with high-energy photons}
\label{sec:indirect}

In the previous section we discussed how the thermal freeze-out of WIMP DM provides motivation for considering the possibility of DM continuing to annihilate today, with an annihilation cross-section -- up to a few caveats that were discussed -- of $\langle \sigma v \rangle \sim 2 \times 10^{-26}$ cm$^3/$s. As was mentioned in the case of the higgsino, generically the tree-level annihilation products are to non-gamma-ray final states, since the DM is electrically neutral, while monochromatic $\gamma\gamma$ and $\gamma Z$ final states can arise at one loop. In this section we discuss how to predict the gamma-ray rates and detection prospects for such signatures with space-based and ground-based gamma-ray detectors.
  
  In other DM models, such as sterile neutrino DM, the DM does not annihilate but rather can decay into SM final states with a decay rate much longer than the age of the Universe.  Just as in the case of DM annihilations, the decays can produce both continuum and monochromatic photon final states.  Though the microphysics is different between annihilation and decay, the astrophysical and observational aspects of the two cases are similar, and so we describe them together.  
  
  DM annihilations and decays may create any stable particle in the final state. In these lecture notes we focus on final-state photons because, unlike charged cosmic rays like {\it e.g.} electrons, positrons, (anti-)protons, amongst other possibilities, the photons propagate along straight lines. In contrast, charged cosmic rays diffuse through the Galactic magnetic fields, meaning that there is less directional information in the signal that can be used to discriminate a putative DM signature from other backgrounds. Still, charged cosmic rays are powerful probes of annihilating and decaying DM and are certainly not worth discounting (see, {\it e.g.},~\cite{AMS:2016oqu,Heisig:2020nse,Cholis:2020twh,Calore:2022stf}). A full description of such signatures, on the other hand, is outside of the scope of these notes.  Neutrinos, on the other hand, behave similarly to photons in that they propagate un-obstructed.  In fact, in many ways neutrinos are simpler than photons: at very high energies photon traveling over extragalactic distances may become attenuated while neutrinos are not. However, it is much easier to detect photons than it is neutrinos, so searches for DM annihilation or decay using photon telescopes are almost always more powerful than searches using neutrino telescopes (see~\cite{Cohen:2016uyg} for an extended discussion of this point in the context of IceCube neutrinos).  Thus, we will also not discuss neutrino signatures in any detail here, though the formalism we develop carries over in a straightforward way to neutrino final states. 
  
  DM annihilation and decay may also be constrained through CMB observables (see the TASI lectures~\cite{Slatyer:2017sev} and references therein for an introduction to this topic).  The basic idea is that energy injected by DM annihilation and decay in the early universe can affect the CMB power spectrum. The CMB probes have the advantage of being robust and relatively model independent, given that the primordial plasma acts as a calorimeter that absorbs (nearly) all of the energy injected by DM annihilation or decay, with little dependence on the final state.  On the other hand, for the models we are primarily interested in (such as those to explain the GCE or heavy, neutralino DM like higgsinos), the gamma-ray probes are stronger, as so we concentrate on those here.  
  
  \subsection{DM annihilation and decay formalism}

Let us now turn to the calculation of the photon flux on Earth from an annihilating DM signal. The case of decaying DM is similar, and we will briefly discuss it afterwards.  Additionally -- motivated by the higgsino DM scenario -- we suppose that the DM is a Majorana fermion $\chi$ to be concrete, so that the DM annihilates with itself. Then, at a given point in space ${\bm x}$ the DM annihilation rate per unit volume is given by $\Gamma_{\rm ann}({\bm x}) = {1 \over 2} n_{\rm DM}^2 \langle \sigma v \rangle$, with $n_{\rm DM} = \rho_{\rm DM}({\bm x}) / m_\chi$.  Clearly, we want to look at DM-rich environments in order to maximize the annihilation rate. This will lead us consider the center of galaxies, and -- in particular -- the center of the Milky Way.  The expression for $\Gamma_{\rm ann}({\bm x})$ may be understood by first considering $n_{\rm DM} \langle \sigma v \rangle$ as the rate (units of 1/s) for a given DM particle to annihilate with another in its vicinity, with the extra factor of $n_{\rm DM}$ then accounting for the number density of DM particles that are undergoing annihilation. The factor of $1/2$ is a symmetry factor to account for identical particles.  

A given DM annihilation event will deposit $2 m_\chi$ of energy into a variety of SM final states. Let us denote $dN_\gamma / dE_\gamma$ as the expected distribution of gamma-rays, per energy $E_\gamma$, produced in a single annihilation event. For example, if $\chi \chi \to \gamma \gamma$, then $dN_\gamma / dE_\gamma = 2 \delta(E_\gamma - m_\chi)$.

The quantity that we are most interested in computing is the photon flux at the detector 
\es{}{
\Phi(E_\gamma, \psi)  = {d N_\gamma^{\rm det} \over dE_\gamma dt d \Omega dA} \,,
}
which is the number of photons $N_\gamma^{\rm det}$ that pass through the detector per unit time $dt$, unit energy $dE_\gamma$, solid angle $d\Omega$, and unit detector area $dA$. Here, $\psi$ is short-hand notation for the set of angular coordinates that tell us where on the sky we are looking. Let us imagine that for a given choice of $\psi$ the DM density falls off with the distance from the detector $r$ as $\rho_{\rm DM}(r)$.  Consider an infinitesimal volume element $d^3x$ at distance $r$. The differential flux $d \Phi$ from this volume element (units of {\it e.g.} cts/GeV/s/cm$^2$) through the detector is given by  
\es{}{
d\Phi  =   \Gamma_{\rm ann} {dN_\gamma \over d E_\gamma} {1 \over 4 \pi r^2} d^3 x \,,
}
where the $1 / (4 \pi r^2)$ factor accounts for the attenuation of the flux with distance.  Said another way, $dA / (4 \pi r^2)$ is the fraction of the emitted photons that cross the detector with area $dA$.  We may write $d^3x = d \Omega r^2 dr$, divide through by $d \Omega$, and identify $\Phi = \int dr \, (d \Phi / d \Omega)$ to derive 
\es{eq:J}{
\Phi(E_\gamma, \psi)  = {\langle \sigma v \rangle \over 8 \pi m_\chi^2} {d N_\gamma \over d E_\gamma} {\mathcal J}(\psi) \,, \qquad {\mathcal J}(\psi) = \int dr \rho_{\rm DM}^2(r) \,,
}
where $ {\mathcal J}(\psi)$ is known as the ${\mathcal J}$-factor, which is independent from the particle physics pre-factors.  Note that $\Phi$ has units of cts/GeV/s/cm$^2$/sr.  Performing the same calculation for DM decay, with DM lifetime $\tau_\chi$, leads to the result
\es{eq:D}{
\Phi(E_\gamma, \psi)  = {1 \over 4 \pi m_\chi \tau_\chi} {d N_\gamma \over d E_\gamma} {\mathcal D}(\psi) \,, \qquad {\mathcal D}(\psi) = \int dr \rho_{\rm DM}(r) \,,
}
where the denominator has a $4 \pi$ instead of an $8 \pi$ because there is no symmetry factor associated with the decay of identical particles in this case. The quantity ${\mathcal D}$ is known as the ${\mathcal D}$-factor. 

Note that some references define the ${\mathcal J}$- and ${\mathcal D}$-factors integrated over $d \Omega$. The way we have defined them is more convenient when looking at signals whose angular extent is large, such as signals from the Milky Way's halo, while the alternate form can be more convenient when studying sources, such as dwarf galaxies, whose angular extent is small.

\begin{figure}[htb]  
\begin{center}
\includegraphics[width=0.49\textwidth]{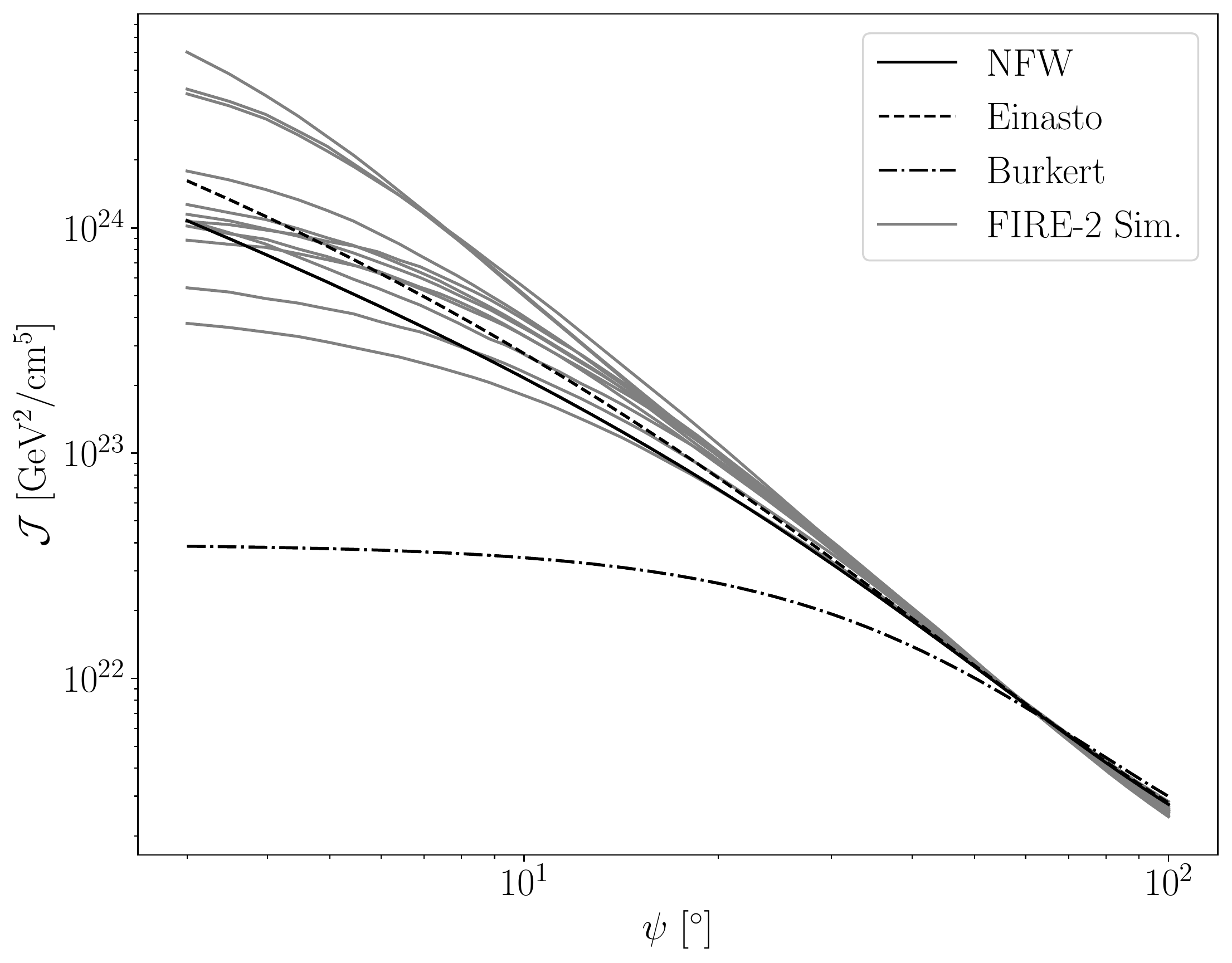}
\includegraphics[width=0.49\textwidth]{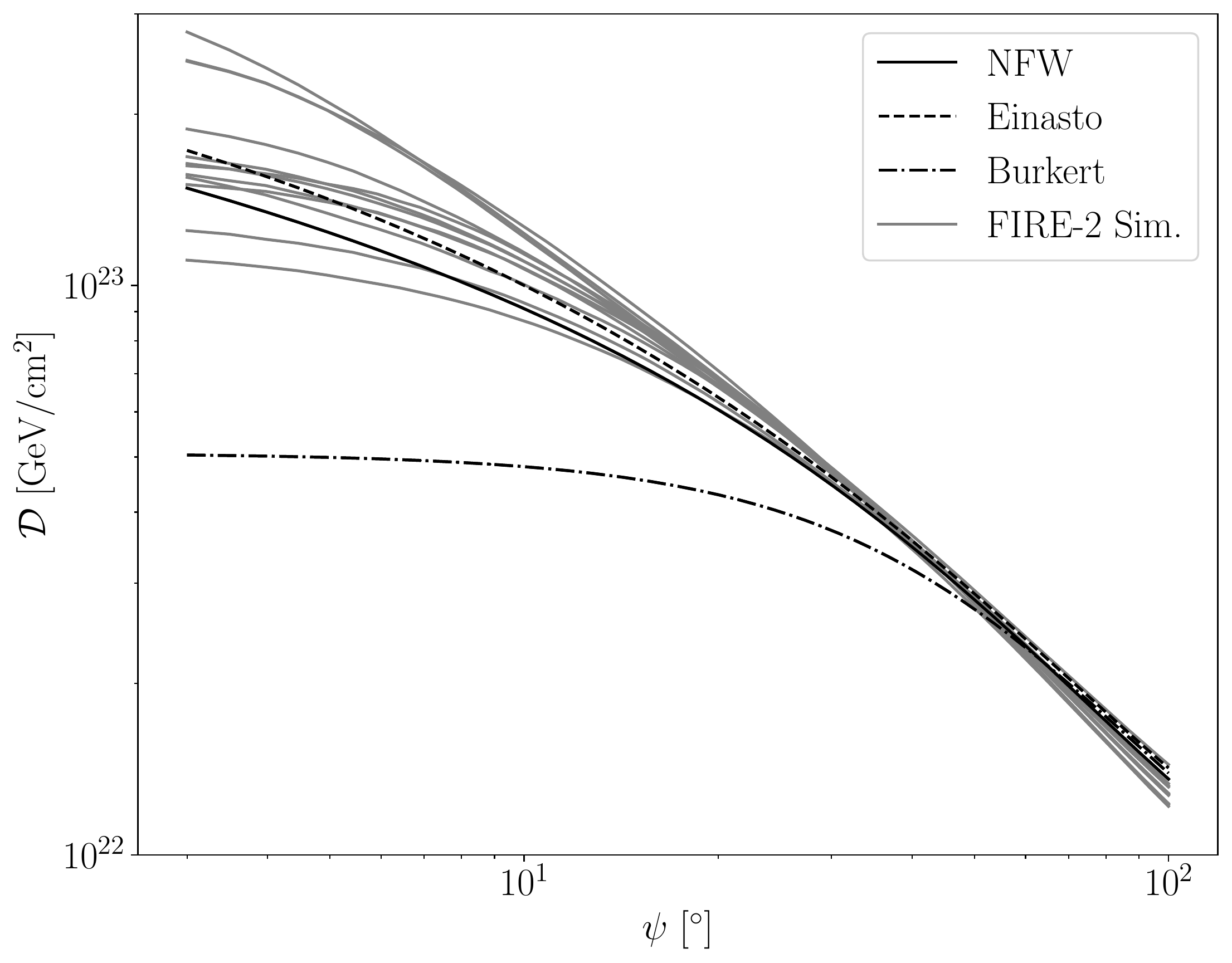}
\caption{The ${\mathcal J}$ and ${\mathcal D}$ factors calculated from the spherically-symmetric DM density profiles in Fig.~\ref{fig:DM}.  The  ${\mathcal J}$ factors have a larger spread because they are proportional to the line-of-sight integral of the DM density squared.  These figures may be reproduced using
\href{https://colab.research.google.com/drive/1MZ6ZeyGYmXlx1pXpeh3NrSani54_zBhr?usp=sharing}{this Colab Jupyter Notebook.}
}
\label{fig:JD}
\end{center}
\end{figure}

Let us now consider a few examples of ${\mathcal J}$- and ${\mathcal D}$-factors.  In Sec.~\ref{sec:DM_MW} we discussed a number of different DM profiles for the MW, some of which are parametric models motivated by $N$-body simulations, such as NFW and Einasto, and others which arise directly and non-parametrically from simulations, like the FIRE-2 profiles. In Fig.~\ref{fig:JD} we use~\eqref{eq:J} and~\eqref{eq:D} to compute the ${\mathcal J}$- and ${\mathcal D}$-factors for these different DM profiles, which are themselves illustrated in Fig.~\ref{fig:DM}.  As in Fig.~\ref{fig:DM}, all of the DM profiles are assumed to be spherically symmetric and are normalized to a common local DM density. The Einasto profile tends to predict the largest ${\mathcal J}$- and ${\mathcal D}$-factors of the parametric models, with Burkert significantly below.  

When discussing MW subhalos, such as dwarf galaxies, and extragalactic objects, it is sometimes convenient to consider the integral of the ${\mathcal J}$ and ${\mathcal D}$ factors over some angular region. This would be the case if one is performing a pure spectral search, where the data is summed spatially within some region of interest (ROI).  Although this notation is not standard, we will define the quantities 
\es{eq:JD_bar}{
\bar { \mathcal J}(\Omega) = {1 \over \Delta \Omega} \int_{\Delta \Omega} d \Omega' \int dr \rho_{\rm DM}^2(r)  \,, \qquad \bar {\mathcal D}(\Omega) ={1 \over \Delta \Omega'} \int_{\Delta \Omega} d \Omega'\int dr \rho_{\rm DM}(r) \,,
}
to be the angular integrals of the ${\mathcal J}$ and ${\mathcal D}$ factors over an ROI $\Omega$, with solid angle $\Delta \Omega$, then divided by $\Delta \Omega$.  That this, these are the average quantities over the ROI. As an illustration, let us fix $\Omega$ to be circular regions of radius $0.5^\circ$ centered around the ultra-faint dSphs given in Tab.~\ref{table:dSphs}.  The values of $\bar {\mathcal J}$ and $\bar {\mathcal D}$ for each of the dSphs, for the best fit NFW parameters given in the table, are provided in Tab.~\ref{table:dSphs}. Comparing these values to the functions for the Milky Way in Fig.~\ref{fig:JD} we see that for both decay and annihilation the Milky Way is by far brighter in terms of the predicted signal towards the center of the Galaxy (with the exception of the Burkert profile for annihilation). The Willman I dwarf galaxy has the largest value of $\bar { \mathcal J}$, but even that value is superseded by the Milky Way's ${\mathcal J}$ factor within the inner $\sim$15$^\circ$ for most of the Milky Way DM density profiles considered. The situation is even more stark for the case of DM decay. For some of the dSphs, the Milky Way still provides a brighter signal even $100^\circ$ away from the GC, bringing about the possibility that the signal from the dSphs would not even be visible above the DM decay signal from the ambient Milky Way halo. 

Why, then, would we look for DM decay or annihilation in dSphs or extragalactic objects, more generally?  For decaying DM, the answer is that, for most instruments, there appears to be very little reason to consider any target other than the Milky Way (see~\cite{Dessert:2018qih} for a discussion of this point).  Even the brightest extragalactic sources in DM decay are subdominant relative to the signal from the Milky Way out to large angles from the GC, where the Milky Way backgrounds are minimal for most energies of interest in decaying DM. For narrow, line-like searches extragalactic objects such as clusters have the advantage that the signal would appear redshifted, which would be a unique fingerprint of decaying DM, but this fingerprint only helps in gaining confidence in a signal ({\it e.g.}, worrying about systematic effects) and not in statistics-limited sensitivity.  For DM annihilation the story is more subtle for multiple reasons.  First, as seen by comparing Fig.~\ref{fig:JD} and Tab.~\ref{table:dSphs}, the ${\mathcal J}$-factors of the brightest dSphs are not too far below the expected ${\mathcal J}$-factor of the Milky Way in the inner $\sim$$10^\circ$. Second, depending on the energy and annihilation channel the dSphs can be much cleaner targets than the central region of the Milky Way, since the center of the galaxy is bright in almost all wavelengths except, perhaps, the multi-TeV regime. We discuss this point further below in the context of the {\it Fermi} GCE.  Third, the ${\mathcal J}$ factors of supermassive, extragalactic objects, such as clusters, are likely enhanced by the so-called boost factors, making them brighter than naively expected based on the main halo density profile alone (see, {\it e.g.},~\cite{Lisanti:2017qlb}). 

The {\it Fermi} GCE is a good case study for the utility of the dSphs and extragalactic halos in constraining models of annihilating DM. The GCE is a spherically symmetric excess of continuum gamma-rays observed around the center of the Galaxy that could be due to annihilating DM~\cite{Hooper:2010mq,Goodenough:2009gk,Daylan:2014rsa,Fermi-LAT:2015sau,Fermi-LAT:2017opo,DiMauro:2021qcf} (see~\cite{Leane:2022bfm} for a recent overview).  Interpreted in the context of annihilating DM the GCE prefers a near-thermal annihilation cross-section $\langle \sigma v \rangle \sim 2 \times 10^{-26}$ cm$^3$/s, and -- depending on the final state -- a DM mass around 10's -- 100 GeV. One of the best fitting models, in terms of the spectral  morphology of the GCE, is that where $\chi \chi \to b \bar b$ is the dominant annihilation channel; in this case, the preferred DM mass is around 30 GeV.  As already mentioned, the DM interpretation of the GCE prefers a contracted gNFW profile with $\gamma_{\rm gNFW} \sim 1.25$. In the context of a given model for Milky Way diffuse emission, the statistical significance of the GCE is usually very high ({\it e.g.}, more than $\sim$10$\sigma$, at which point it no longer really makes sense to talk about statistical significance). That is, there is no doubt that the GCE exists as an excess over certain Milky Way diffuse models (we will discuss these models more below).  The question is whether the GCE is due to DM annihilation or, for example, diffuse mismodeling or mismodeling of other Galactic gamma-ray sources, such as millisecond pulsars~\cite{Abazajian:2010zy,Abazajian:2014fta,Lee:2015fea,Calore:2014xka,Macias:2016nev,Pohl:2022nnd}.  Part of the issue is that in the inner $\sim$$10^\circ$ of the Milky Way the GCE is roughly 10\% of the total gamma-ray emission, with most of the remaining 90\% arising from  ordinary Galactic sources.  This means that in order to be confident in the DM explanation of the GCE we need to understand gamma-ray emission in the center of the Galaxy to better than 10\%, which is challenging. 

In contrast, the dSphs are clean gamma-ray environments, as these sources are not expected to themselves produce appreciable gamma-ray emission. While they do not produce as strong of signals as in the GC, given that the GCE has a high-significance detection already, the target masses and cross-sections for DM explanations of the GCE are within reach of dSphs searches. Moreover, a detection of an excess in the dSphs could be considered a {\it smoking gun} signature of DM, given the relative lack of other confounding backgrounds. Unfortunately, no statistically significant excess of gamma-rays has been observed in the dSphs yet~\cite{Fermi-LAT:2015att,Hoof:2018hyn,Alvarez:2020cmw,DiMauro:2022hue}, though it is unclear to exactly what extent this constrains the DM interpretations of the GCE given the systematic uncertainties on both the dSphs side (mostly related to halo profiles) and on the GCE side (related to the fact that the extracted annihilation cross-section is somewhat degenerate with the DM profile and that the best-fit annihilation cross-section depends on the diffuse model used in the analysis).  For example, according to some of the most recent dSphs analyses with {\it Fermi} data the best-fit cross-section to explain the GCE is essentially on top of the 95\% upper limit from a stacked dSphs analysis~\cite{DiMauro:2022hue}, though within uncertainties the dSphs do not completely rule out the DM interpretation of the GCE.  The DM interpretation of the GCE is also in mild tension with null results from nearby galaxy groups~\cite{Lisanti:2017qlb} and cosmic ray probes (seee~\cite{Leane:2022bfm}  for an extended discussion).

\subsection{Gamma-ray detectors and back-of-the-envelope sensitivities}

Let us now focus on the gamma-ray range, since this of the most interest for searches for WIMP DM. There are two different classes of gamma-ray detectors that are of primary interest, which are illustrated in Fig.~\ref{fig:gamma_ray_ill}:
\begin{figure}[htb]  
\begin{center}
\includegraphics[width=0.8\textwidth]{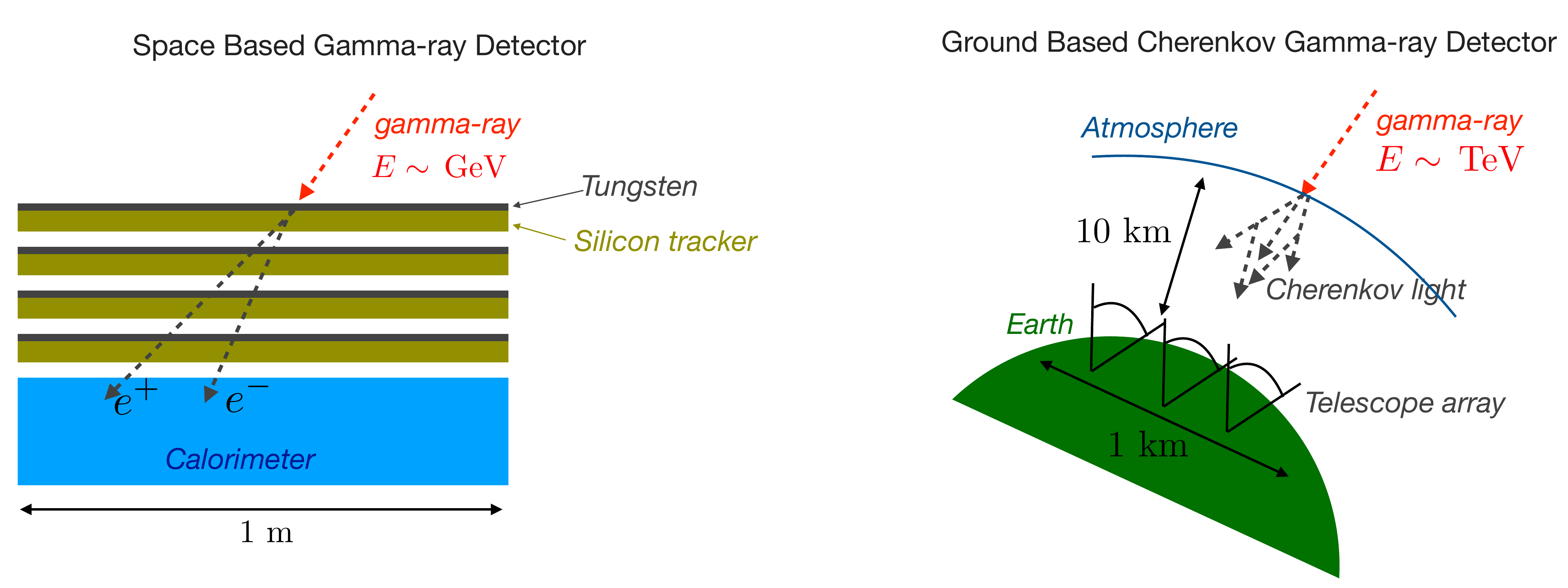}
\caption{Illustrations of the detection concepts for lower-energy ($\sim$100 MeV - TeV) and higher-energy ($\sim$100 GeV and up) gamma-rays in the left and right panels, respectively.  The left panel illustrates, in particular, space-based detector concepts like those employed by the {\it Fermi} Large Area Telescope that work by pair-converting the incoming gamma-ray to electron-positron pairs. These pairs are tracked in the silicon tracking layers to infer the direction of the gamma-ray, while the energy of the gamma-ray is measured with a calorimeter at the bottom of the detector.  The right panel shows the concept behind ground-based Cherenkov detectors, which rely on high-energy gamma-rays pair converting in the atmosphere. The resulting charged particles are then traveling faster than the speed of light in the medium and thus emit Cherenkov radiation, which is detected through a telescope array on Earth.  Examples of such telescopes relevant for DM searches include H.E.S.S. and the upcoming CTA.}
\label{fig:gamma_ray_ill}
\end{center}
\end{figure}
\begin{enumerate}
\item {\bf Space based gamma-ray detectors}: The primary example of a space-based detector is the {\it Fermi} Large Area Telescope onboard the {\it Fermi} Gamma-ray Space Telescope.  This telescope is a pair conversion telescope that works by having alternating layers of a high-$Z$ material (tungsten), which serves to convert the gamma-ray to electron-positron pairs, and then silicon trackers, which track the subsequent electrons and positrons in order to infer the direction of the gamma-ray. The energy of the gamma-ray is measured precisely using a calorimeter at the bottom of the stack of tracking layers. Since space-based gamma-ray detectors must be launched into orbit, their sizes are limited by the payloads of rockets. This explains why {\it Fermi}, along with many other space-based detectors, has an effective area of around $(1 \, \, {\rm m})^2$.  {\it Fermi} has an energy range $\sim$100 MeV to $\sim$2 TeV, with energy resolution $\delta E / E \sim 0.1$ or better and angular resolution around or better than $\sim$0.1$^\circ$ above roughly a GeV.  The cosmic ray and instrumental background for {\it Fermi} is minimal; below a few hundred GeV or so, the backgrounds are almost completely astrophysical. {\it Fermi} can also observe a large fraction of the sky at once, making its exposure map across the full sky nearly isotropic.  The detector has been continuously collecting data since 2008.   {\it Fermi} data is publicly available. 
\item {\bf Ground based Cherenkov gamma-ray detectors:} One of the main downsides of the space-based detectors is the small effective area, which becomes increasingly noticeable at high gamma-ray energies, since gamma-ray fluxes tend to become smaller at increasing energy.  On the other hand, at high gamma-ray energies, roughly above a few hundred GeV, ground-based Cherenkov detectors become available, with existing examples including H.E.S.S., MAGIC, and VERITAS (see~\cite{2022Galax..10...21S} for an overview). In the near-term, the upcoming Cherenkov Telescope Array (CTA)~\cite{CTAConsortium:2017dvg} should come online soon, which will be a game-changing telescope from a DM perspective, as we discuss.  The Cherenkov telescopes work by starting with gamma-ray conversion to electron-positron pairs high up in the atmosphere (10's of km).  These particles are traveling faster than the speed of light in the medium, so they emit Cherenkov radiation (blue light).  That light is collected by an array of telescopes on the ground, which can then infer the direction and the energy of the gamma-ray.  In addition to producing Cherenkov light, the high-energy electron-positron pairs also produce secondary gamma-rays, through {\it e.g.} bremsstrahlung, which in turn can produce more electron-positron pairs, initiating a so-called ``air shower" than can create thousands of particles.  An alternate approach, employed by experiments such as HAWC, is to directly observe air-shower particles that make it to the ground using, for example, water Cherenkov detectors.  Detectors that observe the Cherenkov light directly, such as H.E.S.S., must collect data during the night under pristine ({\it i.e.}, very dark) conditions. On the other hand, detectors such as HAWC do not have this constraint, since they are directly searching for high-energy particles and the backgrounds are not affected by day or night conditions.  With that said, experiments such as HAWC tend to have reduced sensitivity in the low TeV range, since it is difficult for enough particles to make it all the way to the ground (they also generally have worse energy resolution). Since we are mostly interested in low-TeV events for DM searches (think higgsinos and winos), we will focus on the Cherenkov telescopes such as H.E.S.S..  H.E.S.S. achieves energy resolution $\sim$10\%, angular resolution of around $0.1^\circ$, a field of view of approximately $2^\circ$ in radius, and an effective area roughly $0.25 ({\rm km})^2$ at an energy of 1 TeV. CTA will have an improved effective area of around a  $({\rm km})^2$ at 1 TeV (for the CTA South site, which is able to see the GC) and will extend in energy all the way down to a few tens of GeV.  CTA will also have improved energy resolution and a wider field of view than existing telescopes. Note that Cherenkov telescopes are often dominated by cosmic ray backgrounds, since it is difficult to differentiate a gamma-ray event from an isotropic cosmic ray event.
\end{enumerate}

\begin{figure}[htb]  
\begin{center}
\includegraphics[width=0.49\textwidth]{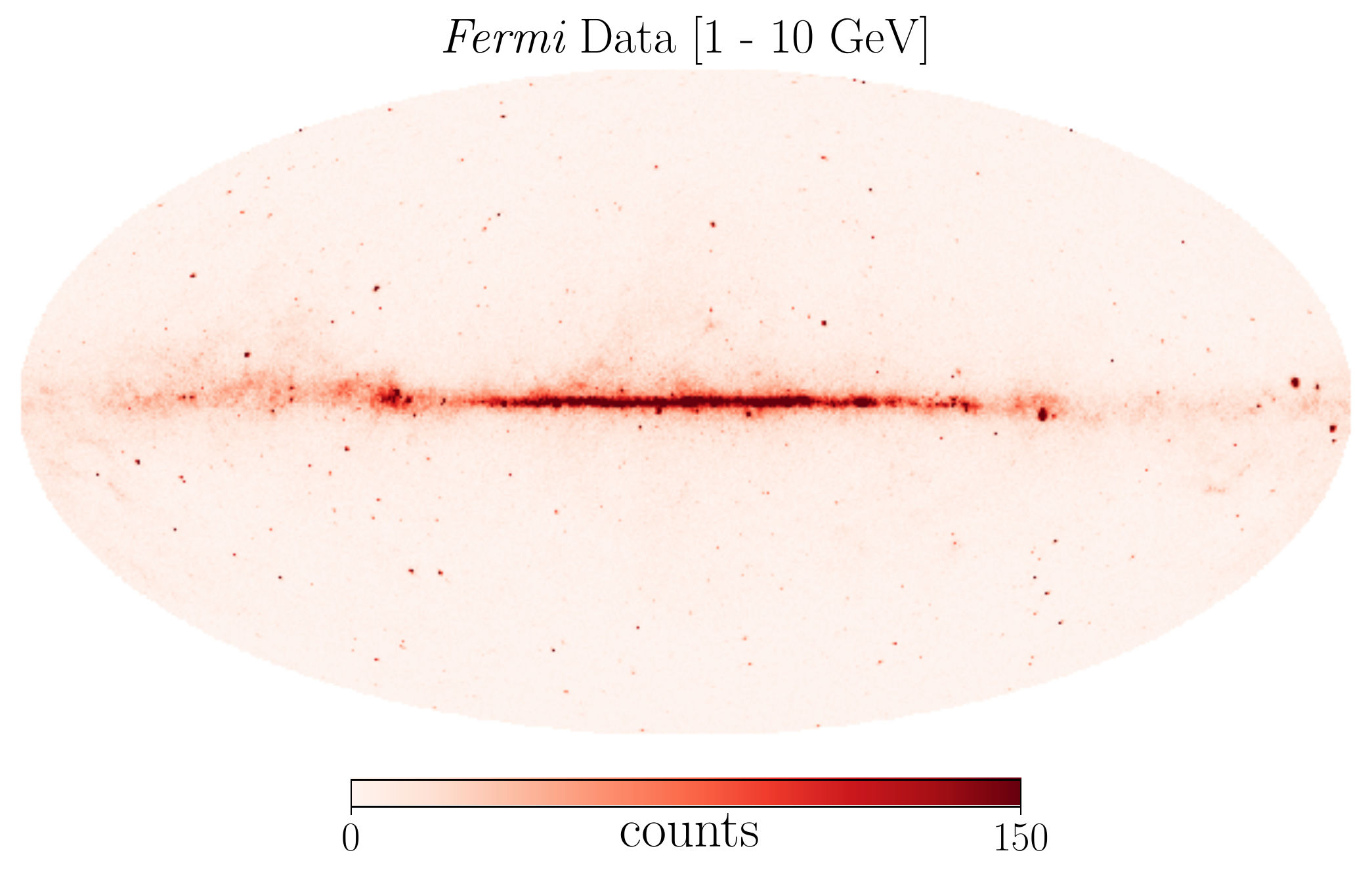}
\includegraphics[width=0.49\textwidth]{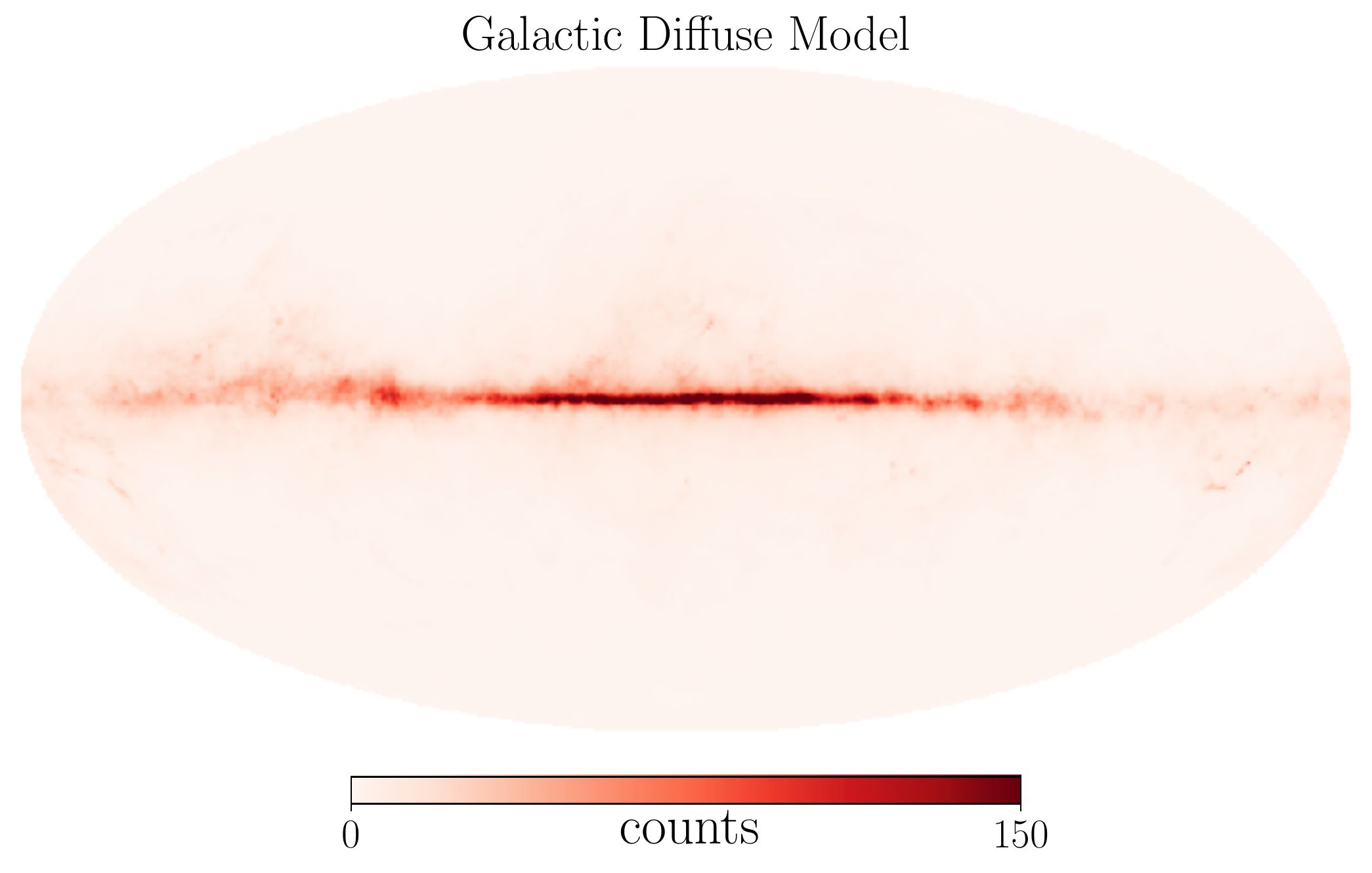}
\includegraphics[width=0.49\textwidth]{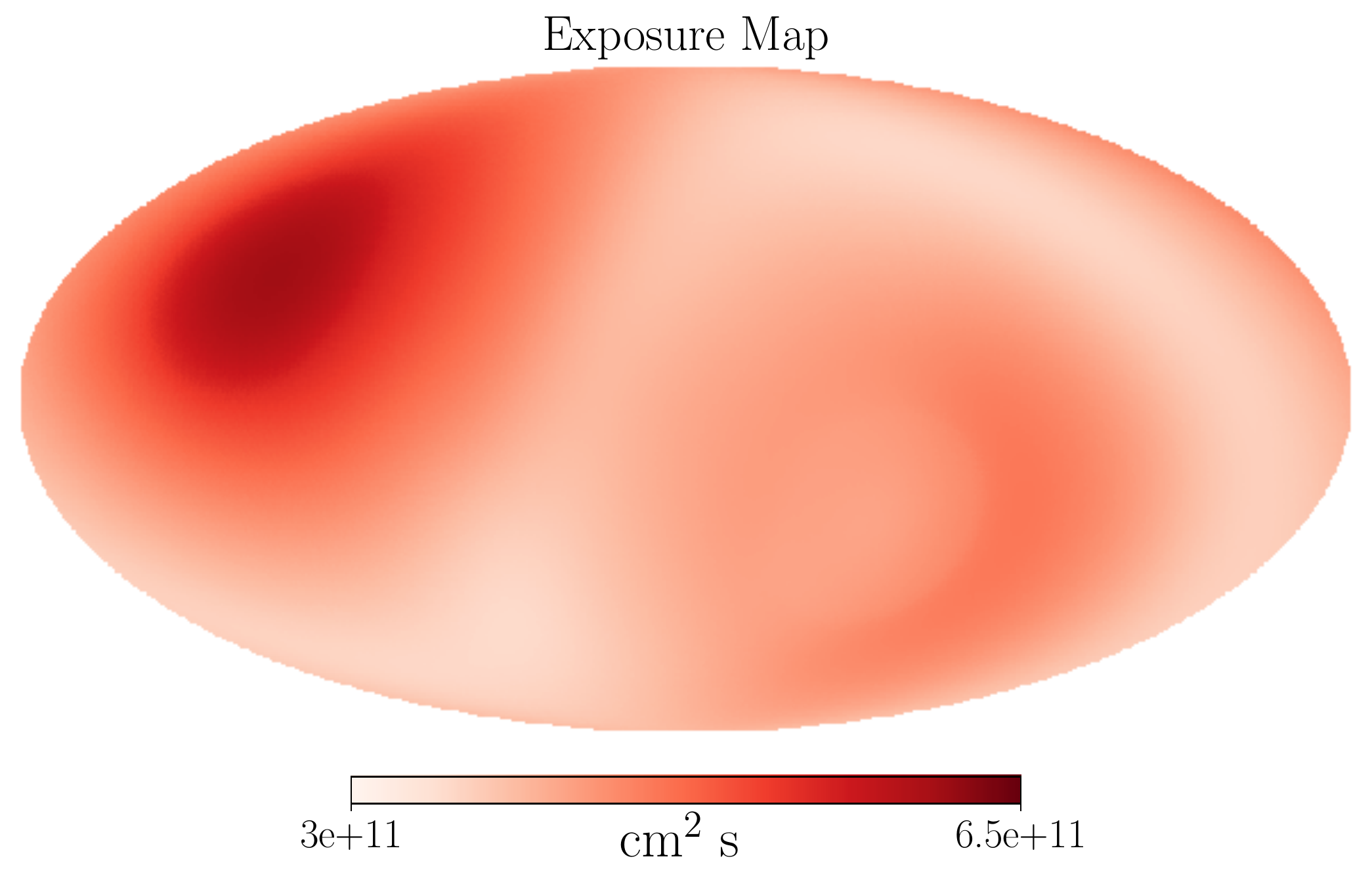}
\caption{(Top left panel) The all-sky {\it Fermi} data summed between 1 and 10 GeV with the selection criterion described in~\cite{Dessert:2022evk}. The Galactic plane cuts across the image, with the GC at the center. (Top right panel) Most of the emission in this energy range arises from Galactic diffuse emission, as shown here by the {\it Fermi} Galactic diffuse model. (Bottom panel) The exposure map, in units of cm$^2$s, for the {\it Fermi} data set described in the top left panel.  The exposure map is roughly isotropic since the detector has a large field of view.  The data are presented in \texttt{healpix} pixelation with ${\rm nside} = 256$.}
\label{fig:fermi_data}
\end{center}
\end{figure}

Having discussed the gamma-ray instruments themselves, we next need to acknowledge that annihilating and decaying DM are not the only sources of gamma-rays in the sky, which complicates searches for new physics since we almost always have to contend with some amount of background.  There are both Galactic and extragalactic gamma-ray sources.  The {\it Fermi} data and dominant background sources are illustrated in Fig.~\ref{fig:fermi_data}.
  The extragalactic emission arises from point sources (PSs) as active galactic nuclei and galaxies with lots of star formation. These are often resolvable as PSs, but there is a diffuse component for any instrument that is constructed from the population of unresolvable sources.  The Galactic sources are predominantly as follows:
\begin{enumerate}
\item {\bf Point sources}: Galactic PSs include supernova remnants, pulsars, and pulsar wind nebula, amongst other sources. The physics that creates the gamma-rays is always non thermal (there are no thermal systems at temperatures near a GeV).  Non thermal emission mechanisms include shocks, jets, and explosions. 
\item {\bf Galactic diffuse emission}:  Galactic emission dominates in the $\sim$MeV - TeV range of primary interest to {\it e.g.} WIMP searches.  There are multiple Galactic emission sources
\begin{enumerate}
\item {\bf $\pi^0$ decay}:  High energy cosmic rays ({\it e.g.}, high energy protons) are accelerated in non-thermal sources such as pulsars and supernova remnants (note that these sources trace star forming regions), and the cosmic rays can then diffuse in the Milky Way before hitting gas. In those collisions one can have processes such as $pp \to \pi^0 + X \to \gamma\gamma + X$, where $X$ stands for additional (mostly hadronic) final states. The ``map" one expects from this gamma-ray source is a convolution of the gas distribution and the cosmic ray distribution. In the GeV range this is the most important source of gamma-rays, with a spectrum $d N / dE \sim 1 / E^\Gamma$, with $\Gamma \sim 2 - 3$. Note that the spectral index is inherited from the index of the parent cosmic rays.
\item {\bf Bremsstrahlung:} Charged high-energy cosmic rays, such as electrons, can emit bremsstrahlung radiation while passing nuclei in gas clouds.
\item {\bf Inverse Compton:} High-energy electrons can up-scatter low-energy photons (such as CMB photon or stellar photons) to gamma-rays. 
\end{enumerate}
\end{enumerate} 

In Fig.~\ref{fig:fermi_data} (top left panel) we illustrate the gamma-ray sky as observed by {\it Fermi} with data collected between 2008 and 2022 between 1 GeV and 10 GeV, with data quality cuts as described in~\cite{Dessert:2022evk}.  This is a Mollweide projection of the full sky, with the Galactic plane visible in the center horizontal line of the image. The Galactic diffuse emission model (technically, we are showing the {\it Fermi} \texttt{gll\_iem\_v07} model), with contributions from $\pi^0$ decay, Bremsstrahlung, and inverse Compton summed, is shown in the top right panel. As is clear by eye, most of the emission in this energy is diffuse emission from the Milky Way, though notably there are a number of bright PSs (most of which are also Galactic in origin) that are also clearly visible.  The bottom panel shows the exposure map for this data set averaged over the energy range, which has units of ${\rm cm}^2 \cdot {\rm s}$. This is the effective area of the detector multiplied by the exposure time. That data corresponds to roughly 15 years, or $\sim$$5 \times 10^8$ s, and the effective area of {\it Fermi} is around a ${\rm few} \times 10^3$ cm$^2$ for the set of quality cuts shown.  The data are presented in \texttt{healpix} pixelation~\cite{Gorski:2004by} with ${\rm nside} = 512$, which corresponds to $12 \times 512^2 = 3145728$ equal-area pixels across the full sky.

We now illustrate a back-of-the-envelope estimate for the sensitivity of {\it Fermi} to a putative DM annihilation signal.  The following numbers are very rough, but they give a general order of magnitude feel for the sensitivity of an instrument like {\it Fermi}.  Further, the approach shown below illustrates how to make a quick estimate of the sensitivity of a given detector to a new-physics signal before investing time and effort into a more precise calculation.  In the inner $\sim$$5^\circ$ of the Milky Way, the astrophysical background has a spectrum $E^2 {d N \over dE} \sim {\rm few} \times 10^{-6}$ GeV/cm$^2$/s/sr at energies $E \sim 10$ GeV. (Recall that $d N / dE \sim 1 / E^\Gamma$, with $\Gamma \sim 2 - 3$, so the above background estimate is reasonable for energies within a factor of a few of 10 GeV.)  In contrast, at high latitudes -- far from the GC -- the background spectrum is around an order of magnitude less this, with $E^2 {d N \over dE} \sim {\rm few} \times 10^{-7}$ GeV/cm$^2$/s/sr.  

The simplest model to consider is that where $\chi \chi \to \gamma\gamma$, since in this case the signal is a sharp line in energy space and not a broad spectrum. Recall that $\chi \chi \to \gamma\gamma$ normally happens at one loop, since DM is usually not electrically charged, and thus the cross-section for this process should be below the thermal cross-section.  The energy resolution of {\it Fermi} is around $\delta E / E \sim 0.1$. Suppose that $m_\chi = 10$ GeV such that our signal would appear in a window of width around 1 GeV centered at 10 GeV.  We start by considering an ROI that consists of the inner $5^\circ$ of the Milky Way. This region has an angular size of $d \Omega \approx 2.4 \times 10^{-2}$ sr, so within a 1 GeV bin around 10 GeV in this ROI we expect to observe around $N_{\rm back} \sim 400$ background event (using the exposure map in Fig.~\ref{fig:fermi_data}).  Now, let us calculate the number of expected signal events as a function of the annihilation cross-section.  We use~\eqref{eq:J} and $d N_\gamma / d E_\gamma = 2 \delta(E_\gamma - m_\chi)$; in conjunction with Fig.~\ref{fig:JD} that tells us to expect ${\mathcal J} \sim 5 \times 10^{23}$ GeV$^2$/cm$^5$ in the ROI assuming a near NFW profile.  We may then estimate the number of signal events as $N_{\rm sig} \sim 10^3 \langle \sigma v \rangle / (2 \times 10^{-28} \, \, {\rm cm}^3 / {\rm s})$.  Let us suppose that we understand our background model perfect. That is, we know that background model should give us, on average, exactly $N_{\rm back} = 400$ events. In any given realization, however, the number of background events that we actually measure will be drawn from a Poisson distribution with mean of 400 events. Since $400 \gg 10$, the Poisson distribution can be approximated by a Gaussian distribution with variance of $400 \, \, {\rm cts}^2$. Thus, if the number of observed counts is more than roughly $400 \, \, {\rm cts} + 2 \times \sqrt{400} \, \, {\rm cts} \approx 450 \, \, {\rm cts}$ then we can start to have a strong suspicion (at the roughly 2$\sigma$ level) that there is evidence for a DM signal. On the contrary, in the absence of a signal we should be able to exclude cross-sections that would be expected to produce less than roughly 50 counts.  In other words, we should be able to set the constraint $\langle \sigma v \rangle \lesssim 10^{-29}$ cm$^3$/s under the null hypothesis.  Indeed, in the recent analysis~\cite{Foster:2022nva} looking for DM annihilation in the GC the 95\% upper limit on this process at $m_\chi = 10$ GeV is around $1.5 \times 10^{-29}$ cm$^3$/s.  Let us now suppose that for $E > 10$ GeV the background scales as $dN / dE \propto 1/E^{2.5}$, which is a reasonable approximation for at least a factor of few in energy, such that the upper limit on the cross-section would be expected to scale with mass as $\langle \sigma v\rangle \propto m_\chi^{5/4}$.  At $m_\chi \sim 100$ GeV this would lead us to expect a limit near $10^{-28}$ cm$^3$/s, which also roughly matches what was found in~\cite{Foster:2022nva}. 

Now suppose that we were to search for DM-annihilation-induced lines from a dwarf galaxy. We take the example of Segue I, since we consider that target more in the next subsection. For an ROI of radius $0.5^\circ$ centered around the target we may estimate, for $m_\chi = 10$ GeV, a number of background counts of around $0.5$ cts, which means that to be confident in a detection (or to set a limit) our signal should produce (or not overproduce) around 3 counts. Note that here we take the background flux to be an order of magnitude smaller than in the GC.  The average ${\mathcal J}$-factor in this ROI is given in Tab.~\ref{table:dSphs}.  We thus expect an upper limit at $m_\chi = 10$ GeV near $\langle \sigma v \rangle \sim 8 \times 10^{-28}$ cm$^3$/s.  This could potentially be improved to closer to $\langle \sigma v \rangle \sim 2 \times 10^{-28}$ cm$^3$/s by considering a more optimal target such as Willman I, though keep in mind that one should also profile over the ${\mathcal J}$ factor uncertainties, which we are not doing here.  These upper limits are not competitive to those from the GC. 

As a last back-of-the-envelope sensitivity estimate, let us convince ourselves that CTA should be sensitive to the thermal higgsino.\footnote{Rough detector performance estimates are taken from \url{https://www.cta-observatory.org/science/ctao-performance/}.}  Let us assume an effective area of $(1 \, \, {\rm km})^2$, an energy resolution $\delta E / E \sim 0.1$, and a total observation time of 100 hr directly at the GC.  CTA should have a roughly flat off-axis effective area until around 2.5$^\circ$ away from the center beam, so we let our ROI extend out to 2.5$^\circ$ from the GC. Motivated by the NFW profile we thus take ${\mathcal J} \sim 10^{24}$ GeV$^2$/cm$^5$, though keep in mind that this close to the GC baryonic feedback is likely important (see Fig.~\ref{fig:JD}).  The background rate, which is dominated by cosmic rays, is estimated at $E = 1$ TeV as $E^2 {d N \over dE} \sim {\rm few} \times 10^{-6}$ GeV/cm$^2$/s/sr.  Thus, in the energy bin of interest we expect around $10^4$ background counts, which means that if we know perfectly the mean expectation for the background model in this energy bin then the number of signal counts should not exceed, roughly, $\sim$$200$ for our 2$\sigma$ limit.  This implies an expected upper limit on the $\chi \chi \to \gamma\gamma$ cross-section around $10^{-28}$ cm$^3$/s.  A more careful analysis, also accounting for the expected increase in flux from endpoint contributions ({\it i.e.}, $\chi \chi \to \gamma + X$, with $X$ additional final states and the final state $\gamma$ at an energy near but slightly below $m_\chi / 2$), projects sensitivity slightly better than this~\cite{Rinchiuso:2020skh}, while the expected cross-section for the gamma-ray line signal for the higgsino is right around $10^{-28}$ cm$^3$/s.  Thus, at least assuming the Milky Way DM profile does not have a sizable core, the higgsino signal should be within reach, though perhaps barely, of CTA.  One point to emphasize is that {\bf CTA will release their data to the community}, like {\it e.g.} {\it Fermi} and other space-based telescopes do, meaning that now is the time to be thinking about what types of analyses are interesting to perform.

\subsection{A case study: Segue I with {\it Fermi} gamma-ray data}
\label{sec:segue}

In the previous subsection we estimated the sensitivity to line-like signals from DM annihilation by using a back-of-the-envelope comparison of the expected background counts to the predicted signal counts in a narrow energy range around the line center. In this subsection we perform a more careful analysis for a more complicated signal, one where the DM annihilates to a continuum spectrum of gamma rays.  We chose an example that is relevant for the DM interpretation of the GCE: a search for $\chi \chi \to b \bar b$ with $m_\chi  = 31$ GeV towards the Segue I dwarf galaxy with $\sim$15 years of {\it Fermi} data.  Recall that the GCE is well fit by annihilating DM to $b \bar b$ with $m_\chi \sim 30 - 50$ GeV and $\langle \sigma v \rangle \sim (1 - 2) \times 10^{-26}$ cm$^3$/s, depending on systematic uncertainties. Our choice of $m_\chi = 31$ GeV is semi-random; to perform a complete analysis one should really repeat this analysis at all $m_\chi$ of interest.  Similarly, our choice of Segue I is also random -- for a full analysis, one should analyze data from all of the relevant dwarf galaxies and combine the results in an appropriate joint analysis. Our simplified goal here is, for this particular choice of $m_\chi$, to derive the 95\% upper limit on $\langle \sigma v \rangle$ from Segue I and to compare that limit with the cross-sections needed to explain the GCE.  The full analysis, described below, is available as a supplementary \texttt{jupyter} notebook, which includes the pre-processed {\it Fermi} data, model, and exposure files.\footnote{See this 
\href{https://colab.research.google.com/drive/1XKnXVonedAZS5wu9cHrqIEwWzlemoCf5?usp=sharing}{Colab Jupyter Notebook.}
}  The {\it Fermi} data is subject to the data quality cuts and exposure time as described in~\cite{Dessert:2022evk}.

\begin{figure}[htb]  
\begin{center}
\includegraphics[width=0.49\textwidth]{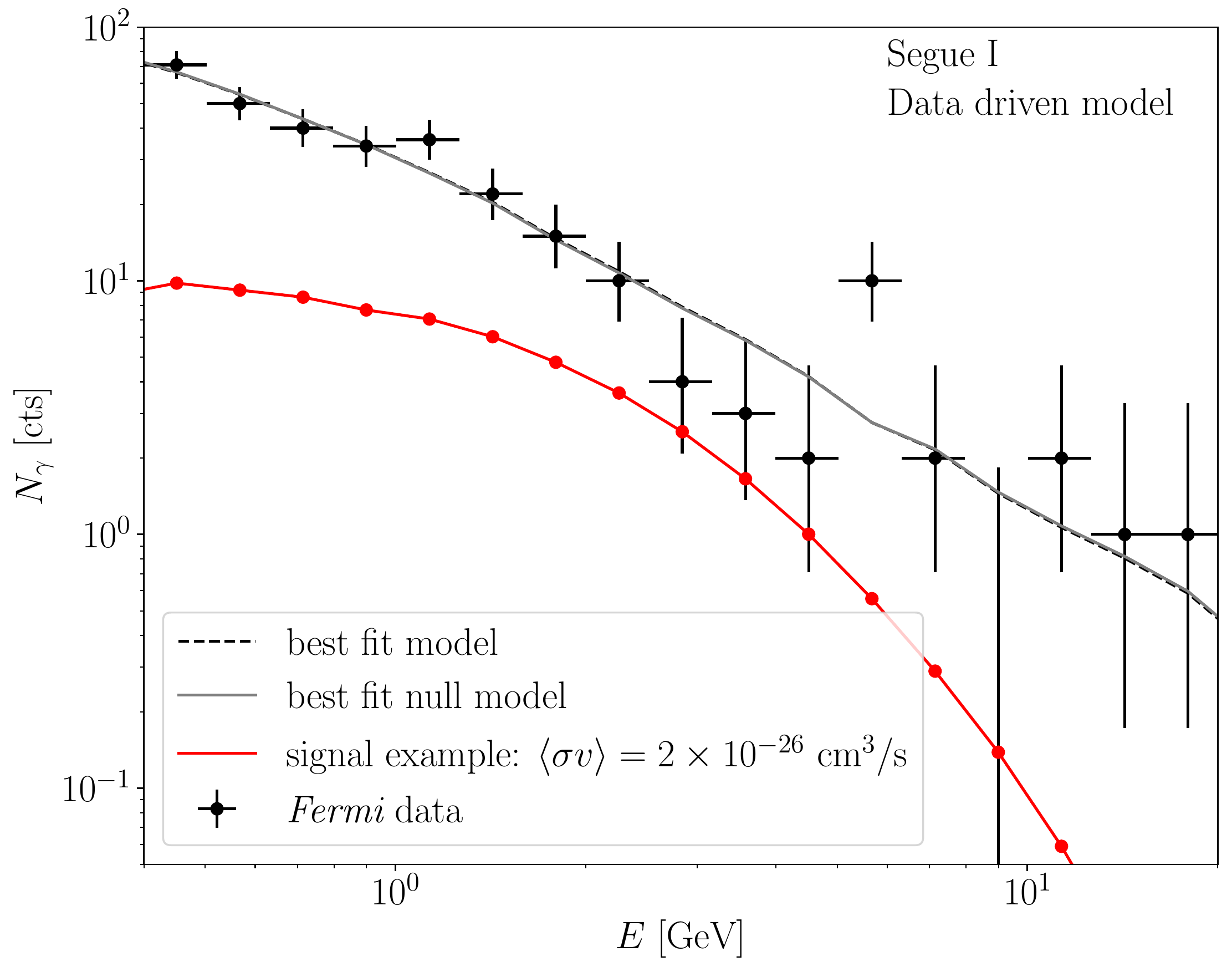}
\includegraphics[width=0.49\textwidth]{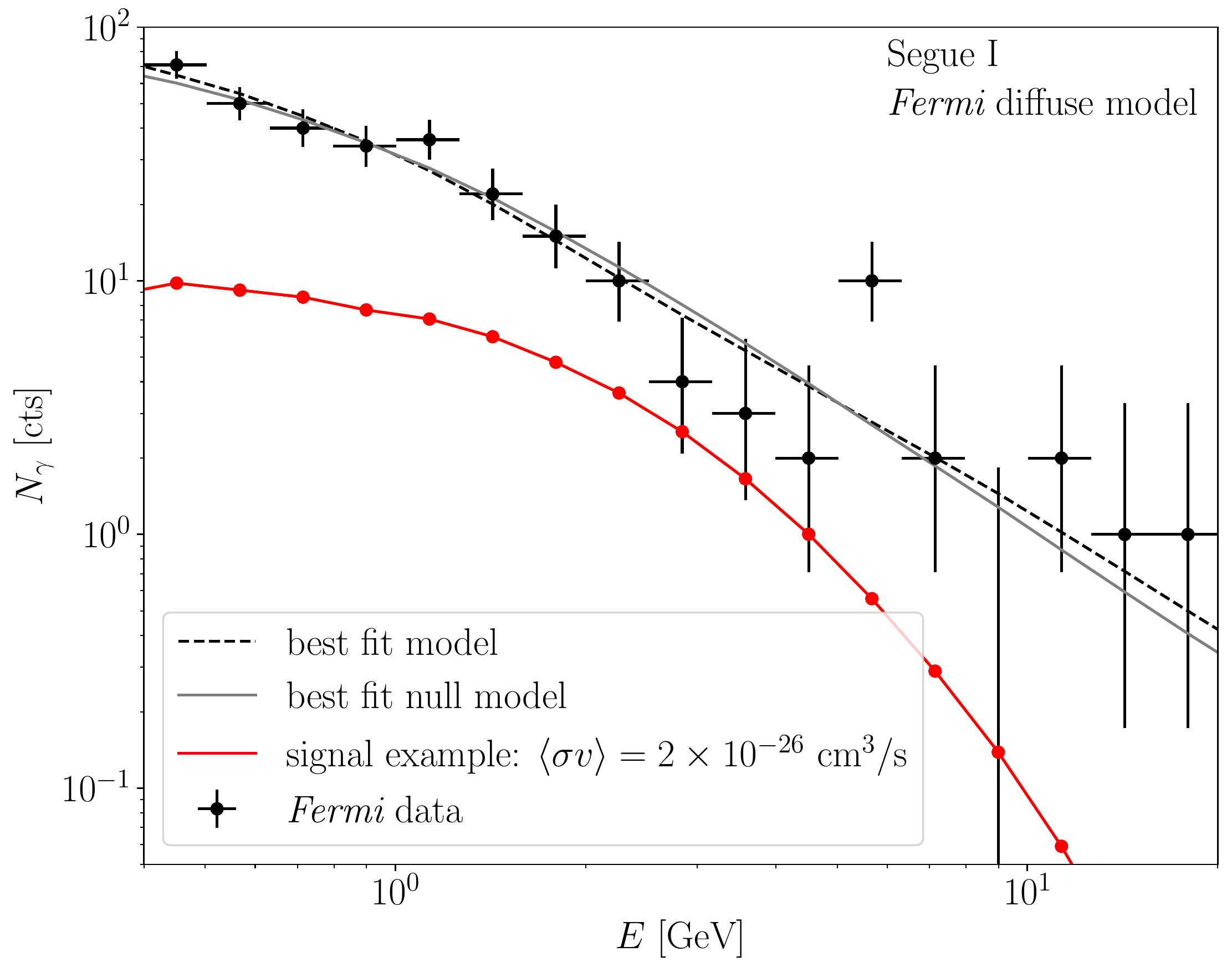}
\caption{Data and models from the example analysis of {\it Fermi} gamma-ray data towards Segue I.  The data is extracted using the selection criterion described in~\cite{Dessert:2022evk} and within a $0.5^\circ$ circle around the center of the dSph. The energy binning is described in the text, and the data are illustrated as counts in these energy bins with Poisson error bars.  The signal example is the same between the two panels and shows the predicted DM annihilation signal in terms of expected counts per energy bin for the given cross-section and for $m_\chi = 31$ GeV. The left panel uses a data driven background model that directly takes the spectral template from data between $0.5^\circ$ and $10^\circ$ from the dSph center, while the right panel uses the {\it Fermi} diffuse emission model instead. Both models give comparable fits and no evidence is found for DM annihilation. }
\label{fig:segue_spectra}
\end{center}
\end{figure}

The Segue I dwarf galaxy is located well away from the Galactic plane, at Galactic coordinates $(\ell, b) \approx (220.50^\circ, 50.45^\circ)$.  We extract spectral data out to $0.5^\circ$ from the source center for the ON data; we also extract OFF data from $0.5^\circ$ to $10^\circ$ from the source center.  The data are binned in 40 logarithmically-spaced energy bins from 200 MeV to 2 TeV, but we only select the bins between $0.5$ GeV and $30$ GeV for our analysis example (bin numbers 4 through 20 inclusive, where we start counting from zero).  The binned count data for the ON data are illustrated in Fig.~\ref{fig:segue_spectra}.
Note that the black data points show the number of counts observed in each energy bin, with the $x$ error bars indicating the bin size and the $y$ error bars the 1$\sigma$ frequentist confidence intervals for the mean number of counts $\mu$ given a single observation of some number $N$ counts.  These error bars are shown to give you a sense of the uncertainty in our ability to infer the true model given the finite counts; they are not actually used in the analysis.

We want to describe the data by a signal plus background model. The signal model is given by the convolution of~\eqref{eq:J} with the exposure (exposure time times effective area) extracted from the instrument response (the exposure file is also available in the supplementary data).  Given that our signal has an energy spread $\delta E / E \gg 0.1$, we can safely ignore the finite energy resolution of the detector. In Fig.~\ref{fig:segue_spectra} we show an example DM annihilation signal with the indicated cross-section forward modeled to detector-level counts. Note that while the DM signal is shown as a smooth curve, the correct way to interpret the curve is the expected number of counts in the appropriate energy bins.  To help with this interpretation, we show red points at the locations of the energy bin centers.  

In computing the signal spectrum from DM annihilation, we need to know the function $dN/dE$ that gives the gamma-ray spectrum from $\chi \chi \to b \bar b$, accounting for the decays of the $b$-quarks. This spectrum is provided in the supplementary data, and it is computed using the easy-to-use code package PPPC 4 DM ID~\cite{Cirelli:2010xx}.  The functionality of PPPC 4 DM ID that we use to construct $dN/dE$ is simply a large interpolation table from pre-computed \texttt{PYTHIA} Monte Carlo simulations of $b \bar b$ showers at different center of mass energies.  That interpolation table may be accessed through a \texttt{Mathematica}\textsuperscript{\textregistered} interface.\footnote{Documentation may be found here: \url{http://www.marcocirelli.net/PPPC4DMID.html}.} 

The signal model is added to the background model to construct our full model for the data. We use two different background models: (i) a data-driven background model, which is illustrated in the left panel of Fig.~\ref{fig:segue_spectra}, and (ii) a physics-based model illustrated in the right panel. The data-driven background model is simply constructed by taking the observed counts data in the OFF region and changing the normalization to account for the smaller angular size of the ON region in addition to the slight change in the exposure map between the two regions.  For the physics-based model we use the sum of the {\it Fermi} 4FGL PS model~\cite{Fermi-LAT:2022byn} in addition to the \texttt{gll\_iem\_v07} {\it Fermi} Galactic diffuse emission model.  This model is provided in the supplementary data.  In both cases, we treat the overall normalization of the  background model as an unconstrained nuisance parameter.

We use the data to constrain the normalization of the signal model, profiling over the nuisance parameter associated with the background model, in the context of a frequentist likelihood.   
See App.~\ref{sec:stats} for a review of frequentist statistics.
Our starting point is to construct the joint Poisson likelihood 
\es{eq:L}{
p({\bf d} | {\mathcal M}, {\bm \theta}) = \prod_{i=1}^{N_e} {\mu_i({\bm \theta})^{d_i} e^{-\mu_i ({\bm\theta})} \over d_i !} \,,
}
where $i$ is an index over the $N_e$ different energy bins.  The data set ${\bf d} = \{ d_i \}_{i=1}^{N_e}$ consists of the set of counts (the $d_i$) observed in each energy bin; note that these are integers.  Our model ${\mathcal M}$ has a parameter vector ${\bm \theta}$, and for a given parameter vector we predict mean counts $\mu_i({\bm \theta})$ in each of the energy bins. We can then recognize the quantity inside of the product in~\eqref{eq:L} as the probability of observing the observed counts $d_i$, given the mean model expectation. It should be relatively intuitive that the best-fit model parameters, which we call $\hat {\bm \theta}$, are found by the values that {\it maximize} the likelihood.  That is, the best fit parameters are the ones for which the observed data set would be most likely, in the context of the model ${\mathcal M}$. 

We will write our parameter vector as ${\bm \theta} = (\mu, A_{\rm bkg})$, where $\mu$ is a parameter that rescales the signal contribution and $A_{\rm bkg}$ is a parameter that rescales the background contribution.  We define $\mu$ such that the cross-section is $\langle \sigma v \rangle = \mu \times (2 \times 10^{-26})$ cm$^3$/s, and $A_{\rm bkg}$ is defined such that for $A_{\rm bkg} = 1$ the background model has its default normalization. Then,
\es{}{
\mu({\bm \theta}) = \mu \,\mu^{\rm sig}_i + A_{\rm bkg} \,\mu^{\rm bkg}_i \,,
}
where $\mu^{\rm sig}_i$ is the mean expected signal model counts for $\langle \sigma v \rangle = 2 \times 10^{-26}$ cm$^3$/s and $\mu^{\rm bkg}_i $ is the default background model contribution in each energy bin.  To compute the upper limit we first compute a quantity called the profile likelihood $q$, found by
\es{eq:q}{
q(\mu) = 2 \log \left[{ p({\bf d} | {\mathcal M}, \hat{\bm \theta}) \over {\rm max}_{A_{\rm bkg}} p\big({\bf d} | {\mathcal M}, (\mu,A_{\rm bkg})\big) }\right]
}
if $\mu > \hat \mu$, with $\hat \mu$ the best-fit signal-strength parameter, and $q(\mu) = 0$ otherwise.  Note that the numerator in~\eqref{eq:q} is the likelihood evaluated at the best-fit parameter vector $\hat {\bm \theta} = (\hat \mu, \hat A_{\rm bkg})$, while in the denominator the likelihood is maximized over $A_{\rm bkg}$ but at fixed signal strength $\mu$.  Wilks' theorem states that in the limit of a large number of data counts we may find the 95\% one-sided upper limit $\mu^{95}$ by solving for $q(\mu^{95}) \approx 2.71$, where importantly we need to account for the possibility that $\hat \mu$ may be at negative values, even though only positive values are physical. In fact, in the large count limit and under the null hypothesis $\hat \mu$ should be negative just as often as it is positive.

In Fig.~\ref{fig:segue_spectra} we show the best-fit signal plus background models and the best fit background-only models for both methods of modeling the background. In both cases, the best fit value of $\hat \mu$ is consistent with $0$ at 1$\sigma$ significance (meaning that $q(\mu)$ changes by less than 1 in going from $\hat \mu$ to zero).  The best-fit value $\hat \mu$ is slightly positive in the case of the data-driven model and slightly negative (by a bit less than 1$\sigma$) for the physics-based model.   Given the lack of evidence in favor of the signal model over the background model, it is relevant to compute the 95\% upper limits.

\begin{figure}[htb]  
\begin{center}
\includegraphics[width=0.7\textwidth]{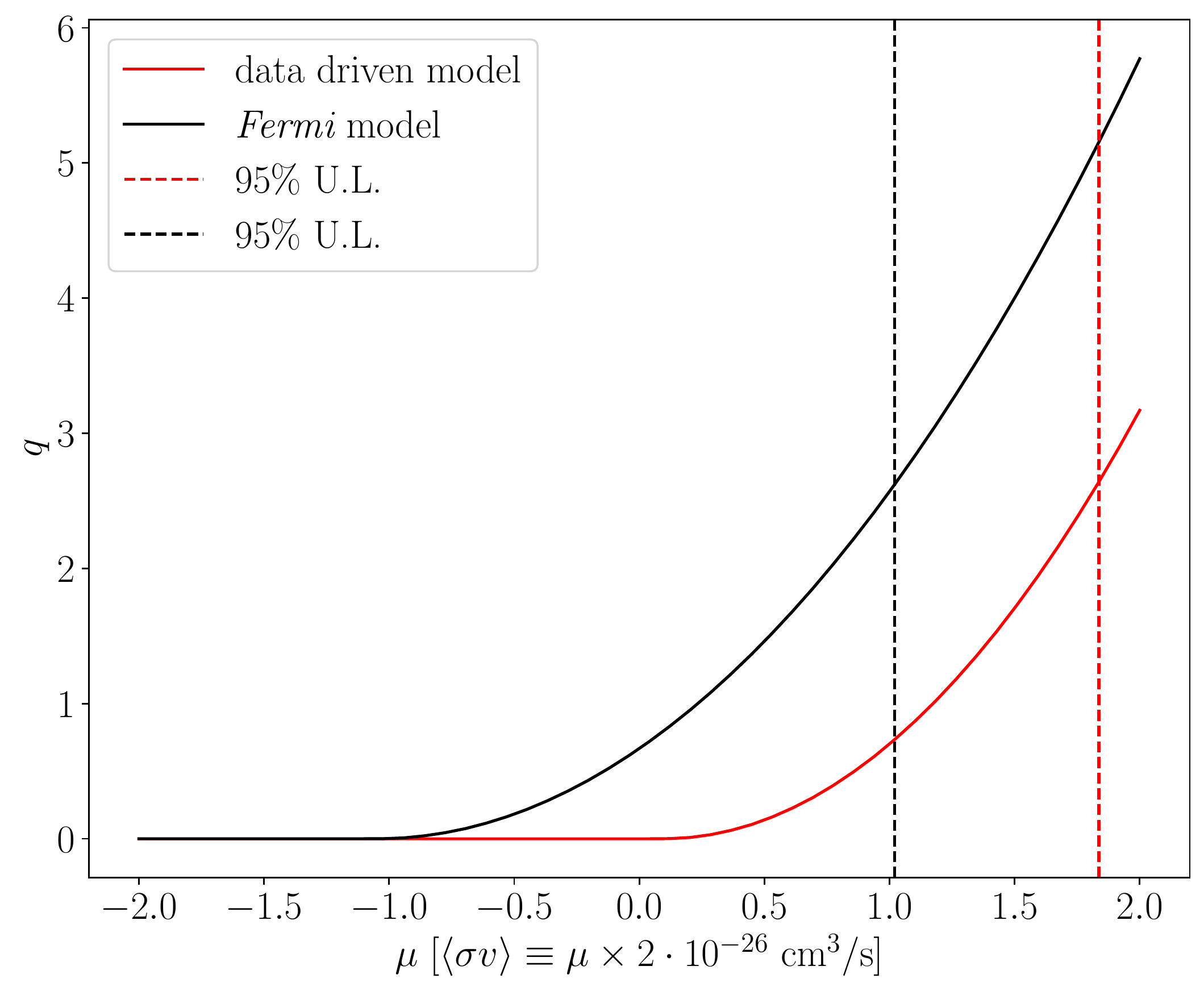}
\caption{The profile likelihood, shown as the test statistic $q$ defined in~\eqref{eq:q}, for the data sets and analyses illustrated in Fig.~\ref{fig:segue_spectra}. The test statistic is shown as a function of parameter $\mu$ that rescales the signal strength as indicated; note, the signal strength is allowed to be negative, even though this is not physical, so that we may make use of asymptotic theorems of frequentist statistics that require us to consider the test statistic in the vicinity of its minimum (zero first derivative), which may be at negative values. The 95\% one-sided upper limits are given, approximately, by where $q(\mu) \approx 2.71$, giving the two upper limits indicated for the data driven background model analysis and the {\it Fermi} diffuse model analysis. The difference between these two upper limits can be thought of a source of systematic uncertainty in this analysis, since both models fit the data almost equally well. }
\label{fig:segue_q}
\end{center}
\end{figure}

In Fig.~\ref{fig:segue_q} we show the profile likelihoods found using both background models.  We also indicate the 95\% one-sided upper limits found in the two analyses by vertical dashed lines.  Using the data-driven background model, for example, we exclude cross-sections greater than $\langle \sigma v \rangle \gtrsim 3.4 \times 10^{-26}$ cm$^3$/s at 95\% confidence; the limit found using the physics-based model is slightly stronger.  These upper limits are very close to the parameters needed to explain the GCE. It would thus not be surprising if by including additional dSphs we could approach an upper limit that would meaningfully constrain the DM explanation of the GCE or provide evidence in its favor. However, keep in mind that here we fixed the ${\mathcal J}$-factor, while in a real analysis one should also profile over its uncertainty.

One question that you may be wondering is: which of the two upper limits that we computed is correct? Unfortunately, there is no right answer to this question. Both analyses that we performed, using the different background models, are valid analyses. Looking at Fig.~\ref{fig:segue_spectra} it also seems that both models are doing a reasonable job of modeling the data, with no clear regions of mismodeling. At a precise level, the two model perform similarly. The difference in twice the log likelihood between the models is less than $0.5$, meaning that there is less than 1$\sigma$ evidence in favor of one model over the other. This is where data analysis becomes part science and part art, since there is no obviously correct way for us to proceed. For example, we could construct a third background model by using the data driven framework but going out to $15^\circ$ instead of $10^\circ$ and get yet another 95\% upper limit; what should we do with this limit? What we are happening upon here is the fact that in any analysis there are both statistical and systematic uncertainties. Here, our systematic uncertainty is related to the fact that we do not have a perfect model for the background emission. We are left to construct approximations to the background emission, either by using a physics-based approach or a data-driven approach or, perhaps, some hybrid approach. If we mis-model the true background emission, then this can affect our inferred best-fit signal parameter and our inferred limit. We want to try our best to make sure that systematics are not so important that we rule out a real signal if one is present or fake a signal if one is not there.

While there is no golden rule for how to deal with systematic uncertainties like those we have encountered here (in contrast to statistical uncertainties, where one simply follows the likelihood framework), there are a few general principles that we can strive to follow. First, try to decide on an analysis approach before actually analyzing the real data. This will prevent you from (perhaps unknowingly) adjusting the analysis to get the results that you want.  Second, when designing your analysis strategy, try to optimize you search such that the statistical and systematic uncertainties will be roughly equal. If you statistical uncertainties are too small, then your answer will be fully limited by systematic uncertainties, which is dangerous since these are hard to quantify. On the other hand, if you are too conservative with your analysis strategy and statistical uncertainties dominate, then you are not getting the full benefit of the data you collected. Roughly matching the two uncertainties can be a good compromise. In our example, one way of increasing the statistical uncertainties and also decreasing the importance of background mismodeling would be to narrow the energy range used in the analysis. Third, plan ahead with the analysis variations that you will do to assess for systematic uncertainties. For example, perhaps you decide ahead of time on a fiducial analysis but then also an ensemble of alternate analyses to look for evidence of mis-modeling and to quantify the systematic uncertainties. In our case, that might be {\it e.g.} constructing the data-driven background model in different ways and/or changing the energy range in the analysis.  Lastly, when in doubt, pick the most conservative result. When you exclude parameter space for DM the idea is that nobody is ever going to bother looking at that parameter space again, since it is excluded. This is how we make progress in the search for DM. With that in mind, we really do not want to be in the situation where someone claims a region of parameter space is excluded but in fact a signal is there, it just was lying below the unaccounted for systematic uncertainties in the analysis. It is best to error on the side of caution.

\section{Axions and axion-like particles}
\label{sec:axions}

Axions are hypothetical ultralight particles that have strong theoretical motivations both to explain outstanding issues with the SM and from UV considerations such as in string theory.  Here, I will use the term axion to refer both to the QCD axion and to more general axion-like-particles (ALPs) that share the QCD axion's interactions with the SM but for the interaction with QCD. There are many excellent reviews of axions, including~\cite{Marsh:2015xka,Graham:2015ouw,DiLuzio:2020wdo} and, recently,~\cite{Hook:2018dlk} in the context of TASI.  We refer the reader to these reviews for alternate perspectives. 

\subsection{The QCD axion and the Strong-{\it CP} Problem}

\begin{figure}[htb]  
\begin{center}
\includegraphics[width=0.5\textwidth]{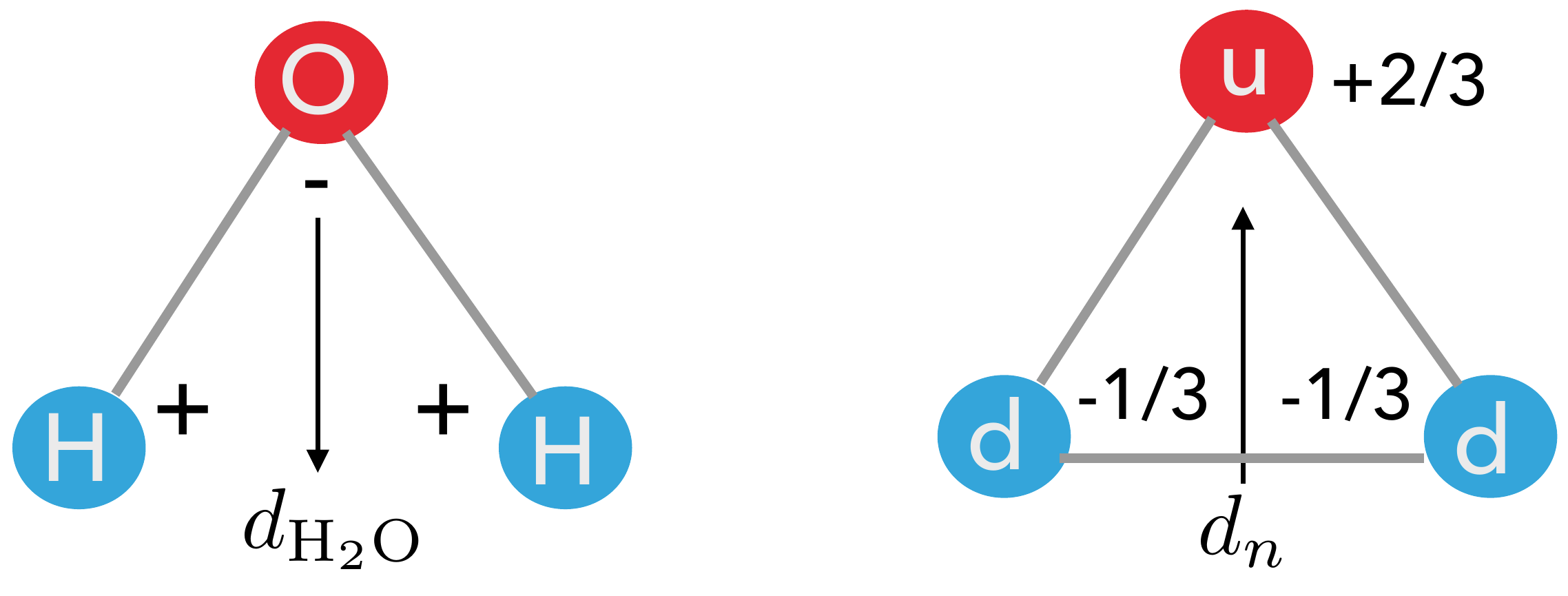}
\caption{A toy illustration of the analogy between the EDM of water (left) and the naive EDM of the neutron (right). The EDM of water may be estimated by one electric charge separated by the length of the water molecule. The same estimate applied to the neutron roughly reproduces the result of a more careful calculation, which depends on the fundamental parameter of nature $\bar \theta$, but the theoretical calculation of the neutron EDM overshoots the observed upper limit on the EDM by 10 orders of magnitude. This implies that $\bar \theta$ is fine-tuned towards zero by at least 10 orders of magnitude, which is known as the strong-{\it CP} problem. Axions may provide a dynamical explanation of the strong-{\it CP} problem, without fine tuning, in addition to explaining the observed DM and arising in the context of theories of quantum gravity such as string theory.}
\label{fig:EDM}
\end{center}
\end{figure}

The strong-{\it CP} problem relates to the non-observation of the neutron electric dipole moment (EDM). To understand the issue behind the neutron EDM, let us begin by understanding the EDM in a simpler system, namely H$_2$O. Consider Fig.~\ref{fig:EDM}, which illustrates in the left panel a particle-physicists-level cartoon of a water molecule. 
Water is electrically neutral, but it has a charge asymmetry, with more negative charges towards the oxygen atom and more positive charges towards the hydrogen atoms.  This asymmetry induces an EDM, whose magnitude is given by the total electric charge that is separated (around $1$ $e^-$ of charge) times the distance of separation (around $0.1$ nm). Thus, we can back-of-the-envelope estimate the EDM $d_{{\rm H}_2{\rm O}} \sim 10^{-8}$ e$\cdot$cm.  The actual EDM of water is measured to be around half as large as this naive estimate, so not bad! Let us now try to apply this logic to the neutron, which -- like water -- is also charge neutral. At face value, the neutron looks analogous to the H$_2$O molecule, as illustrated in Fig.~\ref{fig:EDM}. Given that the size of the neutron is around 1 fm, we may then back-of-the-envelope estimate $d_n \sim 10^{-13}$ e$\cdot$cm.  A more careful calculation yields a result a few orders of magnitude smaller, as we discuss in the following paragraphs. However, no neutron EDM has been observed, and the most stringent upper-limit to-date constrains $|d_n| \lesssim 1.8 \times 10^{-26}$ e$\cdot$cm at 90\% confidence~\cite{Abel:2020pzs}.  This discrepancy is known as the strong-{\it CP} problem. 

More precisely, the SM has a fundamental parameter called ${\bar \theta}$, which can be thought of as an angle ($-\pi \leq \bar \theta < \pi$).  The parameter ${\bar \theta}$ controls the magnitude of the EDM within the SM, and a careful calculation, which we sketch below, yields the result $d_n \approx 3 \times 10^{-16} \bar \theta$ e$\cdot$cm.  A very rough interpretation of the angle $\bar \theta$ is an asymmetry angle between the quarks within the neutron; if $\bar \theta = 0$ then, in some sense, there is perfect alignment between the up quark and the two down quarks such that $d_n = 0$ identically.  More precisely, the term ${\bar \theta}$ enters into the SM through the following term in the QCD Lagrangian: 
\es{eq:CP}{
{\mathcal L}^\slashed{CP} = -{\theta g_s^2 \over 32 \pi^2} G_{\mu \nu}^a \tilde G^{a \mu \nu} \,,
}
where $g_s$ is the strong coupling constant, $G_{\mu \nu}^a$ is the QCD field strength, with $a$ a color index, and $\tilde G$ the Hodge dual field strength: $\tilde G_{\mu \nu} = {1 \over 2} \epsilon_{\mu \nu \rho \sigma} G^{\rho \sigma}$. Here, $\epsilon_{\mu \nu \rho \sigma}$ is the totally antisymmetric tensor known as the Levi-Civita tensor ($\epsilon_{0123} = +1$). The quantity $\theta$ is related to $\bar \theta$, as discussed shortly.  The term in~\eqref{eq:CP} may look familiar, since it is generated by chiral rotations of the quark fields, which are anomalous. (See~\cite{Harvey:2005it} for TASI lectures on anomalies.) Alternatively, we can completely remove~\eqref{eq:CP} from the Lagrangian by the appropriate chiral rotation, though at the expense of introducing a complex phase into the quark mass matrix.  It is then clear that there is some degeneracy between the value of $\theta$ appearing in~\eqref{eq:CP} and the phases of the Yukawa matrices $Y_u$ and $Y_d$ for up- and down-type quarks. Indeed, the invariant combination of parameters is ${\bar \theta} = \theta - {\rm arg} {\rm det} (Y_u Y_d)$, which is the parameter that sets the magnitude of the neutron EDM. The value of {\it e.g.} $\theta$ itself is not physical. 

Let us remove~\eqref{eq:CP} by making the chiral rotation $u \to e^{i \gamma_5 \bar \theta / 4} u$ and $d \to e^{i \gamma_5 \bar \theta / 4} d$, where $u$ and $d$ are the up and down Dirac quark fields, and where we assume -- for simplicity -- that a sequence of chiral rotations has already been performed such that it is ${\bar \theta}$ that appears in~\eqref{eq:CP} and not $\theta$.  We now want to consider how these chiral rotations affect the low-energy chiral Lagrangian, which provides a good description of the strong dynamics below the QCD confinement scale.  

First, we briefly review the chiral Lagrangian.  Let us, in particular, concentrate on the pure pion part of the Lagrangian given by
\es{eq:chpt}{
{\mathcal L} = {F_\pi^2 \over 4} {\rm tr} \left[ (D_\mu U) (D^\mu U^\dagger) \right] + {V^3 \over 2} {\rm tr} \left[ M U + M^\dagger U^\dagger \right] \,,
}
with $F_\pi$ the pion decay constant.
Recall that the chiral Lagrangian is based off the idea of effective field theory (EFT), with the task being to construct the most general Lagrangian consistent with chiral symmetry breaking.  Ignoring the masses of the up and down quarks for a moment, the Lagrangian of QCD has an $SU(2)_L \times SU(2)_R \times U(1)_V$ symmetry (the $U(1)_A$ symmetry is anomalous). After confinement, however, QCD induces non-trivial quark condensates ($\langle \bar q q \rangle \neq 0$), which breaks $SU(2)_L \times SU(2)_R \to SU(2)_{\rm isospin}$. The pions are the Goldstone bosons of this symmetry breaking. They would be exactly massless but for the small quark  masses, which spoils the original symmetry and makes the pions pseudo-Goldstone bosons.\footnote{Note that if the quarks were massless the charged pions would still have small masses from photon loops.} The first term in~\eqref{eq:chpt} is what would be obtained for the chiral Lagrangian in the absence of quark masses, with $U(x) = \exp\left[ 2 i {\pi^a \tau^a \over F_\pi} \right]$, where $\tau^a$ are the $SU(2)$ generators and $\pi^a$ are the three pions ($\pi^0 \equiv \pi^3$, $\pi^{\pm} \equiv {1 \over \sqrt{2}} (\pi^1 \pm \pi^2)$). The covariant derivative showing up in~\eqref{eq:chpt} accounts for the electromagnetic charges of the pions but, importantly, not the QCD interactions. The pions are color neutral, since they emerge as QCD bound states after confinement. The second term includes the effects of the quark masses, which induce non-zero masses for the pions. Here, $M = {\rm diag} (m_u \, \, m_d)$ and $V^3$ is the value of the quark condensate ($\langle \bar u u \rangle = \langle \bar d d \rangle = V^3$). Note that expanding out the mass term in the chiral Lagrangian leads to the pion mass formula, known as the Gell-Mann-Oakes-Renner relation,
\es{}{
m_\pi^2 = {V^3 \over F_\pi^2} (m_u + m_d) \,.
}
Recall that $V \sim \Lambda_{\rm QCD} \sim 250$ MeV, $F_\pi \sim 92$ MeV, and $m_\pi \sim 140$ MeV, from which we can infer $m_u + m_d \sim 10$ MeV.  Lattice QCD calculations find $m_u + m_d \sim 7$ MeV~\cite{ParticleDataGroup:2022pth}, so this is pretty close.

Now let us go back to considering how $\bar \theta$ enters into the chiral Lagrangian after making a chiral rotation of the quark fields. In particular, we have $\langle \bar u u \rangle \to e^{i \bar \theta/2} \langle \bar u u \rangle$, and similar for the $d$ quarks, which implies that $M \to e^{i \bar \theta/2} M$.  We want to understand the vacuum structure of~\eqref{eq:chpt} in the presence of non-zero $\bar \theta$. As can be worked out, $U$ acquires a vacuum expectation value (VEV) for non-zero $\bar \theta$ that corresponds to a VEV of $\pi^0$. In particular, let $\langle U \rangle = {\rm diag} (e^{i \phi} \, \, e^{-i \phi})$.  Substituting this expression into the second term of~\eqref{eq:chpt} and minimizing with respect to $\phi$ one finds that at the VEV $\phi$ is given by
\es{}{
  \tan \phi = {m_d - m_u \over m_d + m_u} \tan { \bar \theta \over 2} \,.
  }
  Now that we have this expression for $\phi$ we can substitute it back into the second term of~\eqref{eq:chpt} to calculate how the vacuum energy depends on $\bar \theta$, finding
\es{eq:vac}{
E(\bar \theta) = - F_\pi^2 m_\pi^2 \sqrt{1 - 4 {m_u m_d \over (m_u + m_d)^2} \sin {\bar \theta \over2}}  \,,
}
which is suggestive because it implies that the theory has the lowest vacuum energy if $\bar \theta = 0$. That is, if $\bar \theta$ was somehow dynamical then it would want to relax to $\bar \theta \to 0$, in which case we would have a dynamical solution to the strong-{\it CP} problem. This is, in fact, exactly how the axion solves the strong-{\it CP} problem! 

Specifically, for the axion effective theory we introduce a massless, pseudo-scalar field $a(x)$ that has the dimension-five interaction with QCD:\footnote{Caution: $a$ is used to denote both the axion field $a(x)$ and the QCD color index.}
\es{eq:axion}{
{\mathcal L} \supset {1 \over 2} \partial_\mu a \partial^\mu a  -{g_s^2 \over 32 \pi^2} { a \over f_a} \, G_{\mu \nu}^a \tilde G^{a \mu \nu} \,.
}
Here, $f_a$ is the UV cut-off of the theory, which we will discuss more shortly and which is referred to both as the Peccei-Quinn (PQ) scale and as the axion decay constant.  For the moment, think of it simply as a UV cut-off that is well above the QCD scale (in fact, we will see that we need $f_a \gtrsim 10^9$ GeV).  Now, we can perform a chiral rotation to remove the combination of Lagrangian terms $\propto$$(a/f_a + \bar \theta) G_{\mu \nu}^a \tilde G^{a \mu \nu}$ but at the expense of modifying the mass matrix in the chiral Lagrangian such that $M \to e^{i({a/f_a + \bar \theta)/2 }}M$. This implies, referring to~\eqref{eq:vac}, that QCD induces a potential for the axion of the form
\es{eq:axion_pot}{
V(a) = - F_\pi^2 m_\pi^2 \sqrt{1 - 4 {m_u m_d \over (m_u + m_d)^2} \sin {1 \over 2} \left( \bar \theta + {a \over f_a} \right) }  \,.
}
Now, when the axion minimizes its potential it dynamically sets $\langle a / f_a \rangle = - \bar \theta$, such that effectively there is no QCD $\theta$-term and the neutron EDM vanishes dynamically.  In the process, if we expand out~\eqref{eq:axion_pot}, we see that the axion acquires a small mass:
\es{eq:ma}{
m_a \approx {F_\pi m_\pi \over f_a} {\sqrt{ m_u m_d} \over m_u + m_d} \approx 6.0 \times 10^{-6} \left( {10^{12} \, \, {\rm GeV} \over f_a} \right) \, \, {\rm eV} \,.
}
Note that $m_a$ is zero if one of the quark masses vanish because in this case the $\bar \theta$ term is unphysical and does not lead to a non-trivial potential.  This is because in that case the $\bar \theta$ term can be completely removed by a chiral rotation of the massless fermion, since there is no mass term that transforms non-trivially under the chiral rotation for the massless fermion. On the other hand, substituting the accepted values $m_u \approx 2.3$ MeV, $m_d \approx 4.8$ MeV, $F_\pi \approx 92$ MeV, and $m_\pi \approx 140$ MeV leads to the result in the second equality in~\eqref{eq:ma}.

Note that the neutron EDM can also be calculated using the chiral Lagrangian with non-trivial $\bar \theta$ dependence through the modified mass matrix $M \to e^{i \bar \theta/2} M$.  This is a relatively textbook-level derivation, so we refer the reader to any number of field theory textbooks that include this calculation, such as {\it e.g.}~\cite{Srednicki:2007qs}.  Here we briefly outline how the calculation proceeds. In the presence of non-zero quark masses the chiral Lagrangian includes terms like 
\es{eq:nucleon}{
{\mathcal L} \supset - c_3 {\rm tr} \left[ M U + M^\dagger U^\dagger \right] \bar N \left( U^\dagger P_L + U P_R\right) N \,,
}
where $c_3$ is a coefficient of the effective theory, $P_{L}$ and $P_R$ are left- and right-handed projection operators, respectively, and $N = (p \, \, n)^T$ is the proton-neutron doublet under $SU(2)_{\rm isospin}$.  Expanding out this Lagrangian for small $\bar \theta$ one finds a term of the form ${\mathcal L} \sim \bar \theta \pi^+ \bar p n$, leaving off other pre-factors other than $\bar \theta$. Through a one-loop process this induces a neutron EDM $d_n \approx 3 \times 10^{-16} \bar \theta$ e$\cdot$cm (see~\cite{Srednicki:2007qs}). Note that the EDM is around two or three orders of magnitude smaller than our naive estimate based on the analogy with water, in part because it arises at one-loop and also in part because of the small quark masses.  That is, by naive dimensional analysis we may guess that 
\es{}{
d_n \sim {\bar \theta \over 4 \pi} {m_u m_d \over m_u + m_d} {1 \over m_\pi^2} \sim 10^{-16} \, \, {\rm e} \cdot {\rm cm} \,,
}
which is remarkably close to the answer obtained from the chiral Lagrangian after doing the loop computation. Here, the $1/(4 \pi)$ is a (rough) loop factor.  The factor ${m_u m_d \over m_u + m_d}$ is motivated by the fact that the EDM should vanish if either quark mass vanishes, and then the $1/m_\pi^2$ is there from dimensional analysis (in the actual computation, this factor arises from a pion propagator). 
Comparing to data, we can conclude that $|\bar \theta| \lesssim 10^{-10}$, which is an incredibly stringent requirement considering that $\bar \theta$ would naturally be expected to take a value anywhere from $- \pi$ to $+\pi$.

As a side-note, at $\bar \theta = 0$ the Lagrangian in~\eqref{eq:nucleon} is responsible for generating a neutron-proton mass difference, since the neutron and proton would have exactly the same masses if not for the quark masses.  The measured neutron-proton mass splitting is $Q_0 = m_n - m_p \approx 1.293$ MeV.  Expanding out to order $\bar \theta^2$ in small $\bar \theta$ one finds an additional mass splitting contribution $\delta Q$ of the form~\cite{Ubaldi:2008nf,Blum:2014vsa}:
\es{eq:Q}{
\delta Q \approx {f_\pi \bar g_{\pi NN} \over 2} \left( {m_d - m_u \over m_d + m_u} \right) \bar \theta^2 \approx 0.37 \, \, \bar \theta^2  \, \, {\rm MeV}  \,,
}
where $\bar g_{\pi NN} \approx 0.023$ is a coupling constant in the chiral Lagrangian.  Given current constraints on $\bar \theta$ we see that it contributes at a negligible level to the neutron-proton mass difference. In principle, the neutron-proton mass difference is very important; the abundance of light elements produced during big bang nucleosynthesis (BBN) is exponentially sensitive to this difference. It is thus interesting to imagine the scenario in which case $\bar \theta$ could have been much larger. From~\eqref{eq:Q}, we see that for $|\bar \theta| \lesssim 1$, roughly, the value of $\bar \theta$ would not qualitatively affect BBN or any other observable. This was looked at carefully in~\cite{Ubaldi:2008nf}, where it was concluded that so long as $|\bar \theta| \lesssim 0.1$, roughly, the universe would look essentially indistinguishable from one with $\bar \theta = 0$. This observation suggests that in a universe where some problems might have an anthropic origin ({\it e.g.}, the cosmological constant problem and perhaps the hierarchy problem), the strong-{\it CP} problem likely does not.  This is because there does not seem to be an environmental reason to prefer $\bar \theta \sim 0$ to the precision at which we measure this to be true. (With that said, one cannot be too sure: perhaps $\bar \theta$ is correlated with a more important observable in the full theory.)  One can take this as additional motivation for finding a dynamical solution to the strong-{\it CP} problem.

On the other hand, unlike the electroweak hierarchy problem the strong-{\it CP} problem admits technically natural solutions that tune $\bar \theta = 0$ in the UV.  This is because within the SM the neutron EDM arises at three-loop order and is well below the upper limit from data. Thus, if $\bar \theta$ is set to zero at high energy scales, the observable EDM generated in the IR will still be well below the upper limit.  In contrast, the Higgs mass runs quadratically. If the Higgs mass is set to zero in the UV, it will acquire a large value in the IR. This is why we say that the strong-{\it CP} problem is not a problem of technical naturalness, while the electroweak hierarchy problem is.  The strong-{\it CP} problem could in principle be solved in the UV, while the hierarchy problem must be solved in the IR to have a natural solution.  Non-axion, UV solutions to the strong-{\it CP} problem typically achieve $\bar \theta = 0$ at high energies through discreet symmetries such as {\it P} or {\it CP}, though these symmetries must be spontaneously broken at intermediate energy scales~\cite{Nelson:1983zb,Barr:1984qx,Choi:1992xp,Babu:1989rb,Barr:1991qx}. The trick is to generate the full {\it P} and {\it CP} violation observed in the SM at the intermediate scale without also inducing $\bar \theta$, which can be difficult without introducing extra fine.  However, this is achievable in a variety of models.    

\subsection{Axion models}

In the previous section we motivated the axion ``from the ground up" as a dynamical solution to the strong-{\it CP} problem. This led to the hypothesis of the Lagrangian in~\eqref{eq:axion}, where the axion has a dimension-five coupling with QCD. Let us now discuss how such an effective theory could come out of UV physics. For an extensive review of axion models, see~\cite{DiLuzio:2020wdo}.  Here, we will work through one model completely and then discuss, at a more heuristic level, a few others.

\subsubsection{Field theory axion models}

Field theory realizations of the axion have it emerging as the pseudo-Goldstone boson of a $U(1)$ symmetry, known as the PQ symmetry, that is broken at a high scale $f_a$.\footnote{This is in contrast to extra-dimensional models, where the axion emerges as the component of a higher-dimensional gauge field; these models are discussed more shortly.} Let us consider the case of
\es{eq:PQ}{
{\mathcal L}_{PQ} = | \partial \Phi |^2 - {\lambda } \left( |\Phi |^2 - {v_a^2 \over 2}\right)^2 \,,
}
where $\Phi(x) = {r(x) + v_a \over \sqrt{2}} \exp \left[ i {a(x) \over v_a} \right]$ is a complex scalar field that we divide into a massive radial component $r(x)$ and the Goldstone mode $a(x)$. The radial mode mass is $m_r=\sqrt{\lambda} v_a$, with $v_a$ the VEV, while the axion is exactly massless  and periodic with period $2 \pi v_a$.  While this sector gives rise to our massless axion field, it does not explain its coupling to QCD. We must couple $\Phi$ to fields charged under QCD in order to acquire such an interaction.

As an example, we consider the  Kim-Shifman-Vainshtein-Zakharov (KSVZ)~\cite{Kim:1979if,Shifman:1979if} model, where we add $N$ identical vector-like fermions $Q_i$, $i = 1, 2, \cdots, N$, with charges $(3,1,0)$ under $SU(3)_c \times SU(2)_L \times U(1)_Y$.  The electroweak representations of the $Q_i$ could be more complicated with only minor modifications to the story, but for simplicity let us start with the case where the vector-like fermions are only charged under the strong force.  We take $\Phi$ to be a SM singlet. Then, we are able to write down the interaction terms between the $Q_i$ and $\Phi$: 
\es{eq:full}{
{\mathcal L} \supset \sum_i \bar Q_i i \slashed{D} Q_i - \sum_i (y_i \bar Q_i Q_i \Phi + {\rm h.c.}) \,,
}
where the $y_i$ are Yukawa coupling constants.  The scalar sector Lagrangian in~\eqref{eq:PQ} is invariant under the $U(1)$ symmetry $\Phi \to e^{i \alpha} \Phi$, and the full Lagrangian -- including the terms in~\eqref{eq:full} -- may also be made invariant by allowing the fermion fields to transform as $Q_i \to e^{i \alpha \gamma_5 / 2} Q_i$.  Of course, this symmetry is anomalous, but let us put that point aside for the moment, as this is exactly how we will end up generating the coupling of the axion to QCD.

Let us now consider the theory at energy scales well below $v_a$, so that we may replace $\Phi \to {v_a \over \sqrt{2}}  \exp \left[ i {a(x) \over v_a} \right]$. We then recognize the quark masses as $m_i = \lambda_i v_a / \sqrt{2}$, allowing us to write 
\es{}{
{\mathcal L} \supset  \sum_i \bar Q_i i \slashed{D} Q_i - \sum_i (m_i \bar Q_i Q_i e^{i {a(x) \over v_a}}+ {\rm h.c.}) \,.
}
We can remove the axion phases by applying an axial transformation to the quark fields, but now we pay attention to the anomaly. That is, we take $Q_i \to e^{-i \gamma_5 a / (2 v_a)} Q_i$, which removes the axion from the equation above but induces the term
\es{}{
{\mathcal L} \supset {g_s^2 \over 32 \pi^2} a {N \over v_a} G_{\mu \nu}^a \tilde G^{a \mu \nu} \,.
}
Referring to~\eqref{eq:axion} we can then identify the decay constant as $f_a = v_a / N$.  We can then integrate out the heavy fermion $Q_i$, such that we are only left with our axion EFT. 

 The integer $N$ is known as the {\it domain wall number}, and in certain cosmological histories for the axion it plays an important role.  The reason is that the QCD-induced potential for the axion is periodic with period $2 \pi f_a$, but in the full UV completion the axion has periodicity $2 \pi f_a N$. In the next section we will discuss topologically-protected field configurations that can arise in axion cosmology known as axion strings, which are characterized by the fact that in going around a circle surrounding a string the axion acquires a $2 \pi f_a N$ phase shift.    In the presence of the QCD-induced potential, however, domains walls form between the strings, and with $N > 1$ these domain walls are stable, which presents a serious cosmological problem (we discuss this problem further shortly).  Note, however, that the domain wall number is only physical in the context of the UV completion of the theory; in the context of the axion EFT it is simply a change in what we mean by $f_a$ by some integer multiple.  
 
 One issue that axion theories face is the so-called ``PQ quality problem"~\cite{Georgi:1981pu,Dine:1986bg,Kamionkowski:1992mf,Barr:1992qq,Holman:1992us}.  For the axion to solve the strong-{\it CP} problem, the PQ symmetry needs to be a really good symmetry, as otherwise the axion will not completely cancel the neutron EDM.  This is completely fine from a field theory perspective, but it is somewhat unexpected in the context of quantum gravity. Quantum gravity is expected to not respect global symmetries (see, {\it e.g.},~\cite{Harlow:2018tng} and references therein).  Some intuition behind this claim comes from the following thought experiment. Let us imagine that a black hole accretes and then ``eats" a chunk of matter charged under some global symmetry. Black holes have no hair, which means that they are only characterized by their mass, spin, and gauge charges.  This means that the resulting back hole will not reflect in any way the global charge that it ate; that charge is destroyed. In particular, if we wait long enough for the black hole to Hawking evaporate away, we will be left with a state with no global charge. Thus, by throwing global charge into the black hole we explicitly violated the global symmetry. Reasoning along this direction motivates us to consider an explicit breaking of the PQ symmetry at the Planck scale:
 \es{eq:PQ_quality}{
 {\mathcal L} \supset \lambda {|\Phi|^{2n} \Phi^m \over m_{\rm pl}^{2n + m - 4} } + {\rm h.c.} \,,
 }
 where $n$ and $m$ are integers, $\lambda$ is a dimensionless coupling constant, and $m_{\rm pl}$ is the Planck scale where quantum gravity is expected to become important. In this case we find, roughly,
 \es{}{
 \langle a \rangle  + \bar \theta \sim \lambda \left( {f_a \over m_{\rm pl}} \right)^{2n + m} \left( {m_{\rm pl} \over \Lambda_{\rm QCD}} \right)^{4} \,,
 }
 with $\Lambda_{\rm QCD}$  the QCD confinement scale and assuming $\lambda$ is real for simplicity. Unless $n$ and $m$ are really large or $\lambda$ is really small, than the right hand side of the equation above will evaluate to a large number, telling us that the axion will not solve the strong-{\it CP} problem.  This seems like a pretty big problem for field-theory-based axion models.

 \subsubsection{Extra dimension and string theory axion models}
 
 On the other hand, who said that $\lambda$ has to be large? Indeed, we motivated operators of the form in~\eqref{eq:PQ_quality} by speculating about what quantum gravity might do, but to actually compute such terms in an EFT one would need an actual theory of quantum gravity. Luckily for us, we do have such a theory: string theory.  We can then ask whether or not axions with good PQ quality appear in string theory? The answer is a resounding yes: axions appear everywhere in string theory, and they can have very high PQ quality (see~\cite{Demirtas:2021gsq} for a recent study of the PQ quality in type IIB string theory compactifications).  The key to high-quality axions in string theory is that the PQ symmetry is secretly tied to a gauge symmetry, and quantum gravity does protect gauge symmetries.  String theory is a ten dimensional theory, but we live in a four dimensional world. Thus, we must compactify the other six dimensions (or 7 for M-theory). String theory has normal gauge field and antisymmetric form fields of various dimensions, which are also gauge fields.  After compactification, these fields can behave like scalars from the perspective of the four-dimensional universe. For example, suppose we have a gauge field $A_M$, with $M= 0, 1, \dots, 9$. Let $\mu = 0, 1, 2,3$ be the indices of four dimensional Minkowski space, with $M > 3$ denoting the compact directions. Then, the $A_M$ with $M > 3$ appears as scalars for a low energy observer making computations well below the compactification scale. The zero-modes of these scalars are massless and can behave as axions from a four-dimensional perspective (see {\it e.g.}~\cite{Svrcek:2006yi} for details).  These axions can still acquire non-local contributions to their potentials, however, for example from QCD instantons.  String theory and gravitational instantons can also contribute to the potential, and these do violate the PQ symmetry. However, such contributions are exponentially suppressed by the instanton action (hence, why I started this paragraph by asking why $\lambda$ necessarily has to be large). Supersymmetry that survives below the Planck scale can provide further parametric suppression of the explicit PQ violating terms. In summary, modern string theory studies using explicit compactifications do find that sufficiently high-quality axions emerge to be candidates for the QCD axion (see~\cite{Demirtas:2021gsq} and references therein).

 To make these concepts more concrete, let us consider a toy five-dimensional orbifold construction that yields a four-dimensional axion and illustrates some of the key points above.  See~\cite{Sundrum:2005jf} for TASI lectures on orbifold and 5D models.  Here, we closely follow the discussion presented in~\cite{Arkani-Hamed:2003xts,Choi:2003wr}.  Let us consider a theory in 5D with the fifth direction, with coordinate $y$, compactified on a circle of radius $R$ modded out by a $Z_2$ symmetry ($S_1 / Z_2$).  Here, the $Z_2$ symmetry is taken to identify $y \to - y$, such that the fundamental domain of the firth coordinate may be taken to be $(0,\pi R)$.  Five-dimensional fields must have definite parity under the $Z_2$. Let us consider a $Z_2$-odd 5D $U(1)$ gauge field $B_M$, with $M = 0,1,2,3,5$ ($\mu = 0,1,2,3$):
 \es{}{
 B_\mu(-Y) = -B_\mu(Y) \,, \qquad  B_5(-Y) = B_5(Y) \,.
 }
 One way to see that $B_\mu$ and $B_5$ must transform differently under $Z_2$ parity is that the one-form $B = B_M dx^M$ must transform homogeneously, and $dx^\mu \to dx^\mu$ while $dy \to - dy$.  The 5D action includes
 \es{eq:5D}{
 {\mathcal L}_{5D} \supset \int d^4x \int_0^{2\pi R} dy \left( {1 \over 4 g_5^2} B_{MN} B^{MN} + \kappa \epsilon^{MNPQR} B_M {\rm tr}\big(F_{NP}F_{QR}\big) \right)\,,
 }
 where $g_5$ is the 5D gauge coupling constant, $\kappa$ is a dimensionless coupling constant, and $F_{NP}$ denotes the field strengths of whatever other gauge fields may be in the theory, with the trace over the gauge indices. Note that the second term above is a Chern-Simons term and arises, for example, at one-loop (through triangle diagrams with three external gauge bosons) by integrating out heavy matter charged under both the $U(1)$ gauge group with field strength $B_{MN}$ and the other non-abelian gauge groups with field strength $F_{MN}$ (for simplicity, let us assume there is only one additional non-abelian gauge group).   

 Upon compactification the zero mode of $B_5$ behaves like an axion from the perspective of the 4D effective theory.
 Moreover, the 5D gauge symmetry of $B_M$ prevents explicit terms being added to the Lagrangian in~\eqref{eq:5D} that would give a potential for $B_5$.  Let us define the 4D axion $a$ through a Wilson loop by
 \es{}{
 e^{2 \pi R i a(x_\mu)} \equiv e^{i\int_0^{2 \pi R} dy \, B_5(x_\mu,y)} \,,
 }
 where we have explicitly separated the 4D coordinates $x_\mu$ and the 5D coordinate $y$.  With this definition the Lagrangian in~\eqref{eq:5D} generates the following terms in the 4D Lagrangian:
 \es{eq:5D_2}{
 {\mathcal L}_{4D} &\supset \int d^4x \left( {2 \pi R \over 2 g_5^2} (\partial_\mu a)^2 + (2\pi R \kappa) a F^a_{\mu \nu} \tilde F^{a \mu \nu} \right) \,, \\
 &=\int d^4x \left( {1 \over 2} (\partial_\mu a)^2 + {g_f^2 \over 32 \pi^2} {a \over f_a} F^a_{\mu \nu} \tilde F^{a \mu \nu} \right) \,,
 }
 where in the last step we have rescaled the axion field to make it canonically normalized and then defined the decay constant through:
 \es{eq:fa_orb}{
 f_a = {1 \over 32 \sqrt{3} \pi^{5/2} \kappa} {1 \over g_5 \sqrt{R}} = {1 \over 64 \pi^3 \kappa g_4} {1 \over R} \,.
 }
 Keep in mind that $g_5$ has dimensions of mass$^{-1/2}$.
 Note that we have also changed the normalization of the non-abelian gauge group with field strength $F$, and coupling $g_f$, to make it canonically normalized, which brings in the factor of $g_f^2$ in~\eqref{eq:5D_2}.  (We change the normalization of $F$ to make contact with previous equations for the axion.)  We have also identified the 4D gauge coupling $g_4$ of the $U(1)$ with the 5D gauge coupling through the relation $2 \pi R / g_5^2 = 1/g_4^2$.

 Note that there is no zero-mode for the 4D gauge field $B_\mu$, since it is odd under the $Z_2$ parity. This implies that the low energy theory has the almost-massless axion but no other light degrees of freedom that descend from $B_M$.  Had we not used the orbifold picture but simply taken the fifth dimension to be an $S_1$, the axion picture would have carried through analogously, but we would also have a massless 4D $U(1)$ theory. 

 As a side-note, suppose we have in mind a model where there is grand unification at the compactification scale $R^{-1}$, such that $M_{\rm GUT} \sim R^{-1}$. Then, we may perform the following estimate for $f_a$. First, we may imagine that $\alpha = g_4^2 / (4 \pi) \sim 1 / 25$, which is the value of the fine structure constant at the GUT scale in supersymmetric unification models. Next, if we have in mind that the Chern-Simons term arises at one-loop, then it is natural to expect $\kappa \sim \alpha / (4 \pi)$. Substituting these values into~\eqref{eq:fa_orb} then allows us to identify $f_a \sim 0.2 \times M_{\rm GUT}$.

 In the theory discussed so far, quantum gravity would not add more terms to the Lagrangian beyond those in~\eqref{eq:5D_2}, with the possible exception of a gravitational Chern-Simons term involving both $B_M$ and the 5D curvature; this term could contribute a potential for $a$ through gravitational instantons. Instantons in $F$ 
 can also contribute to the axion potential.  It may seem odd, however, that any potential at all can be generated for $a$, since such a potential violates the 5D gauge symmetry. What is important to recognize, however, is that the compactification already broke the gauge symmetry, but it breaks it in a special way such that the broken symmetry is only visible to non-local observables that can probe the topology of the spacetime. Locally the theory is gauge invariant, which ensures that the theory is still we behaved in the UV.  We can explicitly see this point by recalling the 5D gauge symmetry 
 \es{}{
 B_M \to B_M + \partial_M \Delta \,.
 }
 A special case of the gauge symmetry is $a \to a + {\rm constant}$, but the Lagrangian in~\eqref{eq:fa_orb} is not invariant under such a gauge transformation precisely because of the instanton configurations in $F$ that give non-vanishing contributions at the infinite boundary of Minkowski space. However, these instanton contributions are exponentially suppressed.

 Another source of PQ breaking in this model could come from 5D matter fields that are charged under the $U(1)$. Suppose that we have a 5D scalar field $\Phi$ with a mass $M$ that is charged with some charge $q$ under the $U(1)$ symmetry. The covariant derivative for this scalar field is $D_M \Phi = (\partial_M + i q \epsilon(y) B_M) \Phi$, where $\epsilon(y) = - \epsilon(-y) = 1$ guarantees that $\Phi$ transforms correctly under $Z_2$ parity. Importantly, note that $B_5$ appears non-derivatively in the covariant derivative, meaning that the symmetry $a \to a + {\rm constant}$ is not guaranteed to be respected by the $(\partial_M \Phi)^2$ term in the 5D Lagrangian. However, the 5D gauge symmetry (and thus the axion shift symmetry) should be respected locally; it is only non-local quantities in the IR that can see the gauge symmetry breaking.  A careful calculation of this effect 
 first decomposes $\Phi$ into Kaluza-Klein modes, computes the 4D quantum effective potential from each mode, and then sums the resulting terms over the Kaluza-Klein modes to derive the axion potential induced by $\Phi$. The result of this calculation is that, approximately,~\cite{Pilo:2003gu,Sundrum:2005jf} 
 \es{}{
 V(a) \sim {1 \over 16 \pi^4} {M^2 \over R^2} e^{-2 \pi M R} \left[ 1 - \cos\left({a \over 1 / (2 \pi g_4 R)}  \right) \right] \,.
 }
 Crucially, the potential is exponentially suppressed by $e^{-2 \pi M R}$ for $M \gg R^{-1}$. This makes sense physically because the scalar field configurations that contribute to this potential must be non-local in order to probe the topology of the space; these are configurations which wrap around the $S_1$. However, the field $\Phi$ is massive, and so in traversing a distance $2 \pi R$ around the circle the propagator is suppressed by $e^{-2 \pi M R}$. (Think here of the Yukawa force resulting from a massive mediator.) Thus, so long as all of the fields charged under the $U(1)$ have masses much larger than $R^{-1}$, their contributions to the explicit PQ violating potential for $a$ can be made sufficiently subdominant to that from QCD.  Lastly, we note that the 5D orbifold theory described here is a toy example but illustrates many of the features found in more complicated string theory compactifications.

 \subsection{The axion EFT} 
 
 One picture that emerged from string theory is that of the ``axiverse"~\cite{Arvanitaki:2009fg}.  The axiverse is the idea that string theory does not just predict a single axion, but in fact string theory compactifications tend to predict large numbers of axion like particles. Since the explicit (non-QCD-induced) masses arise from instantons, the masses are set by exponential factors with the instanton actions in the exponential. It is thus not unexpected that one would obtain a roughly logarithmic distribution of ALP masses over a large mass range, potentially from well below {\it e.g.} $10^{-22}$ eV to values $\sim$$10^{-10}$ eV or higher. Explicit string theory compactifications provide concrete realizations of this picture ({\it e.g.},~\cite{Demirtas:2018akl,Halverson:2019cmy,Mehta:2021pwf}).  From an EFT perspective, let us consider $N$ axions that have the following set of dimension-five interactions with the SM (see, {\it e.g.},~\cite{Foster:2022ajl}):
 \es{eq:axion_EFT}{
{\mathcal L} \supset \sum_{i=1}^N &\left[ {g_s^2 \over 32 \pi^2} {c_s^i a_i \over f_a} G_{\mu \nu}^a \tilde G^{a \mu \nu} + {g_2^2 \over 32 \pi^2} {c_2^i a_i \over f_a} W_{\mu \nu}^a \tilde W^{a \mu \nu}  + {g_1^2 \over 32 \pi^2} {c_1^i a_i \over f_a} B_{\mu \nu} \tilde B^{\mu \nu}  + \right.\\
 &\left. \sum_f C_{aff} {\partial_\mu a_i \over 2 f_a} \bar f \gamma^\mu \gamma_5 f - {1 \over 2} (m_i^0)^2 a_i^2 \right] \,,
 }
 with $f$ SM fermions, $g_1$ and $g_2$ the couplings of $U(1)_Y$ and $SU(2)_L$, respectively, $B_{\mu \nu}$ and $W_{\mu \nu}^a$ the associated respective field strengths, and all of the $c$ parameters dimensionless coupling constants. Note that we have factored out a common dimension-full scale $f_a$.  We assume that the bare masses $m_i^0$, which might arise from {\it e.g.} string instantons, are small relative to that induced by QCD: $\Lambda_{\rm QCD}^2 / f_a \gg m_i^0$.  Then, below the QCD confinement scale the combination of fields $\sum_i c_s^i a_i$ acquires a large mass, such that in diagonalizing the mass matrix the  field
 \es{}{
 a_{\rm QCD} = {\sum_i c_s^i a_i \over \sum_i (c_s^i)^2} 
 }
 becomes an almost pure mass eigenstate with mass $m_a^{\rm QCD} \approx \Lambda_{\rm QCD}^2 / f_a$. This is the QCD axion. The other $N - 1$ axions, however, remain ultralight with masses $\sim$$m_i^0$.  By construction, the other $N - 1$ axion states do not couple to QCD but they still couple to the rest of the SM fields given in~\eqref{eq:axion_EFT}.  We thus arrive at the picture of one (relatively) heavy QCD axion and $N - 1$ ultra-light ALPs.  
 
 For this picture to be consistent with nature, the $m_i^0$ should not be so large that they would misalign the axion field and give rise to a neutron EDM. Given the bound $|\bar \theta| \lesssim 10^{-10}$ this implies $m_i^0 \lesssim 10^{-5} m_a^{\rm QCD}$. Given that we expect $m_a^{\rm QCD} \lesssim 10^{-3}$ eV we could infer that ALPs are best motivated for masses less than around $10^{-8}$ eV, though in reality the axion masses could still be larger than this is there is more non-trivial structure in the theory than assuming all of the dimensionless coefficients are order unity and uncorrelated. 
 
 Most important for this discussion, below the electroweak scale the ALPs acquire couplings to electromagnetism. For a generic axion $a$ we will parameterize the coupling as
 \es{eq:axion_photon_coupling}{
 {\mathcal L} \supset - C_{a\gamma\gamma} {\alpha_{\rm EM} \over 8 \pi} {a \over f_a} F_{\mu \nu} \tilde F^{\mu \nu} = g_{a\gamma\gamma} a {\bf E} \cdot {\bf B} \,,
 }
 where $g_{a\gamma\gamma} \equiv C_{a\gamma\gamma}  {\alpha_{\rm EM} \over 2 \pi f_a}$, $\alpha_{\rm EM} $ is the fine-structure constant, $F_{\mu \nu}$ is the electromagnetic field strength, $C_{a\gamma\gamma}$ is a dimensionless coupling constant, and ${\bf E}$ and ${\bf B}$ are the electric and magnetic fields. Working out the second equality above is a useful exercise.  This interaction plays a crucial role in many astrophysical and laboratory searches for axions.  Note that for the QCD axion $g_{a\gamma\gamma}$ has two contributions, with the most straightforward to understand contribution arising from UV contributions.
 The second contribution, however, appears only in the infrared (IR) and is generated by the mixing of the axion with the neutral pion below the QCD confinement scale. This may be worked out explicitly in the context of the chiral Lagrangian. The axion then acquires, in the IR, the $\pi^0$'s interaction with two photons, which is exactly of the form $\pi^0 F \tilde F$.  Carrying out this calculation yields the result that the IR contribution to $C_{a\gamma\gamma}$, which we call here $C_{a \gamma\gamma}^{\rm IR}$, is $C_{a\gamma\gamma}^{\rm IR} \approx - {2 \over 3} {m_u + 4 m_d \over m_u + m_d} \approx -1.92$.  Working this out from the chiral Lagrangian is also a nice exercise (see the end of~\cite{Svrcek:2006yi} if you are in search of a clean derivation).  Given that $C_{a\gamma\gamma}$ has an IR contribution, this means that axion-photon probes of the QCD axion are robust, despite the QCD axion being defined through its coupling to QCD. Removing the coupling to electromagnetism would require a conspiracy between the UV and IR contributions.

\subsection{Axion dark matter}

A few years after the introduction of the QCD axion as a tool for solving the strong-{\it CP} problem it was realized that axions, if they exist, could also explain the observed DM density~\cite{Preskill:1982cy,Abbott:1982af,Dine:1982ah}.  However, the language that we use when discussing axion DM is fundamentally different from the language that we use when describing heavier DM candidates like WIMPs. The reason is that axions are ultra-light bosons and have high occupation numbers per quantum state, meaning that axion DM is better described by the evolution of classical fields than by the motion of individual particles.  A simple way of estimating the occupation number is to compute the number of particles in a cubic volume of side length equal to the de Broglie wavelength. Locally, the DM velocity is $v \sim 10^{-3}$, which -- taking $\rho_{\rm DM} \sim 0.4$ GeV/cm$^3$, allows us to estimate the occupation number as ${\mathcal N} \sim 10^{27} (10^{-6} \, \, {\rm eV}/m_a)^3$.  Quantum corrections to the classical picture should be suppressed by factors $\sim$$1/\sqrt{\mathcal{N}}$, which are negligible, so from here on we will treat the axion DM as a classical field.

The classical axion DM field is subject to the following equation of motion in the early universe:
\es{eq:axion_EOM}{
\ddot a + 3 H \dot a - {1 \over R^2} \nabla^2 a + {\partial V(a,T) \over \partial a} = 0  \,,
}
where $R$ is the scale factor and the spatial derivatives are with respect to the co-moving coordinates.
This equation may simply be derived from the Euler-Lagrange equation for the axion in the presence of an FRW metric (it is a good exercise to derive this!).  Note that the zero-temperature axion potential $V(a,T=0)$ is given in~\eqref{eq:axion_pot} as calculated in chiral perturbation theory.  We can approximate ${\partial V(a,T) \over \partial a} \approx f_a m_a(T)^2 \sin(a/f_a)$, where $m_a(T)$ is the temperature-dependent axion mass.  For ALPs, we can usually assume that the mass is temperature independent and simply take $m_a(T) = m_a$. For the QCD axion, on the other hand, the mass has important temperature dependence. The temperature dependence arises from the fact that the QCD axion's potential is generated by QCD instantons, and QCD is at finite temperature $T$ in the early universe. For $T \gg \Lambda_{\rm QCD}$, QCD is deconfined and the QCD-induced potential $V(a,T)$ is negligible.  That is, $m_a(T) / m_a \to 0$ in this limit, with $m_a$ the zero-temperature mass.  On the other hand, for $T \ll \Lambda_{\rm QCD}$, we should expect $m_a(T) / m_a \to 1$.  The calculation of $m_a(T)$ is beyond the scope of these lectures, but at high temperatures well above $\Lambda_{\rm QCD}$, the mass may be calculated analytically in the so-called ``dilute instanton approximation," while for $T \approx \Lambda_{\rm QCD}$ lattice QCD results are most accurate (see, {\it e.g.},~\cite{GrillidiCortona:2015jxo,Borsanyi:2016ksw} and references therein for extensive discussions).  
For our purposes it is a reasonable approximation to take~\cite{Buschmann:2019icd}
\es{eq:m_aT}{
m_a(T)^2 = {\rm min} \left[ {\alpha_a \Lambda^4 \over f_a^2 (T / \Lambda)^n} , m_a^2 \right]  \,,
}
where $\Lambda = 400$ MeV, $n > {\rm few}$ is an index that parameterizes the steepness of the mass turn on, and $\alpha_a$ is chosen such that the two expressions above are equal at $T = \Lambda$.  This approximation is reasonable so long as one is not interested in the precise form of $m_a(T)$ right around $T \sim \Lambda$.  Ref.~\cite{Borsanyi:2016ksw}, for example, finds $n \approx 8.2$.

At this point the discussion of axion cosmology bifurcates into two distinct possibilities, depending on whether or not the PQ symmetry is restored after inflation:
\begin{itemize}
\item Option 1: {\bf PQ restored after inflation}: The theory undergoes spontaneous symmetry breaking and develops topological defects known as axion strings, which are singularities in the axion-only picture. The UV theory is restored at the string cores. The axion initial conditions vary from Hubble patch to Hubble patch and these patches come into dynamical contact with each other in the subsequent evolution of the Universe. One must solve, usually numerically, the full non-linear axion equations of motion, including the UV completion, to understand the production of axion DM.
\item Option 2: {\bf PQ not restored after inflation}: Inflation exponentially dilutes the topological defects generated during PQ symmetry breaking, and inflation does what it does best -- it creates homogeneous initial conditions such that all initial Hubble patches that eventually lead into our horizon today have the same initial value for the axion field, which we label as $a(t = 0) / f_a \equiv \theta_i$.  The quantity $\theta_i$ is called the initial misalignment angle.  If $|\theta_i| \ll \pi$, we can linearize the axion field equations. We can also drop the gradient term in~\eqref{eq:axion_EOM}, since the initial conditions are homogeneous. 
\end{itemize}
The distinction between these two cosmological histories is not just relevant for the QCD axion; both possibilities could also play out for ALPs. There are very important differences between the two cosmologies, as we discuss further below.

\subsubsection{PQ not restored after inflation}

Let us begin by assuming that the PQ symmetry is broken before inflation and not restored afterwards. Then, after inflation, we want to solve the equation
\es{eq:a_simple}{
\ddot a + 3 H \dot a + m_a(T)^2 a = 0 \,,
}
where we assume for simplicity $|\theta_i | \ll \pi$ so that we can linearize the $\sin(a/f_a)$ term (note, however, that the dynamics when $\theta_i \sim \pi$ are non-trivial and interesting~\cite{Arvanitaki:2019rax}).  Since $H \sim T^2 / m_{\rm pl}$ at high temperatures when $H \gg m_a$, we can ignore the mass term in the equation of motion, in which case a solution is simply $a = {\rm constant} = \theta_i \times f_a$.  There is also a class of solutions with non-zero initial velocities, for which $a \propto T$ at early times, but this class of solutions does not have the right initial conditions as given by inflation of zero initial velocity (but see~\cite{Co:2021lkc} and references therein).  The axion field is frozen at its initial misalignment angle until $m_a \gtrsim 3 H$, beyond which point the axion field behaves like a damped harmonic oscillator.  

First, consider the case where $m_a(T) \equiv m_a$ does not depend on temperature, which is not the case for the QCD axion but could be the case for an ALP.  Let us assume  that all of the relevant dynamics happen during the radiation-dominated epoch, which is usually the case since the axion should be acting like DM before matter-radiation equality.  Recall that $H = {1 \over 2 t}$ during the radiation dominated epoch. Then, the solution to~\eqref{eq:a_simple} is
\es{}{
{a(t) \over f_a }= \theta_i \left( {2 \over m_a t} \right)^{1/4} \Gamma\left( {5 \over 4} \right) J_{1/4} (m_a t) \,,
}
where $J_{1/4}$ is the Bessel function of the first kind. Expanding the above solution at large values of $m_a t$ we find that at late times, after the axion field has begun oscillating, 
\es{}{
{a(t) \over f_a }\approx \theta_i {1 \over \sqrt{\pi}}\Gamma\left( {5 \over 4} \right)\left({2 \over m_a t}\right)^{3/4} \sin(m_a t + \phi) \propto \left( {T \over \sqrt{m_a m_{\rm pl}} } \right)^{3/2} \sin(m_a t + \phi) \,,
}
where $\phi$ is some unimportant phase. The energy density scales as $\rho_a \sim (\dot a)^2 + m_a^2 a^2 \propto T^3$, which is exactly how non-relativistic matter redshifts. This strongly suggests, and indeed it is the case, that the coherent fluctuations of the axion field behave like cold, non-interacting DM for cosmological purposes, so long as we probe the DM on scales larger than the de Broglie wavelength. 

Having established that at late times axions behave like cold DM, let us compute the relic abundance and determine which values of $\{ \theta_i, m_a, f_a\}$ are needed to match our universe.  If you want to perform the calculation precisely, a good approach is to first compute the axion number density 
\es{eq:n_aTR}{
n_a(T_R) = {\rho_a(T_R) \over m_a} \sim \theta_i^2 f_a^2 m_a \left({T_R \over \sqrt{m_a m_{\rm pl}} }\right)^3
}
 at some reference temperature $T_R$ which is well below where $3 H \approx m_a$ and the axion begins to oscillate.  The number density will simply be diluted by the expansion of spacetime: 
 \es{}{
 n_a(T) = n_a(T_R)\left( {R(T_R) \over R(T)}\right)^3 \,, \qquad T < T_R \,.
 }
 At $T_R$ the entropy density is $s = {2 \pi^2 \over 45} g_{*S}(T_R) T_R^3$. The entropy density scales with the scale factor as $s \propto R^{-3}$, even if $g_{*s}$ changes, which tell us that 
 \es{}{
 \left( {R(T_R) \over R(T_0)} \right)^3 = \left( {g_{*s}(T_0) \over g_{*s}(T_R) } \right) \left({T_0 \over T_R}\right)^3 \,,
 }
 where $T_0 < T_R$ is the temperature at a later epoch, for example today or matter radiation equality.  The energy density at $T_0$ is thus
\es{}{
\rho_a(T_0) = \rho_a(T_R) {g_{*s}(T_0) \over g_{*s}(T_R)} \left( {T_0 \over T_R} \right)^3 \sim \theta_i^2 {f_a^2 m_a^2 \over (m_a m_{\rm pl})^{3/2}} T_0^3 \,,
}
where in the last line we drop factors order unity, including the $g_{*s}$ factors.\footnote{Note that at temperatures well above $10^2$ GeV the entropy degrees of freedom is fixed at $g_{*s} \sim 10^2$, while at $T \sim 1 $ GeV this has dropped to $g_{*s} \sim 80$. During the QCD phase transition the degrees of freedom drops rapidly, such that $g_{*s} \sim 20$ at $T \sim 0.1$ GeV. At temperatures well below $\sim$1 keV $g_{*s}$ asymptotes to a value around 3.}
Just as we did when estimating the WIMP DM abundance, we can make a quick estimate by taking $T_0 = T_{\rm MR}$ to be the temperate of matter-radiation equality.  Then, the ratio $\rho_a(T_{\rm MR}) / T_{\rm MR}^4$ is an estimate  of the relic abundance relative to the observed relic abundance $\Omega_{\rm CDM} = 0.120 \pm 0.001$ by Planck~\cite{Planck:2018vyg}, since at matter-radiation equality the energy density in matter should be equal to that of radiation, which is $\sim$$T_{\rm MR}^4$.
 Thus, the relic abundance of axions $\Omega_a h^2$ today is approximately 
 \es{eq:fuzzy}{
 \Omega_a h^2 \sim 0.1 \theta_i^2 {f_a^2 \over T_{\rm MR} m_{\rm pl}}  \sqrt{{m_a \over m_{\rm pl} }}\sim 0.1 \theta_i^2 {f_a \over 10^{17} \, \, {\rm GeV}} \sqrt{{m_a \over 10^{-22} \, \, {\rm eV} } }\,.
 }
 
 The parameters $\{ m_a , f_a\}$ chosen to illustrate $\Omega_a h^2$ in~\eqref{eq:fuzzy} are chosen because they are the representative parameters for the DM model known as fuzzy DM.  As previously mentioned, fuzzy DM is characterized by having de Broglie wavelength of order galactic scales. The DM field feels a source of quantum pressure at scales below the de Broglie scale, which suppresses structure on small cosmological scales and leads to other non-trivial phenomena such as solitonic cores at the centers of halos (see~\cite{Hui:2016ltb} and reference therein).  Historically, fuzzy DM has been invoked as a solution to a number of small-scale structure problems in the distribution of DM on galactic and dwarf galaxy scales, such as the distribution of DM in dwarf galaxies.  For example, the solitonic cores as the center of fuzzy DM profiles could lead to cored DM density profiles, which matched observations in some dwarf galaxies. These anomalies pointed to the possibility of fuzzy DM with $m_a \sim 10^{-22}$ eV (see~\cite{Hui:2016ltb} for a discussion). However, fuzzy DM is now strongly constrained, with $m_a \lesssim {\rm few} \times 10^{-19}$ eV excluded~\cite{Dalal:2022rmp}, so it is unlikely that fuzzy DM is responsible for any of the currently known discrepancies between the cold DM paradigm and observations.
 
 Let us now turn to the QCD axion, for which the temperature dependence of $m_a(T)$ is crucial and for which $m_a f_a \approx 75$ MeV is fixed by~\eqref{eq:ma}.  Let us assume the temperature dependence in~\eqref{eq:m_aT} and also assume that the axion begins to oscillate at $T_{\rm osc} > \Lambda$, with $T_{\rm osc}$ given approximately by the condition $3 H(T_{\rm osc}) \approx m_a(T_{\rm osc})$. We will again drop factors order unity and simply take $H(T_{\rm osc}) \sim T_{\rm osc}^2 / m_{\rm pl}$. Given $T_{\rm osc}$, we can then compute the number density at this reference time using~\eqref{eq:n_aTR}. The number density will be diluted as $R^{-3}$ for $T \lesssim T_{\rm osc}$, which allows us to calculate the energy density at matter radiation equality just as we calculated it for the case of fixed $m_a$. Concretely, we may write 
 \es{eq:rho_a_T_MR}{
 \rho_a(T_{\rm MR}) = {m_a \over m_a(T_{\rm osc})} \rho_a(T_{\rm osc})  {g_{*s}(T_{\rm MR}) \over g_{*s}(T_{\rm osc})} \left( {T_{\rm MR} \over T_{\rm osc}} \right)^3 \,,
 }
 where $\rho_a(T_{\rm osc}) \approx {1 \over 2} m_a(T_{\rm osc})^2 f_a^2 \theta_i^2$. Following this through leads to the oscillation temperature
 \es{eq:T_osc}{
 T_{\rm osc} \sim \left( m_a m_{\rm pl} \Lambda^n \right)^{1 / (n+2)} \sim {\rm GeV} \left( {10^{12} \, \, {\rm GeV} \over f_a} \right)^{0.1} \,,
 }
 where we use $n \approx 8.2$. Note that we should be careful for $f_a \gtrsim 10^{16}$ GeV, since in this case $T_{\rm osc} \lesssim \Lambda$, in which case it is not appropriate to use the functional form we assumed for $m_a(T)$.  Using the oscillation temperature in~\eqref{eq:T_osc} and assuming the degrees of freedom to be the same between the oscillation temperature and matter radiation equality, which is not a great approximation but which is good enough for an order of magnitude estimate, leads to the result
 \es{eq:Omega_a_QCD}{
 \Omega_a h^2 \sim 0.1 \theta_i^2 \left( {f_a \over 3 \times 10^{11} \, \, {\rm GeV}} \right)^{1 + {1 \over n+2}} \sim 0.1 \theta_i^2 \left( {f_a \over 3 \times 10^{11} \, \, {\rm GeV}} \right)^{1.1}  \,.
 }
 This implies that the correct DM abundance for the QCD axion is naturally achieved, assuming an initial misalignment angle of order unity, for an axion mass around 10 to 20 $\mu$eV.
 
 Suppose that we insist on an axion decay constant at the scale of grand unification (GUT scale) ($f_a \sim 10^{16}$ GeV), perhaps motivated by string theory models~\cite{Svrcek:2006yi} or just the fact that we expect new physics to appear near the GUT scale as the strong and electroweak forces unify.  Taking $f_a \sim 10^{16}$ GeV in~\eqref{eq:Omega_a_QCD} implies that we naively overproduce the observed DM abundance, unless the initial misalignment angle is tuned to be small ($|\theta_i| \sim 5 \times 10^{-3}$).  Before inflation we should think of $\theta_i$ as being randomly distributed across Hubble patches with equal probability of being anywhere from $-\pi$ to $\pi$. Thus, the chance of randomly inflating from a Hubble patch such a small value of $\theta_i$ is low, about one in a thousand.  Other than random chance, there are two mechanisms commonly discussed for accommodating GUT scale QCD axion DM:
 \begin{enumerate}
 \item {\bf entropy dilution:}  Suppose that there is a period of early matter domination that stretches from at least the time at which the axion field starts oscillating to some later time. This epoch must end and reheat the SM to a temperature above a few MeV in order to be compatible with precision BBN. The additional entropy released into the SM during the reheating process, in going from matter domination back to radiation domination, dilutes the axion DM abundance and helps alleviate the fine-tuning problem for GUT-scale axions~\cite{Dine:1982ah,Steinhardt:1983ia,Lazarides:1990xp}.
 \item {\bf anthropic explanation:} A provocative and controversial proposal for explaining the small value of $\theta_i$ needed for GUT-scale axions was put forward in~\cite{Wilczek:2004cr,Tegmark:2005dy} relying on an anthropic argument.  Inflation, an exponentially rapid expansion of space-time, produces homogenous initial conditions where the initial value of the axion field is set by the value $\theta_i = a / f_a$ at the position we inflated from. However, all of space-time expands, and so there are therefore other patches of the universe in which the initial axion field takes on different values. Because the expansion of space is exponential, these patches will not come into causal contact with each other: our cosmological history post inflation only draws on regions of space with the same initial value of the axion field to exponential precision. This means, though, that there are regions of space today which had different initial values of $\theta_i$ and thus which have different DM densities -- we simply cannot access these patches. We therefore have a mechanism for populating a ``multiverse," where the only difference between patches is the DM density (of course, one could make this picture more complicated by having other parameters vary also).  Ref.~\cite{Tegmark:2005dy} suggested, based on a series of physical arguments  regarding {\it e.g.} halo and galaxy formation, that if the DM density was too much larger or smaller than observed in our universe, one might not have the correct conditions to form life. This suggests the possibility that while most of space experiences much larger $|\theta_i|$, our patch has a small $|\theta_i|$ (at the level of one in a thousand) because that is needed for us to be here to make that observation. Of course, it is exceedingly difficult to test such a theory, but it does mean that one should not discount the possibility of GUT-scale axions making up the observed DM abundance.
 \end{enumerate}
 
 \subsubsection*{Cosmological considerations for PQ broken before inflation}
 
 If the PQ symmetry is broken before or during inflation then that means that the axion field undergoes quantum fluctuations during inflation, which are stretched to become isocurvature perturbations.  Isocurvature perturbations are strongly constrained by measurements of the CMB, which in turn leads to constraints (and future discovery potential) for the axion.  Quantum fluctuations of the inflaton during inflation are responsible for generating the curvature perturbations that give rise to structure (for an overview of inflation, see these TASI lecture notes~\cite{Baumann:2009ds}).  Since the inflaton is the source of energy driving inflation, the quantum fluctuations in the inflaton affect all forms of matter: {\it i.e.}, they are fluctuations in the entropy density $\delta s / s$.  Since for matter, $\rho_{\rm matter} \propto T^3$ while for radiation $\rho_{\rm rad} \propto T^4$, this implies for example that for curvature perturbations 
 \es{eq:curvature_pert}{
{ \delta \rho_{\rm matter} \over \rho_{\rm matter} } = {3 \over 4} {\delta \rho_{\rm rad} \over \rho_{\rm rad}} \,,
 }
 where the $\delta \rho$ correspond to the corresponding perturbations in the energies densities of matter and radiation, respectively.  That is, at a fixed time slice one can think of the curvature perturbations as corresponding to common fluctuations in the temperature $\delta T$.  By gauge invariance one can also think of the curvature perturbations as corresponding to fluctuations in the ``clock" from place to place, with some places slightly ahead in their evolution and others slightly behind. The places slightly ahead in time correspond to those where, if we think instead in terms of a constant time slice, they have slightly lower $T$.  
 
 Just like the inflaton, the axion field -- if it exists during inflation -- also undergoes quantum fluctuations in de Sitter space.  In momentum space, let $\delta a ({\bf k})$ be the Fourier transform of the fluctuation $\delta a({\bf x})$. Then (see, {\it e.g.},~\cite{Baumann:2009ds}),
 \es{eq:pert_axion}{
 \langle \delta a ({\bm k}') \delta a({\bm k}) \rangle = (2 \pi)^3 \delta^3({\bm k}' + {\bm k}) P_a({\bm k}) \,, \qquad \Delta_a^2 \equiv {k^3 \over 2 \pi^3} P_a = {H_{\rm inf}^2 \over 4 \pi^2} \,,
 }
 where $H_{\rm inf}$ is the Hubble parameter during inflation.  Note that this two-point function is scale invariant. (As a reminder, during inflation the Universe expands exponentially, such that the scale factor is $R(t) \propto e^{H_{\rm inf} t}$ and $\dot R / R = H_{\rm inf}$ is constant.)  
 In position space, the two-point correlation function is given by
 \es{}{
\langle \delta a({\bm x}') \delta a({\bm x}) \rangle =  \int {d^3 k' \over (2 \pi)^3}  {d^3 k \over (2 \pi)^3} e^{i (k x + k' x')}  \langle \delta a ({\bm k}') \delta a({\bm k}) \rangle 
 }
 The variance of the perturbation $\delta a$ may be computed from the real-space two-point function as
 \es{}{
 \sigma_a^2 \equiv   \langle \delta a({\bm x}) \delta_a({\bm x}) \rangle = \int {dk \over k} \Delta_a^2  = \Delta_a^2 \log(k_{\rm UV} / k_{\rm IR} ) \,,
 }
 where $k_{\rm UV}$ and $k_{\rm IR}$ are UV and IR cut-offs, respectively.  This formula provides another demonstration that the power spectrum is scale invariant: the variance $\sigma_a^2$ has equal contributions from every decade in momentum space. Indeed, the spectrum must be scale invariant since, other than Hubble, there is no other scale in the problem, given that the axion is massless.
 
 Let us, for a moment, refer to the inflaton field as $\phi$. The quantum fluctuations of the inflaton field are of the same form as those of the axion.  That is, the perturbations of the inflaton field $\delta \phi$ obey the same relation as in~\eqref{eq:pert_axion}, again with $\Delta_\phi^2 = H_{\rm inf}^2 / (4 \pi^2)$.  Since the inflaton dominates the energy density of the universe, its perturbations source perturbations in the curvature ${\mathcal R}$: $\delta {\mathcal R} = H \delta \phi / \dot \phi$\footnote{To see this one needs to compute how $\delta \phi$ sources a curvature perturbation, but note that one should be careful with gauge dependence~\cite{Malik:2008im}.}, which implies that the curvature is also subject to a two point function as in~\eqref{eq:pert_axion} but with $\Delta_{\mathcal R}^2 = (H_{\rm inf} / \dot \phi)^2 \Delta_\phi^2$.  The curvature perturbations at a given mode number $k$ exit the horizon during inflation and remain frozen in place until they reenter the horizon later on during the post-inflationary evolution of the universe, when the wavelength of the perturbation becomes sub-horizon scale again. At this point the perturbations are dynamical and lead to structure formation from gravitational collapse. For example, it is these scalar perturbations that lead to the fluctuations observed in the CMB and, later on, to galaxy formation. By fitting the cosmological model to the CMB data on can determine $\Delta_{\mathcal R}^2 \approx 2 \times 10^{-9}$~\cite{Planck:2018vyg}.  
 
 The curvature perturbations source the matter perturbations, with typical fluctuations -- before gravitational collapse -- of $\delta \rho_{\rm matter} / \rho_{\rm matter} \approx \sqrt{\Delta_{\mathcal R}}$.  The perturbations in the axion field also source perturbations in the matter field if the axion becomes the DM, and these perturbations are uncorrelated with the curvature perturbations in the sense that the cross-correlation function between the two perturbations is zero. Moreover, while the curvature perturbations are subject to~\eqref{eq:curvature_pert} the axion perturbations in fact induce a relation closer to $\delta \rho_{\rm matter} = - \delta \rho_{\rm rad}$.  This is because the energy of a non-relativistic axion particle  is given by its rest mass, which is induced by QCD. Thus, at the QCD phase transition places of larger axion number densities draw more energy from the radiation field.  The axion-induced perturbations are known as isocurvature perturbations, and they are strongly constrained by CMB data since they show up differently in the CMB power spectrum than curvature perturbations.  The axion-induced isocurvature perturbations translate to perturbations in the matter energy density $ \delta \rho_{\rm matter}^{\rm iso} $ as 
 \es{eq:rho_iso_axion}{
 \delta \rho_{\rm matter}^{\rm iso} \sim {1 \over 2} m_a^2 \left( \theta_i f_a + \sqrt{\Delta_a} \right)^2 - {1 \over 2} m_a^2 \theta_i^2 f_a^2 \sim m_a^2 \theta_i f_a {H_{\rm inf} \over 2 \pi} \,,
 }
 and thus
 \es{}{
{ \delta \rho_{\rm matter}^{\rm iso} \over \rho_{\rm matter}} \sim {1 \over \pi}{ H_{\rm inf} \over |\theta_i |f_a} \,.
 }
 Planck measurements constrain the variance in ${ \delta \rho_{\rm matter}^{\rm iso} \over \rho_{\rm matter}} $ to be less than around 1\% of $\Delta_{\mathcal R}^2$~\cite{Planck:2018jri}, which implies that  ${ \delta \rho_{\rm matter}^{\rm iso} \over \rho_{\rm matter}}  \lesssim 5 \times 10^{-6}$.  We want to combine this constraint with~\eqref{eq:rho_iso_axion}, but to do so it is useful to use~\eqref{eq:Omega_a_QCD} to solve for the value of $|\theta_i|$ needed to produce the correct DM density. Doing this and re-arranging the inequality leads to the result
 \es{eq:iso_curvature_constraint}{
 H_{\rm inf} \lesssim 2 \times 10^9 \, \, {\rm GeV} \left( {f_a \over 10^{16} \, \, {\rm GeV}} \right)^{{n+3 \over 2 n + 4}} \approx 2 \times 10^9 \, \, {\rm GeV} \left( {f_a \over 10^{16} \, \, {\rm GeV}} \right)^{0.45}  \,.
 }
 This is a stringent result with dramatic cosmological implications. For example, if high-scale inflation is discovered, {\it e.g.} through primordial B-modes in the CMB polarization, then this would effectively rule out GUT-scale axions. Indeed, this has already happened.  In 2014 the BICEP2 collaboration claimed a discovery of B-mode polarization pointing to $H_{\rm inf} \sim 10^{14}$ GeV~\cite{BICEP2:2014owc}.  This result generated a period of soul searching and model building in the axion community (see, {\it e.g.},~\cite{Visinelli:2014twa,Kearney:2016vqw}), though it turns out that the BICEP2 result was not correct.  There is currently no evidence for primordial B-mode polarization in the CMB, and at present $H_{\rm inf} \lesssim 6 \times 10^{13}$ GeV~\cite{Planck:2018jri}.
 
 Lastly, we note that the axion isocurvature perturbations are also non-Gaussian~\cite{Kawasaki:2008sn}. Loosely speaking, the non-Gaussianity may be seen by noting that expanding~\eqref{eq:rho_iso_axion} one term further, so that $\delta \rho_{\rm matter}^{\rm iso} \sim m_a^2 f_a^2 \delta a + {1 \over 2} m_a^2 \delta a^2$, then while $\delta a$ obeys Gaussian statistics, $\delta a^2$ does not. For example, if we compute a three-point function, known as the bi-spectrum, between density perturbations, we find $\langle \delta \rho_{\rm matter}^{\rm iso} \delta \rho_{\rm matter}^{\rm iso} \delta \rho_{\rm matter}^{\rm iso} \rangle \sim \langle \delta a \delta a \delta a \delta a \rangle \sim   \Delta_a^4$, since the four-point function of $\delta a$ decomposes into a sum of the product of two-point functions.  (In contrast to the four-point function, note that three-point functions of zero-mean Gaussian random fields vanish.)   Working this through, from the absence of observed isocurvature non-Gaussianity one can then set a constraint on the scale of Hubble during inflation, though for GUT scale axions the constraint from the isocurvature two-point function is currently stronger. (See~\cite{Kawasaki:2013ae} for further discussion.)

 \subsubsection{PQ restored after inflation}
   
   Let us now imagine that the coin flips the other way, and the PQ symmetry is restored after inflation.  Then, in the subsequent thermal evolution of the universe we pass through the PQ phase transition, generating different initial misalignment angles from Hubble patch to Hubble patch. These Hubble patches come into causal contact with each other in the evolution of the Universe, such that our horizon today includes in its history earlier patches that had many different misalignment angles. Are we then supposed to average over all possible initial misalignment angles? The answer is more subtle than that, because {\it e.g.} the gradients that we threw away in~\eqref{eq:axion_EOM} will clearly become important, and we threw these gradients away when writing down~\eqref{eq:a_simple}. We also linearized the equations of motion in writing down~\eqref{eq:a_simple}, but with PQ broken after inflation there will be regions of space where the non-linearities are important, since the axion field will have maximal field excursion over the relevant region of space.  The biggest complication, though, comes from the fact that the axion-only equation of motion gives rise to solutions known as axion strings that have singularities at their cores. We will define the strings more precisely below, but roughly speaking that are characterized by the fact that going in a circle around a string core the axion will acquire a phase shift $\sim$$2 \pi f_a$ and come back to where it started, since the axion is a periodic field. Now imagine shrinking that circle around the string core. Clearly, the field becomes singular at the string core, since for a string with radius $\epsilon$ going in a circle of circumference  $2 \pi \epsilon$ needs to result in a field excursion $\sim$$2 \pi f_a$, and thus as $\epsilon \to 0$ we see that the derivative of the axion field diverges.  The apparent singularity at the core of the string is a sign that new physics comes into play to resolve the singularity. As we show below, the PQ symmetry is restored at the string core and the radial mode, while frozen out elsewhere, is dynamical within a distance $\sim$$1/f_a$ of the string core.  The axion strings that develop after the PQ phase transition are currently thought to be the dominant source of axion DM with the PQ symmetry broken after inflation~\cite{Gorghetto:2020qws,Buschmann:2021sdq}, so let us take a moment to understand better their properties.
   
\begin{figure}[htb]  
\begin{center}
\includegraphics[width=0.5\textwidth]{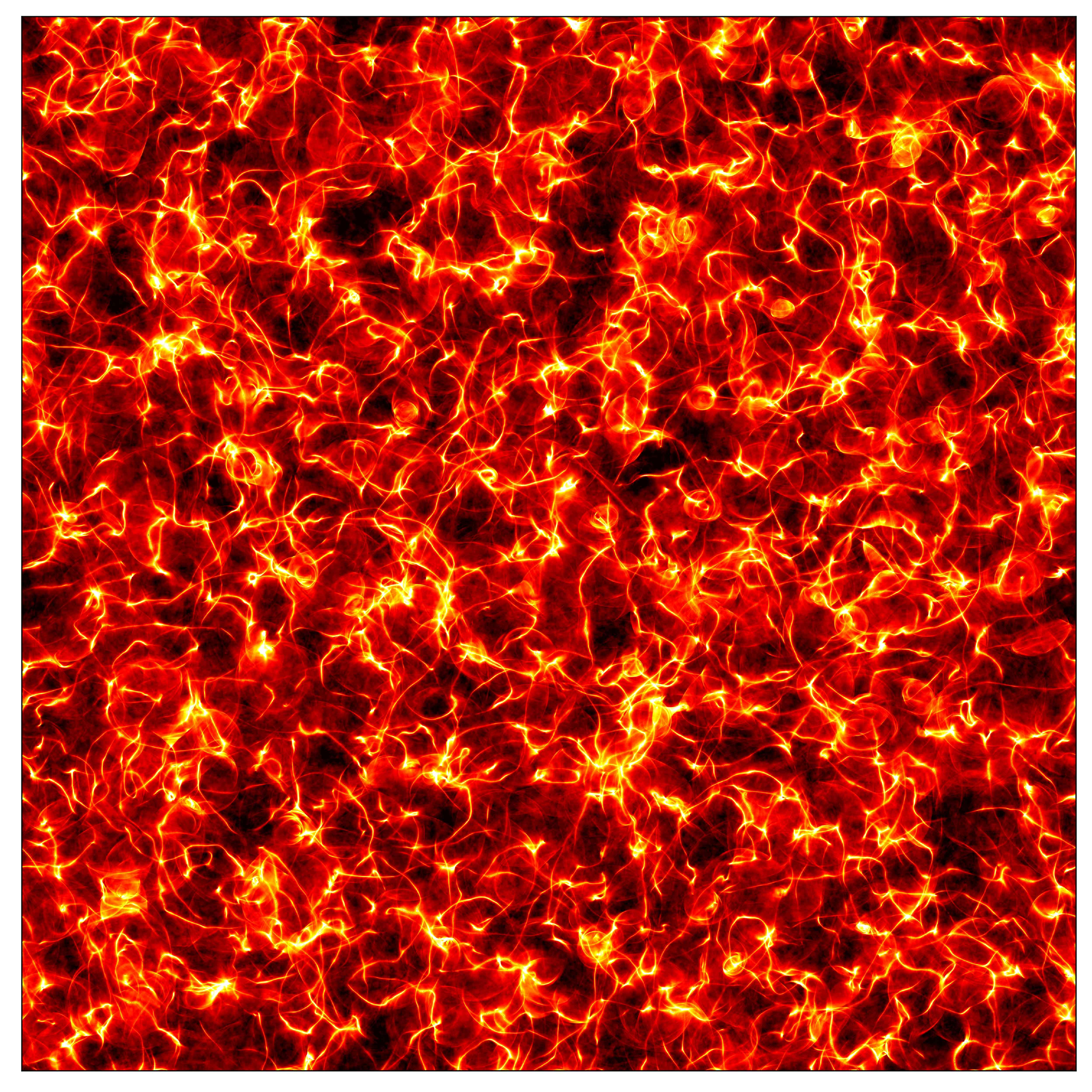}
\caption{ An illustration of axion strings and their axion radiation from a large-scale axion-cosmology simulation along the lines of that in~\cite{Buschmann:2021sdq}.  As the universe expands, the axion string network evolves to maintain the so-called scaling solution, with roughly one string per Hubble patch. The network evolves by emitting radiation in the form of axions. In the early universe these axions are relativistic, but at the QCD phase transition all of the axions acquire a mass (from QCD instantons) and the majority of the axions stop in-place and behave like cold DM from thereafter. (Figure courtesy Malte Buschmann.)}
\label{fig:axions}
\end{center}
\end{figure}

   In Fig.~\ref{fig:axions} we illustrate a snapshot of an axion simulation, based on the framework in~\cite{Buschmann:2021sdq}. The axion strings are clearly visible, which evolve by emitting relativistic axions that go on to form the DM abundance, as we discuss more below.

 Including the radial mode, we need to solve the differential equation
 \es{eq:EOM_string}{
 \ddot \Phi  + 3 H \dot \Phi - {\nabla^2 \over R^2} \Phi + {m_r^2 \over v_a^2} \Phi \left( |\Phi|^2 - {v_a^2 \over 2} \right) = 0 \,,
 }
 where $\Phi$ is the complex scalar field that undergoes PQ symmetry breaking, as given in the Lagrangian in~\eqref{eq:PQ}, and where $m_r = \sqrt{2 \lambda} v_a$ is the radial mode mass.  Here, $v_a$ is the VEV of the PQ field.  Note that we are interested in the string dynamics at temperatures above the QCD phase transition, so we ignore the QCD-induced potential. Physically, the string network will evolve dynamically from the PQ phase transition until the QCD phase transition. At the QCD phase transition we need to add the temperature-dependent potential for the axion to~\eqref{eq:EOM_string}. At this point in the cosmological history domain walls develop between the strings. The cosmology then depends drastically on whether or not the domain wall number is equal to or greater than unity. If the domain wall number is equal to one, then throughout the full $2 \pi v_a$ excursion of the axion field, the QCD axion potential, which is periodic with period $2 \pi v_a$, only encounters a single minimum. In this case, the domain walls quickly collapse, producing axions in the process (see, {\it e.g.},~\cite{Hiramatsu:2012gg,Buschmann:2019icd,OHare:2021zrq} for simulations of this collapse).   However, if   the domain wall number $N > 1$, then the QCD-induced potential is period with period $2 \pi v_a / N$, which means that during the full field excursion the axion field sees $N$ unique minima of the potential. This means that certain regions of space will collapse towards one minimum while other regions will collapse towards another. The domain walls that develop in this case are stable, which is bad because they rapidly come to dominate the energy density of the universe and thus mess up all of the precision cosmological history from BBN onwards. As a result, we insist on domain wall number $N = 1$ in this case, such that $f_a = v_a$. (Alternatively, we could {\it e.g.} add explicit PQ symmetry breaking to destabilize the domain walls for $N > 1$~\cite{Harigaya:2018ooc}, but we will not consider that possibility here.)  Depending on your taste, the requirement $N = 1$ might give you some pause about the scenario of the PQ symmetry broken after inflation.  
 
 With $N = 1$ fixed, let us now consider an infinitely straight, static string in flat spacetime ($H = 0$ and $R = 1$ in~\eqref{eq:EOM_string}) stretched in the $z$ direction with the core of the string at $r = 0$ in radial coordinates.  We make the ansatz  
 \es{eq:Phi}{
 \Phi(x) = {f_a \over \sqrt{2}} g(m_r r) e^{i \theta} \,,
 }
 such that the axion field -- which is the phase factor $a = \theta f_a$ -- acquires a $2 \pi f_a$ phase shift as we go in a circle around the string core. The function $g(m_r r)$ should go to zero at $r = 0$ and to one at $r \to \infty$. Substituting the ansatz for $\Phi$ into~\eqref{eq:EOM_string} we find a non-linear differential equation for $g$:
 \es{}{
g''(x) +  {1 \over x} g'(x) - {1 \over x^2} g(x) + {1 \over 2} g(x) \left[ g^2(x) - 1 \right] = 0 \,.
 }
 In general, this differential equation should be solved numerically. However, the most interesting part of the solution is that at large $x$, where one can show that $g(x) = 1 - {1 \over x^2} + {\mathcal O}(1/x^4)$.  Now, let us compute the tension in the string (energy per unit length), which is given by 
 \es{}{
 \mu = \int d^2x \langle H \rangle &= \int d^2x \left[ |\nabla \Phi|^2 + {1 \over 2} {m_r^2 \over f_a^2} \left( |\Phi|^2 - {f_a^2 \over 2} \right)^2 \right] \\
 &= \pi f_a^2 \left[ \int_0^{\rm  m_r r_{\rm IR}} dx {1 \over x} g^2(x) + \int_0^\infty dx\, x^2 \left( (g'(x))^2 + {1 \over 4} (g^2(x) - 1)^2 \right) \right]  \\
&= \pi f_a^2 \log( m_r r_{\rm IR} C) \,.
 }
 In the second line we use the fact that since $g(x) \to 1$ as $x \to \infty$, the first term -- which arises from the derivative with respect to $\theta$ in $|\nabla \Phi|^2$ -- diverges logarithmically. We thus assume an IR spatial cut-off $r_{\rm IR}$. The second term in the second line above arises from the potential and the derivative-with-respect-to-$r$ terms in $|\nabla \Phi|^2$. Since $g(x) \to 1 - {1 \over x^2}$ at large $x$, this term converges and integrates to a constant, which we absorb into a constant within the logarithm $C$ in the last line.  Physically, for strings in an expanding universe the natural IR cutoff is given by Hubble. Thus, we expect $\mu = \pi f_a^2 \log(m_r / H C)$ in an expanding universe. 
 
 The strings evolve to roughly maintain the so-called scaling solution during the evolution of the network. The important point to note is that the strings may shrink by emitting radiation in the form of axions.  Let us fix the cosmology such that the strings evolve in a radiation dominated era, since that is likely the important epoch between PQ symmetry breaking and the QCD phase transition. The free string network would have an energy density that falls off as $\rho_{\rm string}^{\rm free} \propto {1 / R^2}$. To see this, consider a grid of infinitely straight strings pointing in the ${\bf \hat z}$ direction; then, it is only the distances in the ${\bf \hat x}$ and ${\bf \hat y}$ directions that determine the energy density.  The interacting string network, on the other hand, redshifts as radiation: $\rho_{\rm string}^{\rm interacting} \propto {1 /R^4}$.
 
 To make these concepts more precise, let us define the number of strings per Hubble patch $\xi$:
 \es{}{
 \xi(t) \equiv \lim_{L \to \infty} {\ell_{\rm tot}(L) \over L^3} t^2 \,,
 }
 where $\ell_{\rm tot}(L)$ is the total string length within a box of co-moving side-length $L$.   In terms of $\xi$, the energy density in the string network may be written as 
 \es{eq:rho_string}{
 \rho_{\rm string} = {\xi(t) \mu  \over t^2} \,,
 }
 as this is simply the string length is a large box of side-length $L$ multiplied by the string tension and then divided by the volume of the box.  In the scaling solution $\xi$ is constant or at most logarithmically varying with time.  Some intuition behind this result is that the strings are frozen on super-horizon scales but dynamical on scales smaller than Hubble, where they lose energy and shrink by emitting axions. The network loses energy in such a way such that $\xi$ is constant. At a more mathematical level (following an argument in~\cite{Buschmann:2021sdq}), it has been shown by various simulations and semi-analytic arguments that axion string loops emit energy as $dE / dt = - \alpha f_a^2$, where $\alpha \sim {\mathcal O}(10)$ is a constant, regardless of the loop size. While this might be surprising, it arises because smaller loops, despite their size, have more curvature than larger loops. The curvature causes the strings to emit more axions per unit length in such a way that all loop sizes emit the same total amount of energy per unit time. Let us consider the strings as consisting of a number density $n_c$ of correlation lengths, with each correlation length behaving as an independent string loop. Then, we may write the axion emission rate per unit volume as $\Gamma_a = n_c \alpha f_a^2$. Alternatively, we may compute $\Gamma_a$ by taking a time derivative of~\eqref{eq:rho_string} minus the contribution from the free string network and noting that any energy loss must go into axions. Neglecting all factors order unity and order $\log(m_r / H)$ we can then identify:
 \es{eq:energy_balance}{
 {\dot \xi f_a^2 \over t^2} + {\xi \mu \over t^3} \sim n_c f_a^2 \,.
 } 
 We may relate $n_c$ to $\xi$ through $n_c = {\xi \over L_c} H^2 \sim {\xi / t^3}$, where $L_c$ is the correlation length, which should scale as $H^{-1}$ since that is the scale over which the strings become dynamical. Thus, $n_c f_a^2 \sim \xi f_a^2 / t^3$ and therefore, referring to~\eqref{eq:energy_balance}, we infer that $\dot \xi = 0$. Carrying out this argument more precisely, including the $\log(m_r / H)$ factors in the tension, one can motivate that $\xi \propto c_1 \log(m_r / H)$ at large values of $\log(m_r / H)$ for some constant $c_1$~\cite{Buschmann:2021sdq}.  Modern simulations of the axion string network confirm the logarithmic scaling of $\xi$ and estimate $c_1 \approx 0.25$~\cite{Gorghetto:2020qws,Buschmann:2021sdq}, though this scaling is still debated (see, {\it e.g.},~\cite{Hindmarsh:2021vih} for an alternative scaling proposal). 
 
 The string network maintains the approximate scaling solution by emitting relativistic axions. This happens until the QCD phase transition, when the network collapses.  Also at the QCD phase transition, the axions rapidly acquire a mass, which ``stops them in their tracks." The axions go from being relativistic to highly non-relativistic, at least most of them, in a short period of time around the moment when $3 H(T_{\rm osc}) \approx m_a(T_{\rm osc})$.  Referring back to~\eqref{eq:rho_a_T_MR} from the constant misalignment case, we can approximate the DM density from axions produced by the string network as 
 \es{eq:rho_a_string}{
 \rho_a(T_{\rm MR}) \approx m_a n_a(T_{\rm osc})  {g_{*s}(T_{\rm MR}) \over g_{*s}(T_{\rm osc})} \left( {T_{\rm MR} \over T_{\rm osc}} \right)^3 \,,
 }
 where $n_a(T_{\rm osc})$ is the axion number density generated by string decay at $T_{\rm osc}$. Here we assume, as a rough approximation, that the string network evolves as in the $m_a(T) = 0$ case up until $T_{\rm osc}$ and that for $T < T_{\rm osc}$ the network has disappeared.  We can then compute $n_a(T_{\rm osc})$ by 
 \es{eq:n_a}{
 n_a(T_{\rm osc}) = \int_{\rm k_{\rm IR}}^{k_{\rm UV}} {dk \over k} {d \rho_{\rm string} \over dk} \,,
 }
 where $k_{\rm UV}$ and $k_{\rm IR}$ are the UV and IR cut-offs in momentum, respectively. We expect $k_{\rm UV} \sim m_r$ while $k_{\rm IR} \sim H  \sqrt{\xi}$.  The reason that $k_{\rm IR}$ is expected to scale with $\sqrt{\xi}$ is that the IR cut-off is really the typical distance between strings, and as there are more strings per Hubble the string become closer together by a factor $\sim$$\sqrt{\xi}$. Between the IR and UV cutoffs there is no scale in the problem, suggesting that between these scales we might expect $d \rho_{\rm string} / dk \propto 1/k^q$ for some index $q$. Recent simulations suggest $k \approx 1$~\cite{Buschmann:2021sdq}, though this is highly debated~\cite{Gorghetto:2020qws}.  Let us assume $q = 1$ for our estimate, though do keep in mind that future simulations could lead to different results. The case $q = 1$ corresponds to a conformal spectrum, since in that case the axion energy density receives equal contributions from each decade in momentum. In order for the energy density to be properly normalized this means that 
 \es{}{
{d \rho_{\rm string} \over dk} \approx {1 \over \log(m_r / H)} \rho_{\rm string} {1 \over k} \approx {1 \over \log(m_r / H)}  \xi \mu {H^2 \over k} \sim \pi f_a^2 \xi {H^2 \over k} \,,
}
using $H = 1/(2 t)$ and dropping all factors order unity.  Integrating the expression in~\eqref{eq:n_a} we then find 
\es{}{
n_a(T_{\rm osc}) \approx {1 \over \log(m_r / H)}  \sqrt{\xi} \mu H \approx \sqrt{c_1} \pi f_a^2 H(T_{\rm osc}) \sqrt{\log(m_r / H)} \,,
}
which, substituting into~\eqref{eq:rho_a_string} and taking $m_r \approx 10^{10}$ GeV and $T_{\rm osc} \approx 1$ GeV, so that $\log(m_r / H) \approx 65$, leads us to estimate that at fixed $f_a$ and $\theta_i \sim 1$ the DM density from strings is larger than that from the misalignment angle by an amount $\sim \sqrt{c_1} \pi \sqrt{\log(m_r / H)} / 3 \sim  4$ (hint: if you are trying to reproduce this, use the assumption $3 H(T_{\rm osc}) \approx m_a(T_{\rm osc})$), with all other parametric dependences the same.  Referring to~\eqref{eq:Omega_a_QCD} we then estimate that in this case the correct DM abundance is achieved for $m_a \sim 70$ $\mu$eV, matching the more careful calculation results that predict $m_a \in (40,180)$ $\mu$eV, with $m_a \sim 65$ $\mu$eV if $q = 1$~\cite{Buschmann:2021sdq}.

 \subsubsection*{Cosmological considerations for PQ broken after inflation}
 
 There are two main cosmological considerations in the case where the PQ symmetry is broken after inflation:
 \begin{itemize}
 \item Small-scale isocurvature perturbations lead to axion minihalos 
 \item High reheat temperature implies a thermal population of axions, which contributes to $N_{\rm eff}$
 \end{itemize}
 The first consideration is unique to the case of the PQ symmetry being broken after inflation, while the second may, or may not, be relevant if PQ is broken before inflation but is certainly relevant if PQ is broken after inflation. We briefly discuss both of these effects below, as they are observational handles that may one day lead to evidence for the axion model.
 \\
 
 \noindent
 {\bf Axion minihalos}
 \newline
 
 Looking back at {\it e.g.} Fig.~\ref{fig:axions}, it is not surprising that the axion-string-domain-wall network would leave behind large perturbations in the DM density distribution on scales of order the cosmological horizon at the QCD phase transition. At distance scales much larger than the horizon these perturbations are white-noise. Let us denote $\delta = \delta \rho_{\rm matter} / \rho_{\rm matter}$ to be the dimensionless size of the density perturbations induced by the axion isocurvature.  Further, let us denote by $L_{\rm osc}$ the comoving size of the cosmological horizon  at $T_{\rm osc}$: $L_1 = 1 / (R(T_{\rm osc}) H(T_{\rm osc}))$.  At $T_{\rm osc}$ we expect 
 $\delta \sim {\mathcal O}(1)$ 
 on scales of order $L_1$. However, on length scales much larger than $L_1$ the density perturbations should be uncorrelated, since these Hubble patches are causally disconnected.  Thus, if we were to compute the variance $\sigma(k)^2$ of $\delta$ averaged over length scales $2 \pi / k \gg L_1$, with $k$ the comoving wave number, then we expect $\sigma(k)^2 \propto k^3$, since the number of uncorrelated perturbations within the spatial volume grows like $1/k^3$ at small $k$ and the variance is proportional to one over the number of uncorrelated patches within the volume. Writing 
 \es{eq:pert_axion2}{
 \langle \delta({\bf k}') \delta({\bf k}) \rangle = (2 \pi)^2 \delta({\bf k'} + {\bf k}) {2 \pi^2 \over k^3} \Delta_\delta^2 \,,
 }  
 as in~\eqref{eq:pert_axion}, we expect now $\Delta_\delta^2 = \Delta_0^2 (k L_1)^3$ for $k L_1 \ll 1$, since this gives rise to the expected variance at small $k$. Note that for $k \gtrsim L_1$ the density perturbations are highly non-Gaussian. At small $k$, though, power spectra for which $\Delta_\delta^2 \propto k^3$ are called white-noise power spectrum. We expect $\Delta_0^2 \sim 1$, since we want perturbations order unity at $k \sim (2 \pi)/L_1$; for example, Ref.~\cite{Eggemeier:2019khm} concludes $\Delta_0 \sim 0.1$.  Recall that the curvature perturbations from the inflaton field that give rise to the observed structure in the universe and the fluctuations in the CMB have $\Delta^2 \sim 10^{-9}$, meaning that the axion-induced white noise fluctuations will dominate on small scales with $k \gtrsim 5 \times 10^{-3} L_1^{-1} \approx 7 \, \, {\rm pc}^{-1}$, which are scales much too small to probe in {\it e.g.} CMB or large scale structure. (This is roughly the distance between us and the nearest star.) Note that here we use $T_{\rm osc} \sim\,\, {\rm GeV}$ such that $L_1 \approx 10^{-3}$ pc. 
 
 The density perturbations generated by the axion will roughly be frozen in place until matter-radiation equality, when they will begin to collapse. After matter-radiation equality they will undergo hierarchical structure formation, just like the normal curvature perturbations on larger scales, and form a spectrum of so-called ``minihalos" with masses spanning from a minimum value $M_{\rm min}$ to increasingly high masses.  Following the general procedure of hierarchical mergers one expects $dN_{\rm MH} / d M_{\rm MH} \propto 1 / M_{\rm MH}^{\rm 3/2}$ for minihalo masses $M_{\rm MH} \gg M_{\rm min}$, where $N_{\rm MH}$ is the number of subhalos as a function of halo mass (see, {\it e.g.},~\cite{Eggemeier:2019khm} and references therein).  The minimum halo mass is roughly set by the amount of matter within a Hubble volume at the oscillation time $T_{\rm osc}$. This scale gives the characteristic mass of the first subhalos that form after matter-radiation equality.  We can estimate this mass as $M_{\rm MH} \sim {4 \over 3} \pi L_1^3 \Omega_{\rm DM} \rho_c \sim 10^{-16}$ $M_\odot$, where $\rho_c \approx 1.3 \times 10^{-7}$ $M_\odot/$pc$^3$ is the critical density and $\Omega_{\rm DM} \approx  0.27$ is the DM energy fraction today~\cite{Planck:2018vyg}.
 
 The axion minihalos may have a number of interesting implications for direct and indirect searches.  For example, we can try to directly search for the minihalos using gravitational probes, such as gravitational lensing by the minihalos of light from highly magnified stars~\cite{Dai:2019lud} and pulsar timing searches~\cite{Ramani:2020hdo} (see, {\it e.g.},~\cite{Xiao:2021nkb}).  Note that pulsar pulses are extremely stable, and that the searches using pulsar timing exploit this fact by looking for small changes to the pulsar pulse timings due to the gravitational effects of transiting DM minihalos.  The axion minihalos could also lead to non-gravitational signals, for example by colliding with highly magnetic neutron stars (NSs)~\cite{Edwards:2020afl}, or they could provide sources for the much more dense, hypothetical axion stars~\cite{Braaten:2015eeu}. 
  \\
 
 \noindent
 {\bf Cosmic axion microwave background and $N_{\rm eff}$}
 \newline

 If the PQ symmetry is broken after inflation then the reheat temperature from inflation must be greater than, very roughly, around $10^{10}$ GeV.   Such a high reheat temperature necessarily means, as we illustrate below, that a thermal population of axions was once present and that this population underwent relativistic freeze-out at some intermediate temperature. Depending on the reheat temperature after inflation if the PQ symmetry was broken before inflation, a thermal axion population may also be present in that cosmological history as well (but it is not guaranteed, since the reheat temperature could be low in that case).  Apart from entropy dilution effects, the effective temperature of the thermally-produced axion population would be roughly the same as that of the CMB today, and so we refer to this population of relic, relativistic axions as the cosmic axion microwave backgrounds (CAMB).  The CAMB is a very important target for the next generation of cosmological surveys because it contributes to $N_{\rm eff}$, the effective number of neutrino degrees of freedom.\footnote{Various relativistic cosmic axion backgrounds may also be within reach of direct detection experiments~\cite{Dror:2021nyr}.} This is simply because the relic axions behave like non-interacting radiation in the early universe, the same way that neutrinos do.  See the TASI 2022 lecture notes~\cite{Green:2022bre} for a discussion of how $N_{\rm eff}$ may be constrained from cosmological observables. Here, we simply sketch an argument for the conditions under which a CAMB should be present and how, roughly, it should contribute to $N_{\rm eff}$.
 
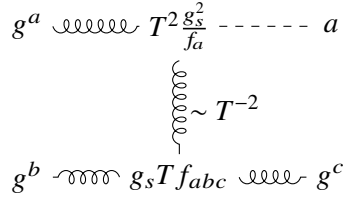
\begin{figure}[htb]  
\begin{center}
\begin{tikzpicture}
  \begin{feynman}
    \vertex (a){$g^a$};
    \vertex [right=2cm of a] (b){$T^2 \frac{g_s^2}{f_a}$};
    \vertex [below=2cm of b] (c){$g_s T f_{abc} $};
    \vertex [below=2cm of a] (d){$g^b$};
    \vertex [right=2cm of b] (e){$a$};
    \vertex [right=2cm of c] (f){$g^c$};
    \diagram* {
      (a) -- [gluon] (b),
      (b) -- [gluon, edge label =\(\sim T^{-2}\)] (c),
      (c) -- [gluon] (d),
      (b) -- [scalar] (e),
      (c) -- [gluon] (f)
    };
  \end{feynman}
\end{tikzpicture}
\caption{One of the diagrams that contributes to thermal axion production in the early universe in scenarios where the reheat temperature after inflation is high enough for axions to be in thermal equilibrium with the SM plasma.  Here, for example, two gluons $g^a$ and $g^b$ fuse to form an axion $a$ and a gluon $g^c$. We indicate the parametric dependence of each vertex as well as the internal gluon propagator. The gluon-gluon-axion vertex brings a factor of $1/f_a$, and then beyond this factor all other mass scales are set by the temperature $T$. }
\label{fig:axion_fo}
\end{center}
\end{figure}

 Let us suppose that the axions scatter with the SM with rate $\Gamma_a(T)$. Then, just as in the case of non-relativistic freeze-out discussed in Sec.~\ref{sec:freeze_out},  the condition for the axion to be in thermal equilibrium at a temperature $T$ with the SM is that $\Gamma_a(T) \gtrsim 3 H(T)$, such that each axion particle scatters, on average, at least once with the SM bath during a Hubble time. Consider temperatures $T \gg \, {\rm GeV}$ such that we may treat QCD perturbatively. Then, focusing on the coupling of the axion to gluons given in~\eqref{eq:axion}, the axion can be produced and annihilated in interactions with thermal gluons ($gg \to ga$ and $ag \to gg$, for gluons $g$).  See, for example, the diagram illustrated in Fig.~\ref{fig:axion_fo}, which can describe the emission or absorption of a thermal axion. 
Let us roughly estimate the rates $\Gamma_a$ for these processes. The relevant momentum scale for all particles is $p \sim T$.  Note that the gluon also acquires a thermal mass in the plasma $m_g = g_3 T$, with $g_3$ the strong coupling constant, which regulates IR divergences.  The momentum and coupling factors, with $f_{abc}$ the $SU(3)$ structure constants, of the various parts of the diagram are illustrated in Fig.~\ref{fig:axion_fo}. This leads to the estimate 
\es{}{
|i \,M|^2 \sim {g_s^6 |f_{abc}|^2 \over 64 \pi^4} {T^2 \over f_a^2} \,,
}
with the factors of $2$ above somewhat arbitrary. We then estimate the thermally-averaged cross-section for axion absorption  or emission as 
\es{}{
\langle \sigma v \rangle \sim {|i\, M|^2 \over 16 \pi s} \sim {|i\, M|^2 \over 64 \pi T^2} \sim {\alpha_s^3 (N_c^2 - 1) \over 64 \pi^2} {1 \over f_a^2} \,,
}
The rates of  {\it e.g.} axion production and absorption are then approximately $\Gamma_a = n_g \langle \sigma v \rangle \sim {\alpha_s^3 \over 8 \pi^2} {T^3 \over f_a^2} \sim 2 \times 10^{-5} (T^3 / f_a^2)$, with $n_g \sim T^3$ the number density of gluons.  A more accurate calculation, keeping track of all the factors of $2$ and treating the system within the context of thermal field theory, yields a similar but slightly larger result~\cite{Salvio:2013iaa} (see also~\cite{Graf:2010tv,Baumann:2016wac}):
\es{}{
\Gamma_a \approx 1.5 \times 10^{-4} {T^3 \over f_a^2} \,.
} 
Approximating the freeze-out temperature by $3 H(T_{\rm fo}) \sim 3 T_{\rm fo}^2 / m_{\rm pl} \approx \Gamma_a(T_{\rm fo}) $, we find
\es{}{
T_{\rm fo} \sim 10^8 \, \, {\rm GeV} \left( {f_a \over 10^{11} \, \, {\rm GeV}} \right)^2 \,.
}
 If the reheat temperature after inflation is larger than $T_{\rm fo}$, then a thermal axion population would have been produced after inflation, and that population would still exist today, having free-streamed since its decoupling at $T_{\rm fo}$.  On the other hand, if the reheat temperature is below $T_{\rm fo}$ then such a thermal population would never have been produced. If the PQ symmetry is broken after inflation, then the reheat temperature must be above $f_a$ and thus also above  $T_{\rm fo}$; in this cosmological scenario, we expect a thermal population of axions. On the other hand, if the PQ symmetry is broken before inflation then we may or may not have a thermal axion population, depending on $f_a$, but we certainly cannot have such a population if $f_a$ is at the GUT scale ($f_a \gtrsim 10^{15}$ GeV), since in that case  our calculation yields $T_{\rm fo} > f_a$.  On the other hand, if $f_a \sim 10^{11}$ GeV then it is perfectly consistent to have a reheat temperature between $10^8$ GeV and $10^{11}$ GeV and produce thermal axions but not restore the PQ symmetry, while also satisfying the isocurvature constraint~\eqref{eq:iso_curvature_constraint}.  An axion population with a freeze-out temperature well above the electroweak scale ($T_{\rm fo} \gg 200$ GeV) produces a contribution to the effective neutrino degrees of freedom around $\Delta_{\rm Neff} \approx 0.027$, which may be in reach of future surveys (see, {\it e.g.},~\cite{Baumann:2016wac,Green:2022bre}).

\section{Astrophysical probes of axions}
\label{sec:axions_indirect}

\begin{figure}[htb]  
\begin{center}
\includegraphics[width=1.0\textwidth]{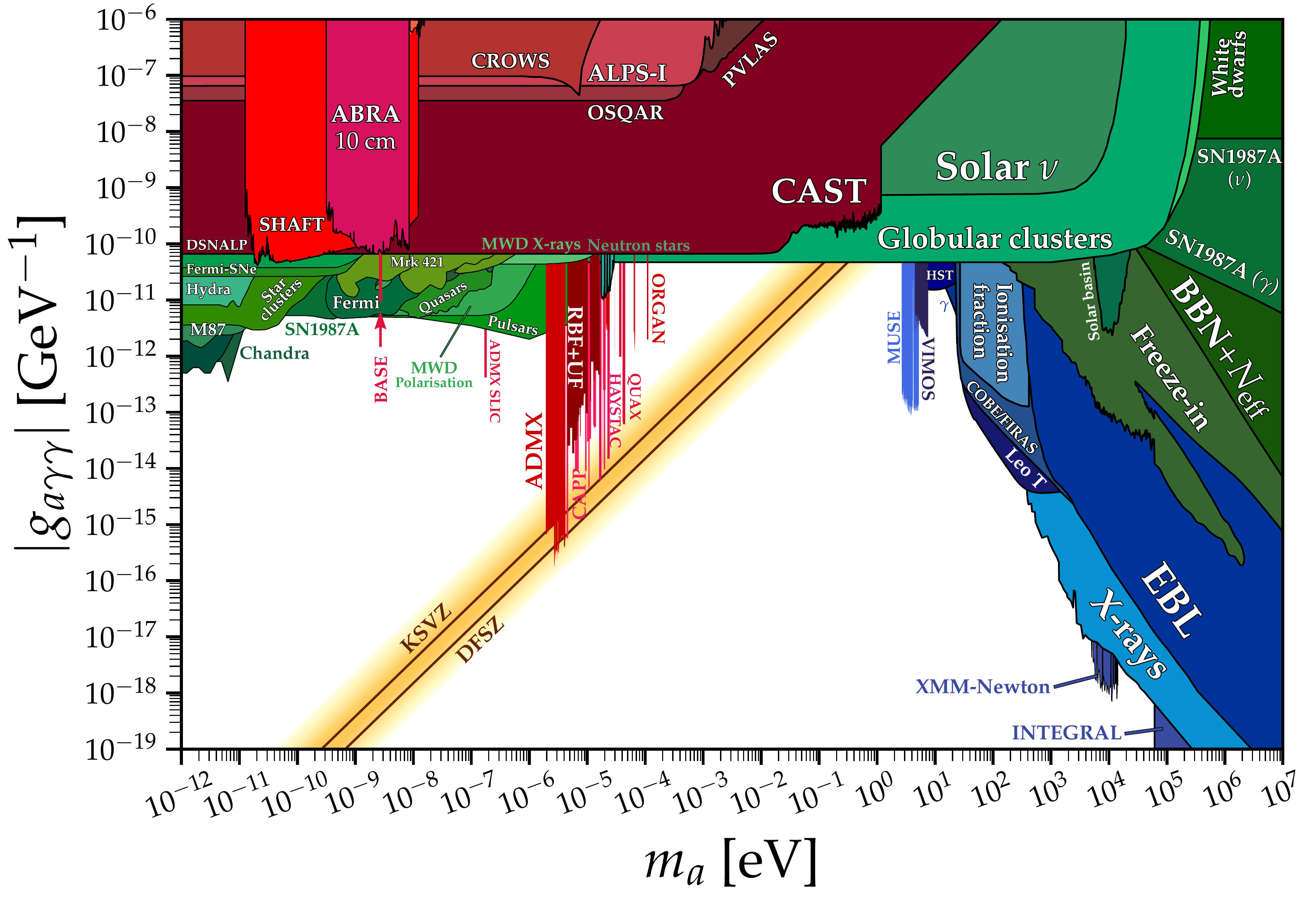}
\caption{Current upper limits on the axion-photon coupling as a function of the axion mass from astrophysical and laboratory probes, with the QCD axion band indicated.  We discuss many of the astrophysical probes in this section, including spectral modulation searches, MWD polarization, supernova constraints, NS radio searches, CAST, and MWD $X$-ray searches, amongst other concepts. Figure compiled from~\cite{AxionLimits}.}
\label{fig:axion_limits}
\end{center}
\end{figure}

We now turn to astrophysical probes of the QCD axion and ALPs, first discussing probes only involving gravity and then turning to a sequence of probes  involving the axion-photon and axion-mater couplings.
Many of the probes we discuss have current upper limits illustrated in Fig.~\ref{fig:axion_limits}.

\subsection{Black hole superradiance}

One of the most intriguing classes of astrophysical probes of axions only involves gravity.  These include, for example, the isocurvature probes that we already discussed. At lower redshifts, another promising detection pathway for axions is a process called black hole superradiance~\cite{Arvanitaki:2009fg,Arvanitaki:2010sy,Brito:2015oca} (see~\cite{Baryakhtar:2022hbu} for a broad overview of the recent literature on this topic).  The basic idea of superradiance can be understood from the following basic concept from classical mechanics. Imagine a ball hitting the edge of a rotating cylinder. If the ball is moving slower than the rotating point of the cylinder on impact, then it can gain kinetic energy and angular momentum, about the center of the cylinder, assuming that there is a source of dissipation ({\it i.e.}, friction), which allows the ball to change velocity. Similarly, a classical axion field surrounding a black hole may gain energy and angular momentum from the black hole, assuming that the black hole is rotating faster than the axion cloud.  The event horizon of the black hole provides the source of dissipation. In order for the superradiance to be efficient, the Compton wavelength of the axion should be comparable to the black hole Schwarzschild radius $r_s$:
\es{eq:ma_SR}{
m_a \sim {1 \over r_s} = {1 \over 2 G M_{\rm BH}} \sim 7 \times 10^{-11} \, \, {\rm eV} \left( {1 \, \, M_\odot \over M_{\rm BH}} \right) \,,
}
with $M_{\rm BH}$ the black hole mass and $G$ Newton's gravitational constant.  Also, the angular velocity
 should be less than the black hole's angular velocity $\Omega_{\rm BH}$, so that the axion field may extract energy and angular momentum from the black hole.  Note that the gravitational potential
\es{eq:SH}{
V(r) = - {G M_{\rm BH} m_a \over r}  \equiv - {\alpha_{\rm SR} \over r} 
}
 of the black hole provides a form of reflection or trapping, such that the axion waves continue to return to the black hole and extract more energy and momentum, until the black hole has spun down to the point that superradiance is no longer possible. 
 The closer $m_a$ is to $r_s^{-1}$, the more efficient this trapping is and thus the shorter the superradiance time-scales.  The axion superradiant cloud that is populated is quantized into energy levels similar to the hydrogyen atom, with -- comparing to~\eqref{eq:SH} -- the fine structure constant replaced by $\alpha_{{\rm SR}} = G M_{\rm BH} m_a$.  Unlike for hydrogen, though, since the axions are bosons the energy levels become populated with exponentially large numbers of axions.
 
 The rough story of the life of a black hole in the presence of superradiance is that if a black hole is born with a large spin, then it will rapidly spin down via superradiance if an appropriate-mass axion is in the spectrum.  Note that the superradiance growth is exponential, so the axion does not have to be any appreciable fraction of the DM to source the superradiance instability; even quantum fluctuations in the axion field are sufficient to begin the superradiance process. This means that if one finds a rapidly-spinning black hole of a given mass, one can use this to set constraints on axions that might exist within the spectrum of nature, regardless of whether or not those axions make up any fraction of the DM. In addition, transitions between the energy levels of the gravitational atom formed around the black hole can give continuous gravitational wave signatures that could be in reach of current- and next-generation gravitational wave observatories.
 
 In fact, black hole superradiance already disfavors axions with masses roughly in the range $ {\rm few} \times 10^{-13} < m_a \lesssim 10^{-11}$ eV~\cite{Arvanitaki:2010sy,Stott:2018opm,Baryakhtar:2020gao}.  This constraint is highly relevant, since it means that QCD axions with Planck-scale and slightly lower decay constants, above roughly $5 \times 10^{17}$ GeV, are disfavored.  We can understand this constraint by going slightly more into detail in the superradiance requirements, timescales, and rates. (Our presentation here follows closely~\cite{Baryakhtar:2020gao}.)  Just like for hydrogen, the states of the gravitational atom that are populated by superradiance are described by three quantum numbers $\{n,\ell,m\}$, with $n$ the principle quantum number, $\ell$ the total angular momentum, and $m$ the angular momentum in the azimuthal direction $\phi$, which we define to be the direction in-line with the spin of the black hole. Concentrating only on the time evolution and the evolution in $\phi$, the axion mode evolves as $a \sim e^{i (m \phi - \omega t )}$. The azimuthal phase velocity is then $v_p = \omega / m$. Comparing to the black hole angular frequency $\Omega_{\rm BH}$ ({\it i.e.}, requiring $v_p < \Omega_{\rm BH}$), we see that the superradiance condition should be $\omega < m \Omega_{\rm BH}$.  The frequency of the superradiant states depends only on $n$, just like in hydrogen,
 \es{}{
 \omega \approx m_a \left( 1 - {\alpha_{\rm SR}^2 \over 2 n^2 } \right) \,,
 }
 to leading order in $\alpha_{\rm SR}$.  Thus, for $m_a > m \Omega_{\rm BH}$ black hole superradiance into the mode with azimuthal quantum number $m$ is not possible.  On the other hand, this seems to imply that regardless of $m_a$ we can always have the superradiance condition satisfied by populating a state with a high enough quantum number $m$. While this is true, it also turns out that the rate for forming the superradiance cloud scales as $\Gamma_{\rm SR} \propto m \Omega_{\rm BH} \alpha_{\rm SR}^{4 \ell + 5}$ (recall that $\ell \geq m$).  Thus, the high-$m$ states have suppressed superradiant rates and are less relevant for astrophysical applications. The first state which is relevant for superradiance is that with $\{n, \ell,m\} = \{2,1,1\}$; for this state, the superradiance rate is $\Gamma_{\rm SR} \approx {a_\star \over 24} \alpha_{\rm SR}^8 m_a$ to leading order in $\alpha_{\rm SR}$. (Note that including the next-leading term in $\alpha_{\rm SR}$ one actually sees that $\Gamma_{\rm SR}$ vanishes when $\alpha_{\rm SR}$ is large enough that the superradiance condition is saturated.)  Here, $a_\star < 1$ is the dimensionless spin of the black hole.  In terms of $a_\star$ the angular velocity of the BH is 
 \es{}{
 \Omega_{\rm BH} = {1 \over 2 G M_{\rm BH}} \left( {a_\star \over 1 + \sqrt{1- a_\star^2}} \right) 
 \,.
 } 
 
 Let us now consider an example black hole and understand how its properties translate to constraints on the QCD axion.  Suppose that we had a $M_{\rm BH} = 20$ $M_\odot$ black hole with $a_\star$ very close to $1$; what would this tell us about the allowed valued of $m_a$? First, we need $m_a < 3.5 \times 10^{-12}$ eV in order to have superradiance at all (for $a_\star = 1$).  In this case, $\alpha_{\rm SR} < 0.5$, with $\alpha_{\rm SR}$ largest for the highest allowed value of $m_a$. The superradiance  rate is $\Gamma_{\rm SR} \approx {1 \over 1 \, \, {\rm s}} \left( {m_a \over 3.5 \times 10^{-12} \, \, {\rm eV}} \right)^9$, though this is technically only valid for $m_a \ll 3.5 \times 10^{-12} \, \, {\rm eV}$.  Thus, the rate drops rapidly for $m_a$ below the critical value, meaning that for masses well below $\sim$$10^{-12}$ eV superradiance will not be efficient enough to lead to constraints on $m_a$, since if we observe a black hole with large spin then it is possible that superradiance simply has not had enough time to spin down the black hole.  On the other hand, while superradiance is highly efficient for $m_a \sim 3.5 \times 10^{-12}$ eV, the black hole would immediately spin down such that its value of $a_\star$ puts it below the superradiance threshold  ($m_a > m \Omega_{\rm BH}$), meaning that superradiance would shut off and the black hole would still have a large value of $a_\star$. Thus, the uncertainty on the measured value of $a_\star$ will, in practice, limit the highest obtainable $m_a$.  By considering the next excited state $(3,2,2)$ we can in principle double the reach in $m_a$, though the superradiance rate becomes further suppressed.  Given the existence of $\sim$$10$ $M_\odot$ black holes with spin measurements near unity, we can rule out QCD axions with masses in the range, roughly, $10^{-13} \, \, {\rm eV} < m_a < 10^{-11} \, \, {\rm eV}$~\cite{Arvanitaki:2010sy,Stott:2018opm,Baryakhtar:2020gao}. 
 
 Another very important consideration for superradiance is also the strength of the axion self interactions~\cite{Baryakhtar:2020gao}, which can destabilize the superradiant cloud and shut off the superradiance process, thereby limiting the ability for axions to spin down black holes.  Expanding the potential in~\eqref{eq:axion_pot} about its minimum we see that the axion has quartic interactions $\propto$$a^4$, which allow {\it e.g.} $3 \to 1$ processes.  These processes can shuffle energy between the different states of the gravitational atom and also take energy away to infinity.  For the QCD axion the self-interactions are relatively weak for Planck-scale decay constants, but still they are border-line relevant for superradiance in the $(3,2,2)$ state for $m_a \sim 10^{-11}$ eV axions from $\sim$$10$ $M_\odot$ black holes and likely limit the ability to go to much higher masses~\cite{Baryakhtar:2020gao}, where the self interactions are stronger.  Also note that by going to higher mass black holes we can constrain the existence of ALPs with much lower masses, so long as their self-interactions are sufficiently week. For example, referring to~\eqref{eq:ma_SR}, spinning black hole with $M_{\rm BH} \sim 10^6$ $M_\odot$ can probe the existence of ALPs with $m_a \sim 10^{-16}$ eV.  Superradiance thus provides a promising probe of the axiverse, which will become even more exciting in the era of gravitational wave astronomy, as gravitational waves can both be emitted by the superradiant axion clouds and also provide pathways by which the masses and spins of black holes can be mapped out.   %For ALPs  

\subsection{Probes only involving axion-photon mixing}

Most terrestrial and astrophysical probes of axions involve the coupling to photons, as given in~\eqref{eq:axion_photon_coupling}, since macroscopic electromagnetic fields are easily manipulated in the laboratory and found in astrophysical environments (in these lecture notes we focus on the astrophysical applications). 

\subsubsection{Magnetic white dwarf linear polarization from axions}
\label{sec:MWD}

\begin{figure}[htb]  
\begin{center}
\includegraphics[width=0.5\textwidth]{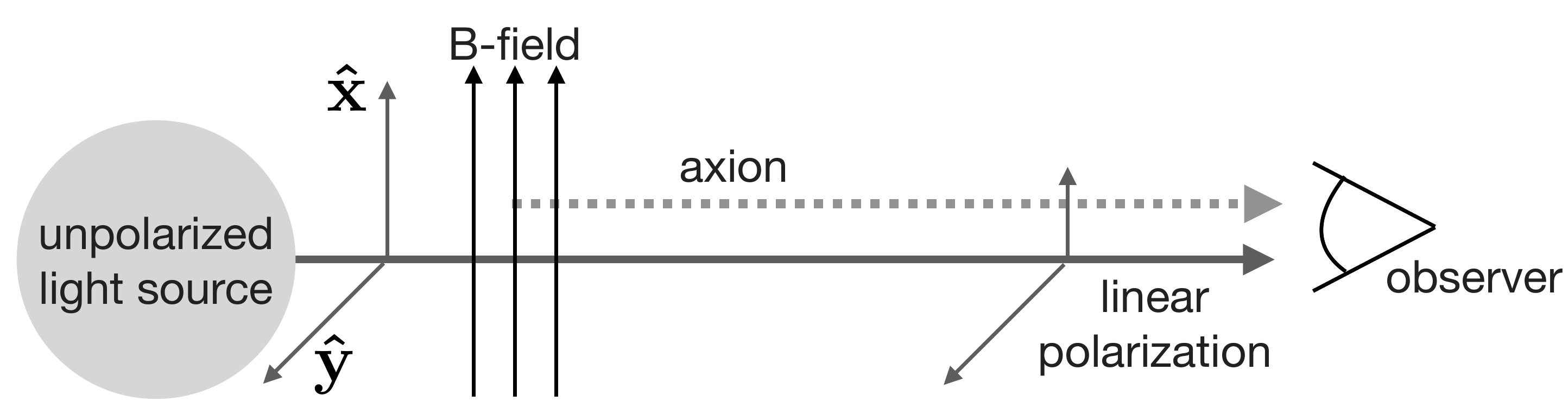}
\caption{An illustration of how axions may induce linear polarization in an otherwise unpolarized light source (such as a MWD) if the photons traverse a region of magnetic field. The photons polarized parallel to the magnetic field may convert to axions, which are unobserved, while the photons polarized normal to the magnetic field are unaffected. (Figure adapted from~\cite{Dessert:2022yqq}.)}
\label{fig:ill_pol}
\end{center}
\end{figure}

One of the most straightforward applications of the axion-photon coupling is to magnetic white dwarf (MWD) polarization searches~\cite{Gill:2011yp,Dessert:2022yqq}, which we briefly sketch below.  The operator ${\mathcal L} \supset g_{a\gamma\gamma} a {\bf E} \cdot {\bf B}$ allows a dynamical axion, propagating through a static magnetic field region, to rotate into a photon, which is polarized parallel to the magnetic field, since the operator has ${\bf E} \cdot {\bf B}$. Similarly, a photon propagating through a magnetic field region may rotate into an axion, so long as the photon is polarized parallel to ${\bf B}$.  Imagine, then, we have an initially unpolarized light source, as illustrated in Fig.~\ref{fig:ill_pol}.  Suppose that the light propagates in the ${\bf \hat z}$ direction to the observer, who measures the degree of linear polarization. In the absence of an axion, the observer will measure no linear polarization. In the presence of an axion with non-zero $g_{a\gamma\gamma}$, however, along with a region of magnetic field, taken to point in the ${\bf \hat x}$ direction, a linear polarization will be induced in the ${\bf \hat y}$ direction. The reason is that the photons polarized parallel to ${\bf \hat x}$ can convert to axions, which are unobserved, and this decreases the intensity of that polarization direction, as sketched in Fig.~\ref{fig:ill_pol}. 

Let us now estimate the magnitude of the axion-induced linear polarization.  First, we need the equations of motion for the mixing of axions with photons. Let us start with~\eqref{eq:axion_photon_coupling} and work in Weyl gauge ($A_0 = 0$, with $A_i = - {\bf A}_i$, such that $A^\mu = (0, {\bf A})$).  Recall that in this gauge ${\bf E} = - \partial {\bf A} / \partial t$ and ${\bf B} = {\bf \nabla} \times {\bf A}$.  It is a nice exercise to use the Euler-Lagrange equations to show that in flat spacetime 
\es{eq:axion_photon_eom}{
\ddot a - \nabla^2 a + m_a^2 a &= - g_{a\gamma\gamma} \dot {\bf A} \cdot ({\bf \nabla} \times {\bf A}) \\
\ddot {\bf A} - \nabla^2 {\bf A} + {\bf \nabla} ( {\bf \nabla} \cdot {\bf A}) &= g_{a\gamma\gamma} \left[ \dot a {\bf \nabla} \times {\bf A} - {\bf \nabla} a \times \dot {\bf A} \right] \,.
} 
The last term in the second line above, involving ${\bf \nabla } a \times \dot {\bf A}$, is higher-order in $g_{a \gamma \gamma}$ for our applications, since we only consider here the case of an external ${\bf B}$-field.  Initially, before entering the magnetic field region, we start with initial conditions for which $a = 0$. The induced axion-field is first order in $g_{a\gamma\gamma}$, which implies that the contribution of the ${\bf \nabla } a \times \dot {\bf A}$ term in~\eqref{eq:axion_photon_eom} to ${\bf A}$ is second order in $g_{a\gamma\gamma}$.  

Let us take the electromagnetic waves to propagating in the ${\bf \hat z}$ direction, as in Fig.~\ref{fig:ill_pol}.
To make further progress, we make a WKB-type approximation to reduce the second order differential equations to a first order equations. Free electromagnetic fields propagating in the ${\bf \hat z}$ direction are plane waves: ${\bf A} \sim e^{-i \omega t + i k z} \sim e^{i \omega(z-t)}$. Let us thus write 
\es{}{
{\bf A}(t,z) = i {\bf A_s}(z) e^{i \omega(z-t)} \,, \qquad a(t,z) = a_s(z) e^{i \omega(z-t)} \,,
}
with the $i$ in front of ${\bf A_s}$ a convenient phase factor,
and make the approximation that $|\partial_z A_s| / \omega \ll 1$ and $|\partial_z a_s| / \omega \ll 1$. This approximation is equivalent to the statement that the amplitude modulations of the axion and electromagnetic waves are slow relative to the oscillation frequency of the carrier wave.   With this approximation, we can for example write 
\es{}{
\partial_z^2 {\bf A} \approx  e^{i \omega(z-t)} \left[ \omega^2 {\bf A_s} + 2 \, \omega \partial_z {\bf A_s} \right] \,,
}
where the term proportional to $\partial_z^2 {\bf A_s}$ is subdominant and neglected.  We thus find the Schr\"odinger-type equation 
\es{eq:axion_mixing_eom}{
 \left[i \partial_z + {\bf \Delta} \right]{\bf {\mathcal  A}} = 0 \,, 
 }
with ${\bf{ \mathcal A}} = (a, A_x,A_y,A_z)^{\rm T}$, dropping the subscript ``$s$" labels, and 
\es{eq:Delta_matrix}{
{\bf \Delta} = \left(
\begin{array}{cccc}
\Delta_{aa} & \Delta_{ax} & 0 & 0 \\
\Delta_{ax}& \Delta_{xx} & 0 & 0 \\
0 & 0 & 0 & 0 \\
0 & 0 & 0 & 0 \\
\end{array}
\right) \,, \qquad \Delta_{aa} = -{m_a^2 \over 2 \omega} \,, \qquad \Delta_{ax} = {g_{a\gamma\gamma} B \over 2} \,, \qquad \Delta_{xx} = 0 \,.
}
Of course, for now it is a bit overkill to include the $A_y$ and $A_z$ components in this equation, but for more general applications these terms can also be included in the mixing.  Also, we define the term $\Delta_{xx}$, even though it is zero in this application, since it is relevant when we start to include a plasma mass for the photon. In that case, there would also be non-trivial $\Delta_{yy}$, $\Delta_{zz}$, along with other terms that we will discuss shortly. 

The $\Delta_{ax}$ term mixes the axion and photon, while the $\Delta_{aa}$ term accounts for the difference in the dispersion relations between the massless photon and the massive axion.  The latter point means that for the same energy, the two waves will have different momenta. The photon has $k_\gamma = \omega$, while the axion has $k_a = \sqrt{\omega^2 - m_a^2} \approx \omega - {m_a^2 \over 2 \omega}$, for $\omega \gg m_a$. Thus, we can identify $\Delta_{aa} = \Delta k$ as the momentum mismatch between the axion and photon at fixed frequency. 

Let us now suppose that at $z = 0$ we have the initial conditions ${\bf \mathcal{A}} = (0,A_0 {\hat x_1},A_0 {\hat x_2},0)^T$, as appropriate for an unpolarized electromagnetic wave (equal amplitudes in the ${\bf \hat x}$ and ${\bf \hat y}$ directions) propagating in the ${\bf \hat z}$ direction.  Here, ${\hat x_1}$ and ${\hat x_2}$ are treated as uncorrelated, complex Gaussian random fields with mean zero and unit variance: $\langle |{\hat x_{1,2}}|^2 \rangle$ = 1, with the cross-correlation vanishing and with the average taken over the thermal ensemble of photons. The amplitude for ${\bf A}_y$ is unchanged after propagating a distance $L$, but working to second-order in perturbation theory, using the formalism of perturbation theory in time-dependent quantum mechanics, 
\es{eq:A_x_second_order}{
A_x(z = L) \approx A_0 {\hat x_1}\left[ 1 - \int_0^L ds \, \Delta_{ax}(s) \int_0^s ds' \, \Delta_{ax}(s') e^{i \int_0^{s'} ds'' \left(\Delta_{xx} - \Delta_{aa} \right)} \right] \,. 
}
Note that we have allowed for the possibility that $B$ is a function of position, which is generically going to be the case, and we have also considered the possibility that $\Delta_{xx}$, which is zero for our current application, depends on position as well.  For a constant B-field ($B(s) = B_0$) and $\Delta_{xx} = 0$ the equation above may be integrated exactly. Let us now define the linear polarization as 
\es{}{
L_p \equiv {\sqrt{\langle  |A_x|^2 - |A_y|^2  \rangle^2 + \langle 2 {\rm Re}\big( A_x A_y^{*} \big)  \rangle^2 }  \over \langle |A_x|^2 + |A_y|^2 \rangle } \,,
}
in which case 
\es{}{
L_p \approx {\Delta_{ax}^2 \over \Delta_{aa}^2 } \left[ 1 - \cos(L \Delta_{aa} ) \right] \,.
}
If the axion and photon waves do not become out of phase after propagating over a distance $L$ ($| L \Delta_{aa} | \ll 1$), then we say the conversion is resonant and 
\es{eq:res_conv}{
L_p \approx {1 \over 8} g_{a\gamma\gamma}^2 B_0^2 L^2 \,, \qquad (| L \Delta_{aa} | \ll 1) \,,
}
while if the two waves do become out of phase ($|L \Delta_{aa}| \gg 1$), the linear polarization is rapidly oscillating with $L$ but has magnitude around 
\es{eq:L_p_large_ma}{
L_p \sim  g_{a\gamma\gamma}^2 B_0^2 {\omega^2 \over m_a^4} \,, \qquad  (| L \Delta_{aa} | \gg 1)  \,.
}
That is, at very low axion masses the axion and photon are in-phase and the linear polarization $L_p$ does not depend on $m_a$, but as $m_a$ increases to the point where the axion and photon become out of phase, the linear polarization becomes increasingly suppressed at large $m_a$. This is a generic feature of axion-photon mixing and is one of the reasons that astrophysical constraints are typically ``flat" as a function of axion mass up to some critical mass, after which point they are drastically reduced in sensitivity, as found in Fig.~\ref{fig:axion_limits} for many of the constraints (and in particular MWD polarizations).

Note, also, that the resonant conversion probability in~\eqref{eq:res_conv} is very much a quantum or wavelike feature. Imagine a beam of particles that decays with some rate $\Gamma$. After propagating over a distance $L$, the number of particles that have decayed will be proportional to $\Gamma \times L$, linear in $L$. On the other hand, the conversion probability formula in~\eqref{eq:res_conv} is quadratic in $L$, since the linear mixing occurs at the level of the waves, while the observable involves the amplitude squared.  

Let us now put in some numbers for MWD polarization. A typical MWD has a magnetic field $B_0 \sim 500$ MG that extends over a distance of order the MWD radius $R_{\rm MWD} \sim 0.01 R_\odot$. (Note that the MWD magnetic fields are well measured by {\it e.g.} Zeeman splitting of lines in the MWD atmospheres.)  Current data constrains the linear polarization from multiple MWDs to be less than around 1\%, though such constraints could certainly improve in the future~\cite{Dessert:2022yqq}.  The frequencies relevant are in the optical (let us take $\omega \sim $~eV for definiteness) given the surface temperatures of typical MWD.  Thus, we find  
\es{}{
L_p \approx 0.5 \% \left( {g_{a\gamma\gamma} \over 10^{-12} \, \, {\rm GeV}^{-1} } \right)^2 \left( {B_0 \over 500 \, \, {\rm MG}} \right)^2 \left( {L \over 0.01 \, \, R_\odot} \right)^2 \,, \qquad m_a \lesssim 2 \times 10^{-7} \, \, {\rm eV} \,,
}
leading to constraints of around $|g_{a\gamma\gamma}| \lesssim 10^{-12} \, \, {\rm GeV}^{-1}$ at low axion masses, which is roughly what was found in {\it e.g.}~\cite{Dessert:2022yqq}.

A natural question to ask next is why not consider the polarization of NSs, which have much larger magnetic field strengths. It turns out, however, that NSs have too large of field strengths, such that the conversion probability between axions and photons is suppressed by non-linear corrections to Maxwell's equations due to the Euler-Heisenberg Lagrangian.  Here, we briefly sketch how to include the effects of the Euler-Heisenberg terms.  The Euler-Heisenberg Lagrangian arises from considering the field theory at energies below the electron mass where we integrate out the electron.  At one loop the following terms are induced in the effective Lagrangian for electromagnetism:
\es{eq:EH}{
{\mathcal L}_{\rm EH} = {\alpha_{\rm EM}^2 \over 90 m_e^4} \left[ \left(F_{\mu \nu} F^{\mu \nu} \right)^2 +{ 7 \over 4} \left(F_{\mu \nu}\tilde  F^{\mu \nu} \right)^2 \right] = 2 {\alpha_{\rm EM}^2 \over 45 m_e^4} \left[ ({\bf E}^2 - {\bf B}^2)^2 + 7 ({\bf E} \cdot {\bf B})^2 \right]  \,.
}
While the exact numerical values in front of the coefficients above require actually performing the one-loop computation, we can easily understand the parametric dependencies.  The Lagrangian arises from considering the diagram with four external photons connected to an internal electron loop. There are thus four electron propagators that enter into the scattering matrix $i {\mathcal M}$, meaning that we expect $i {\mathcal M} \propto {1 \over m_e^4}$, since the external momenta are small relative to $m_e$. Each of the four vertices has a factor of the electric charge $e$, so we also expect $i {\mathcal M} \propto \alpha_{\rm EM}^2$. There needs to be four external gauge fields, which means that when we consider the equivalent term ({\it i.e.}, the Euler-Heisenberg Lagrangian) in the effective theory where the loop has been contracted to a point, that term will need to have four gauge fields $A_\mu$. The only gauge invariant combinations of four $A_\mu$ are $\left(F_{\mu \nu} F^{\mu \nu}\right)^2$ and $\left(F_{\mu \nu} \tilde F^{\mu \nu}\right)^2$.  Working out the numerical pre-factors by performing the loop computation leads to the Lagrangian in~\eqref{eq:EH}.  

Consider a dynamical photon propagating through a region of external magnetic field ${\bf B}$. Separating the dynamical and static magnetic field components in~\eqref{eq:EH} we see that the Euler-Heisenberg Lagrangian generates mass-type terms for the photon, but that these mass terms (or, more appropriately, modifications to the dispersion relations) affect the different polarization components differently.  In our example application, the component  $\Delta_{xx}$ that was zero in~\eqref{eq:Delta_matrix} is modified to $\Delta_{xx} = \Delta_{{\rm EH} ,||} = (7/2) \xi \omega$, with $\xi = (\alpha_{\rm EM} / 45 \pi) (B / B_{\rm crit})^2$, with $B_{\rm crit} \equiv m_e^2 / e \approx 4.4 \times 10^{13} \, \, {\rm G}$ the critical field strength.  Referring to~\eqref{eq:A_x_second_order}, we see that the Euler-Heisenberg term is relevant if, roughly, $L \xi \omega > 1$, which -- for $\omega \approx 1 \, \, {\rm eV}$ and $L \approx 0.01 R_\odot$ -- corresponds to $B \gtrsim 600$ MG.  This justifies us dropping the Euler-Heisenberg term in our simple estimate before for MWD polarization, but if the fields become much larger than 500 MG we certainly need to include the effects of this term. Let us, then, consider the case when $|\Delta_{xx}| \gg |\Delta_{aa}|$ and $|L \Delta_{xx}| \gg 1$. In this case, $L_p \sim \Delta_{ax}^2 / \Delta_{xx}^2 \sim 1 / B^2$; that is, going to larger $B$ actually reduces the axion-photon conversion process, since the change in the photon dispersion relation suppresses the conversion more than the increase in $\Delta_{ax} \propto B$ increases the conversion.   For this reason, polarization probes of axions using NSs are likely less powerful than those using MWDs~\cite{Dessert:2022yqq}.

\subsubsection{Axion-induced spectral modulation from extragalactic sources}

In addition to modifying the polarization properties of the electromagnetic waves, axions also modify the intensity, since part of the electromagnetic beam is lost to unobserved axion radiation.  The intensity modulation is difficult to observe unless there are reasons that the change in intensity varies rapidly with frequency.  Such spectral modulations are indeed expected from a variety of extragalactic $X$-ray sources (see, {\it e.g.},~\cite{Berg:2016ese,Conlon:2017qcw,Conlon:2017ofb,Schallmoser:2021sba} and references therein).

Let us consider a closely-related object to the axion-induced linear polarization, which is the photon survival probability $P_{\gamma \to \gamma}$.  For simplicity, we suppose that the electromagnetic wave of frequency $\omega$ propagates, as in the previous subsection, through a medium of length $L$ with constant magnetic field ${\bf B} = {B}_0 {\bf \hat x}$.  Now, however, we include the plasma frequency in the dispersion relation for the photon. The plasma frequency $\omega_p$ can be thought of as a mass term for the photon, and referring back to~\eqref{eq:Delta_matrix} it adds a contribution to $\Delta_{xx}$ of the form $\Delta_{xx} = - \omega_p^2 /(2 \omega)$. (Note that in this section we will be considering $B \ll B_{\rm crit}$, so the Euler-Heisenberg term will not be relevant.) The plasma frequency for a cold medium with non-relativistic free electron number density $n_e$ is given by
\es{eq:omega_p}{
\omega_p \approx \sqrt{ {4 \pi \alpha_{\rm EM} n_e \over m_e}} \approx 1 \times 10^{-12} \, \, {\rm eV} \sqrt{ {n_e \over 10^{-3}/{\rm cm}^3} } \,.
}
The parametric form for the plasma frequency is relatively straightforward to understand, as it arises from the Coulomb restoring force between electrons (or electrons and ions) in the plasma. In equilibrium the typical separation between electrons in the plasma is $d \propto 1/n_e^{1/3}$. Suppose we have a perturbation that displaces an electron a distance $\delta \ll d$ from its equilibrium position. In practice, the source of this perturbation is the propagation electromagnetic wave, which exerts an electric force on the charge. That charge now experiences a restoring force that pushes it towards its equilibrium position. The restoring force is 
\es{}{
F_{\rm restore} \sim  {e^2 \over d^2} - {e^2 \over (d + \delta)^2} \sim - {e^2 \delta \over d^3} \sim  -\alpha_{\rm EM} n_e \delta \,,
}
with the sign such that the force pushes the charge back towards equilibrium.
Setting $F_{\rm restore} = m_e \ddot \delta$, we see that the charge oscillates with frequency $\omega_p \sim \sqrt{ \alpha_{\rm EM} n_e / m_e}$, as in~\eqref{eq:omega_p}.  Accounting for this oscillation as a contribution to the electrical conductivity it is straightforward to see that electromagnetic waves propagates with the modified dispersion relation $\omega^2 = k^2 + \omega_p^2$, meaning that $\omega_p$ acts as a mass for the photon.   

Returning to the problem at hand, using~\eqref{eq:axion_mixing_eom} and~\eqref{eq:Delta_matrix} it is straightforward to show that 
\es{eq:P_gg}{
P_{\gamma \to \gamma} =  1 - {1 \over 2}  {\Theta^2 \over 1 + \Theta^2} \sin^2 \left( {\Delta m^2 L \over 4 \omega} \sqrt{ 1 + \Theta^2 }  \right) \,,
}
where $\Delta m^2 \equiv m_a^2 - \omega_{\rm pl}^2$ and $\Theta \equiv 2 B_0 g_{a\gamma\gamma} {\omega \over \Delta m^2}$. We also assume that the initial photon is unpolarized, so that there is only a fifty percent chance that it ends up in the ${\bf \hat x}$ state that mixes with the axion.  Let us now assume that $m_a \ll \omega_{\rm pl}$ ($m_a \ll 10^{-12}$ eV for $n_e \sim 10^{-3}$ cm$^{-3}$), then we see that if ${\omega_{\rm pl}^2 L  \over 4 \omega} \gtrsim 1$ the survival probability will oscillate rapidly with frequency $\omega$.

\begin{figure}[htb]  
\begin{center}
\includegraphics[width=0.5\textwidth]{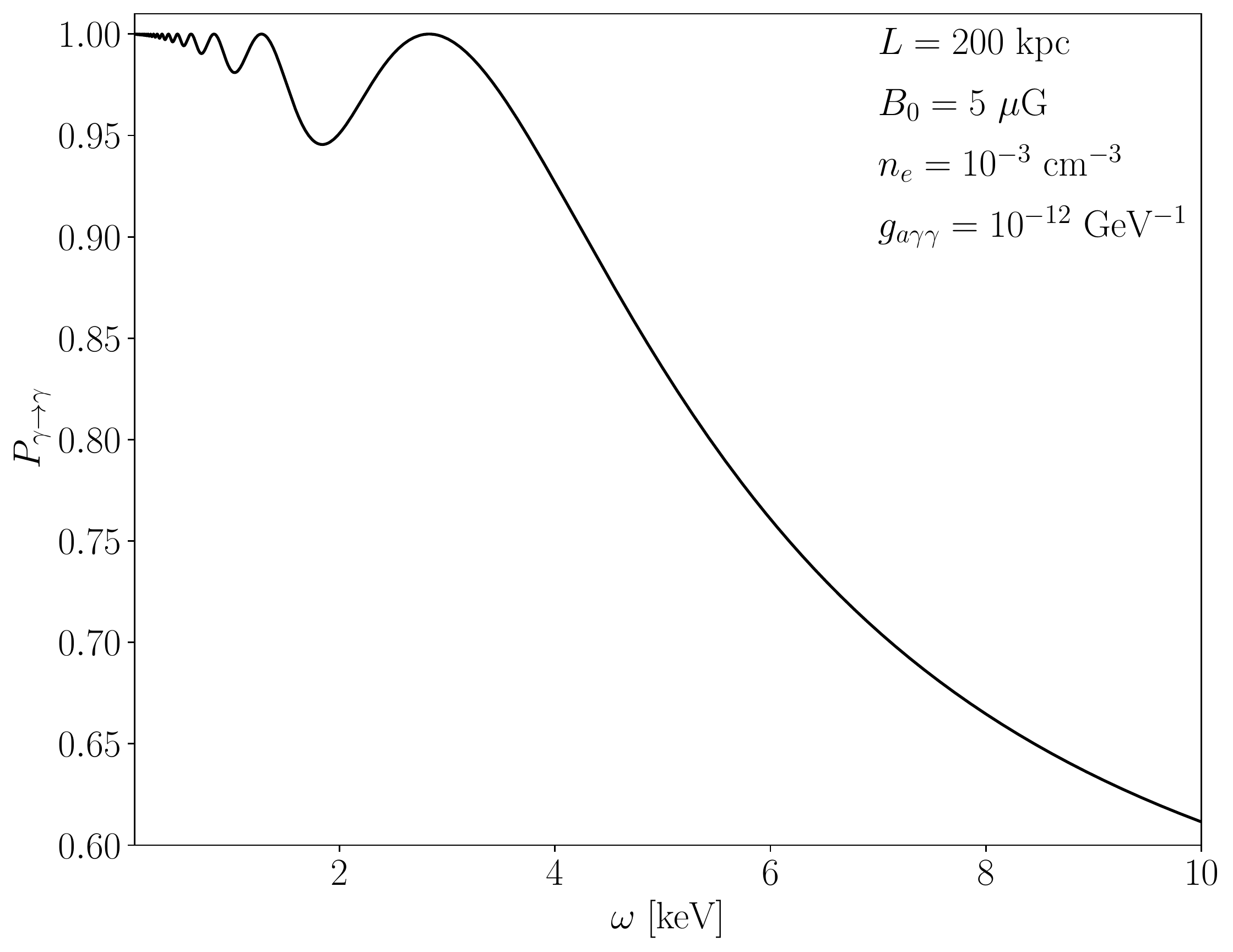}
\caption{An example of spectral modulations, illustrated through the photon-photon survival probability $P_{\gamma\to\gamma}$, as a function of the photon frequency $\omega$  This analysis is a toy example illustrating the signal one may expect for $X$-rays propagating through a galaxy cluster. The simplified model for a cluster here is a box of length $L$ with constant magnetic field $B_0$ and electron number density $n_e$, which gives the photon a plasma mass. In reality, there are many magnetic field domains and a non-trivial profile of $n_e$. The spectral modulations can be searched for on top of the otherwise smooth spectra of sources such as bright central galaxies.  This figure may be reproduced \href{https://colab.research.google.com/drive/16jRto0zfEqxC98MP_gfF0RXIy8WAbmYy?usp=sharing}{with this Colab  Jupyter Notebook}. }
\label{fig:p_agg}
\end{center}
\end{figure}

Consider now $X$-rays propagating out from the central region of a galaxy cluster. The intracluster medium is often modeled as a series of disconnected domains. For example, for the Coma cluster the domain sizes are estimated as $L \sim 10$ kpc (the cluster size is around a Mpc), the free electron density is $n_e \sim 10^{-3}$ cm$^{-3}$, and the characteristic magnetic field strengths in the domains are estimated as $B_0 \sim 5$ $\mu$G (see~\cite{Schallmoser:2021sba} for more precise values).  The magnetic field changes direction and orientation by an amount ${\mathcal O}(1)$ from domain to domain, which suppresses the conversion probability relative to what one would assume taking a single domain of the same total length.  The magnetic field magnitude also fluctuates between domains, and $n_e$ falls going away from the cluster center.  As a rough illustration of what one might find, consider Fig.~\ref{fig:p_agg}.  We take a single domain of $L = 200$ kpc with $B_0 = 5$ $\mu$G, $n_e = 10^{-3}$ cm$^{-3}$, and fix $g_{a\gamma\gamma} = 10^{-12}$ GeV$^{-1}$.  The survival probability is computed using~\eqref{eq:P_gg}. This is not a realistic picture of what  $P_{\gamma \to \gamma}$ may look like, since it approximates the cluster as a single domain, but it illustrates the oscillatory features that arise due to the $\sin$ term in~\eqref{eq:P_gg}. With multiple domains, the oscillations at keV frequencies can become even more frequent (see, {\it e.g.},~\cite{Berg:2016ese,Conlon:2017qcw,Conlon:2017ofb,Schallmoser:2021sba}).  The axion-photon coupling $g_{a\gamma\gamma}$ controls the strength of the oscillations. At lower frequencies than shown, $P_{\gamma \to \gamma}$ is essentially fixed at unity, while at frequencies much higher than shown $P_{\gamma \to \gamma}$ asymptotes to its minimum value.\footnote{For non-zero axion masses the oscillations appear at higher energies (see, {\it e.g.},~\cite{Fermi-LAT:2016nkz}).} It is only in the vicinity of the $X$-ray band that one finds oscillations.

To search for axion-induced spectral modulations one wants to observe the $X$-ray spectrum of a source deep within a cluster, such as Coma.  The predicted spectra of galaxies within these clusters are generically relatively smooth in the $X$-ray band. For example, the spectra may be modeled as a power-law with hydrogen absorption at low energies and various, known atomic absorption lines. Suppose the astrophysical model for the galaxy emission flux is $F_0(\omega; {\bm \theta})$, with ${\bm \theta}$ the vector of nuisance parameters for the parametric astrophysical model. Then, in the presence of axions one constructs the model $F_1(\omega; g_{a\gamma\gamma}, {\bm \theta}) = F_0(\omega; {\bm \theta}) \times P_{\gamma \to \gamma}(\omega, g_{a\gamma\gamma})$. The model $F_1$ may then be compared to data to constrain $g_{a\gamma\gamma}$ (and also $m_a$, which has been set to zero in the discussion so far). The issue, though, is that we can only make statistical predictions for $P_{\gamma \to \gamma}(\omega, g_{a\gamma\gamma})$ because of the random domains; a given realization will have a unique modulation pattern. See~\cite{Berg:2016ese,Conlon:2017qcw,Conlon:2017ofb,Schallmoser:2021sba} for methods that have been developed to deal with this complication, with the most up-to-date limits constraining $g_{a\gamma\gamma} \lesssim {\rm few} \times 10^{-13}$ GeV$^{-1}$ for $m_a \ll 10^{-12}$ eV, though these constraints are subject to systematic uncertainties on {\it e.g.} magnetic field assumptions.

\subsubsection{Narrow-band radio lines from axion dark matter conversion in neutron stars}
\label{sec:axion_NS}

The axion probes involving $g_{a\gamma\gamma}$ that we have discussed so far have the advantage of not requiring the axion to be DM, which is useful because while {\it e.g.} string theory constructions point to the possible existence of ultra-light axions with nontrivial $g_{a\gamma\gamma}$, these particles would generically be expected from the misalignment mechanism to have a negligible DM abundance. On the other hand, if we assume that the axion is DM then it is not surprising new probes become available to us, one of which we discuss now. This probe will be especially promising, since it will target $\sim$$\mu$eV axions potentially reaching down to QCD axion-level $g_{a\gamma\gamma}$, where we do expect the axion to have a sufficient DM abundance.  With that in mind, for the remainder of this subsection we assume that the axion is 100\% of the DM. 

\begin{figure}[htb]  
\begin{center}
\includegraphics[width=0.5\textwidth]{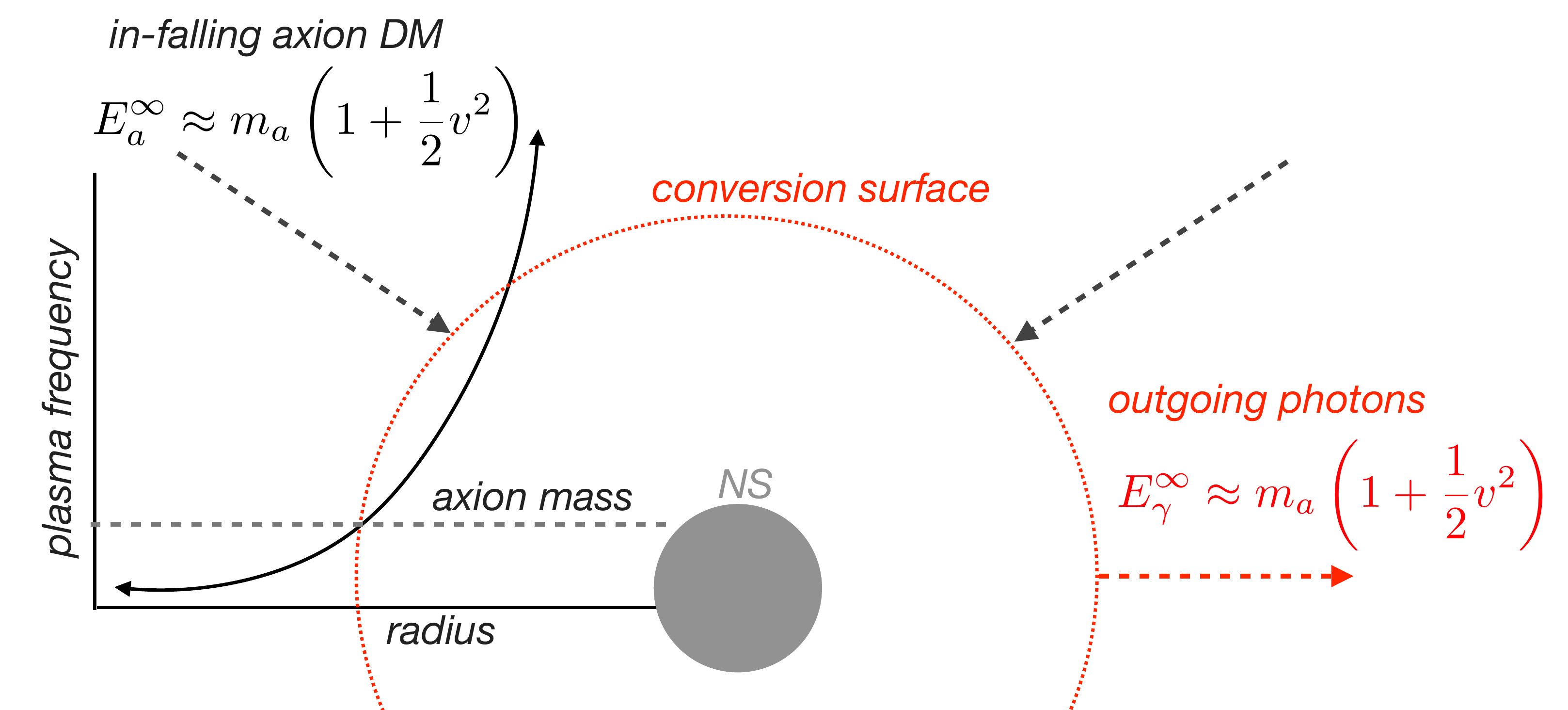}
\caption{Radio telescopes searches for axion DM search for ambient DM axions falling into the NS magnetosphere, by gravity, and then converting to outgoing radio photons in the strong magnetic fields surrounding the NS. The energy of the photons, far away from the NS, is equal to the energy of the incoming axions, asymptotically far in the past, so long as energy is conversed during the conversion process.  This implies that for Galactic NSs the signals may be as narrow as $\delta f / f \sim 10^{-6}$, though energy non-conservation processes can broaden this further (see the text).  The conversion of the axions and photons is stymied by the fact that axions and photons have different dispersion relations. However, the NS is surrounded by a plasma, which gives the photon an effective mass.  The photon mass falls off monotonically from the NS, meaning that there exists a surface, called the conversion surface, where the axion mass equals the photon mass and the conversion may occur resonantly.  The axion-photon conversion occurs predominantly in the vicinity of this conversion surface.    }
\label{fig:axion_NS_ill}
\end{center}
\end{figure}

The general idea, outlined in~\cite{Pshirkov:2007st,Huang:2018lxq,Hook:2018iia,Safdi:2018oeu,Leroy:2019ghm,Battye:2019aco,Battye:2021xvt,Foster:2022fxn} and references therein, is that axion DM may fall into the magnetospheres surrounding NSs and then convert to radio-frequency photons in the strong magnetic fields surrounding the stars.  The radio photons free stream to Earth, where they may be detected with our radio telescopes.  The radio signals should be narrow in frequency space, since the axions that fall into the NSs are initially non relativistic and thus the energy of outgoing photons should be $\sim$$m_a$, as illustrated in Fig.~\ref{fig:axion_NS_ill}.  For NSs in the Milky Way the DM velocity dispersion is $v \sim 10^{-3}$, meaning that we expect $\delta E_a^\infty / E_a^\infty \sim 10^{-6}$, with $E_a^\infty$ the axion energy asymptotically far away from the NS's gravitational field. If energy is conserved during the conversion process, then the energy of the outgoing photons, asymptotically far from the NS, should have the same energy dispersion $\delta E_a^\infty / E_a^\infty \sim 10^{-6}$, with central energy $E_a^\infty = m_a \left( 1 + v_{\rm los} \right)$, where $v_{\rm los}$ is the line-of-sight velocity of the NS. In reality, there are non-energy-conserving processes in the NS magnetosphere that can further broaden the line, as we discuss briefly later in this subsection.

We might guess that the Euler-Heisenberg term is going to be important for this process, since the magnetic fields of NSs can surpass $10^{15}$ G, and we already found the Euler-Heisenberg term was important for $\sim$$10^9$ G fields around MWDs. However, it turns out that this is not the case, and in fact the Euler-Heisenberg term is not relevant at all for axion DM conversion in NS magnetospheres.  To see this, let us refer back to~\eqref{eq:Delta_matrix}, and estimate $\Delta_{aa} = - m_a^2 / (2 \omega) \approx - m_a $, using $\omega \approx m_a$. (Note that in this subsection that axion mass will play a crucial role.)  On the other hand, the Euler-Heisenberg term contributes with $\Delta_{{\rm EH},||} = (7/2) \xi \omega \approx (2\times 10^{-4}) m_a (B / B_{\rm crit})^2$. Thus, for $B \lesssim 2 \times 10^{15}$ G the Euler-Heisenberg term is a subdominant contribution to the matrix ${\bf \Delta}$ and may be neglected. The crucial point, here, is that $\Delta_{aa}$ grows in importance with decreasing $\omega$, while $\Delta_{{\rm EH},||}$ decreases in that limit.

The magnetic fields around NSs are well described by dipoles, for which the magnetic field magnitude falls off as $B = B_0 (r_{\rm NS}/r)^3$, where $B_0$ is the field value at the surface and $r_{\rm NS} \approx 10$~km is the NS radius. Naively, we may then estimate the axion-to-photon conversion probability as $p_{a \to \gamma} \sim g_{a\gamma\gamma}^2 B_0^2  L^2$, for some length scale $L$.  Referring back to~\eqref{eq:L_p_large_ma} we may guess that $L \sim m_a / \omega^2 \sim 1/ m_a$ (since $| r_{\rm NS} \Delta_{aa} | \gg 1$), in which case $p_{a \to \gamma} \sim g_{a\gamma\gamma}^2 B_0^2  m_a^{-2}$ would be too small to be of practical interest.  As we motivate below, on the other hand, a more careful calculations yields the result $L \sim \sqrt{ r_{\rm NS} m_a^{-1} }$, in which case, very roughly, $p_{a \to \gamma} \sim g_{a\gamma\gamma}^2 B_0^2  r_{\rm NS} \, m_a^{-1}$ becomes large enough to be phenomenologically relevant.

The key feature in understanding axion-photon conversion in NS magnetospheres is the plasma frequency profile exterior to the NS surface in the magnetosphere.  Following Goldreich and Julian (GJ)~\cite{1969ApJ...157..869G}, we can derive a straightforward estimate for the free electron density, and in-turn $\omega_p$, in the magnetosphere. First, let us start with the assumption that the co-rotating magnetosphere is cold and in equilibrium.\footnote{Of course, the idea of the co-rotating magnetosphere cannot be the full story.  Further out from the NS the magnetosphere would rotate with increasing speed, such that at the so-called light cylinder the magnetosphere would rotate at the speed of light.} This means, by definition, that the charged particles in the co-rotating magnetosphere should not feel any net force that would change their co-rotating positions.  Naively, however, the charges (let us assume they are electrons) feel a Lorenz force from the NS's magnetic field, whose magnitude is $F \sim e \Omega r B(r)$, with $\Omega$ the angular frequency of the NS and $B(r)$ the magnitude of the magnetic field at the charges radius $r$ from the NS center. Note that we leave off geometric factors in this estimate.  Depending on where the charge is relative to the rotation and the magnetic field, the Lorentz force will either try to push the charge towards the NS surface or move it away from the surface. This Lorentz force must be compensated by an electrostatic force that is equal and opposite. Such a force is generated by hypothesizing a distribution of co-rotating charges, with number density $n_e$, exterior to the NS.  The electrostatic force will have magnitude, approximately, $F \sim e^2 n_e r$, and it will be either towards or away from the NS's surface depending on whether or not that region of the magnetosphere is electron or ion (or possibly even positron) dominated. Equating these two forces we infer that $n_e \sim \Omega B / e$.  Performing a more careful calculation one finds the GJ formula~\cite{1969ApJ...157..869G}
\es{}{
n_e \approx {2 {\bf \Omega} \cdot {\bf B} \over e } \,,
}
for $r \ll \Omega^{-1}$. Here, ${\bf \Omega}$ is the NS's angular velocity vector.  More careful calculations and simulations reveal a number of subtleties and situations where the GJ formula does not apply, for example in the open-field regions of pulsars where there is significant charge acceleration leading to observed pulsar radiation, but in general the GJ formula sets the correct scale, at least relative to what is found in simulations in the quieter parts of the magnetosphere with closed fields lines. One disclaimer to note about the GJ model is that formally it only solves for the charge distribution and not the distribution of total particles. In principle, there could be a charge-symmetric distribution of {\it e.g.} electrons and positrons, which would leave the charge distribution unchanged but affect the plasma density.  Such charge symmetric populations may be present in younger and more active NSs, though simulations of older and quieter NSs find that such a symmetric population is likely not important.  Note that at least for standard pulsars with field strengths below the critical strength the charges that make up the magnetosphere arise in the first place by being stripped off the NS surface.

The GJ model suggests that, at a given radius, the plasma frequency falls off as $\omega_p \approx \omega_p^0 ( r_{\rm NS}/r)^{3/2}$, with $\omega_p^0$ the plasma frequency at the NS surface.  Typical values for $\omega_p^0$ are 
\es{}{
\omega_p^0 \approx \sqrt{ {8 \pi \alpha_{\rm EM} \Omega B_0 \over e \, m_e}} \approx 5 \, \, \mu {\rm eV} \sqrt{ { 1 \, \, {\rm s} \over P_{\rm NS}} {B_0 \over 10^{14} \, \, {\rm G}} } \,,
}
with $P_{\rm NS}$ the NS's period.  Now imagine an outgoing axion, which has fallen through the NS and is heading outwards. As it propagates, it sees a monotonically decreasing plasma frequency. At the radius $r_c$, defined such that $\omega_p(r_c) = m_a$, the axion and the photon have the same dispersion relations. Resonant conversion between the axion and the photon may occur in the vicinity of $r_c$, leading to the enhanced conversion probability. Similarly, in-falling axions may also convert to photons at the conversion radius, and those photons are then reflected and propagate outwards to infinity.  The set of all conversion radii across the magnetosphere is referred to as the {\it conversion surface}, as illustrated in Fig.~\ref{fig:axion_NS_ill} (note that generically the conversion surface is not spherical and, furthermore, not even a surface). 

Consider an outwards propagating axion on a radial trajectory.  Going back to~\eqref{eq:axion_mixing_eom} and again applying the trick of using first-order time-dependent perturbation theory from quantum mechanics, we can approximate the axion-to-photon conversion probability of the axion having converted to a photon by radius $r$ as 
\es{eq:pagg_NS}{
p_{a\to\gamma}(r) \sim \left| \int_0^r dr' {B(r') g_{a\gamma\gamma} \over v_c} e^{-i  {\int_0^{r'} dr'' \left[ m_a^2 - \omega_p^2(r'') \right] \over 2 m_a v_c}} \right| \sim g_{a\gamma\gamma}^2 B(r_c)^2 r_c m_a^{-1} \,, 
}  
where we have again left off geometric factors, and where $v_c$ is the velocity of the DM at the conversion surface, where we approximate the conversion predominantly takes place.  The factors of $v_c$ can be found by re-deriving the matrix ${\bf \Delta}$ in~\eqref{eq:Delta_matrix} in the case of a non-relativistic wave, for which {\it e.g.} $k = v_c \omega$ in the vicinity of the conversion surface.  To evaluate the integral in~\eqref{eq:pagg_NS} one may use the method of stationary phase, since the exponential oscillates rapidly except in the vicinity of where the axion and photon dispersion relations are matched.  It is also appropriate to compute an asymptotic conversion probability $p_{a \to \gamma}^\infty \equiv v_c \lim_{r \to \infty} p_{a\to \gamma}(r)$, where the factor of $v_c$ accounts for the decrease of the amplitude of the electromagnetic waves as they propagate away from the conversion surface. By approximating all axion trajectories as having a conversion probability as in~\eqref{eq:pagg_NS}, which -- as we discuss below -- is not always the best approximation, we may roughly estimate the radio power from a NS from axion-photon conversion as 
\es{eq:P_NS_radio}{
{\mathcal P} &\approx 4 \pi r_c^2 \times 2 p_{a \to \gamma}^\infty \rho_{\rm DM}(r_c) v_c \sim 8 \pi r_c^3 {v_c^3 \over v_0} \rho_{\rm DM}^\infty g_{a\gamma \gamma}^2 B(r_c)^2 m_a^{-1} 
\\
&\sim 5 \times 10^{21} \, \, {{\rm erg} \over s} \left({g_{a\gamma\gamma} \over 10^{-12} \, \, {\rm GeV}^{-1}} {m_a \over 5 \, \, \mu{\rm eV} }\right)^2 \sqrt{{B_0 \over 10^{14} \, \, {\rm G}} }\left( {P \over 1 \, \, {\rm s}} \right)^{3/2} {\rho_{\rm DM} \over 0.4 \, \, {\rm GeV}/{\rm cm}^3} {200 \, \, {\rm km/s} \over v_0} \,,
}
where we use the approximate relation for the conversion radius 
\es{}{
r_c \approx r_{\rm NS} \left( {\omega_0 \over m_a} \right)^{2/3} \,,
}
keeping in mind that we need $r_c > r_{\rm NS}$, in addition to $v_c \approx \sqrt{2 G M_{\rm NS} / r_c}$.  Note that $v_0^2$ is the velocity dispersion of the asymptotic DM velocity distribution, which enters because$\rho_{\rm DM}(r_c) \sim (v_c / v_0) \rho_{\rm DM}^\infty$ by gravitational capture.

As an illustration, consider a nearby NS, like the magnificent seven NSs, at $d \sim 1$ kpc from Earth. Let us also assume that the signal at $f = {m_a \over 2 \pi} \approx 1 \, \, {\rm GHz} \left( {m_a \over 5 \,\, \mu{\rm eV}} \right)$ has a width $\delta f / f \sim 10^{-6}$ set by the asymptotic energy dispersion of the DM.  A convenient quantity to use when quantifying the sensitivity of radio telescopes is the flux density $S$, which is equal to the flux at Earth (in units of erg/s/cm$^2$) divided by the bandwidth of the signal in Hz. Fixing the parameters to the fiducial ones in~\eqref{eq:P_NS_radio}, and taking $m_a = 5 \, \, \mu{\rm eV}$, we then estimate that for our 1 kpc source the flux density is $S \sim 5 \times 10^{-26} \, \, {\rm erg/cm}^2/{\rm s}/{\rm Hz} \left( {g_{a\gamma\gamma} / 10^{-12} \, \, {\rm GeV}^{-1}} \right)^2 \sim 5 \times 10^{-3} \, \, {\rm Jy}  \left( {g_{a\gamma\gamma} / 10^{-12} \, \, {\rm GeV}^{-1}} \right)^2 $, where we have converted to the radio astronomy unit the jansky (Jy), defined as $1 \, \, {\rm Jy} \equiv 10^{-23} \, \, {\rm erg}\cdot {\rm s}^{-1} \cdot {\rm cm}^{-2} \cdot {\rm Hz}^{-1}$.  At this frequency and bandwidth modern radio telescopes can achieve sensitivity to $S \sim 10^{-4}$ Jy, meaning that such telescopes should be able to probe axion-photon couplings already around or smaller than $g_{a\gamma\gamma} \sim 10^{12} \, \, {\rm GeV}^{-1}$, which is indeed the case (see, {\it e.g.},~\cite{Foster:2020pgt,Foster:2022fxn}).

The description above is based off of analytic estimates of the axion-induced radio power done up to $\sim$2018~\cite{Pshirkov:2007st,Huang:2018lxq,Hook:2018iia,Safdi:2018oeu}.
A large body of work since $\sim$2018, however, has shown that the picture presented above is overly simplistic in a number of important ways, and any serious attempt to model the axion flux from NSs should account for these more recent modifications to the formalism (see, {\it e.g.},~\cite{Leroy:2019ghm,Battye:2019aco,Battye:2021xvt,Foster:2022fxn} and reference therein).  Here, we list two of the important modifications. First, the bandwidth is broadened beyond $\delta f / f \sim v_0^2 \sim 10^{-6}$ by energy non-conserving processes during the conversion process. For example, the outgoing photons are refracting and reflecting in a moving medium, which imparts a frequency shift. As a simple example of this, consider the case where the in-falling axions convert to photons, which reflect off of the conversion surface. The conversion surface is moving, since the NS is rotating, and this leaves a frequency shift in the outgoing photon proportional to the appropriately projected speed of the conversion surface. A second important aspect of the story that needs to be accounted for is the de-phasing between the axion and photon for non-radial trajectories. To maintain resonant conversion the axion and photon waves must remain in-phase over the distance $\sim$$\sqrt{m_a^{-1} r_c}$. However, for non-radial trajectories the photon will be refracted  while the axion will not, leading to the two waves becoming out of phase at distances shorter than $\sim$$\sqrt{m_a^{-1} r_c}$, which suppresses the conversion.  

Lastly, we note that the radio searches are receiving attention at the moment in light of the upcoming square kilometer array (SKA) radio telescope array, which will push us to unprecedented sensitivity to faint radio sources. The most promising target appears to be the GC region of the Milky Way~\cite{Safdi:2018oeu}, since there is more DM there and since there are many NSs in that region, but more work is needed to determine if SKA will be sensitive enough to detect a few $\mu$eV QCD axion.

\subsection{Probes only involving axion production in stars}

Many of the strongest probes of axions involve the production of the particles within stars. These include, for example, probes using NSs and supernovae (SN)~\cite{Raffelt:1996wa,Raffelt:2006cw,Hamaguchi:2018oqw,Buschmann:2021juv}, WD stars~\cite{Raffelt:1985nj,Blinnikov:1994eoa}, horizontal branch (HB) stars~\cite{Ayala:2014pea,Dolan:2022kul}, and red giants (RG)~\cite{Capozzi:2020cbu,Straniero:2020iyi}.  The references~\cite{Raffelt:1996wa,Raffelt:2006cw,Irastorza:2021tdu} are excellent starting points in understanding how axions are produced within different types of stars and how this subsequent energy loss can lead to observable signatures. In this section, we take a more simplistic approach and perform a rough estimate of axion production within the Sun. We then comment on the modifications needed to calculate axion production in other types of stars.

\begin{figure}[htb]  
\begin{center}
\includegraphics[width=0.4\textwidth]{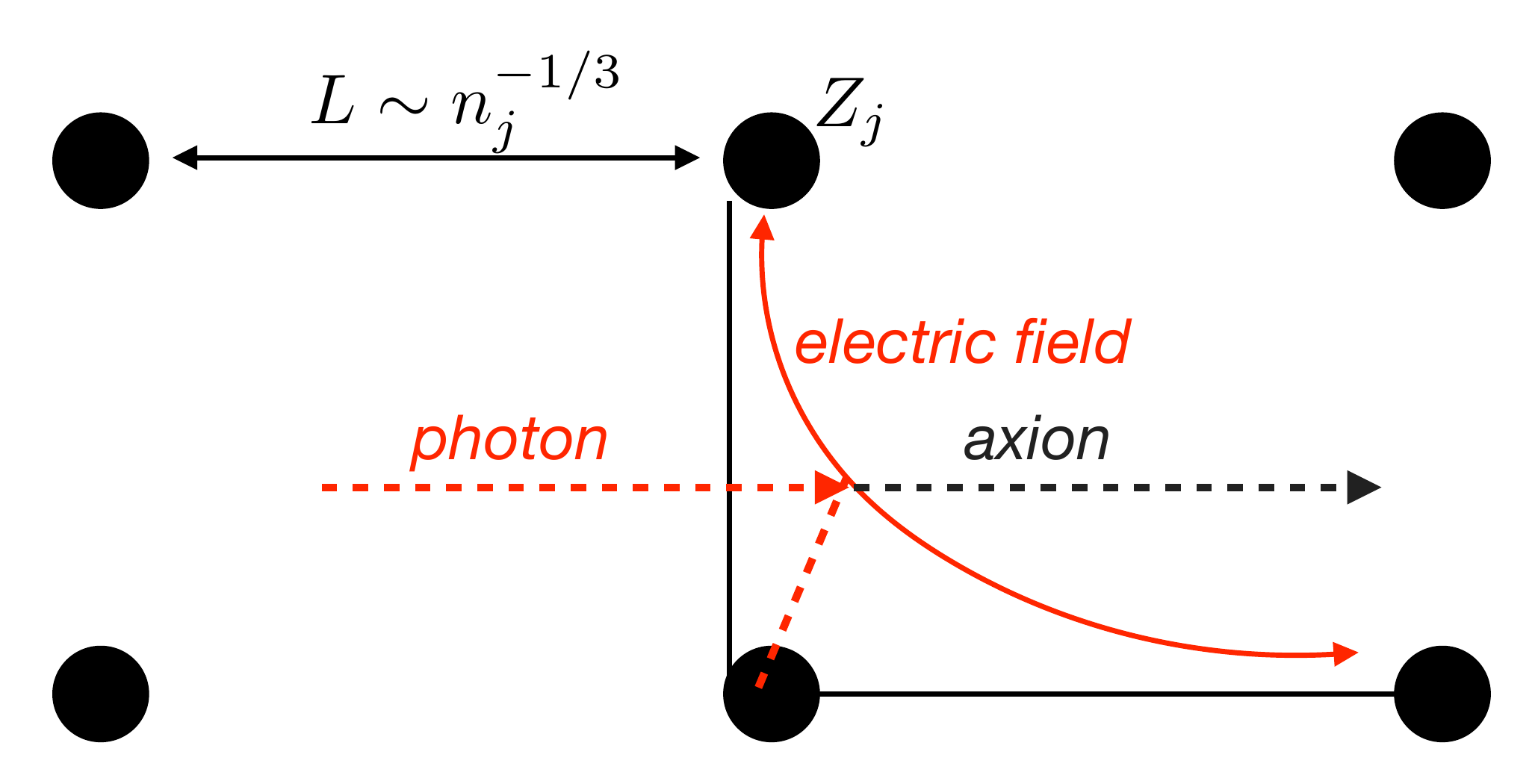}
\caption{A toy illustration for how photons can convert to axions in the hot plasma of a stellar interior. The ions and electrons in the plasma source electric fields, and the thermal photons may resonantly convert to axions in the electric fields over length-scales of order the coherence length for the electric fields, which is typically set geometrically by the number density of ions and electrons as illustrated.  The converted axions then provide a new pathway for stellar cooling, may be detected in laboratory experiments such as CAST in the case of the Sun, or may lead to novel $X$-ray signatures if the axions subsequently reconvert to photons in astrophysical magnetic fields outside of the star.}
\label{fig:star_ill}
\end{center}
\end{figure}

Let us consider the solar core as a hot ionized gas at a temperature $T \sim 0.7$ keV~\cite{Raffelt:1996wa}.  Consider an ion species $j$ with electric charge $Z_j$ and number density $n_j$. The ions source a Coulomb field that is coherent over the inter-ion spacing $L \sim n_j^{1/3}$ (see Fig.~\ref{fig:star_ill}).
Within the volume between ions the typical Coulomb field strength is $E \sim Z_j e / L$.  Propagating photons can resonantly convert to axions through the Coulomb field. The conversion probability $p_{\gamma \to a}$ after propagating a distance $L$ should be of order $p_{\gamma \to a} \sim g_{a\gamma\gamma}^2 E^2 L^2 \sim g_{a\gamma\gamma}^2 4 \pi Z_j^2 \alpha_{\rm EM}^2 n_e^{2/3}$.  Now suppose that a relativistic photon propagates a distance $\Delta t$ in a time $\Delta t$, with $\Delta t \gg L$. The photon effectively sees $N_L = \Delta_t / L$ different domains of coherent electric field. We can thus compute the rate at which the photon is expected to convert to an axion as 
\es{}{
\Gamma_{\gamma \to a} \sim {N_L \over \Delta t} p_{\gamma \to a} \sim g_{a\gamma\gamma}^2 4 \pi Z_j^2 \alpha_{\rm EM}^2 n_e \,.
}
This naive estimate is remarkably similar to the result from a more careful calculation~\cite{Raffelt:2006cw}, which yields 
\es{eq:full_solar_rate}{
\Gamma_{\gamma \to a} \approx {g_{a\gamma\gamma}^2 T \kappa_s^2 \over 32 \pi} \left[ \left(1 + {\kappa_s^2 \over 4 \omega^2} \right) \log \left( 1 + {4 \omega^2 \over \kappa_s^2} \right) - 1 \right] \,,
}
where 
\es{eq:kappa_s}{
\kappa_s^2 \equiv {4 \pi \alpha_{\rm EM} \over T} \left( \sum_j Z_j^2 n_j + n_e \right) \,,
}
allowing for the possibility of multiple ion species with number densities $n_j$, in addition to an electron population with number density $n_e$.  Note that the pre-factor in~\eqref{eq:full_solar_rate} is independent of $T$, but the $T$ dependence is put into~\eqref{eq:kappa_s} since this combination is roughly $T$-independent ($(\kappa_s / T)^2 \approx 12$ in the Sun~\cite{Raffelt:2006cw}). 

To derive~\eqref{eq:full_solar_rate} precisely, we can compute the matrix element for photon-axion conversion off of an external photon field, in the limit where the ion mass is much heavier than the temperature such that the ion does not recoil. The differential cross-section for this process is 
\es{eq:diff_xsec_prim}{
{d \sigma_{\gamma \to a} \over d \Omega} = {g_{a\gamma\gamma}^2 Z^2 \alpha_{\rm EM} \over 8 \pi} { | {\bf k_a} \times {\bf k_\gamma}|^2 \over ({\bf q}^2 + \kappa_s^2)^2} \,,
}
where ${\bf q}$ is the momentum transfer between the photon and the axion (${\bf q} = {\bf k_\gamma} - {\bf k_a}$), with ${\bf k_\gamma}$ and ${\bf k_a}$ the photon and axion momentum vectors, respectively.  Here, $\kappa_s$ enters because it is the Debye-H\"uckel screening length in a plasma at high temperatures (at low temperatures, the screening length asymptotes to the value $\sim$$n_e^{-1/3}$ that we discussed previously, known as the Thomas-Fermi screening length).  The transition rate $\Gamma_{\gamma \to a}$ is then given by the total cross-section $\sigma_{\gamma \to a}$ multiplied by the number density of targets. Performing the integral of~\eqref{eq:diff_xsec_prim} for the total cross-section then leads to~\eqref{eq:full_solar_rate}.  See~\cite{Raffelt:1985nk} for a further discussion of the screening effects. 

 The total axion luminosity $L_a$, in {\it e.g.} units of erg/s, is estimated as $L_a \sim V_{\rm core} \cdot T \cdot n_\gamma \cdot \Gamma_{\gamma \to a}$, where $n_\gamma \approx {2 \zeta(3) \over \pi^2} T^3$ is the photon number density and $V_{\rm core} = {4 \over 3} \pi r_{\rm core}^3$ is the iso-thermal core volume, with $r_{\rm core} \approx 1.7 \times 10^5$ km for the Sun.  Using~\eqref{eq:full_solar_rate} with $\omega \sim T$ we thus estimate 
\es{eq:L_a_sun}{
L_a \sim 3 \times 10^{30} \left( {g_{a\gamma\gamma} \over 10^{-10} \, \, {\rm GeV}^{-1}} \right)^2 {{\rm erg}\over s} \sim 10^{-3} \, \, L_\odot  \left( {g_{a\gamma\gamma} \over 10^{-10} \, \, {\rm GeV}^{-1}} \right)^2   \,.
}
If $g_{a\gamma\gamma}$ becomes too large, then $L_a$ is large enough that the standard solar model would begin to differ from data. For example, if $g_{a\gamma\gamma} \sim 10^{-6}$ GeV$^{-1}$ one can estimate that the Sun would only live for around 1000 years in total, which is clearly not correct.  At a more precise level, but studying solar neutrino production in the solar neutrino model one can constrain $L_a \lesssim 0.04 L_\odot$, or $g_{a\gamma\gamma} \lesssim 5 \times 10^{-10}$ GeV$^{-1}$~\cite{Raffelt:2006cw}.

We should check that the plasma frequency within the solar core $\omega_p$ is actually below the temperature, such that it makes sense to talk about a thermal population of photons.  Given the mass and size of the solar core (and there being around 1.5 nucleons per electron in the core), we can estimate an electron number density $n_e \sim 6 \times 10^{25}$ cm$^{-3}$, giving a plasma frequency through~\eqref{eq:omega_p} of $\omega_p \sim 250$ eV.  This plasma frequency is below the temperature, meaning that there are propagating, non-Boltzmann suppressed photons, and the estimate we performed above is likely correct up to corrections of order $\omega_p / T$.  

Note that $\omega_p$ is very close to $T$ for the Sun, which means that modest changes to the core composition could easily push $\omega_p$ higher and thus effectively shut off the Primakoff axion production mechanism described above of thermal photons scattering into axions using $g_{a\gamma\gamma}$.  This is precisely what happens as stars on the main sequence exit onto the RG branch. On the main sequence stars burn H in their cores, generating He, but eventually this H fuel source burns out. When this happens, a degenerate, dense core of He develops in the solar core, surrounded by an increasingly small H-burning shell. This stage of stellar evolution characterizes the RG branch. On the RG branch, the plasma frequency in the stellar core is higher than the temperature, so Primakoff production of axions does not take place.  However, eventually the temperature within the core rises to become high enough to burn He into carbon, starting what is called the Helium flash, where the core rapidly heats up and expands as it ignites. This new stellar phase, called the HB phase, is characterized by a He burning core. Since the He burning core is hotter and less dense, its plasma frequency now drops again below the temperature, meaning that Primakoff production is allowed. Thus, axions are produced by $g_{a\gamma\gamma}$ in the HB phase but not in the RG phase.  Axions can provide an efficient source of energy loss for HB stars, which decreases their lifetime since they burn their helium fuel more rapidly.  A powerful way for quantifying this affect is by measuring $R = N_{\rm HB} / N_{\rm RG}$ is globular clusters, which are clusters of relatively old and low-mass stars with similar evolutionary tracks and initial conditions, where $N_{\rm HB}$ is the number of HB stars and $N_{\rm RG}$ is the number of RG stars. As $g_{a\gamma\gamma}$ increases, $R$ is expected to decrease. By comparing the theoretical prediction for the $g_{a\gamma\gamma}$ dependence of $R$ with data one finds the constraint $g_{a\gamma\gamma} \lesssim 6.6 \times 10^{-11}$ GeV$^{-1}$~\cite{Ayala:2014pea}.

While more compact stars, such as RG stars, WDs, and NSs, cannot produce axions through the Primakoff mechanism, since the plasma frequency is higher than the temperature, they can still efficiently produce axions through their derivative couplings to fermions (see~\eqref{eq:axion_EFT}).  As a concrete example, let us consider WDs, which are supported by electron degeneracy. In WDs electrons can scatter off of ions by Coulomb exchange and emit an axion in the final state by Bremsstrahlung.  That is, $e^- Z \to e^- Z a$, where $Z$ is an ion of charge $Z$. A typical white dwarf has a mass $\sim$$0.6$$M_\odot$, a radius $\sim$$0.01$$R_\odot$, and a temperature $T \sim {\rm keV}$.  They also typically have twice as many nucleons as electrons, and so a generic electron number density is around $n_e \sim 4 \times 10^{29}$ cm$^{-3}$, giving a Fermi energy $E_f = {1 \over m_e}   (3 \pi^2 n_e)^{2/3} \sim 0.2$ MeV, meaning that the electrons are almost relativistic. Approximating the electrons as non-relativistic, the Fermi momentum is $p_f \sim 0.4$ MeV.
The fact that $T \ll p_f$ is important because the electrons must scatter into free states, since they are fermions, but all states with momentum less than $p_f$ are filled. Thus, the only electrons that can take place in the scattering process are those with momentum difference of order $T$ from $p_f$, since these can have the chance of scattering into a free state. This brings about the so-called Fermi suppression factor of $(T / p_f)^2 \sim 6 \times 10^{-6}$ for the scattering electron. Skipping the details (but see {\it e.g.}~\cite{Raffelt:1996wa}), the luminosity in axions emitted from a WD is, roughly, 
\es{eqref:L_a_WD}{
L_a \approx 3 \times 10^{-4} L_\odot \left( {g_{aee} \over 10^{-13} } \right)^2 \left( {M_{\rm WD} \over 1\, \, M_\odot} \right) \left( {T \over 1 \, \, {\rm keV}} \right)^4 \,,
}
where we have defined $g_{aee} \equiv C_{aee} m_e / f_a$, referring back to~\eqref{eq:axion_EFT}.  Let us understand the factors of $T$ in~\eqref{eqref:L_a_WD}.  First, the cross-section $\sigma$ for $e^- Z \to e^- Z a$ is proportional to $T$ ($\sigma \propto T$), since there is an electron spin flip associated with the electron-electron-axion vertex, which brings down an axion energy (and the axion energy is of order $T$).  Second, since we are calculating a luminosity we need to integrate the differential axion transition rate times the energy of the axions, and the axion energies are of order $T$. The remaining two powers of $T$ come from the Fermi suppression discussed above.  Axion emission in WDs can also be an appreciable source of energy loss, which accelerates WD cooling. But comparing the predicted luminosity function of WDs to data one can thus constrain the coupling $g_{aee}$~\cite{Raffelt:1985nj,Blinnikov:1994eoa}. 

In passing, we note that NSs can also emit axions from a similar process to that in WDs, but now the axions are emitted by bremsstrahlung during nucleon-nucleon scattering ($N N \to NN a$), which is mediated by pion exchange, through {\it e.g.} the fundamental axion-quark or axion-gluon couplings.  The physics is similar to electron bremsstrahlung, but in this case $L_a \propto T^6$, since both nucleons are degenerate and thus each is subject to a $(T/p_f)^2$ Fermi suppression factor. NS and SN cooling by axion production currently provides the strongest constraint on high-mass QCD axions, disfavoring -- roughly -- QCD axions above $\sim$20 meV (see, {\it e.g.},~\cite{Buschmann:2021juv}).

\subsection{Probes involving both axion production in stars and axion-photon mixing}

A number of terrestrial and astrophysical searches combine the concepts of axion production in stars with axion-photon mixing in external magnetic fields. The most canonical search along these lines is employed by the CERN Axion Solar Telescope (CAST) (see~\cite{CAST:2017uph} and references therein).  The basic idea behind CAST is to look for the solar axions at energies $\sim$keV converting into $X$-ray photons within a terrestrial magnetic field. The CAST experiment consists of a beam pipe (literally an LHC beam pipe) with length $L \sim 10$ m, radius $r \sim 3$ cm, and transverse magnetic field $B_0 \sim 10$ T.  The detector is pointed at the Sun (clearly with shielding between the Sun and the detector to block the ordinary $X$-rays!) around 1 hr per day, leading to approximately $t_{\rm exp} = 1000\, \,  {\rm hours}$ of exposure time in the final CAST analysis~\cite{CAST:2017uph}. For low mass axions the conversion probability is $p_{a \to \gamma} \sim g_{a\gamma\gamma}^2 B_0^2 L^2 \sim 10^{-16} (g_{a\gamma\gamma} / 10^{-10} \, \, {\rm GeV}^{-1} )^2$. While this conversion probability may seem small, it is compensated by the large flux of axions incident on the detector, $F_a = L_a / (4 \pi d^2)$ with $d$ the distance to the Sun, which may be estimated using~\eqref{eq:L_a_sun}.  Combining the conversion probability with the axion flux, along with the cross-sectional area $A$ of the detector, we estimate the number $N_a$ of axion-induced $X$-rays that CAST observes as
\es{}{
N_a \approx {F_a \over \omega} A p_{a \to \gamma} \times t_{\rm exp} \sim 10^4 \left( {g_{a\gamma\gamma} \over 10^{-10} \, \, {\rm GeV}^{-1}} \right)^4 \,,
}
where we have approximated all of the axions having energy $\omega \sim {\rm keV}$.  If this was a zero-background experiment we would expect CAST to be sensitive to axions with $g_{a\gamma\gamma} \sim 10^{-11}$ GeV$^{-1}$ or above, but in reality CAST does have a small background rate and an efficiency factor below unity, so their low-axion-mass limit is slightly worse.  Including all of the factors order unity and the background, their upper limit becomes $g_{a\gamma\gamma} \gtrsim 6.6 \times 10^{-11}$ GeV$^{-1}$ at low $m_a$~\cite{CAST:2017uph}.

What defines low $m_a$ for CAST? Going back to the discussion in Sec.~\ref{sec:MWD}, low-mass axions are those for which $|\Delta_aa| L  \sim {m_a^2 L \over \omega} \lesssim 1$. For axion masses above this critical value, which is around $m_a \sim 6$ meV for CAST, the conversion probability becomes increasingly suppressed like $m_a^{-4}$.  As we discovered in Sec.~\ref{sec:axion_NS}, one way of addressing the dispersion mismatch issue is to give the photon a plasma frequency. CAST has done this, in fact, by filling the beam pipe with a gas and adjusting the density in order to scan over different plasma masses and, thus, $m_a$.

In addition to converting to photons in a terrestrial experiment, axions produced in stars may also convert to photons in astrophysical magnetic fields. There are two different classes of searches that one can do along these lines: (i) searches for axions converting to photons in the stellar magnetic fields of the stars that are producing the axions, and (ii) looking for axions converting in the Galactic or extragalactic magnetic fields between the star and Earth.  One example of the latter process is the search for a gamma-ray burst coincident with the neutrino burst from SN1987a~\cite{Payez:2014xsa}.  SN1987a was a nearby type II SN that went off around 50 kpc from Earth in 1987; it was a core-collapse SN that should have led to the creation of a proto-NS.  That NS has a large initial temperature on the order of 10's of MeV, which is large enough such that there is a non-negligible population of thermal photons with energies above the plasma frequency.  These thermal photons undergo Primakoff production of axions in the proto-NS core; those axions then escape the SN, due to their weak interactions. The axions, however, may reconvert to photons in the Milky Way's magnetic field. The coherent part of the magnetic field of the Milky Way is on the order of $\sim$$\mu$G, with the size of the Milky Way $\sim$10 kpc. Thus, we naively predict conversion probabilities $p_{a \to \gamma} \sim g_{a\gamma\gamma}^2 B_0^2 L^2 \sim 10^{-4} \left( {g_{a\gamma\gamma} \over 10^{-12} \, \, {\rm GeV}^{-1}} \right)^2$, though doing this more carefully leads to slightly smaller conversion probabilities~\cite{Payez:2014xsa}.  Luckily, at the same of the SN 1987a explosion a gamma-ray telescope, the Gamma-Ray Spectrometer  of the Solar Maximum Mission, was taking data and would have been sensitive to the gamma-ray signal, though it observed no excess gamma-ray counts. The absence of a gamma-ray signal implies a limit $g_{a\gamma\gamma} \lesssim 5 \times 10^{-12}$ GeV$^{-1}$ for low mass axions. We should all be thinking about how to be ready to discover or further constrain axions the next time a nearby SN goes off! Other examples of searches for axion-photon conversion in Galactic magnetic fields include searches for $X$-rays from Primakoff production in super star clusters~\cite{Dessert:2020lil} and Betelgeuse~\cite{Xiao:2020pra}.

In order for it to be favorable for an axion to convert in a stellar magnetic field instead of the Galactic magnetic fields, the stellar magnetic fields should be very large. This essentially reduces case (i) listed in the previous paragraph to MWDs and NSs~\cite{Raffelt:1987im,Fortin:2018ehg,Buschmann:2019pfp,Dessert:2021bkv}.  In these cases, the probes are of mixed coupling combinations, such as $g_{aee} g_{a\gamma\gamma}$, though in terms of those coupling combinations such searches, which are usually carried out in the $X$-ray band, tend to be the strongest available.

\section*{Acknowledgements}

I thank the organizers of TASI 2022 -- Jiji Fan, Stefania Gori, and LianTao Wang -- for the invitation to give these lectures, and I think the students of TASI 2022 for the stimulating discussions.  I also thank Joshua Benabou, Quentin Bonnefoy, Malte Buschmann, Christopher Dessert, Joshua Foster, Soubhik Kumar, Kevin Langhoff, Mariangela Lisanti, Claudio Manzari, Toby Opferkuch, Nadav Outmezguine, Yujin Park, Nick Rodd, Christiane Scherb, Tracy Slatyer, Linda Xu, and Sam Witte for helpful discussions and Linda Xu for help compiling App.~\ref{app:xsec}.  I am supported in part by the DOE Early Career Grant DESC0019225.  

\appendix

\section{Cosmology -- a rapid review}
\label{sec:cosmo}

In these lecture notes we focus on aspects of the particle nature of DM. The starting point to any such endeavor, however, is an understanding of the successful picture of the big bang cosmology, where the DM is described by a perfect fluid.  There are numerous, excellent reviews of cosmology that any student of particle theory would be wise to study, including~\cite{Trodden:2004st} from TASI 2002 and 2003,~\cite{Baumann:2009ds} from TASI 2009, and~\cite{Green:2022bre} from TASI 2022.  The summary presented here is rapid and only meant as a quick ``look-up table" for the relevant equations needed when {\it e.g.} computing WIMP DM freeze-out or the axion DM abundance.

The metric of spacetime in the early universe is described by the flat Friedmann-Robertson-Walker (FRW) metric 
\es{}{
ds^2 = -dt^2 + R^2(t) \times \underbrace{\left[dr^2 +r^2 d\Omega^2\right]}_{\text{``co-moving coordinates"}} \,,
}
where $R(t)$ is the time-dependent scale factor that describes the expansion of the Universe and where the co-moving spatial coordinates are indicated in a spherical basis.  The expansion of the Universe is described by the Hubble parameter $H \equiv \dot R / R$; the value of the Hubble parameter today is the Hubble constant $H_0 \approx 67.4 \pm 0.5$ km/s/Mpc~\cite{Planck:2018vyg}, though see {\it e.g.}~\cite{DiValentino:2021izs} for a summary of recent controversy related to $H_0$.   Suppose that the Universe is filled with a homogeneous fluid with energy density $\rho$. Then, Einstein's equation yield the Friedmann equations
\es{eq:Friednmann}{
H^2 = {1 \over 3 m_{\rm pl}^2} \rho \,, \qquad \dot H + H^2 = - {1 \over 6 m_{\rm pl}^2} \left( \rho + 3 p\right) \,,
}
where $m_{\rm pl} \approx 2.4 \times 10^{18}$ GeV is the reduced Planck mass and $p$ is the pressure of the fluid.  Note that the reduced Planck mass is related to Newton's gravitational constant by
\es{}{
G = {1 \over 8 \pi m_{\rm pl}^2} \approx 6.7 \times 10^{-39} {1 \over {\rm GeV}^2} \,.
}
The Planck mass $M_{\rm pl}$ (not reduced) is defined simply by $G = 1/M_{\rm pl}^2$, giving the value $M_{\rm pl} \approx 1.2 \times 10^{19}$ GeV.
For a perfect fluid the pressure is related to the density by the equation of state parameter $w$, $p = w \rho$, which yields the solution $R(t) = R_0 t^{ {2 \over 3(w+1)}}$ and, accordingly, $H(t) = {2 \over 3(1+w)} {1 \over t}$.  Non-interacting, non-relativistic matter ({\it e.g.}, cold DM) is pressureless ($w = 0$), while a gas of relativistic particles has $w = 1/3$ such that the stress-energy tensor is traceless.   A cosmological constant has $w = -1$.

The difference between matter and radiation may be understood as follows.  Imagine a gas of non-interacting particles in an expanding universe. The number density will scale with the scale factor like $n \propto 1 / R^3$ as the Universe expands ($R$ increases), since the physical number density is being diluted by a factor of $R$ in each spatial direction. For non-relativistic matter, the energy of each particle is simply given by its rest mass, such that the energy density redshifts as $\rho \propto n \propto 1 / R^3$. However, for ultra-relativistic matter ({\it i..e.}, radiation) the energy of each particle is proportional to its frequency, which also redshifts with the expanding Universe like $1 / R$, giving $\rho \propto 1 / R^4$.  The cosmological constant describes the energy density of free space and this thus constant as space expands: $\rho \propto R^0$.       
 
The early universe is filled with a hot plasma of SM -- and perhaps BSM -- particles, which sequentially freeze-out of the plasma as the Universe expands and cools down. Let us suppose that a particle $X$ is kept in thermal equilibrium through scattering processes with the rest of the plasma with rate $\Gamma_X$. The particle $X$ ``freezes out" of equilibrium when $\Gamma_X$ becomes much less than $H$ ({\it i.e.}, if $\Gamma_X \ll H$ a particle $X$ cannot scatter during the time it takes for the Universe to double in size, making it increasingly hard for that particle to ever scatter again).  Typically, $\Gamma_X = n \langle \sigma v \rangle$, where $n$ is the number density of particles in the thermal plasma that $X$ can scatter off of, with cross-section $\sigma$, and the average is over the relative velocity $v$ given the relative velocity distribution.

In general, regardless of whether the application is to the early universe primordial plasma or to air particles in the room you are currently in, the phase-space distribution of a thermal species is described by the Fermi-Dirac (Bose-Einstein) distribution for fermions (bosons):
\es{}{
f({\bm p}) = {1 \over e^{E/T} \pm 1} \,,
}
with $E = \sqrt{{\bm p}^2 + m^2}$, $m$ the particle mass, and with the sign in the denominator positive for the Fermi-Dirac distribution and negative for the Bose-Einstein distribution.  The number density, energy density, and pressure may be computed by computing the appropriately weighted integrals over the phase-space density (see, {\it e.g.},~\cite{Trodden:2004st} for an extended discussion).  For example, the number density is given by 
\es{eq:n}{
n = g \int {d^3 p \over (2 \pi)^3} f({\bm p}) \,,
}
where $g$ is the number of spin degrees of freedom for the particle (two for a photon, two for an electron or positron, and so forth).  The expression for the energy density $\rho$ is analogous to~\eqref{eq:n}, but with an extra insertion of $E$ within the integral.  Performing these integrals for relativistic particles ($E^2 = {\bm p}^2$) leads to the following expressions for the number and energy densities of bosons and fermions: 
\es{eq:n_rel}{
n_{\rm rel}(T) = {\zeta(3) \over \pi^2} g T^3 \, \quad {\rm (bosons)} \,, \qquad n_{\rm rel}(T) ={3 \over 4} {\zeta(3) \over \pi^2} g T^3 \, \quad {\rm (fermions)} \\
\rho_{\rm rel}(T) = {\pi^2 \over 30} g T^4 \, \quad {\rm (bosons)} \,, \qquad \rho_{\rm rel}(T) ={7 \over 8} {\pi^2 \over 30} g T^4 \, \quad {\rm (fermions)} \,,
}
with $\zeta$ the Riemann zeta function  ($\zeta(3) \approx 1.20$).
On the other hand, non-relativistic particles (both bosons and fermions), with $E = m$, have exponentially suppressed number densities in thermal equilibrium:
\es{eq:n_NR}{
n_{\rm non-rel}(T) = g \left( {m T \over 2 \pi} \right)^{3/2} e^{-m /T} \,.
}
The intuition behind this expression is that a massive particle with $m \gg T$ in thermal equilibrium may annihilate into lighter thermal plasma particles but cannot be created, since the plasma with kinetic-energy-per-particle $T$ does not have enough energy to create particles of mass $m$.  Thus, the number density of particles in thermal equilibrium is rapidly depleted for $m \gg T$.  This point plays a crucial role in the process of thermal freeze-out. 

Let us now briefly consider the radiation dominated universe, which is the one more relevant for early-universe DM cosmology, well before matter-radiation equality. (We know from early-Universe probes like the CMB that DM must behave like DM before matter radiation equality, which implies that DM production mechanisms typically take place in a radiation-dominated universe.)  In this context it makes sense to define the total number of relativistic degrees of freedom as 
\es{eq:energy_dof}{
g_* \equiv \sum_{{\rm particles} \, \, i} f_i g_i \left( {T_i \over T} \right)^4 \,,
}
where $f_i =1$ for bosons and $7/8$ for fermions and where particles and anti-particles are counted separately.  If all particles are in thermal equilibrium then $T_i = T$, the temperature of the thermal bath, but on the other hand it is possible that some particles may freeze-out of the thermal bath and then maintain thermal-like phase space distributions but with $T_i \neq T$. For example, the temperature of the thermal bath evolves with the scale factor as $T \propto 1/R$, but if a particle $i$ falls out of equilibrium and becomes non-relativistic, then the effective temperature of that sector redshifts more rapidly as $T_i \propto 1/R^2$ (since the kinetic energy in that case, which determines the effective temperature, scales like $v^2$).  Then, the total energy density of the Universe is given by $\rho = {\pi^2 \over 30} g_* T^4$. Referring to the  first of the Friedmann equations we may then infer that 
\es{eq:Hubble_rad}{
H = { \pi \sqrt{g_*} \over 3 \sqrt{10} } {T^2 \over m_{\rm pl}} \,,
}
which also gives a relation between time and temperature through $H = 1/(2t)$. (Note that at temperatures well above the electroweak phase transition $g_* \approx 100$.)

In calculations of the DM density we often make use of the entropy density 
\es{eq:entropy}{
s = {2 \pi^2 \over 45} g_{*S} T^3 \,,
}
where $g_{*S}$ is the entropy degrees of freedom, closely related to the energy degrees of freedom in~\eqref{eq:energy_dof}, but with a slightly different temperature dependence:
\es{eq:entrop_dof}{
g_{*S} \equiv \sum_{{\rm particles} \, \, i} f_i g_i \left( {T_i \over T} \right)^3 \,.
}
Note that at high and low temperatures $g_*$ and $g_{*S}$ asymptote the same asymptotic values of around 100 and 3, respectively.   The entropy density is a very useful quantity because one can prove that the comoving entropy density is constant. That is, the physical entropy scales with the scale factor as $s \propto R^{-3}$ with the scale factor.  This fact is used in the calculation of the relic DM density in a variety of scenarios, including axions and WIMPs. That is because if we calculate the number density $n$ of some relic at a high temperature $T_*$ and we assume that for $T < T_*$ that particle redshifts like cold DM ($n \propto R^{-3}$), then we know that for $T < T_*$ the quantity $n / s$ is constant, allowing us to infer the number density of our relic candidate today given the calculation of the number density at $T_*$.

Lastly, let us briefly review the important cosmological epochs in the early Universe. It is common to refer to epochs by their cosmological redshift $z$, defined through
\es{}{
{R(t_0) \over R(t)} = 1 + z \,,
}
with $z$ the redshift of the epoch at time $t$ and $t_0$ denoting today.  We can also talk about the temperature of the photon bath as a clock, since this temperature is monotonically decreasing in the standard cosmology (though one needs to be careful about epochs where the entropy degrees of freedom change, since it is the comoving entropy density that is actually conserved).  In particular, apart from times of changing entropy degrees of freedom, $T \propto 1/R$, such that
\es{}{
T(z) \approx T_0 (1 + z) \,, 
}
with $T_0$ the temperature of the CMB today.\footnote{Note that the CMB is no longer in thermal equilibrium, but after decoupling it maintains the thermal distribution with a redshifting temperature parameter $T$.} Below, we highlight a few important cosmological epochs:
\begin{itemize}
    \item $z = 0$: The temperature of the CMB today is $T_0 \approx 2.35 \times 10^{-4}$ eV.
    \item $z \approx 1100$: This is the epoch of recombination ($T \approx 0.3$ eV).  Recall that the binding energy of the hydrogen ground state is $\sim$$13.6$ eV.  Thus at temperatures below around 10 eV the process $p + e^- \longleftrightarrow H 
+ \gamma$, with $H$ denoting hydrogen and $\gamma$ photons, begins to deplete the free protons and electrons by forming the energetically favorable bound hydrogen atoms. By $z \sim 1100$ there are almost no free electrons and protons left (by relative abundance). This is important because it implies that the photons are now decoupled and free stream essentially unperturbed until their detection as the CMB today.
    \item $z = z_{\rm eq} \approx 3402$: this is the epoch of matter-radiation equality. That is, going off of~\eqref{eq:Friednmann} we can write
    \es{}{
    H^2(z) = H_0^2 \left[ \Omega_r(1+z)^4 + \Omega_m(1+z)^3 + \Omega_\Lambda\right] \,,
    }
    with $\Omega_r$, $\Omega_m$, and $\Omega_\Lambda$ the energy density fractions today in radiation, matter, and cosmological constant, respectively. 
 Matter-radiation equality occurs at $\Omega_r(1+z_{\rm eq})^4 = \Omega_m(1+z_{\rm eq})^3 $.  This corresponds to a photon bath temperature of $T \approx 0.799$ eV. 
 \item $z \sim 10^{10}$: After the QCD phase transition at $T \sim 200$ MeV, the quarks and gluons are confined to neutrons and protons.  However, the nucleons are still kept in thermal equilibrium with the rest of the SM plasma through weak interactions. The neutron and proton freeze-out of equilibrium, non relativistically, at a temperature $T \sim 0.8$ MeV.  Below this temperature the comoving number densities of neutrons and protons are fixed, though the nucleons fuse to form light nuclei such as helium and deuterium, until $T \sim 0.1$ MeV.  The epoch of BBN marks our first observational handle of the Universe.  Before BBN the standard cosmological picture assumes radiation domination, but there is plenty of room to modify the cosmological history prior to BBN, for example through a period of early matter domination or a low reheat temperature from inflation. However, from BBN onwards the Universe must be radiation dominated (until matter-radiation equality) with little to no extra entropy dilution in order to be consistent with the combination of precision BBN and precision CMB analyses.  
\end{itemize}

\section{Frequentist statistics for particle and astro-particle physics}
\label{sec:stats}

In this section we briefly review some of the foundational principles for frequentist statistics that are needed in any non-trivial analysis of laboratory or astrophysical data to search for new physics. Here, we emphasize the basic and foundational concepts; see~\cite{Cowan:2010js,doi:https://doi.org/10.1002/9783527653416.ch4,Cousins:2018tiz} for additional details and results.  There are also many excellent introductions to Bayesian statistics. While Bayesian statistics is sometimes used in particle physics, at the moment most analyses that search for DM tend to rely on frequentist statistics. Part of the reason for this is simply that this is, at the moment, the convention of the field, but also Bayesian statistics force you to chose a prior for the DM model of interest. It can be hard to write down a principled prior for particle DM model parameters, and care needs to be taken in the Bayesian framework to make sure that, for example, the upper limit inferred from the data is actually a reflection of the data and not the prior. On the other hand, there are principled choices of noninformative priors that one can make in the context of Bayesian inference. At the least, it is probably true that if your result depends strongly on whether you use Bayesian or frequentist statistics, then probably you should proceed by collecting more data or improving your modeling framework.  With that philosophy in mind, from here on out we only discuss the frequentist approach.

Let us suppose that we perform the following experiment: we draw a real number $x$ from a probability distribution $p(x) ={1 \over \sqrt{2 \pi} \sigma} \exp\left[ -{ (x - \mu_{\rm true})^2  \over 2 \sigma^2} \right]$, where $\mu_{\rm true}$ is the expected value and $\sigma^2$ is the known variance.  If we only perform one draw from this distribution to get a single value $x$, how can we determine the confidence interval for $\mu$ and the 95\% one-sided upper limit for $\mu$? While this example is simplistic, it will illustrate many of the features that arise in more complicated statistical analyses for DM signals.  First, given the data ${\bm d} = x$ and the model ${\mathcal M}$, which consists of the Gaussian distribution with variance $\sigma^2$ (we fix $\sigma$ at its true value for simplicity), we write down the likelihood
\es{}{
p({\bm d}| {\mathcal M}, \mu) = {1 \over \sqrt{2 \pi} \sigma} \exp\left[ -{ (x - \mu)^2  \over 2 \sigma^2} \right] \,.
}
While it may look as if we have simply rewritten the probability distribution, the likelihood has a distinct interpretation.  In particular, the likelihood is evaluated for the observed and fixed data set ${\bm d}$ as a function of the model parameter $\mu$ of the model ${\mathcal M}$. It should be unsurprising that the best-fit value for $\mu$, which we denote by $\hat \mu$, is that 
 which maximizes the likelihood. That is, the ``best-fit" value of $\mu$ is that which makes the observed data most likely in the context of the parametric model of interest. In our example, this implies that $\hat \mu = x$.  Note that $\hat \mu$ will never actually equal $\mu_{\rm true}$, so our estimate for $\mu$ is only meaningful in so much as we can compute confidence intervals.  Note that if we had $N$ data point such that ${\bm d} = \{x_i\}_{i=1}^N$, all drawn from the same distribution, then the likelihood would be given by the product
 \es{}{
p({\bm d}| {\mathcal M}, \mu) = \prod_{i=1}^N {1 \over \sqrt{2 \pi} \sigma} \exp\left[ -{ (x_i - \mu)^2  \over 2 \sigma^2} \right] \,.
}
For simplicity, however, in our example we restrict ourselves to $N=1$.

 To compute a confidence interval for $\mu$ we start by computing the profile likelihood ratio
 \es{}{
 \lambda(\mu) = {p({\bm d}| {\mathcal M}, \mu) \over p({\bm d}| {\mathcal M}, \hat \mu)} \,.
 }
 In analyses that are more nontrivial, in particular those with nuisance parameters, the numerator would denote the likelihood evaluated at fixed signal parameter $\mu$ but maximized over the nuisance parameters, while the denominator would be maximized over both the signal parameter and the nuisance parameters. Nuisance parameters are model parameters that also must be modeled to correctly describe the data but which are not the parameter of interest to be constrained. For example, if we wanted to we could treat $\sigma$ as a nuisance parameter, though in our simple analysis we keep this parameter fixed at its true value.  An equivalent but computationally easier quantity to keep track of instead of the profile likelihood ratio is the test statistic 
 \es{}{
 t(\mu) = - 2 \log \lambda(\mu) \,.
 }
 In our example
 \es{}{
 t(\mu) = {(x - \mu)^2 \over \sigma^2} \,.
 }
 The best-fit value $\hat \mu$ is that which minimizes the test statistic. Let us then formally define the confidence interval at confidence $1-\alpha$ as the set of all $\mu$ such that $t(\mu) < c$ for some threshold quantity $c$ determined by $P\big(t(\mu) < c\big) = 1 - \alpha$, with $P\big(t(\mu) < c\big)$ denoting the probability that $t(\mu)$ is less than $c$, assuming the data is generated from the null hypothesis.  

 To make progress, let us define $f\big(t(\mu) \big)$ as the probability distribution function (pdf) of observing $t(\mu)$.  That is, $f\big(t(\mu) \big)$ is the pdf of $(\mu - \hat \mu)^2 / \sigma^2$, where $\mu$ is a random variable drawn from the normal distribution with variance $\sigma$ and mean $\hat \mu$. To find the pdf $f$ it is convenient to first compute the cumulative distribution function (cdf) $\Phi$ for the normal distribution:
 \es{}{
 \Phi(\mu) \equiv \int_{- \infty}^\mu d \mu' {1 \over \sqrt{2 \pi} \sigma} \exp\left[ -{ (\mu' - \hat \mu)^2  \over 2 \sigma^2} \right] = {1 \over 2} {\rm erfc} \left[ {\hat \mu - \mu \over \sqrt{2} \sigma}\right] \,,
 }
 where ${\rm erfc}$ is the complementary error function (${\rm erfc}(z) = 1 - {\rm erf}(z)$, with ${\rm erf}$ the error function).  The cdf for $t$, which we call $F$, is then given by
 \es{}{
 F(t) =\Phi(\hat \mu + \sigma \sqrt{t}) -\Phi(\hat \mu - \sigma \sqrt{t}) = {\rm erf}\left[ \sqrt{t \over 2}\right] \,.
 }
 Note that the two terms above arise because when we invert $t = (\mu - \hat \mu)^2 / \sigma^2$ to find $\mu$ in terms of $t$ there are two possibilities, depending on whether $\mu$ is above or below $\hat \mu$.
 The pdf for the test statistic $t$ is then computed through a derivative of the cdf:
 \es{}{
 f\big(t(\mu) \big) = F'(t) = {e^{-t/2}\over \sqrt{2 \pi} t} \equiv \chi^2_1(t) \,,
 }
 with $\chi^2_1$ the chi-square distribution with one degree of freedom.  We may also identify $P(t(\mu) < c) = F(c)$, such that $c$ is given by $c = F^{-1}\left[1 - \alpha\right] $, where $F^{-1}$ is the inverse cdf for the chi-square distribution with one degree of freedom.  In particular, we find
 \es{}{
 c = 2 \left( {\rm erf}^{-1}(1 - \alpha) \right)^2 
 }
 as the threshold value for the test statistic in order to obtain a confidence interval at confidence $1 - \alpha$.  For 68\% confidence ($\alpha \approx 0.32$) this implies $c =1$ (that is, the 1$\sigma$ confidence interval is the set of all $\mu$ near $\hat \mu$ where $t(\mu) <1$).  The 2$\sigma$ ($\sim$95.5\%) confidence interval is given by the set of $\mu$ where $t(\mu) < 4$.  (Note that by definition, we say the $N$$\sigma$ confidence interval is that where $t(\mu) < N^2$.)

 A closely related quantity to the confidence interval is the one-sided upper limit. The one-sided upper limit is most naturally computed through the test statistic for upper limits, defined as
 \es{}{
q(\mu) = \left\{ 
\begin{array}{cc}
t(\mu) & \hat \mu \leq \mu \,, \\
0 & \hat \mu > \mu \,.
\end{array}
\right.
 }
 The cumulative distribution function for this test statistic is
 \es{}{
 F_q(t) = \Phi(\hat \mu + \sigma \sqrt{t}) =  {1 \over 2} {\rm erfc} \left[ - {\sqrt{t \over 2}}\right] \,.
 }
 Thus, at confidence $1- \alpha$ we may constrain the parameter $\mu$ to the interval given by $q(\mu) < c$, with $c$ determined through the relation
 \es{}{
 c = 2 \left( {\rm erfc}^{-1}\left[2 (1 - \alpha)\right]  \right)^2\,.
 } 
 Note that this value of $c$ is found by setting $F_q(c) = 1 - \alpha$ and solving for $c$, assuming $c>0$.
 For example, 95\% one-sided upper limits are found by setting $c \approx 2.71$, which is a very commonly used value.

 Let us now turn to the computation of the evidence of the signal model over the null hypothesis.  That is, in addition to constraining the model parameters, such as the signal strength, we would also like to know if the data prefers our signal model (for us, our DM model) over the null hypothesis of no signal contribution.  We again illustrate this computation in the context of our Gaussian example. Without loss of generality we take ${\mu}_{\rm true} = 0$ to describe the null hypothesis, while a signal contribution would be  ${\mu}_{\rm true} > 0$. For example, perhaps $\mu_{\rm true}$ rescales the annihilation cross-section in a DM annihilation search. In that case, $\mu < 0$ would be unphysical, but we still consider such values to make sure that we find the point of maximum likelihood.\footnote{See~\cite{Cowan:2010js} for details of how to work with the restriction $\mu > 0$, which is a valid restriction but which complicates the applications of the analytic distributions for the test statistics.} Still, since only $\mu > 0$ is physical, we perform a one-sided hypothesis test. 

 The discovery test statistic for a one-sided test is defined as 
 \es{}{
q_0 = \left\{ 
\begin{array}{cc}
t(0) & \hat \mu \geq 0 \,, \\
0 & \hat \mu < 0 \,.
\end{array}
\right.
 }
 This test statistic describes how unlikely the null hypothesis ($\mu = 0$) is given the signal model. That is, the discovery test statistic allows us to reject the null hypothesis at a quantifiable confidence level.  The probability of observing a discovery test statistic as high or higher than $q_0$ is
 \es{}{
 p = \int_{q_0}^\infty f(q_0 | \hat \mu) = {1 \over 2}  {\rm erfc} \left( \sqrt{q_0 \over 2} \right) \, \, \qquad (q_0 > 0) \,.
 }
 Note that here $p$ is the $p$-value and it denotes the probability of the finding a discovery test statistic as high or higher than that of the null hypothesis in the context of the signal model.  In many fields significance is reported by the $p$-value, but particle physicists often instead report significance using language such as ``a $Z$$\sigma$ detection," for some number $Z$.  The number of $\sigma$ is quantified through the $Z$-score, defined as 
 \es{}{
 Z = \Phi^{-1}(1 - p) = \sqrt{2} \, {\rm erfc}^{-1} \left[ 2 p \right] = \sqrt{q_0} \,,
 }
 with $\Phi(z) = {1 \over 2} {\rm erfc}(-z / \sqrt{2})$ the cumulative distribution function of the standard ($\sigma = 1$, $\mu_{\rm true} = 0$) normal distribution and $\Phi^{-1}$ its inverse.  Here, $Z$ is interpreted as the number of standard deviations above the true value for which it would be as likely to draw a random value for $x$ (at or above this value) as it is to observe a discovery test statistic with a $p$-value $p$.  For example, a 3$\sigma$ discovery ($Z = 3$, $q_0 = 9$) means that the null hypothesis is as likely as it is to draw a value from a Gaussian that happens to be at or more than three standard deviations upwards from the mean; the associated $p$-value (or probability of this happening) is $p \approx 1.3 \times 10^{-3}$.  Importantly, keep in mind that the formulae above are for a one-sided test, which is the case typical in DM searches; the analogous formulae for two-sided tests have additional factors of $2$ and may be derived by following the logic above but without the restriction $q_0 = 0$ for $\hat \mu < 0$.  
 
 In particle physics, for one-sided tests, the threshold of a 5$\sigma$ discovery is often implemented. Note that this corresponds to a $p$-value of $p \approx 2.9 \times 10^{-7}$, which is exceedingly unlikely by random chance alone. However, there are two important caveats to keep in mind with this threshold. First, one should always account for the so-called look-elsewhere effect through the trials factor  when performing multiple, independent searches for new physics. An illustrative example of the look-elsewhere effect is found in axion DM searches.  Consider, for example, the searches for axion DM from radio telescope observations discussed in Sec.~\ref{sec:axion_NS}. Here, the signal appears at the frequency $\omega = m_a$, but it is narrow at the level $\delta \omega / \omega \sim 10^{-6}$ (here, let us assume for definiteness that $\delta \omega / \omega = 10^{-6}$, though physically that signal may be slightly broadened).  We do not know the mass of the axion, so we must search over many different possibilities. For example, suppose we search for evidence of axions in radio spectral data over one order of magnitude in frequency space. This corresponds to, roughly, $10^6$ different, independent analyses. In that ensemble of $10^6$ different analyses the likelihood that one of them will have a discovery significance at or above 5$\sigma$ is approximately $10^6 \times 2.9 \cdot 10^{-7} \sim 0.3$, which is exceptionally high! This means it would not be surprising at all to find one or more test mass points that exceed our 5$\sigma$ discovery threshold since we searched over so many independent mass points. The key to making progress here is to distinguish the local significance, which is the one we have been discussing so far, from the global significance, where the global significance accounts for the number of independent analyses that are being performed. In this case, a 5$\sigma$ discovery threshold for the global analysis would correspond to a local significance threshold with $p$-value such that $10^6 \times p \sim 2.9 \cdot 10^{-7}$, which gives $p \sim 2.9 \cdot 10^{-13}$ and which corresponds to a $\sim$7.2$\sigma$ local significance threshold ($q_0 \approx 52$).  The correction applied above is known as the Bonferroni correction. 

 The second important caveat is that this discussion has only focused on statistical uncertainties, while in reality almost every analysis will have a combination of statistical and systematic uncertainties. A systematic uncertainty is some unknown aspect of the data that does not vary randomly from data point to data point. For example, suppose that we are looking for a signal over a background model that is poorly determined. We would have a systematic uncertainty associated with mismodeling the background model.  An example of this type of systematic uncertainty is discussed in more detail in Sec.~\ref{sec:segue}. While statistical uncertainties are usually straightforward to account for, it is much harder to quantify and account for systematic uncertainties, since by definition they are aspects of the data that we do not fully understand.  When discussing systematic versus statistical uncertainty it is useful to distinguish the concepts of {\it accuracy} and {\it precision}. Here, {\it accuracy} refers to systematic uncertainties: does the result of your analysis give the correct result ({\it e.g.}, does your recovered model parameter match the true model parameter within the quoted uncertainties).  Precision, on the other hand, refers to statistical uncertainties: more precise measurements have smaller error bars. Ideally, we want our inference of model parameters to be both accurate and precise. In practice -- as discussed more in Sec.~\ref{sec:segue} -- it can be a good goal to try to compromise accuracy and precision, such that one designs analyses where the expected statistical uncertainty is roughly at the level of the expected systematic uncertainty. If the precision is better than this, the analysis will be systematics-limited, and systematic uncertainties are hard to quantify. If, on the other hand, the precision is worse, then one is not fully exploiting the data set. Of course, such a compromise is not always possible: often one is statistics limited, for example, simply due to the finite size of the data set.

 Let us now return to the discussion of statistical uncertainties.  Often in particle physics we want to project our sensitivity to a putative new-physics model. For example, we may want to compute the expected upper limit on a cross-section, given some future measurement. Not only are we interested in the expected upper limit, but often times we would like to know the percentiles of the expected upper limit so that we can make the so-called ``Brazil band plots," which show the 1$\sigma$ and 2$\sigma$ expected ranges for, for example, the 95\% one-sided upper limit.  One way to make progress here is to simply perform a large number of toy analyses, where one creates toy data sets under, {\it e.g.}, the null hypothesis and then analyses these toy data sets with the combined signal and background model. By performing a large number of such toy analyses one will construct a distribution of 95\% one-sided upper limits, from which one can find the mean expected upper limit and various percentiles.  However, there is a very convenient shortcut to this procedure called the Asimov method~\cite{Cowan:2010js}.  

 Using the Asimov procedure we can compute, for example, the expected 95\% one-sided upper limit under the null hypothesis without having to perform Monte Carlo simulations. (Note that one can also use this method in a straightforward way to find the expected discovery significance of a model.)  Instead, we use the Asimov data set ${\bf d}_A$, which is equal to the expectation under the hypothesis of interest, with no statistical variance added in. For example, suppose we want to compute the expected 95\% one-sided upper limit on $\mu$ under the null hypothesis in a scenario where the null hypothesis is described by $\mu_{\rm true} = 0$ in our Gaussian example, with variance $\sigma^2$. The Asimov data set for the null hypothesis is simply $x = 0$. The Asimov test statistic $t_A$ is then simply 
 \es{}{
 t_A(\mu) = {\mu^2 \over \sigma^2} \,,
 }
 and we thus determine that the expected 95\% one-sided upper limit is $\mu^{95\%}_{\rm expected} \approx \sqrt{2.71} \sigma$.  On the other hand, no actual analysis will report this precise upper limit, unless that data set happens to have precisely $x = 0$. It is thus useful to also know the expected ranges for the upper limit. This, too, may be computed using the Asimov framework, but we refer the reader to~\cite{Cowan:2010js} for details. Instead, here we simply quote the result that:
 \es{eq:asimov_ul}{
 \mu^{1 - \alpha}_{\rm expected}( \pm N\sigma) &= \hat \mu_{\rm expected} + \sigma \left[ \Phi^{-1}(1-\alpha) \pm N \right]  \,,\\
 \mu^{95\%}_{\rm expected}( \pm N\sigma) &= \hat \mu_{\rm expected} + \sigma \left[ \sqrt{2.71} \pm N \right]  \,. 
 }
 In the top line above, we give the general result for the upper limit at $1-\alpha$ confidence, while in the bottom line we specify to 95\% confidence ($\alpha = 0.05$). Here, $\pm N \sigma$ refers to the expected interval at $\pm N \sigma$ containment. For example, the 95\% upper limit should be within the range $\big(\mu^{95\%}_{\rm expected}( -\sigma),\mu^{95\%}_{\rm expected}( +\sigma)\big)$ approximately 68\% of the time. (Note, by {\it e.g.} $N\sigma$ containment we mean the mass of the standard normal distribution contained within $N\sigma$ of either side of the mean.)  The quantity $\hat \mu_{\rm expected}$ is the expected best-fit value of the signal parameter, which was taken to zero in our simple example.

 One non-trivial aspect of~\eqref{eq:asimov_ul} is that, for example, the expected lower edge of the 2$\sigma$ containment region is actually below $\hat \mu_{\rm expected}$! For example, if $\hat \mu_{\rm expected} = 0$ then 95\% of the time the upper limit should be in the range $\sigma \times (-0.35,3.6)$.  This is strange because it means that some fraction of the time our upper limit will rule out the true value, which here is $\mu = 0$.  Of course, this is expected since our upper limit is only at 95\% confidence, but this is physically an unpleasant feature of the upper limits in the context of new-physics searches since if $\mu$ describes a cross-section then we are saying we rule out all physical cross-section values if our upper limit happens to be below zero.  There are multiple ways forward here~\cite{Cousins:2018tiz}. One principled approach is to power constrain the limits, which is to not let the limits go below the lower edge of the expected $1\sigma$ containment region for the 95\% one-sided upper limit~\cite{Cowan:2011an}.

 While we have illustrated this frequentist formalism for our simple Gaussian likelihood example, all of the test statistics are still well defined for an arbitrary likelihood. One modification is that the profile likelihood ratio should be profiled over possible nuisance parameters, as mentioned previously:
 \es{}{
 \lambda(\mu) = {p({\bm d}| {\mathcal M},\{\mu,\hat {\hat {\bm \theta}}_{\rm nuis}\}) \over p({\bm d}| {\mathcal M},\{\hat \mu, {\hat {\bm \theta}}_{\rm nuis}\})} \,.
 }
 Here, we assume that our model ${\mathcal M}$ has the signal parameter of interest $\mu$ in addition to a set of nuisance parameters ${\bm \theta}_{\rm nuis}$ that may, for example, describe other aspects of the background or signal model that we are less interested in constraining but still need to describe. The notation $\{\hat \mu, {\hat {\bm \theta}}_{\rm nuis}\}$ means that the likelihood is evaluated at the best-fit combined signal parameter and nuisance parameter vector, while $\{\mu,\hat {\hat {\bm \theta}}_{\rm nuis}\}$ means we maximize the likelihood over the nuisance parameters at fixed signal parameter $\mu$.  After defining the profile likelihood, the test statistics have the same definitions as before.  However, our interpretations of the test statistics in terms of the precise threshold values $c$ taken to obtain various confidence levels for upper limits and significances for detection require the profile likelihood to be chi-square distributed with one degree of freedom. If the profile likelihood is not chi-square distributed then you need to either work out the analogous formulae for whatever distributions the test statistics are subject to or, more often than not, compute the distributions numerically through Monte Carlo. 

 Luckily, there is a very powerful theorem that can be invoked to make use of the machinery described above in a broad range of circumstances, even when the likelihood is not constructed from normal distributions.  Wilks' theorem states that as the data size is taken to infinity, then subject to a few caveats that are unimportant for this discussion the profile likelihood asymptotes to being chi-squared distributed with one degree of freedom.  This means that even if the likelihood is not constructed from Gaussian probability distributions, so long as there is a ``large" amount of data in the analysis, chances are the profile likelihood is reasonably described by the chi-square distribution under the null hypothesis.   A closely related concept to Wilks' theorem is the central limit theorem, which states that averages over large numbers of random variables are described to increasingly good precision by the normal distribution, even if the underlying random variables are themselves not normally distributed, as the number of random variables is taken to infinity.  The key question in invoking Wilks' theorem, of course, is what does a ``large" amount of data mean. As an example, let us consider the Poisson distribution: $p(k| \mu) = \mu^k e^{-\mu} / k!$, with $k$ the number of counts and $\mu$ the expected number of counts. In the large $\mu$ limit the Poisson distribution may be approximated by the normal distribution (this is an example of the central limit theorem): $p(k| \mu) \approx {1 \over \sqrt{2 \pi \mu}} e^{-(k - \mu)^2 / (2 \mu)}$.  That is, at a large number of expected count $\mu$ the Poisson distribution is approximately described by the normal distribution with mean $\mu$ and variance $\mu$. However, the degree to which this description is adequate depends on how far beyond the mean we are probing. For example, suppose that $\mu = 9$ such that with the Gaussian approximation the variance is $3$. If we only probe 1$\sigma$ away from the mean, then corrections to the Gaussian approximation can be expected to be of order $\sigma / \mu \sim 1/3$, which may be acceptable.  However, if we probe more than three standard deviations away from the mean then clearly issues will arise, since for example the Poisson distribution will not return negative counts while the Gaussian approximation will return negative counts if probed sufficiently far from the mean.  For most applications, Poisson distributions are reasonably approximated by normal distributions if the expected number of counts is much larger than $\sim$10, since we often do not go much further than 3$\sigma$ away from  the mean.  The Poisson distribution is one of the most commonly used building blocks in constructing likelihoods in particle physics, since we often conduct counting experiments; for example, in gamma-ray astro-physics we count photons.

\section{Minimal dark matter annihilation cross-sections}
\label{app:xsec}

In this Appendix we provide more details of the calculation that go into~\eqref{eq:xsec_higgsino}, which is the thermal annihilation cross-section for higgsino annihilation in the early universe (see also~\cite{Kowalska:2018toh}). Note that this cross-section includes co-annihilations. On the contrary, annihilations today relevant for indirect detection only involve the lightest Majorana state $\chi$, since the heavier ones have already decayed to $\chi$ and SM states. 

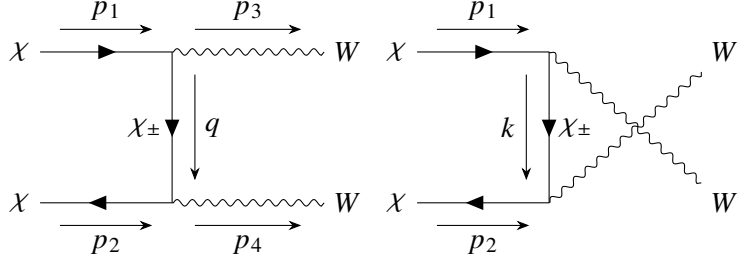
\begin{figure}
\begin{center}
\begin{tikzpicture}
  \begin{feynman}
    \vertex (a){$\chi$};
    \vertex [right=2cm of a] (b);
    \vertex [below=2cm of b] (c);
    \vertex [below=2cm of a] (d){$\chi$};
    \vertex [right=2cm of b] (e){$W$};
    \vertex [right=2cm of c] (f){$W$};
    \diagram* {
      (a) -- [fermion,  momentum=\(p_{1}\)] (b),
      (b) -- [fermion, ,  momentum=\(q\), edge label' =\(\chi_\pm\)] (c),
      (d) -- [anti fermion,  momentum'=\(p_{2}\)](c),
      (b) -- [photon,  momentum=\(p_{3}\)] (e),
      (c) -- [photon,  momentum'=\(p_{4}\)] (f)
    };
  \end{feynman}
\end{tikzpicture}
\begin{tikzpicture}
  \begin{feynman}
    \vertex (a){$\chi$};
    \vertex [right=2cm of a] (b);
    \vertex [below=2cm of b] (c);
    \vertex [below=2cm of a] (d){$\chi$};
    \vertex [right=2cm of b] (e){$W$};
    \vertex [right=2cm of c] (f){$W$};
    \diagram* {
      (a) -- [fermion,  momentum=\(p_{1}\)] (b),
      (b) -- [fermion,  momentum'=\(k\), edge label =\(\chi_\pm\)] (c),
      (d) -- [anti fermion,  momentum'=\(p_{2}\)] (c),
      (b) -- [photon] (f),
      (c) -- [photon] (e)
    };
  \end{feynman}
\end{tikzpicture}
\end{center}
\caption{Example annihilation channels for $\chi \chi \to WW$ relevant for higgsino annihilation today and in the early universe.}
\label{fig:app_0}
\end{figure}

There are many Feynman diagrams that contribute to~\eqref{eq:xsec_higgsino}, as shown in Figs.~\ref{fig:feyn_1},~\ref{fig:feyn_2},~\ref{fig:feyn_3},~\ref{fig:feyn_4},~\ref{fig:feyn_5},~\ref{fig:feyn_6}, and~\ref{fig:higgsino_ann}.
We will not calculate the cross-sections associated with all of these processes but rather illustrate the calculations for an example processes. In particular, we will consider one $t$/$u$ channel process, $\chi \chi \to WW$, with $\chi$ the lightest Majorana state. The $t$ and $u$ channel diagrams contributing to this process are illustrated in Fig,~\ref{fig:app_0}. 
The associated scattering matrix is
\es{eq:app_chichi_WW}{
    i \mathcal{M}_{\chi\chi \to WW} = \frac{g^2}{4} \bar v_2 \left(   \frac{ \gamma_\mu (\qs - m_\chi) \gamma_\nu  \epsilon^\mu_3  \epsilon^\nu_4 }{t^2 - m_\chi^2 } +  \frac{ \gamma_\mu (\ks - m_\chi) \gamma_\nu  \epsilon^\mu_4  \epsilon^\nu_3 }{u^2 - m_\chi^2 }  \right) u_1  \,,
}
with $p_1$ and $p_2$ the incoming momenta, $p_3$ and $p_4$ the outgoing momenta, $\epsilon_3^\mu$ and $\epsilon_4^\mu$ the polarizations of the outgoing vector bosons, $q$ the momentum exchange, and $u_1$/$v_2$ the wavefunctions for the incoming fermions.  We then square the matrix element, average over the initial fermion spins, and sum over the final polarization states of the vector bosons to compute
\es{}{
    \sum \overline{ \left| \mathcal{M} \right|^2}_{\chi\chi \to WW} &= \frac{g^4}{64} \left[ \frac{\mathrm{Tr}[( \Ps{2} - m_\chi ) \gamma_\lambda (\ks + m_\chi) \gamma_\mu  (\Ps{1} + m_\chi) \gamma_\alpha (\ks + m_\chi) \gamma_\beta ]}{(u - m_\chi^2)^2} \left(g_{\mu \alpha} - \frac{p_{4}^\alpha p_4^\mu}{m_W^2} \right) \left(g_{\lambda \beta} - \frac{p_{3}^\lambda p_3^\beta}{m_W^2} \right) \right.  \\
    & \left.  + \frac{\mathrm{Tr}[( \Ps{2} - m_\chi ) \gamma_\lambda (\qs + m_\chi) \gamma_\mu  (\Ps{1} + m_\chi) \gamma_\alpha (\qs + m_\chi) \gamma_\beta ]}{(t - m_\chi^2)^2} \left(g_{\mu \alpha} - \frac{p_{3}^\alpha p_3^\mu}{m_W^2} \right) \left(g_{\lambda \beta} - \frac{p_{4}^\lambda p_4^\beta}{m_W^2} \right)  \right. \\
    & \left.  + \frac{\mathrm{Tr}[( \Ps{2} - m_\chi ) \gamma_\lambda (\ks + m_\chi) \gamma_\mu  (\Ps{1} + m_\chi) \gamma_\beta (\qs + m_\chi) \gamma_\alpha ]}{(t - m_\chi^2)(u - m_\chi^2)} \left(g_{\mu \alpha} - \frac{p_{3}^\alpha p_3^\mu}{m_W^2} \right) \left(g_{\lambda \beta} - \frac{p_{4}^\lambda p_4^\beta}{m_W^2} \right)  \right. \\
    & \left. + \frac{\mathrm{Tr}[( \Ps{2} - m_\chi ) \gamma_\lambda (\qs + m_\chi) \gamma_\mu  (\Ps{1} + m_\chi) \gamma_\beta (\ks + m_\chi) \gamma_\alpha ]}{(t - m_\chi^2)(u - m_\chi^2)} \left(g_{\mu \alpha} - \frac{p_{4}^\alpha p_4^\mu}{m_W^2} \right) \left(g_{\lambda \beta} - \frac{p_{3}^\lambda p_3^\beta}{m_W^2} \right) \right]. 
}
We can then compute the Mandelstam variables $s$, $t$, and $u$, giving 
\es{}{
    s &= 4 m_\chi^2  (1+ v^2)  \\
    t & = (m_\chi^2 + m_W^2 - 2(m_\chi^2 (1+v^2 ) - m_\chi v \cos \theta \sqrt{m_\chi^2 (1 + v^2) - m_W^2 })  \\ 
    u & = (m_\chi^2 + m_W^2 - 2(m_\chi^2 (1+v^2 ) + m_\chi v \cos \theta \sqrt{m_\chi^2 (1 + v^2) - m_W^2 }) \,, \\
}
where $v$ is the velocity, which will be taken to zero shortly. Expanding in small velocity and in $m_\chi \gg m_W$ then gives
\begin{equation}
\sum \overline{ \left| \mathcal{M} \right|^2}_{\chi\chi \to WW}  \simeq \frac{g^4 m_\chi^2 (m_\chi^2 - m_W^2)}{ \left( 2 m_\chi^2 - m_W \right)^2}  + \mathcal{O}(v^2)  \simeq \frac{g^4}{4} + \mathcal{O}(v^2, \frac{m_W^2}{m_\chi^2}) \,.
\end{equation}
Substituting this result into the formula for the velocity averaged annihilation cross-section gives
\begin{equation}
    \langle \sigma v \rangle_{\chi\chi \to WW} = \frac{\left| \mathcal{M} \right|^2}{8 \pi s } \sqrt{1 - \frac{m_W^2}{m_\chi^2}} \simeq \frac{g^4}{128 \pi m_\chi^2}.
\end{equation}
By dimensional analysis we could have guessed the final answer, up to the ${\mathcal O}(1)$ factors.

\subsection{Accounting for the full cross-section}

The remaining diagrams can be computed through a similar fashion as the example described above. It is useful to keep in mind that the 1 charged Dirac fermion and 2 neutral Majorana fermions contribute a total of  $4 \times 1 + 2 \times 2 = 8$ degrees of freedom.  That is, the higgsino in the early universe is made up of equal parts $\chi, \tilde \chi, \chi_+,$ and $\chi_-$.  The diagrams shown in Figs.~\ref{fig:feyn_1},~\ref{fig:feyn_2},~\ref{fig:feyn_3},~\ref{fig:feyn_4},~\ref{fig:feyn_5},~\ref{fig:feyn_6}, and~\ref{fig:higgsino_ann} then evaluate to (note: here we express results in terms of $g$ and $\tan \theta_w$ instead of {\it e.g.} $g$ and $g'$):
\begin{itemize}
    \item Figs.~\ref{fig:feyn_1} and~\ref{fig:feyn_4}: $\chi \chi$ and $\tilde \chi \tilde \chi$ annihilation
    \begin{equation}
        \sigma_{\chi\chi} = \frac{g^4}{256 \pi m_\chi^2} \left( 3 + 2\tan^2 \theta_w + \tan^4\theta_w \right) = \sigma_{\tilde \chi \tilde \chi}
        \label{eq:today}
    \end{equation}
    \item Fig.~\ref{fig:feyn_2}: $\chi \tilde \chi$ annihilation
    \begin{equation}
        \sigma_{\chi \tilde\chi} =  \frac{g^4}{512 \pi m_\chi^2} \left(29 + 21 \tan^4 \theta_w \right)
    \end{equation}
    \item Fig.~\ref{fig:feyn_3} and~\ref{fig:feyn_5}: $\chi \chi_{\pm}$ and $\tilde \chi \chi_{\pm}$  annihilation
    \begin{equation}
        \sigma_{\chi \chi_+} = \frac{g^4 }{128 \pi m_\chi^2} \left( 6 + \tan^2 \theta_w \right)    = \sigma_{\chi \chi_-} = \sigma_{\tilde \chi \chi_+} = \sigma_{\tilde\chi \chi_-}
    \end{equation}
    \item Fig.~\ref{fig:feyn_6}: $\chi_+ \chi_{-}$ annihilation
    \begin{equation}
        \sigma_{\chi_+ \chi_-} = \frac{g^4 }{ 512 \pi m_\chi^2} \left(29 + 4 \tan^2 \theta_w + 44 \tan^4 \theta_w \right)
    \end{equation}
    \item Fig.~\ref{fig:higgsino_ann}: $\chi_{\pm} \chi_{\pm}$ annihilation
    \begin{equation}
        \sigma_{\chi_+ \chi_+} = \frac{g^4 }{64 \pi m_\chi^2}  = \sigma_{\chi_- \chi_-}
    \end{equation}
\end{itemize}
The effective cross section of the entire higgsino multiplet is then the average of all $4 \times 4 = 16$ possibilities, giving
\begin{equation}
    \langle  \sigma v \rangle  =  \sum_{i,j \in  \chi , \tilde \chi, \chi_+, \chi_-}  \frac{\langle \sigma v \rangle_{ij} }{16} = \frac{g^4}{512 \pi m_\chi^2} \left( 21 + 3 \tan^2 \theta_w + 11 \tan^4 \theta_w \right). 
\end{equation}

Note that the cross section during the present day is simply that in~\eqref{eq:today}.  Also relevant for the present day is the one-loop annihilation to $\gamma\gamma$, which is suppressed relative to the tree-level processes but gives a more discernible signal in indirect detection experiments.

\bibliography{refs.bib}

\end{document}